\documentclass[fleqn,10pt]{wlscirep}
\usepackage[utf8]{inputenc}
\usepackage[T1]{fontenc}

\usepackage{nameref}
\usepackage{caption}
\usepackage{subcaption}
\usepackage{hhline}
\usepackage{pdflscape}

\definecolor{degree-colour}{RGB}{152,78,163}
\definecolor{closeness-colour}{RGB}{77,175,74}
\definecolor{betweenness-colour}{RGB}{228,26,28}
\definecolor{pers-pagerank-colour}{RGB}{55,126,184}

\newcommand{\supplementarysection}{%
  \setcounter{figure}{0}
  \let\oldthefigure\thefigure
  \renewcommand{\thefigure}{S\oldthefigure}
  \setcounter{table}{0}
  \let\oldthetable\thetable
  \renewcommand{\thetable}{S\oldthetable}
  \section*{Supplementary Material}
  \setcounter{section}{0}
  \let\oldthesection\thesection
  \renewcommand{\thesection}{S\oldthesection}
}

\title{Early warning signals for predicting cryptomarket vendor success using dark net forum networks}

\author[1,*]{Hanjo D. Boekhout}
\author[2,3]{Arjan A.J. Blokland}
\author[1]{Frank W. Takes}
\affil[1]{Leiden University, Institute of Advanced Computer Science, Niels Bohrweg 1, 2333 CA Leiden, Netherlands}
\affil[2]{Leiden University, Institute of Criminal Law and Criminology, Steenschuur 25, 2311 ES Leiden, Netherlands}
\affil[3]{Netherlands Institute for the Study of Crime and Law Enforcement (NCSR), De Boelelaan 1077, 1081 HV Amsterdam, Netherlands}

\affil[*]{h.d.boekhout@liacs.leidenuniv.nl}

\begin{abstract}
In this work we focus on identifying key players in dark net cryptomarkets that facilitate online trade of illegal goods.
Law enforcement aims to disrupt criminal activity conducted through these markets by targeting key players vital to the market's existence and success.
We particularly focus on detecting successful vendors responsible for the majority of illegal trade.
Our methodology aims to uncover whether the task of key player identification should center around plainly measuring user and forum activity, or that it requires leveraging specific patterns of user communication.
We focus on a large-scale dataset from the Evolution cryptomarket, which we model as an evolving communication network.
Results indicate that user and forum activity,  measured through topic engagement, is best able to identify successful vendors.
Interestingly, considering users with higher betweenness centrality in the communication network further improves performance, also identifying successful vendors with moderate activity on the forum.
But more importantly, analyzing the forum data over time, we find evidence that attaining a high betweenness score comes before vendor success.
This suggests that the proposed network-driven approach of modelling user communication might prove useful as an early warning signal for key player identification.
\end{abstract}
\begin{document}

\flushbottom
\maketitle

\thispagestyle{empty}

\section*{Introduction}
\label{sect:introduction}
The dark net, a part of the internet that requires specific software or authorization to access~\cite{darkwebhow}, hosts a myriad of online fora that are increasingly a hotbed for criminal behavior and radicalisation~\cite{nadini2022emergence, chainanlysis2021}.
Dark net fora can, both theoretically and empirically, be split in those functioning as meeting places for the exchange of criminal information and those where criminal goods and services are traded, i.e., criminal marketplaces.
These fora and marketplaces can serve up to hundreds of thousands of users.
They are often moderated and organized in a professional manner, with cryptocurrencies, such as Bitcoin, serving as currency and are therefore referred to as \emph{cryptomarkets}~\cite{martin2014lost, shortis2020drug}.
To efficiently coordinate its activities disrupting these cryptomarkets, law enforcement aims to target key players that are vital to these market's existence and success~\cite{fonhof2018characterizing, shortis2020drug}.

Key players include the administrators and moderators responsible for the existence and proper functioning of the cryptomarket.
However, they also include the more successful vendors that are responsible for the majority of the trade conducted on the cryptomarket.
Identifying which users function as administrators can often be as easy as looking at the titles assigned to them on the cryptomarkets' forums.
Similarly, if sales statistics were shared on the cryptomarket, currently successful vendors would be easily identifiable.
We study the Evolution cryptomarket, a dataset with more than half a million posts and over four thousand vendors. It is moreover one of the few cryptomarkets that recorded sales information.
However, many cryptomarkets do not record sales information and at best provide a label to vendors independent of their success.
Furthermore, it is nearly impossible to identify those vendors whose success is yet to come.
Yet, if law enforcement wishes to disrupt future sales, it is exactly these future successful vendors that they would need to identify in order to dissuade them from continued participation in the cryptomarket.
Therefore, in this paper we focus on identifying key players in the form of both current and future successful vendors.

Existing research studying the workings of cryptomarkets and aimed at assisting law enforcement in identifying key players, often uses methods such as topic modelling or sentiment analysis~\cite{munksgaard2016mixing, van2017new, moeller2017flow, armona2018measuring, kamphausen2019digital, li2021demystifying}.
These methods rely on (combinations of) commonly used words and sentence structures in the forum message contents.
However, the rise of the use of message encryption in criminal communication, calls for the development of methods not reliant on knowledge of message content.
In this work, we aim to develop a method to identify key players based on the temporal structure of their communication network alone; thus, ignoring message content entirely.

\emph{Communication networks} model the interaction between entities within communication systems, such as mobile phone~\cite{li2014comparative, cavallaro2020disrupting, reisch2022monitoring}, face-to-face~\cite{lee2022universal}, and social media communication~\cite{wang2023evidence, lu2014network}; but also communication through online fora~\cite{fonhof2018characterizing, jo2023stage}.
Online fora, including those associated with cryptomarkets, usually consist of \emph{topics}, which may be grouped by subject.
Each topic is started by one user with a first message, also called a \emph{post}, and allows the set of users with access to respond by placing their own posts.
This activity can be considered a form of indirect communication from the posters to those users who placed posts on the same topic before them.
We can model this indirect communication using what we call a user-to-user communication network that directly connects users that posted in the same topic.
At the very least, a link in such a network represents a shared interest in the same topic as well as a level of familiarity with one another due to the likelihood of having seen each others' posts.
At best, a link can signify direct communication between two users that are, by means of forum posts, responding to one another.
Thus, links represent potential social ties formed on a dark net forum.
In this work, we leverage the structure of these communication networks without relying on knowledge of message content, with the goal of identifying and predicting successful vendors.

To find important users in a (criminal) network, one of the most commonly used approaches is to apply network centrality measures, which rank users based on their position in the network~\cite{liu2012criminal, fonhof2018characterizing, cavallaro2020disrupting, reisch2022monitoring, lee2022universal, wang2023evidence}.
Different network centrality measures often imply different roles a given user plays within a network.
In this paper, we explore four different measures: degree, harmonic closeness centrality, betweenness centrality and PageRank.
This allows us to grasp what type of role, as defined by the user's structural position in the network, may be more suited to the task of identifying key players in cryptomarkets.
The nuances of the interpretation of centrality measures can vary depending on whether we account for edge weights, i.e., the strength of social ties, and edge directions, i.e., who responds to whom.
Therefore, we consider for each measure whether the direction and strength of social ties matters, for identifying (successful) vendors for law enforcement applications.

Beside network measures, several intuitive straightforward measures can be obtained directly from the forum data.
We consider three such measures: post activity, topics started, and (started) topic engagement.
We henceforth refer to these measures as forum \emph{activity indicators}.
The rationale behind these three activity indicators, relies on vendors' tendency to start topics to promote their listings~\cite{li2021demystifying, familmaleki2015analyzing} and the concept of \emph{name recognition}.
Name recognition, also called brand awareness in a marketplace context, has been linked to improved trust~\cite{chen2003interpreting} and market outcomes~\cite{huang2012brand} (e.g., more sales).
Furthermore, Duxbury \& Haynie~\cite{duxbury2018network} concluded that trustworthiness is a better predictor of vendor selection than product diversity or affordability.

In this paper, we investigate to what level employing network measures computed on user-to-user communication networks are useful in identifying both current and future successful vendors on cryptomarkets.
We look at three law enforcement applications, each increasingly more useful to law enforcement practitioners.
We investigate whether (1) network measures can be used to distinguish vendors and their level of success; if (2) rankings induced by network measures can narrow down the user base to a significantly smaller set of potentially relevant users for law enforcement to investigate; and, to what extent (3) the top ranked users include successful vendors and other key players.
Furthermore, we study the Evolution cryptomarket at different points in time, i.e., we look at at various \emph{snapshots} of the communication network.
By doing so we simulate law enforcement investigating the state of the cryptomarket at those specific points in time, while subsequent data, i.e., at that point future data, shows how the cryptomarket would progress without intervention.
Consequently, we propose a methodology with the potential to serve as an \emph{early warning signal} for future vendor success on cryptomarkets.

The remainder of this paper is structured as follows.
In the Results section we shortly describe the dataset and measures used before reporting on our results.
The results and their implications for law enforcement are discussed in the Discussion section.
Finally, the Methods section provides more in-depth descriptions on the dataset and network extraction as well as the activity indicators, network measures, and evaluation metrics used in this work.
\section*{Results}
\label{sect:results}
In this section we first discuss our dataset and the (network) measures for identifying key players.
Next, we report and interpret results for the task of distinguishing vendors from non-vendors and predicting the levels of vendor success.
Then, we explore to what extent the rankings induced by (network) measures can reduce the set of users for law enforcement to investigate, while still including the greatest share of successful vendors.
Finally, we look at the set of top ranked users for the most promising network centrality measure and activity indicator at a specific point in time.
We do so to establish how well represented key players are among these top ranked users.
\subsection*{Data}
\label{subssect:results-data}
In this study we focus on the cryptomarket \emph{Evolution}.
Evolution was active from January 2014 until March 2015, when it closed due to an exit scam.
At the time, it was one of the most popular cryptomarkets~\cite{shortis2020drug}.
It formed a combination of a carding forum, where card information (e.g., credit/debit/ID/etc.) is traded, and an underground drug market~\cite{evolutionbackground}.

We obtained raw data of the Evolution marketplace and forum from the dark net market archives~\cite{dnmArchives}.
From this, we extracted a structured dataset, established a method of linking the market and forum data, and subsequently extracted communication network(s).
The extraction and linking process, the resulting dataset, and various statistics on the dataset and its completeness, are presented in Boekhout et al.~\cite{boekhout2023largescale}.
Parameters of the network extraction procedure control respectively the bounds on when two posts constitute a social tie ($\delta_o$ and $\delta_t$) and the strength of the social tie ($\omega_{lower}$, $t_{lim}$, and $\omega_{first}$).
For the communication network(s) studied in this work, the same extraction procedure and parameters were used as those used in Boekhout et al.~\cite{boekhout2023largescale}, i.e., $\delta_o = 10$, $\delta_t = 1$ month, $\omega_{lower} = 0.2$, $t_{lim} = 7$ days, and $\omega_{first} = 0.5$.
We demonstrate the robustness of our findings for each of these parameters in Supplementary Material Section S1.

The cryptomarket Evolution observed two notable changes in user and post activity.
In the initial months up to May 2014, the cryptomarket underwent steady growth in terms of both post activity and the number of active users.
However, monthly post activity stabilised from May until October (see Figure~\ref{fig:activity-trend}).
Notably, May saw a change in the vendor ranking system, which assigns textual labels to vendors that are visible on the marketplace to potential customers and imply a level of success and trustworthiness.
Obtaining a label representing greater success and trustworthiness as a vendor now required sufficient positive feedback, but most important for us, the new ranking system also reported on the exact number of sales a vendor had made up to that point.
The second major change to the cryptomarket came in early November 2014, as a by-product of the closure of six cryptomarkets following the joint international law enforcement operation dubbed ``Onymous''~\cite{shortis2020drug}.
After this disruption, Evolution showed a significant increase in overall activity until its closure.

Both the communication networks and current \& future sales counts were extracted on a monthly basis using data up to the end of each month, including all data prior to the given month.
As such, we obtained 15 network snapshots (starting from January 2014 up to March 2015).
Note that we rely on all data prior to the given month and not only the most recent month(s), because a vendor's reputation plays an important role in their success and is based not just on the most recent activity.
In fact, the build up reputation is such a vital aspect that it is predominantly the successful vendors with a large number of sales and high reputation who choose to migrate and maintain their identity in new cryptomarkets after market closures~\cite{norbutas2020reputation}.
Details on the network extraction process and the computation of monthly sales statistics are provided in the Methods section.
\subsection*{Network measures \& activity indicators}
\label{subsect:results-measures}
Each considered network measure captures a different role a user may play within the user-to-user communication network.
To cover a wide range of user roles that may be important to vendor success, we report on four centrality measures: (1) in-degree; (2) bidirectional harmonic closeness centrality; (3) directed weighted betweenness centrality; and (4) directed weighted PageRank.
The \emph{in-degree} of a user indicates the number of different users that posted (shortly) after them on the same topic(s). Thus, it can serve as a proxy of how many users have seen one or more of their posts and thus to some extent their level of name recognition.
The bidirectional \emph{harmonic closeness centrality}~\cite{rochat2009closeness} is a measure of a user's ability to reach the entire network, following paths regardless of link direction. High harmonic closeness centrality indicates that it should be relatively easy to reach and therefore potentially be visible to the entire user base.
The directed weighted \emph{betweenness centrality}~\cite{freeman1977set, brandes2001faster} computes how often a user lies on shortest paths connecting other nodes, taking into account both the direction and strength of social ties. High betweenness nodes often lie `between' communities. As such, it may be a good measure of how well a (potential) vendor reaches different, otherwise separated, communities of customers.
Finally, the directed weighted \emph{Pagerank}~\cite{page1999pagerank} computes the probability that a random walker that infinitely traverses a network ends up at a given node, taking into account both the direction and strength of social ties.
High PageRank centrality is often an indicator of being well connected to other important users.
Duxbury \& Haynie~\cite{duxbury2018network} found that buyers were more likely to continue ordering with vendors within the same community.
As such, a close connection with other key players, as indicated by a high PageRank value, can be indicative of a high perceived trust, positively affecting sales.
Finally, we note that links in the communication network are temporally independent unless they rely on the same post(s), i.e., the link $(a,b)$ is not dependent on the existence of link $(b,c)$ unless they were formed based on the same post by user $b$.
As such, were we to consider only time-respecting paths (e.g., as introduced by Kempe et al.~\cite{kempe2000connectivity}), which require temporally dependent links, we would not adequately capture the social aspect of the network, i.e., the desired concepts of familiarity and shared interest. 
Therefore, we focus on `static' network measures.

To evaluate the network measures we compare them against three activity indicators, which serve as our baselines.
These activity indicators can be computed directly from the forum data, so without aforementioned communication network extraction, are intuitively meaningful in the context of cryptomarket vendor success and also do not require knowledge of message content.
We consider: (1) post activity; (2) topics started; and (3) topic engagement.
\emph{Post activity} refers to the number of posts a user has placed on the forum.
It relies on the idea that greater activity means greater visibility, which in turn leads to greater name recognition.
\emph{Topics started} determines the number of topics a user started and \emph{topic engagement} subsequently computes the sum of all posts placed within those topics, regardless of who posted them.
These measures rely on the fact that the more topics a user has started and the more engagement those topics received, the greater the likelihood that they are a (successful) vendor.
This is supported by Armona~\cite{armona2018measuring} previously concluding that a similar measure of vendor forum sentiment could be indicative of higher demand for a vendor on the Agora cryptomarket.
Whereas their measure relied entirely on forum post texts (and thread titles) for the selection of posts and computation of the sentiment, our activity indicators can be determined entirely independent of post content.
Again, the increased visibility through starting topics also boosts name recognition.

Further details on the computation and interpretation of the measures is provided in the Methods section.
\subsection*{Distinguishing vendors and their level of success}
\label{subsect:results-distinguish}
\begin{figure}[t]
  \centering
    \begin{subfigure}[b]{0.46\textwidth}
      \centering
      \includegraphics[width=\textwidth]{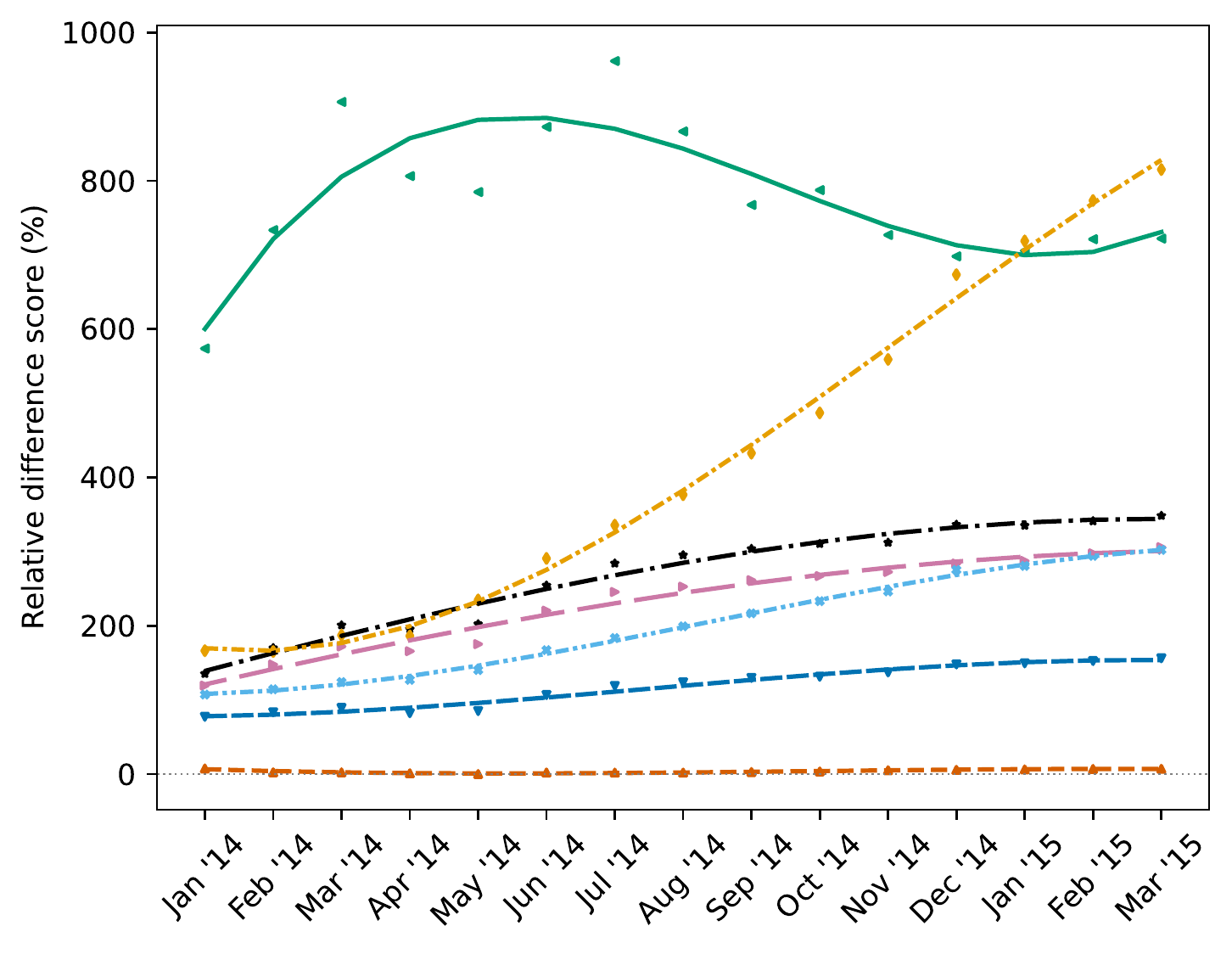}
      \caption{Relative difference score}
    \end{subfigure}
    ~
    \begin{subfigure}[b]{0.44\textwidth}
      \centering
      \includegraphics[width=\textwidth]{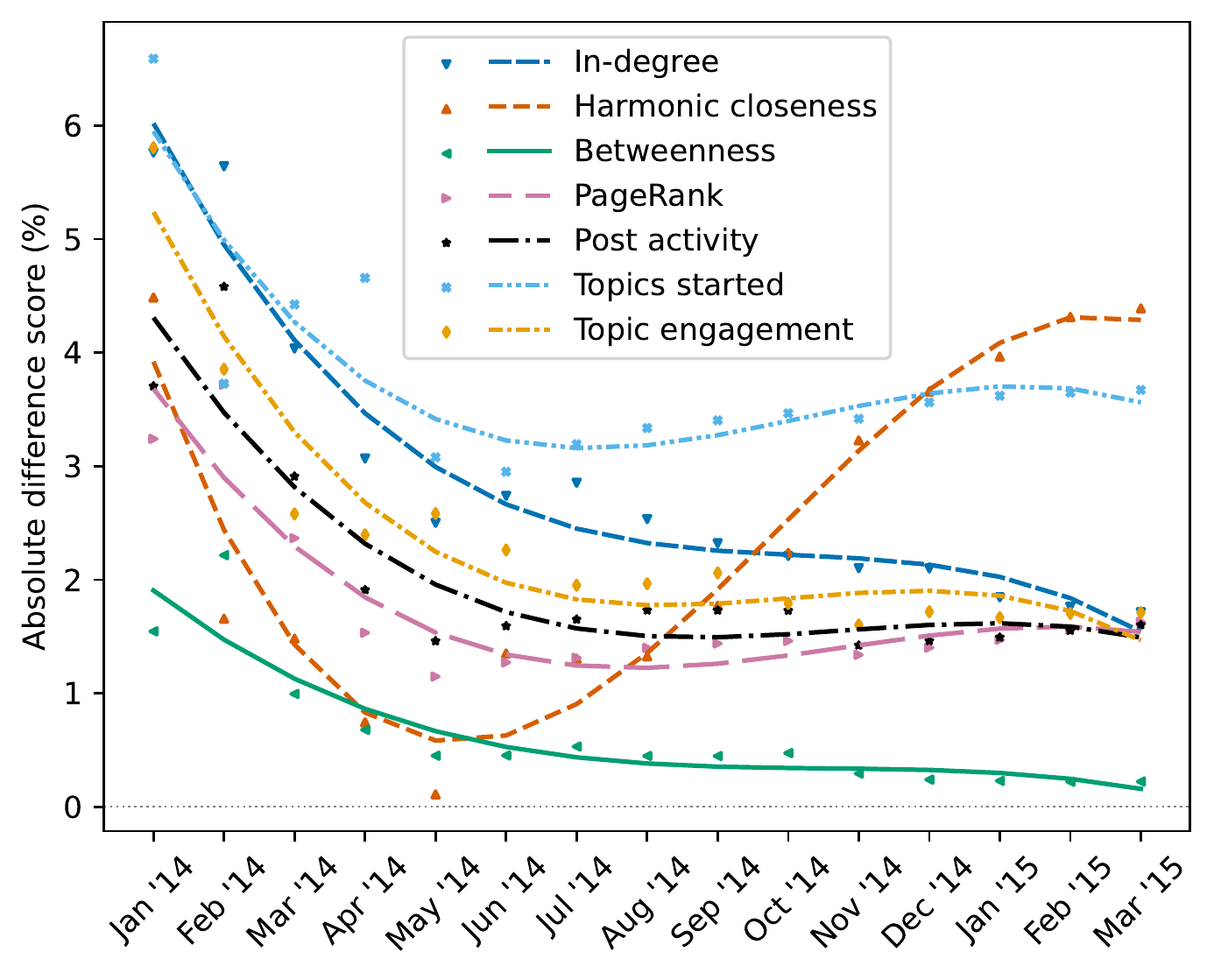}
      \caption{Absolute difference score}
    \end{subfigure}
  \caption{The relative (a) and absolute (b) difference score between vendors over non-vendors. Positive scores indicate that vendors achieve higher normalized network centralities or activity indicators than non-vendors on average.} \label{plots:avgvalue-main-1}
\end{figure}
\begin{figure}[t]
  \centering
    \begin{subfigure}[b]{0.44\textwidth}
      \centering
      \includegraphics[width=\textwidth]{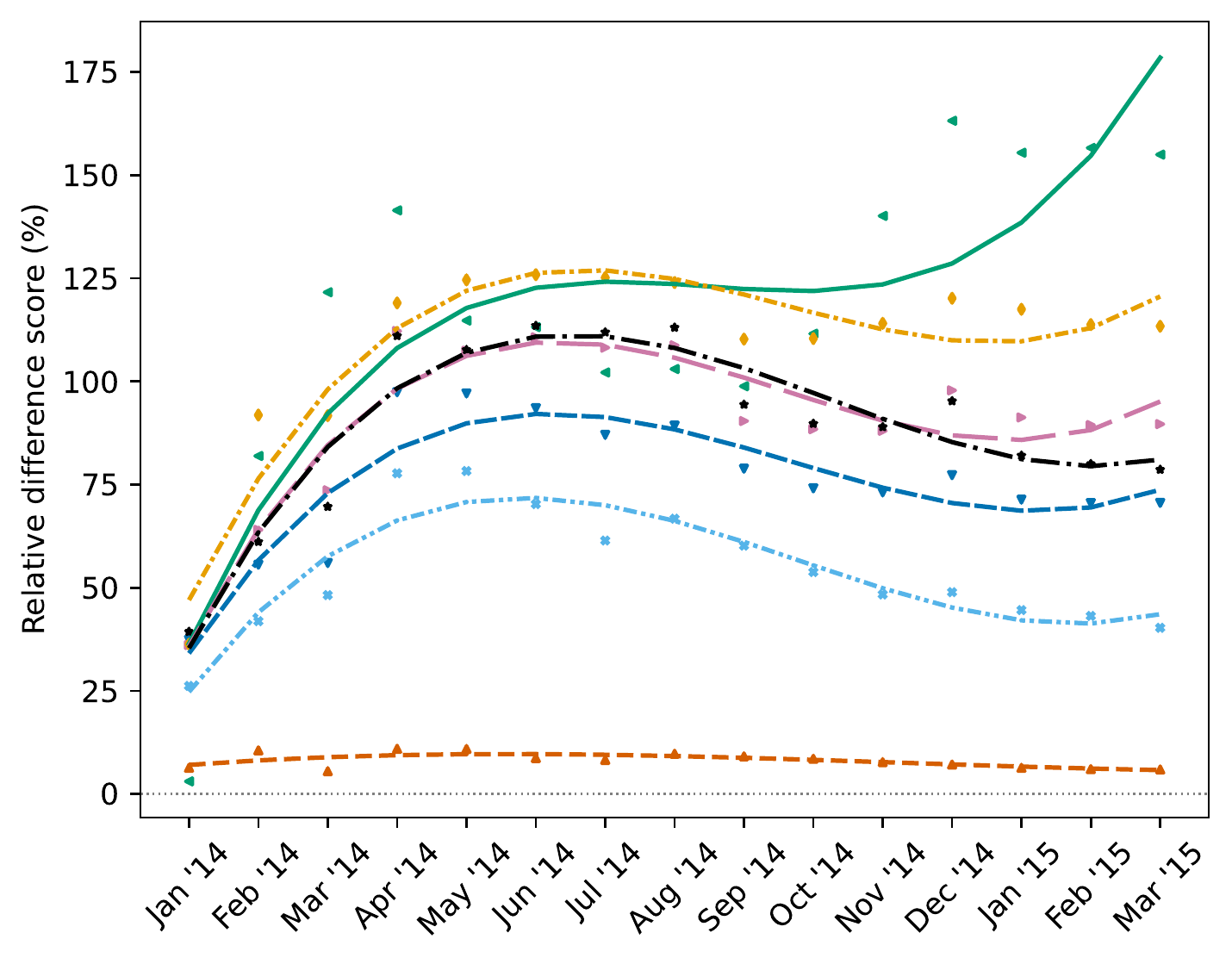}
      \caption{Relative difference between the top percentile and all vendors, current success}
    \end{subfigure}
    ~
    \begin{subfigure}[b]{0.44\textwidth}
      \centering
      \includegraphics[width=\textwidth]{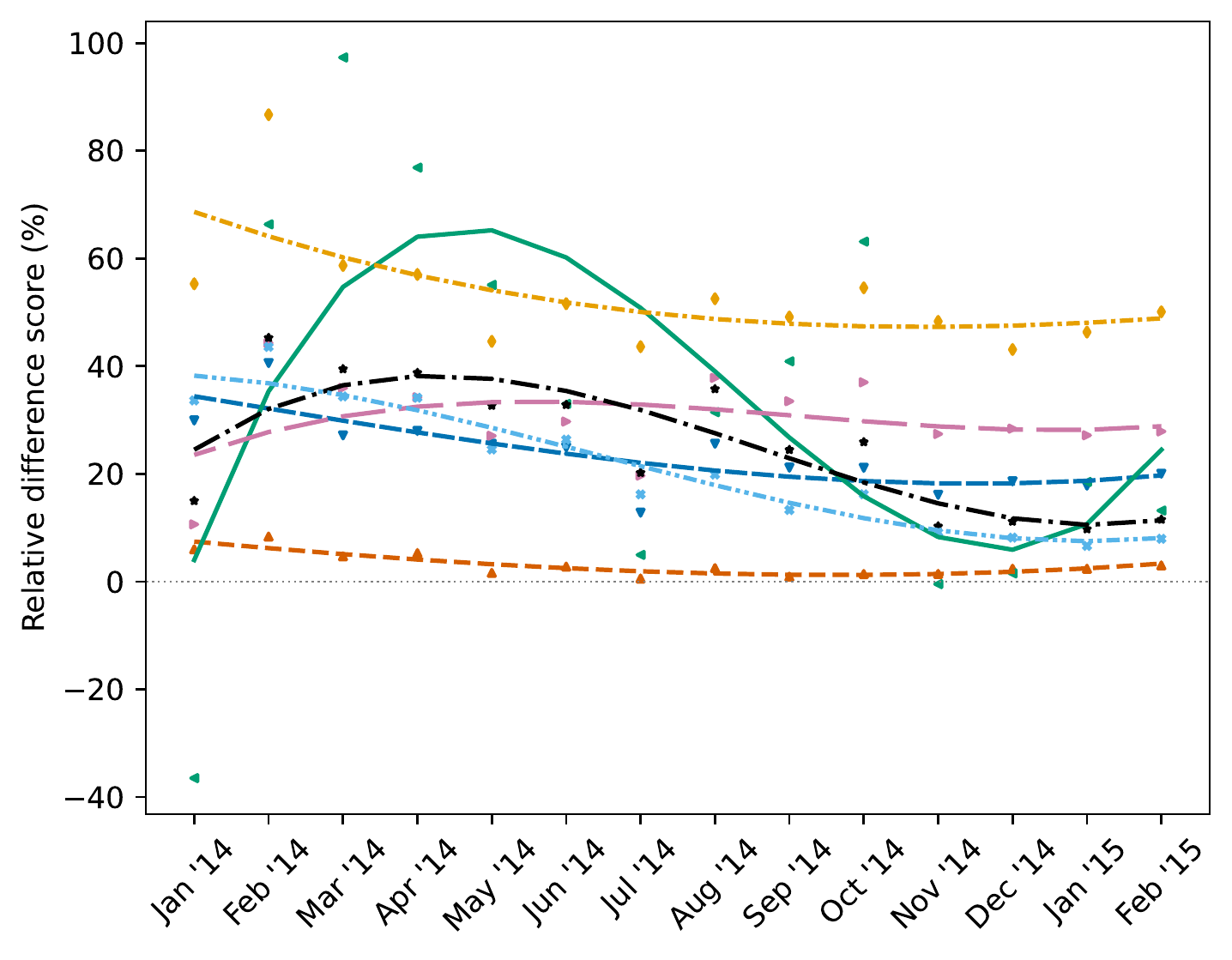}
      \caption{Relative difference between the top percentile and all vendors, future success}
    \end{subfigure}
    \begin{subfigure}[b]{0.44\textwidth}
      \centering
      \includegraphics[width=\textwidth]{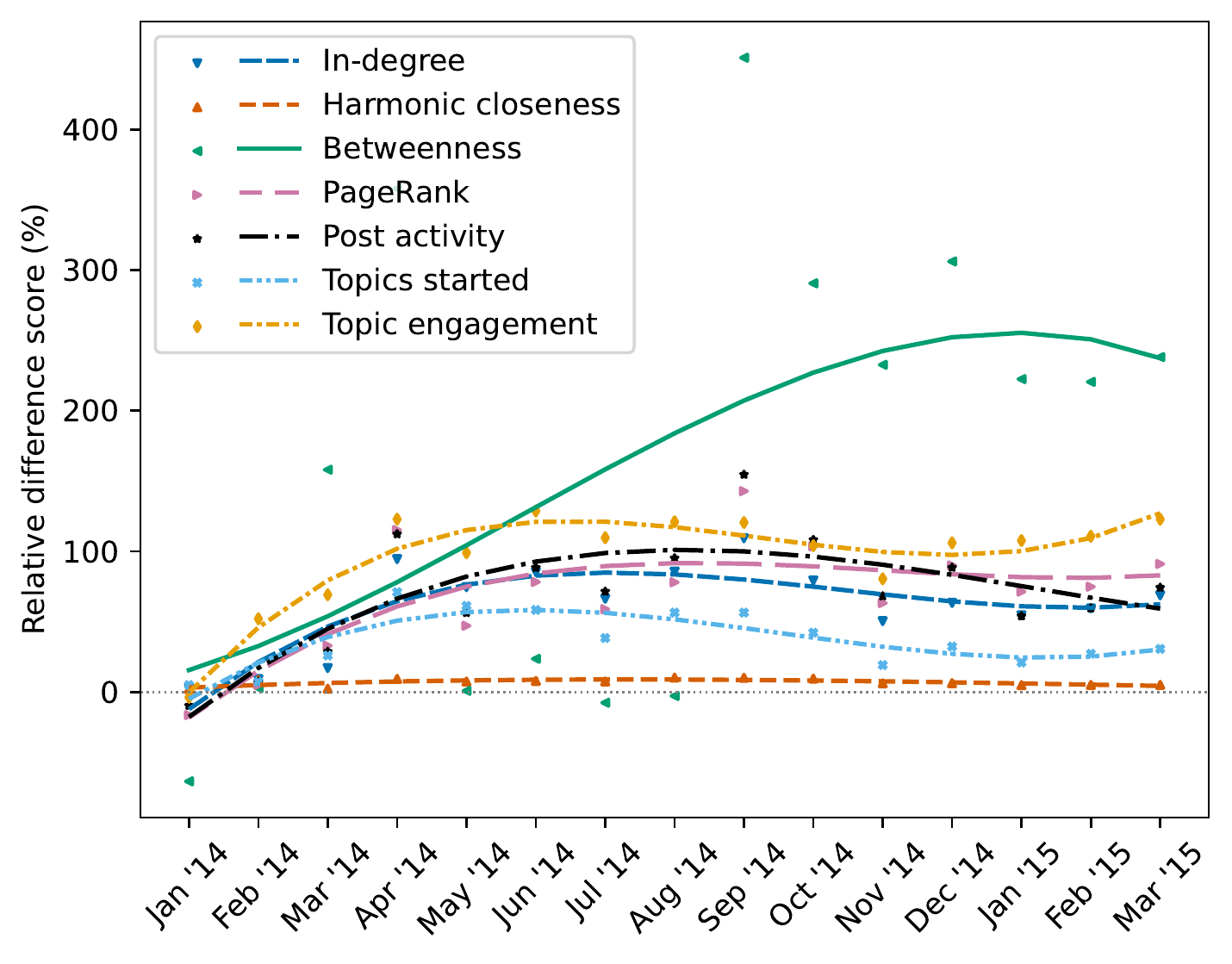}
      \caption{Relative difference between the top and sub-top vendor percentile, current success}
    \end{subfigure}
    ~
    \begin{subfigure}[b]{0.44\textwidth}
      \centering
      \includegraphics[width=\textwidth]{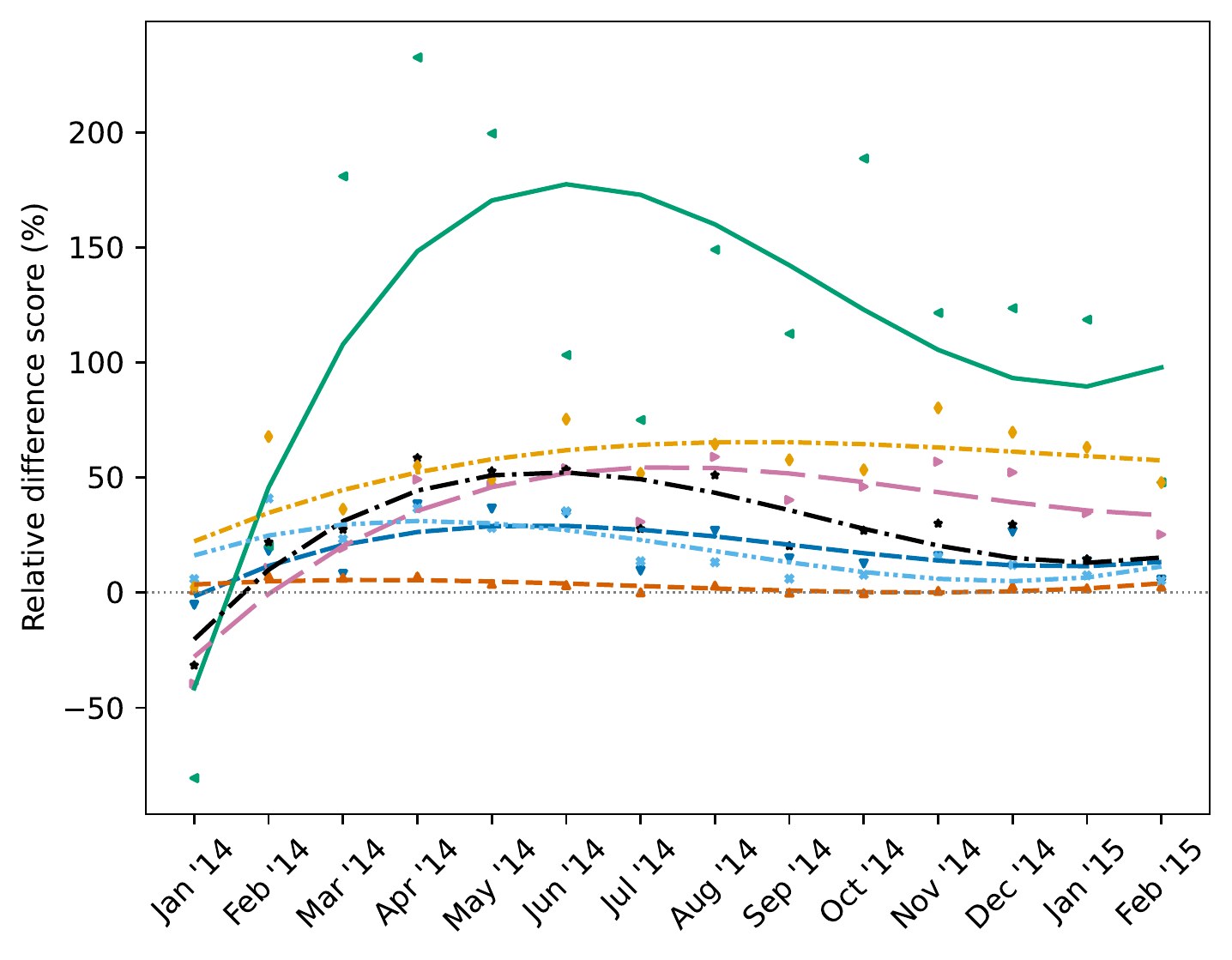}
      \caption{Relative difference between the top and sub-top vendor percentile, future success}
    \end{subfigure}
  \caption{The relative difference score between the top percentile and all vendors and between the top and sub-top percentiles. Positive scores indicate on average higher normalized network centrality or activity indicators for the more ``successful'' group.} \label{plots:avgvalue-main-2}
\end{figure}
To predict vendor success, we must determine if it is possible to distinguish between vendors and non-vendors, as well as between various levels of success.
We look at the average network centralities and activity indicators for groups of users, in an attempt to distinguish groups with greater success.
To this end, we divided, for each month, all active vendors, i.e., all users that are or will become vendors with at least one post already posted at that time, into five groups of success percentiles, each including respectively the top 0-20\%, 20-40\%, etc. of vendors in terms of sales.
We refer to these groups as \emph{vendor percentiles}.
Separate vendor percentiles are formed for current and future success.
We refer to the most and second most successful percentiles as the \emph{top} and \emph{sub-top percentile}, respectively.
The \emph{non-vendors}, consisting of regular forum users and those vendors with no recorded sales at all, form a separate sixth group.

First, we computed for each month the mean normalized value for each measure for the groups of all vendors and all non-vendors, using min-max normalization.
From this the \emph{relative} and \emph{absolute difference scores} between vendors and non-vendors was computed for each of the four network measures and three activity indicators (see Methods section for more details on their computation).
The resulting scores are depicted in Figure~\ref{plots:avgvalue-main-1}.
In these figures, lines give a third polynomial approximation of the trend based on the monthly centralities and activity indicators.
Here, the third polynomial is used to try to account for the two aforementioned events that took place in the Evolution cryptomarket~\cite{boekhout2023largescale}.
Dashed lines are used for the network measures and dotted lines for the activity indicators.

Figure~\ref{plots:avgvalue-main-1} show that, for all measures, vendors have higher network centralities and activity indicators than non-vendors.
Furthermore, they show that although the relative difference score for betweenness centrality of vendors over non-vendors is quite significant (600-1000\%), the corresponding absolute difference score is the smallest of all these measures.
This indicates that betweenness has relatively small values overall with some extremely high outliers.
On the contrary, harmonic closeness centrality has low relative difference scores but nominal absolute difference scores.
Since these effects are expected to disappear when inducing a ranking from the actual values, it is less the size of the difference scores than the fact that they are positive that are an indicator of (useful) predictive power.
After all, the ranking induced by the centralities and activity indicators is more useful to law enforcement practitioners than the actual values.
Thus, the exclusively positive values in Figure~\ref{plots:avgvalue-main-1}, indicate the potential of all network measures and activity indicators to distinguish vendors from non-vendors.

Next, we investigate whether these measures can also distinguish between vendors' levels of success.
To assess this, we looked at the relative difference scores between the top percentile and all vendors (Figures~\ref{plots:avgvalue-main-2}a,b) and between the top and sub-top percentile (Figures~\ref{plots:avgvalue-main-2}c,d) for both current and future success.
Figure~\ref{plots:avgvalue-main-2}a shows that for all measures the currently most successful vendors have on average higher network centralities and activity indicators.
After the first month and with the exception of July and August 2014 for betweenness centrality, Figure~\ref{plots:avgvalue-main-2}c demonstrates this also holds when comparing the top with the sub-top percentile.
Interestingly, trend changes for most measures follow cryptomarket developments.
For example, up until May the difference score increases monthly, similar as to how the level of activity on the cryptomarket increased during this period.
The following period, up to the November 2014 ``Onymous'' disruption~\cite{shortis2020drug}, shows stable but slightly decreasing difference scores for most measures.
Finally, after this disruption, we see a small increase in difference scores again.

When we consider future success, Figure~\ref{plots:avgvalue-main-2}b shows again positive difference scores between the top vendor percentile and all vendors.
However, they are noticeably lower than for current success.
Similarly, Figure~\ref{plots:avgvalue-main-2}d shows mostly positive difference scores when comparing with the sub-top percentile, but with lower scores.
Thus, for both current and future success the network centralities and activity indicators show the potential to distinguish vendors' level of success.

Notably, betweenness centrality shows trends that differ from the all other measures.
In particular, for current success we see clearly higher difference scores in the last months.
On the contrary, for future success the final months show lower difference scores than before.
This behaviour is likely due to the delay between successful vendors establishing themselves in the network and reaping the benefits in terms of sales.
In other words, high betweenness centrality is expected to be more a prelude to than a consequence of vendor success.
Thus, these results show the potential of betweenness centrality as an early warning signal for future vendor success.

In short, for all measures vendors show positive difference scores over non-vendors and less successful vendors.
Thus, rankings induced by these measures are expected to rank successful vendors (relatively) higher.
Therefore, the induced rankings have the potential to assist law enforcement by allowing them to focus investigative efforts on higher ranked users.
Furthermore, betweenness centrality was shown to have potential as an early warning signal, as high betweenness appears to precede vendor success.
Finally, among the remaining network measures and activity indicators, topic engagement consistently showed the highest difference scores.
This suggests that topic engagement may provide the best predictions of vendor success.
\subsection*{Detecting vendors in the user base}
\label{subsect:results-recall}
\begin{figure}[t]
  \centering
    \begin{subfigure}[b]{0.44\textwidth}
      \centering
      \includegraphics[width=\textwidth]{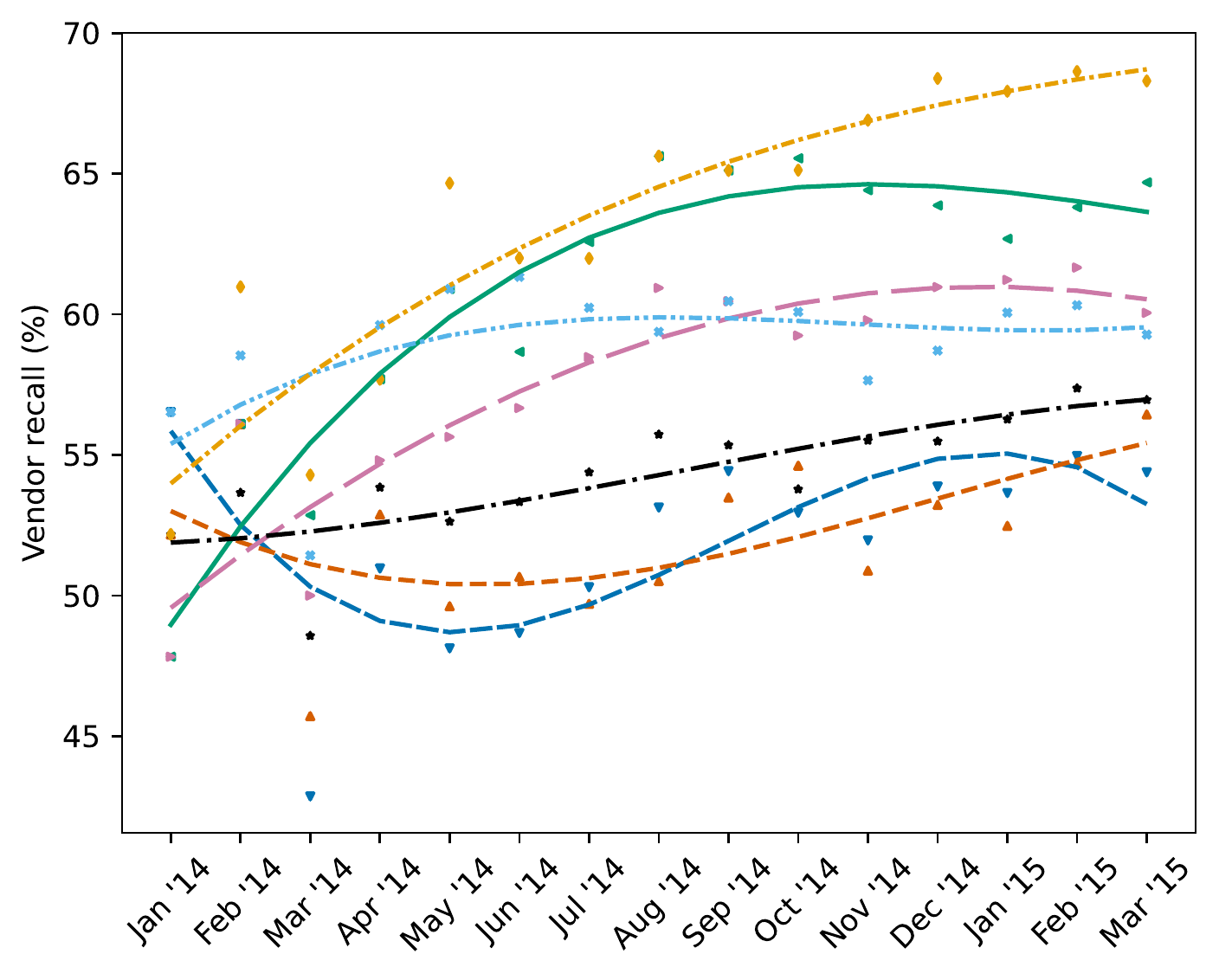}
      \caption{Current success}
    \end{subfigure}
    ~
    \begin{subfigure}[b]{0.44\textwidth}
      \centering
      \includegraphics[width=\textwidth]{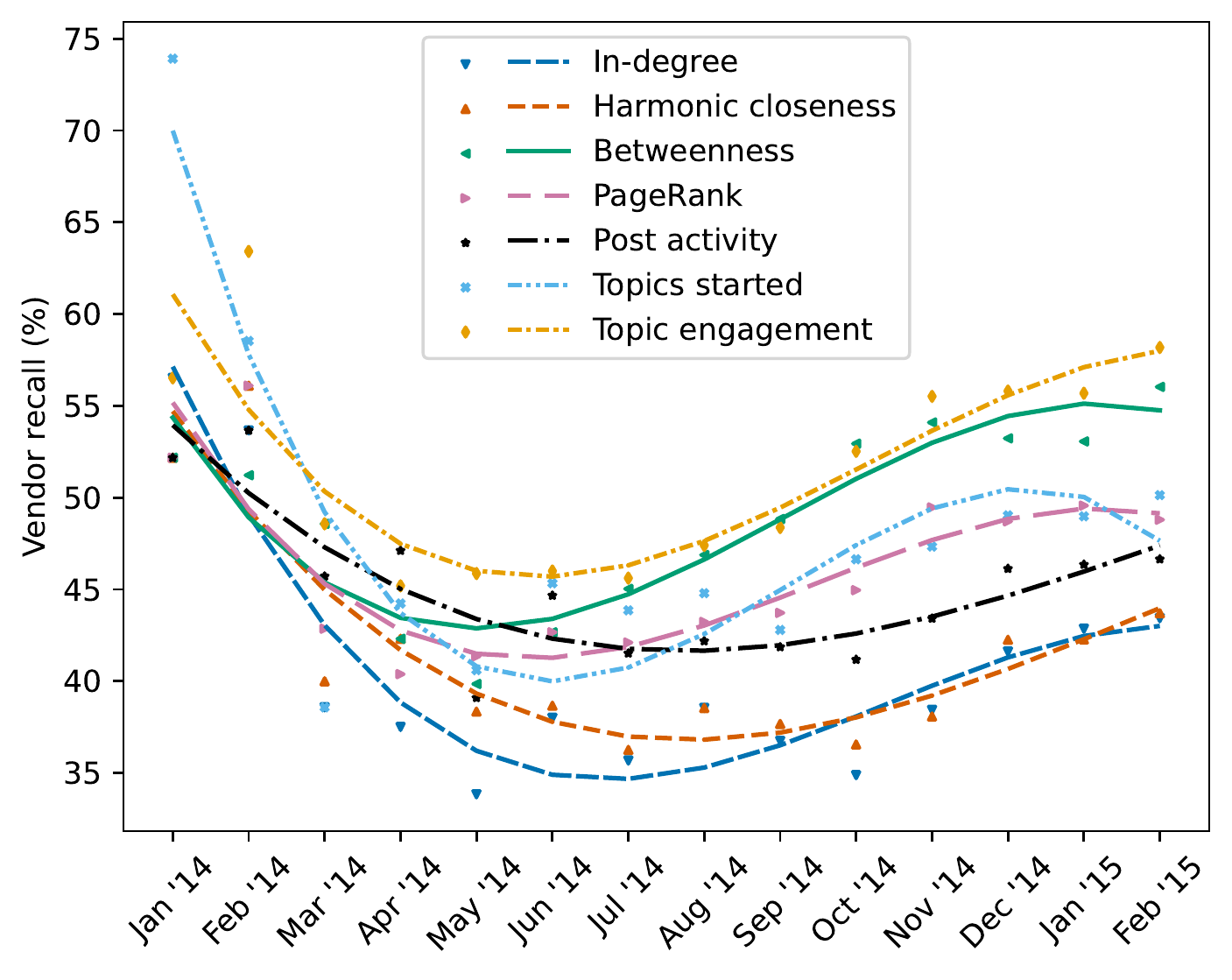}
      \caption{Future success}
    \end{subfigure}
  \caption{Monthly vendor recall of top vendor percentile (top 0-20\% vendors in terms of sales) among the top 20\% of all users based on the network measures and activity indicators. Plots cover recall for both current and future success.
  Higher vendor recall indicates a greater portion of the top vendor percentile was found.} \label{plots:recall-vendor}
\end{figure}
In their efforts to disrupt cryptomarkets, law enforcement has access to limited personnel and resources.
One method employed by law enforcement to deal with this limitation, is to reduce the set of users to investigate based on a ranking induced by some measure.
Rankings that after such a reduction still include many users of interest, are of course preferable.
In the previous section, we established the predictive potential of the network measures and activity indicators for predicting (successful) vendors.
Now, with the specific law enforcement perspective of aiming to find as many (hard to identify) vendors as possible, we want to explore how this predictive potential translates to the task of reducing the set of users to investigate.
To do this, we consider what we call the \emph{vendor recall}.
The vendor recall computes what percentage of users among the top vendor percentile (the top 20\% of vendors) is also among the top percentile of all users, i.e., among the top 20\% of all users when ranked on a given network measure or activity indicator (see Methods section for further details).
Note that we focus on the top percentile, instead of the absolute top vendors, as this aligns with the law enforcement intervention method of dissuading continued participation in the cryptomarket.
Since this intervention method is known to be ineffective for the absolute top vendors and comes at a relatively low cost to law enforcement, it is more suited to targeting larger groups of vendors.
For this reason we also prioritize reporting the recall of vendors over sales.
Monthly vendor recalls are plotted in Figure~\ref{plots:recall-vendor} for current (a) and future (b) success, respectively.
As noted before, lines in these plots are third polynomial approximations of the trend.

Figure~\ref{plots:recall-vendor} shows that, for both current and future success, degree and closeness centrality generally have a worse vendor recall than any of our activity indicators.
From May onwards, PageRank outperforms post activity and performs on par with the topics started indicator.
Meanwhile, from July onwards, betweenness centrality consistently outperforms both the post activity and topics started activity indicators and performs (nearly) on par with topic engagement.
Overall, the topic engagement indicator most consistently achieves high performance in terms of vendor recall.
These observations tell us two things.
First, network centrality measures require the communication network to have developed and stabilised sufficiently before achieving reliable vendor recall.
During the initial months the communication network and its structure are still undergoing significant changes.
Consequently, we also see large fluctuations in vendor recall for the network measures between these months.
Second, network measures do not strictly improve on our best activity indicator(s) in terms of vendor recall.
\begin{table}[t]
  \centering
  \caption{Mean (and standard deviation) of the monthly overlap between network centrality based and activity indicator based detected vendors for the top vendor percentile (top 0-20\% of vendors in terms of sales) as shown in Figure~\ref{plots:recall-vendor}. (Abbreviations of activity indicators: pa = post activity, ts = topics started, and te = topic engagement.)} \label{table:overlap}
  \begin{tabular}{|r||ccc|c||ccc|c|} \hline
      & \multicolumn{4}{|c||}{Current success} & \multicolumn{4}{|c|}{Future success} \\ \hline
      & pa & ts & te & pa $\cup$ te $\cup$ ts & pa & ts & te & pa $\cup$ te $\cup$ ts \\ \hhline{|=========|}
    & \multicolumn{8}{|c|}{\% of indicator detected vendors also found by centrality (higher is better)} \\ \hline
    In-degree           & 90.3 & 77.9 & 78.4 & $72.3$ $\pm 4.49$ & 83.8 & 70.4 & 72.1 & $66.0$ $\pm 4.60$ \\
    Harmonic closeness  & 88.9 & 77.1 & 75.5 & $70.3$ $\pm 4.04$ & 83.9 & 72.1 & 70.6 & $66.2$ $\pm 7.09$ \\
    Betweenness         & 89.8 & 83.1 & 85.4 & \textbf{80.4} $\pm 5.23$ & 82.5 & 77.4 & 79.5 & $74.1$ $\pm 4.33$ \\
    PageRank            & 95.7 & 83.2 & 86.5 & $79.6$ $\pm 2.78$ & 90.9 & 76.4 & 82.4 & \textbf{74.6} $\pm 4.14$ \\ \hline
    & \multicolumn{8}{|c|}{\% of centrality detected vendors also found by indicator (lower is better)} \\ \hline
    In-degree           & 94.1 & 88.1 & 94.8 & $98.7$ $\pm 2.04$ & 93.6 & 83.1 & 91.9 & $97.8$ $\pm 4.04$ \\
    Harmonic closeness  & 92.5 & 87.1 & 91.2 & $95.9$ $\pm 2.27$ & 91.3 & 82.9 & 87.7 & $95.4$ $\pm 3.90$ \\
    Betweenness         & 80.4 & 80.7 & 89.0 & \textbf{94.4} $\pm 2.25$ & 75.9 & 75.6 & 83.8 & \textbf{90.6} $\pm 2.88$ \\
    PageRank            & 90.6 & 85.4 & 95.0 & $98.8$ $\pm 1.14$ & 89.0 & 79.1 & 92.3 & $97.0$ $\pm 2.02$ \\ \hline
  \end{tabular}
\end{table}

Despite achieving the best vendor recall, topic engagement is only able to detect up to 2/3rd of the most successful vendors for current success and even fewer for future success.
Thus, there may still be a significant number of successful vendors that are not detected by the activity indicators that may be included by network measures.
To investigate this, we analyse the overlap of detected vendors between the network measures and activity indicators.
Table~\ref{table:overlap} shows the average monthly overlap of each network measure with each individual activity indicator and the union of detected vendors by all activity indicators.
We see that PageRank and betweenness centrality detect the greatest share of vendors also found by the activity indicators, detecting on average approximately 80\% of all current vendors and 75\% of all future vendors found.
However, respectively nearly 99\% and 97\% of all vendors detected by PageRank are also found by the activity indicators.
As such, PageRank is not able to identify many new vendors.
On the contrary, the activity indicators find respectively only 94\% and 90\% of the vendors included by betweenness centrality.
Notably, individual indicators find far fewer.
Thus, betweenness centrality is able to detect the largest share of successful vendors not included by any of the activity indicators.
Therefore, reducing the set of users for law enforcement to investigate using betweenness centrality may provide a fresh perspective.
\begin{figure}[t]
  \centering
      \begin{subfigure}[b]{.44\textwidth}
        \centering
        \includegraphics[width=\textwidth]{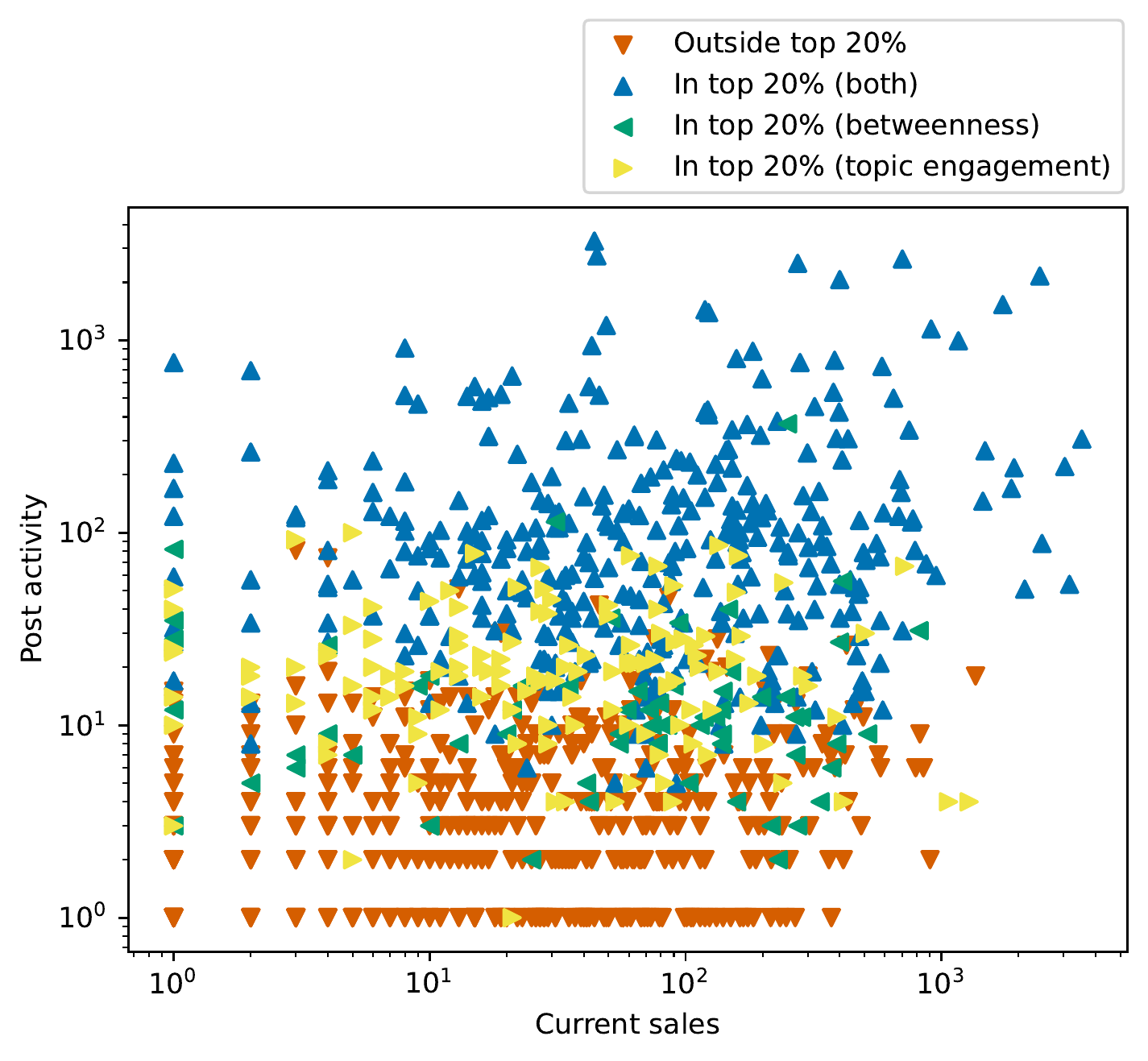}
        \caption{Current success}
      \end{subfigure}
      ~
      \begin{subfigure}[b]{.44\textwidth}
        \centering
        \includegraphics[width=\textwidth]{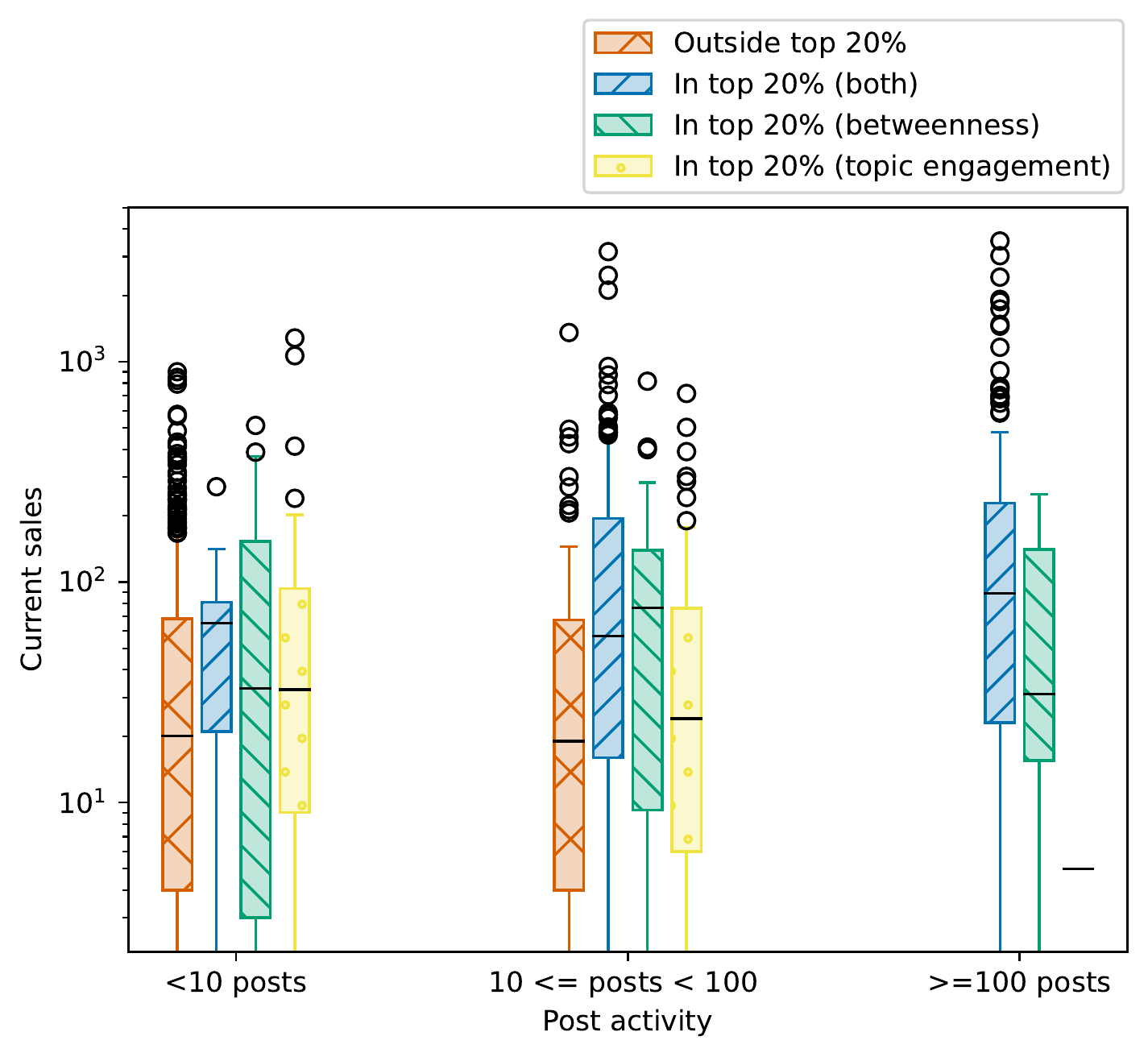}
        \caption{Current success}
      \end{subfigure}
      \begin{subfigure}[b]{.44\textwidth}
        \centering
        \includegraphics[width=\textwidth]{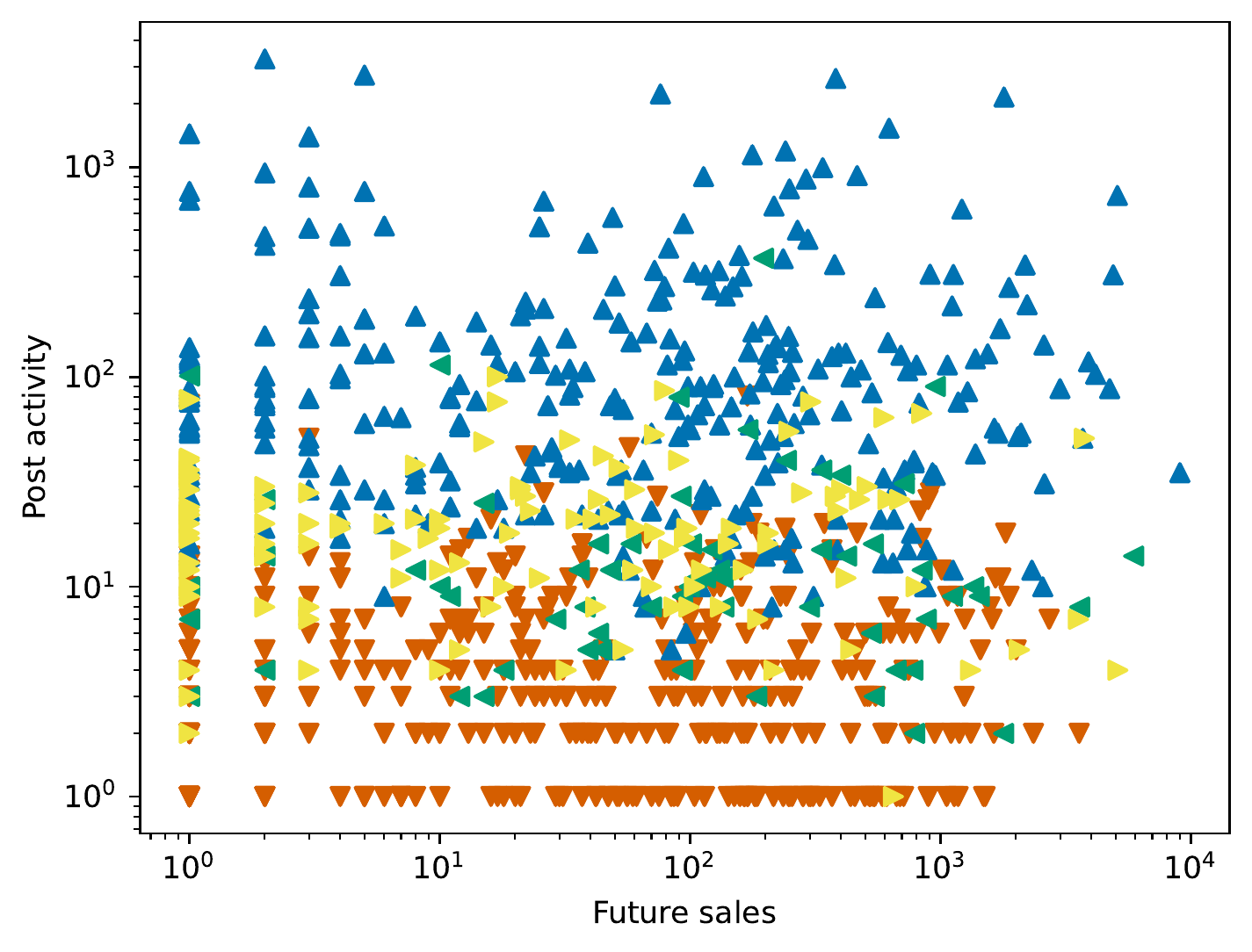}
        \caption{Future success}
      \end{subfigure}
      ~
      \begin{subfigure}[b]{.44\textwidth}
        \centering
        \includegraphics[width=\textwidth]{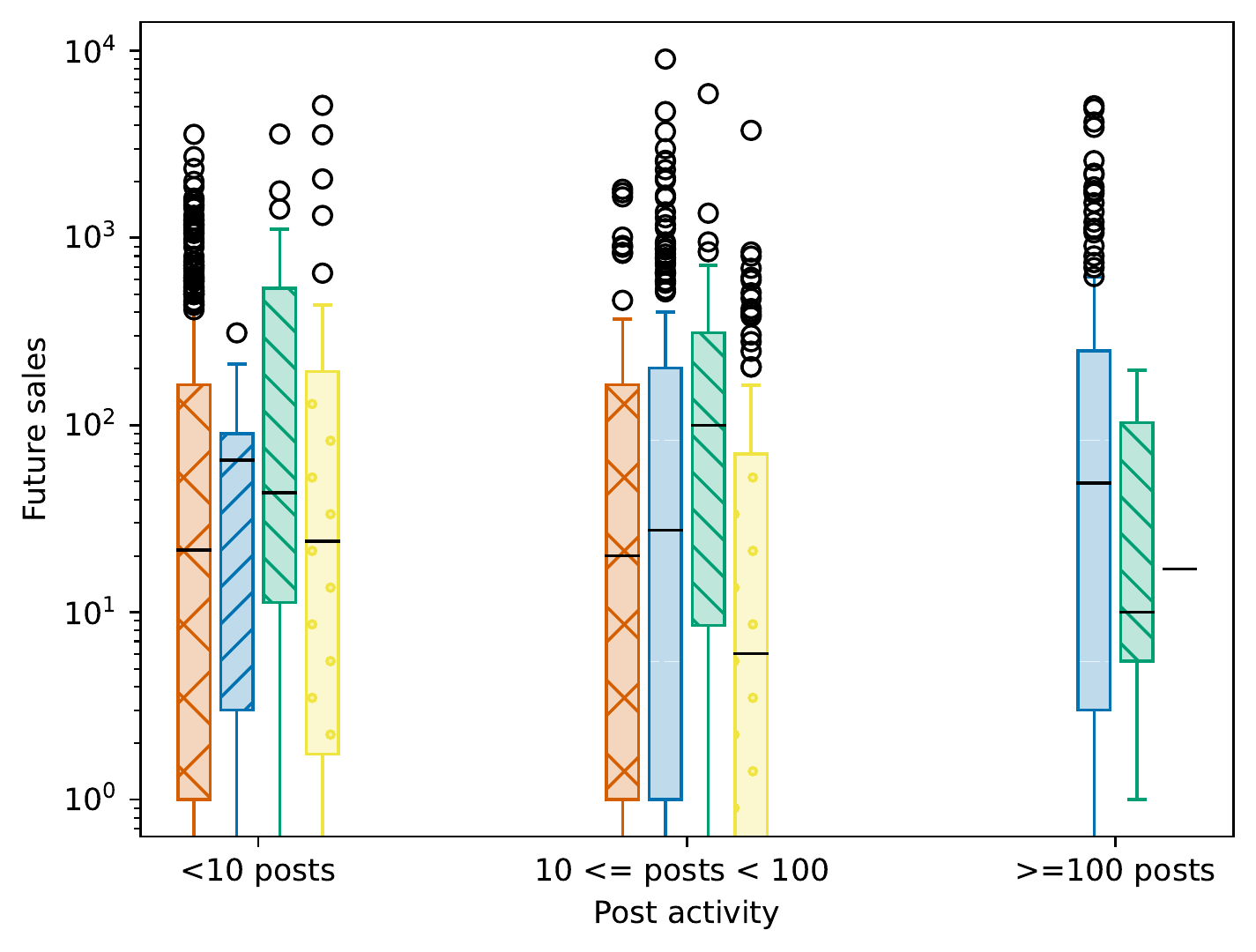}
        \caption{Future success}
      \end{subfigure}
    \caption{Sales and post activity of recalled (in top 20\%) and non-recalled (outside top 20\%) users for topic engagement, betweenness centrality, and their intersection for September 2014, for current (a,b) and future success (c,d), respectively.} \label{fig:sales-and-activity}

\end{figure}

Despite finding additional vendors, the union of all successful vendors detected by betweenness centrality and all activity indicators only finds around 75\% and 65\% of the top percentile for current and future success, respectively.
This means there is still a significant segment of the most successful vendors that would not be found for any of these measures.
One possible explanation for scoring low on any of these measures is simply low posting activity.
To assess whether this holds for the successful vendors that do not score high enough to be detected, we look at what we call the \emph{post activity recall} of the top vendor percentile in Supplementary Material Section S2.
The post activity recall is the percentage of the top vendor percentile's total post activity, for a given month, that is associated with those vendors detected with vendor recall (see Methods section for further details).
We find that for both current and future success, the vast majority of post activity is associated with the vendors with high network centrality and activity indicators.
As such, low post activity can be considered the main reason for the relatively low vendor recalls we observe.
After all, though over 30\% of successful vendors are not found, they are responsible for less than 10\% of the post activity of the entire group (in most cases even less).
Indeed, Figures~\ref{fig:sales-and-activity}a,c show that any vendor with activity above a certain threshold is always among the detected vendors, while most vendors with very few sales are not.
Specifically, it demonstrates that for both topic engagement and betweenness centrality for September 2014, this threshold is below 100 posts (as confirmed by Figures~\ref{fig:sales-and-activity}b,d).
This also holds for the other centrality measures, as demonstrated  in Supplementary Material Section S3.
We note that vendors with low post activity are also much less likely to be found using other methodologies.
Therefore, applying the methods discussed in this paper is likely not to miss vendors that other methodologies might have found.
Thus, the relatively low vendor recall achieved by betweenness centrality and topic engagement should not discourage law enforcement practitioners from using them.

Figure~\ref{fig:sales-and-activity} further indicates that vendors are overall more likely to be identified the greater their respective success.
This is also demonstrated through \emph{sales recall}, which measures what percentage of sales of the entire top percentile the detected vendors are responsible for (see Methods section for further details), in Supplementary Material Section S2.
There we show that the sales recall is generally between 10-20\% higher than the corresponding vendor recall. 
This indicates that the detected vendors are, on average, the more successful vendors.
In Supplementary Material Section S3 we show that this finding also holds for other months and network measures.
Furthermore, from Figures~\ref{fig:sales-and-activity}a,c it appears that the vendors found by topic engagement, and not betweenness centrality, are generally slightly more active and less successful compared to those found by betweenness centrality and not topic engagement.
Indeed, Figures~\ref{fig:sales-and-activity}b,d confirm that, for vendors with between 10 and 100 posts, those found exclusively by betweenness are generally more successful.
Notably, the effect seems to be even stronger for future success and this is moreover confirmed to hold for other months in Supplementary Material Section S3.
This observation once more highlights the potential of betweenness centrality as an early warning signal.

Throughout this section we have considered a single threshold at which we cut-off the rankings, namely 20\% of all users.
In Supplementary Material section S4, we investigate the performance of the measures at different thresholds. 
We observe that for low false positive rates, up to around 20\%, our findings hold. 
For higher false positive rates however, topic engagement clearly outperforms all other measures.
However, given the limited resources of law enforcement, it is unlikely that such large user samples would ever be considered for investigation.
After all, the resources required to investigate even 20\% of users would likely exceed those available to law enforcement.
Additionally, we find that topic engagement is the best measure for predicting vendors, regardless of their level of success.

To summarise, topic engagement provides the best single measure recall performance.
Meanwhile, betweenness centrality identifies the greatest share of vendors that do not score high for any of the activity indicators.
Additionally, betweenness centrality detects the most vendors of all network measures.
As such, betweenness centrality is the network measure most likely to be of use to law enforcement for detecting vendors in the user base.
Furthermore, betweenness centrality uniquely finds relatively more successful vendors among those with moderate activity. 
Notably, this effect is stronger for future success, further demonstrating its potential as an early warning signal.
\subsection*{Key player identification}
\label{subsect:results-keyidentification}
\begin{table}[t]
  \centering
  \caption{Top 25 users for betweenness and topic engagement for September 2014. Titles are determined as their most significant observed over the entire dataset (Administrator > Market Moderator >  Moderator > Public Relations >  Vendor > Banned > Troll > Member).} \label{table:topusers}
  \begin{tabular}{r|ll|rr||ll|rr|}
    & \multicolumn{4}{c||}{Betweenness} & \multicolumn{4}{c|}{Topic engagement} \\ \cline{2-9}
    &  &  & \multicolumn{2}{c||}{\#sales} &  &  & \multicolumn{2}{c|}{\#sales} \\ \cline{4-5} \cline{8-9}
    Rank & username & title & current & future & username & title & current & future \\ \hline
     1 & penissmith     & Troll            &   45 &    5 & Yasuo          & Vendor           & 2420 & 1790 \\
     2 & wefinance      & Banned           &   44 &    2 & themostseekrit & Moderator        &    0 &    0 \\
     3 & themostseekrit & Moderator        &    0 &    0 & Grandeur       & Vendor           & 1735 &  622 \\
     4 & FRIM           & Vendor           &  703 &  381 & First          & Banned           &  119 &    1 \\
     5 & scrufffe       & Member           &    0 &    0 & SingularLee    & Member           &    0 &    0 \\
     6 & Yasuo          & Vendor           & 2420 & 1790 & penissmith     & Troll            &   45 &    5 \\
     7 & Scattermind    & Public Relations &  400 &    0 & kalashnikov    & Vendor           & 3029 & 2214 \\
     8 & LudoTilMortem  & Market Moderator &  274 &    0 & wefinance      & Banned           &   44 &    2 \\
     9 & Kimble         & Administrator    &    0 &   76 & FRIM           & Vendor           &  703 &  381 \\
    10 & leon-trotsky   & Troll            &    0 &    0 & moka           & Vendor           &    0 &    0 \\
    11 & ScoobyJew      & Moderator        &    0 &    0 & JoeBloggs      & Member           &   17 &    0 \\
    12 & elmachico777   & Vendor           &   49 &  240 & highasakite    & Vendor           &  158 &    3 \\
    13 & Cypher         & Vendor           &  123 &    3 & ucard          & Vendor           &  910 &  177 \\
    14 & evilsmile      & Banned           &    0 &    0 & mountainhigh9  & Vendor           &    0 &    0 \\
    15 & Trippyy        & Moderator        &    8 &  464 & Scattermind    & Public Relations &  400 &    0 \\
    16 & Grandeur       & Vendor           & 1735 &  622 & misterbitcoin  & Vendor           &    0 &    0 \\
    17 & nerotic        & Member           &    0 &    0 & kesh           & Vendor           &   16 &    4 \\
    18 & sinordos       & Member           &    0 &    0 & alphawolf89    & Vendor           &  378 &   94 \\
    19 & d33poutside    & Administrator    &   43 &    2 & SkypeMan       & Vendor           &  585 & 5078 \\
    20 & moka           & Vendor           &    0 &    0 & fbgduck55      & Troll            &    0 &    0 \\
    21 & johnjones      & Member           &    0 &    0 & DonaldTrump    & Member           &    0 &    0 \\
    22 & misterbitcoin  & Vendor           &    0 &    0 & IronHeart      & Vendor           &  144 &  148 \\
    23 & Gold           & Vendor           &   21 &  216 & fnufnu         & Member           &    0 &    0 \\
    24 & Sportlife      & Vendor           &  649 &  268 & Verto          & Administrator    & 1163 &  338 \\
    25 & maaadcity      & Member           &    0 &    0 & ScoobyJew      & Moderator        &    0 &    0 \\ \hline
  \end{tabular}
\end{table}
In the previous section we determined that betweenness centrality and topic engagement are the measures with the greatest vendor recall performance.
That is to say, they are likely to have the most successful vendors among the top ranked users when ranked on these measures.
Here we look at the top scoring users to investigate to what extent the top scoring users are indeed key players in the cryptomarket.
To this end, we report the top 25 users, their member title, and their current and future sales for September 2014 for these measures in Table~\ref{table:topusers}.

We see that among the top 25 users in betweenness centrality and topic engagement there are ten (i.e., 40\%) that occur in both rankings.
Furthermore, we observe that for both measures over half of the top 25 users have current and/or future sales (56\% and 64\% respectively).
The probabilities of this happening randomly are more than a million times smaller ($3.47 \times 10^{-7}$ and $3.44 \times 10^{-9}$ respectively).
Note, not all users with sales also have the corresponding ``Vendor'' member title.
The reason for this is twofold: first, more important titles such as ``Administrator'' and ``Moderator'' supersede the ``Vendor'' title; and second, the ``Vendor'' title did not exist before September leading to some older vendors with few future sales not to be labelled as such.
This also illustrates a potential pitfall of relying too much on forum member titles for key player identification.

Of the users with sales, twelve are among the top percentile for current sales and eight are among the top percentile for future sales. 
Respectively three (\emph{kalashnikov}, \emph{Yasuo}, and \emph{Grandeur}) and one (\emph{SkypeMan}) of them are in fact in the top 10 current and future sales.
This suggests, these two measures are suitable for predicting potential successful vendors.
Notably, \emph{Trippyy}, who is included in the top 25 for betweenness centrality, is the only user that is a member of the top percentile for future sales, but not a member of the top percentile for current sales. 
Note, that \emph{Trippyy}'s member title in September was still ``Vendor''.
Additionally, betweenness centrality appears to include a greater proportion of vendors for whom the majority of their sales are yet to come.
On the other hand, we observe that the inclusion of \emph{kalashnikov} and \emph{SkypeMan} for topic engagement means that it captures a substantially greater total of future sales among the top 25 users.
If our goal were to identify the absolute top vendors specifically, these results may be interpreted to imply that topic engagement is the better choice of measure.
However, we must caution that sales volume does not equate to trade volume.
After all, the trade volume associated with a single sale can differ between listings and we are not able to differentiate between which sales came from which listings. 
Therefore, 100 sales could represent a larger total trade volume than 1000 sales.
As such, we can not conclusively say whether the inclusion of \emph{SkypeMan} by topic engagement is indeed the better choice compared to the inclusion of \emph{Trippyy} by betweenness centrality.
This uncertainty is another reason why we put a greater emphasis on vendor recall than sales recall in this work, and why we focus on the top vendor percentile instead of the pure top vendors in terms of sales.
Regardless, the results in Table~\ref{table:topusers} are a concrete example of how these measures can potentially serve as early warning signals for future vendor success.

In addition to vendors, we also find users with other important positions on the forum, such as ``Administrator'' and ``Moderator'', among the top 25 for both measures.
In fact, betweenness centrality and topic engagement combined include three out of the four users to have held the title ``Administrator'' among their top users.
Furthermore, the only missing administrator became inactive within a month of the founding of the cryptomarket.
Thus, we can say that all active administrators were found.
Additionally, betweenness centrality identifies five out of nine users to have held the title of ``Moderator'' and who registered before the end of September 2014 (four out of seven if we exclude users who obtained the title after September, including \emph{d33poutside}).
The probability of this happening randomly is more than 250 million times smaller ($2.07 \times 10^{-11}$ ($2.27 \times 10^{-9}$)).
On the other hand, topic engagement includes two out of nine (two out of seven) with probabilities of this randomly occurring that are just over 700 times smaller ($3.10 \times 10^{-4}$ ($1.82 \times 10^{-4}$))
Thus, these measures are suited to predicting key players beyond just successful vendors.
Though neither measure perfectly identifies only key players, they provide an excellent way of identifying individuals to investigate further manually.
\section*{Discussion}
\label{sect:discussion}
The identification of key players in cryptomarkets such as successful vendors and administrators, is a vital step in law enforcement interventions.
Whereas it can be easy to identify administrators due to titles given to these users, it may be harder to identify successful vendors.
It is especially difficult to identify those vendors whose success is yet to come.
These tasks might be further complicated when encryption is used for message contents.
The results presented in this work showed that network measures computed on the user-to-user communication network and three forum activity indicators, not reliant on knowledge of message content, are useful in predicting (future) successful vendors.
Specifically, the topic engagement indicator and betweenness centrality showed the best performance.

Our results showed that, on average, it is possible to distinguish between vendors and non-vendors using both network centrality and the activity indicators.
Additionally, we found that more successful vendors have on average higher centralities and activity indicators than less successful vendors.
This holds for both current and (to a slightly lesser extent for) future success.
However, it is important to remember that these findings are about the average case; perfect delineations cannot be made.
Even so, they indicate that the rankings induced by the measures have predictive potential for vendor success and may be useful to law enforcement activities.

To reduce the workload for law enforcement, it can be beneficial to reduce the set of users that need to be manually investigated.
We found that the measures of betweenness centrality and topic engagement included the greatest proportion of successful vendors when applying such a reduction (up to two thirds of the successful vendors when reducing to 20\% of the users).
Additionally, results showed that the vast majority (up to 98\%) of post activity of the most successful vendors was produced by those included and that they were the relatively more successful vendors.
As such, most successful vendors that are not retained by these measures are simply not very active on the forum.
We note that the network centrality measures appear to require the communication network to have sufficiently developed and stabilised for good predictive performance.
We found that betweenness centrality was the only network measure that was able to detect a substantial set of successful vendors that were not found by any of the activity indicators.
Thus, there are vendors that may not be the most active, start the most topics or get the most engagement on their topics, but that are able to establish themselves in the structure of the communication network such that they lie on many shortest paths. 
High betweenness vendors may, for example, be connecting buyers of distant locations and/or diverse goods.
However, the question of why (certain) vendors achieve high betweenness scores, remains an open question to be addressed in future work through methods such as topic modelling.
In short, while topic engagement showed the best overall performance, betweenness centrality could provide the greatest added value to law enforcement activities for reducing the set of users to investigate.

The results highlight that the same measures are almost as effective at recognizing those that will do well in the future.
This can partly be explained by those vendors that are already quite successful and will simply continue to do well.
However, results indicate that the top ranked users by betweenness and topic engagement in fact include several vendors whose majority of sales are yet to come.
Additionally, for vendors that are moderately active, betweenness centrality was shown to be more effective at finding vendors with high future sales.
Furthermore, results suggest that high betweenness centrality may (often) precede sales success.
As such, beyond predicting current success, the proposed approach can provide early warning signals for future success.
However, how early we may be able to predict future success remains an open question for future work.

Finally, we highlight some possible limitations of this work.
First, this study focused on a single, somewhat older, dark web cryptomarket.
As such, the extent to which our findings can be generalized to other cryptomarkets or Dark Web marketplaces, is an open question, as other markets may show unique characteristics not represented in this study.
Regardless, it is worth noting that most fora and marketplaces appear to be operated in a similar fashion, i.e., with the fora being used to advertise and discuss vendors and their listings.
Moreover, betweenness centrality has been shown to perform well in similar criminal network related settings before.
Therefore, we expect our findings may very well hold up for other cryptomarkets and Dark Web marketplaces.
Second, although Supplementary Material Section S1 demonstrates that our findings are generally robust with respect to variations in parameter choices during the communication network extraction, the performance of the network measures are sensitive to these parameter choices, implying that parameter tuning is likely needed depending on the considered data and precise setting.
Since our findings would be applied in a setting where sales information is unknown, especially in the case of future sales, it would be infeasible to automatically search for optimal parameter values.
Third, we note that, by virtue of not having access to hidden and missing data, this study focused only on the visible public communication on this cryptomarket.
In Boekhout et al.~\cite{boekhout2023largescale} we estimated that roughly 8\% of posts were on hidden parts of the forum or were otherwise missing from the scraped data.
Additionally, any off-market private communication is not included in the analysis.
These missing links may impact the extent to which key players can be identified through user-to-user communication networks.
Finally, we note that our analysis is hardly exhaustive in terms of considered network measures.
Although we did experiment with a wider selection of measures, none of which outperformed betweenness centrality, it is possible that another (specialized) network measure would provide better performance.
Despite this, we note that the network measures reported on in this paper cover a wide range of network interpretations relevant to the cryptomarket forum setting.
Therefore, we believe that the results reported in this paper are a good account of what can be achieved with network measures.
\section*{Methods}
\label{sect:methods}
In this section we discuss our dataset, followed by a description of how the communication networks were extracted.
Next, we discuss the rationale and computation of our activity indicators and the four network measures employed, in the context of finding key players in cryptomarkets.
\subsection*{Dataset}
\label{subsect:method-dataset}
As previously discussed in the Data section, we use the data presented in Boekhout et al.~\cite{boekhout2023largescale}.
This dataset consists of data on the forum and the market, as well as data that links forum users to market users, i.e., vendors.
For the forum data, we rely almost exclusively on the post and user data, ignoring more general information about topics and fora.
For the market data, we rely exclusively on the vendors data.
This vendors data includes their sales statistics at specific moments in time.
However, in most cases, these moments in time are not conveniently at the end of each month.
As such, the \emph{current sales} of a vendor at the end of a given month were estimated based on their average daily growth in the number of sales between the most recent sales information available before and after the change of month.
For the months after the last available sales information, the final sales total is used.
We note that, since sales statistics are only ever compared within the perspective of that same month, we can include as much future data as available instead of always looking ahead the same number of months.
\emph{Future sales} of a vendor were therefore determined as the difference between their current sales, for a given month, and the last available sales information.

Figure~\ref{fig:activity-trend} shows the total and monthly post activity and number of active users and vendors.
Here, \emph{active} users and vendors are those with at least one post up to and including the given month, where for the monthly active users we require at least one post that month.
Throughout our results we relied on the total sets of active users and vendors for each month.
\subsection*{Network extraction}
\label{subsect:methods-network-extraction}
Along with the dataset, we also utilise the communication network extraction method proposed in Boekhout et al.~\cite{boekhout2023largescale}.
This extraction method creates nodes for all active users and adds an edge connecting nodes for any posts by a pair of users that are in the same topic and adhere to certain parameters.
The direction of these edges are from the user who placed the later post to the user who placed to earlier post.
Additionally, edges are formed from every user who placed a post in a topic to the user who placed the first post in the topic.
All edges are weighted to indicate the strength of the social tie implied by the edge.

\begin{figure}[t]
  \centering
  \begin{minipage}{.58\textwidth}
    \centering
    \includegraphics[width=\textwidth]{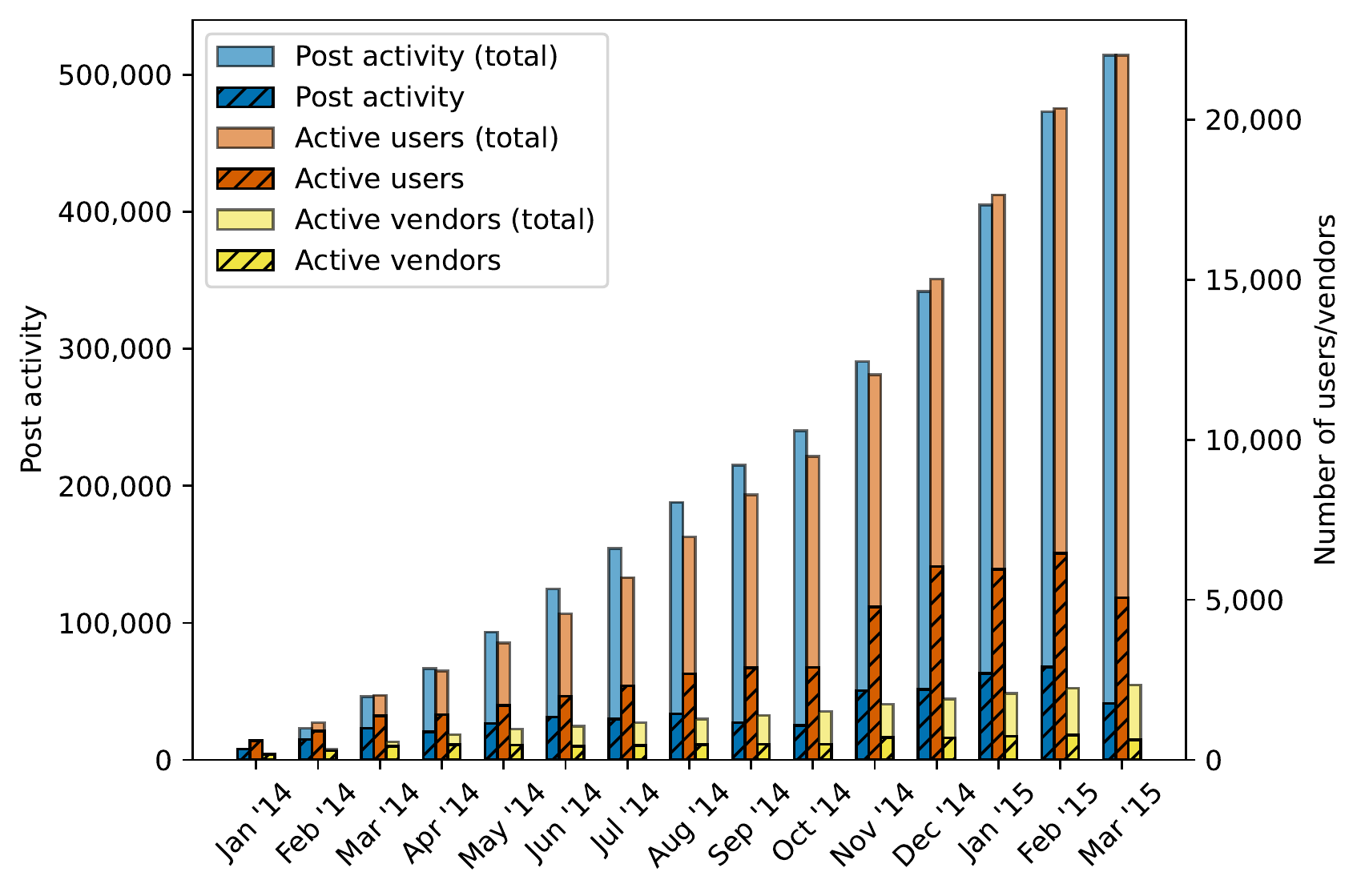}
    \captionof{figure}{Post activity and active users over time.}
    \label{fig:activity-trend}
  \end{minipage}%
  \begin{minipage}{.4\textwidth}
    \centering
    \includegraphics[width=\textwidth]{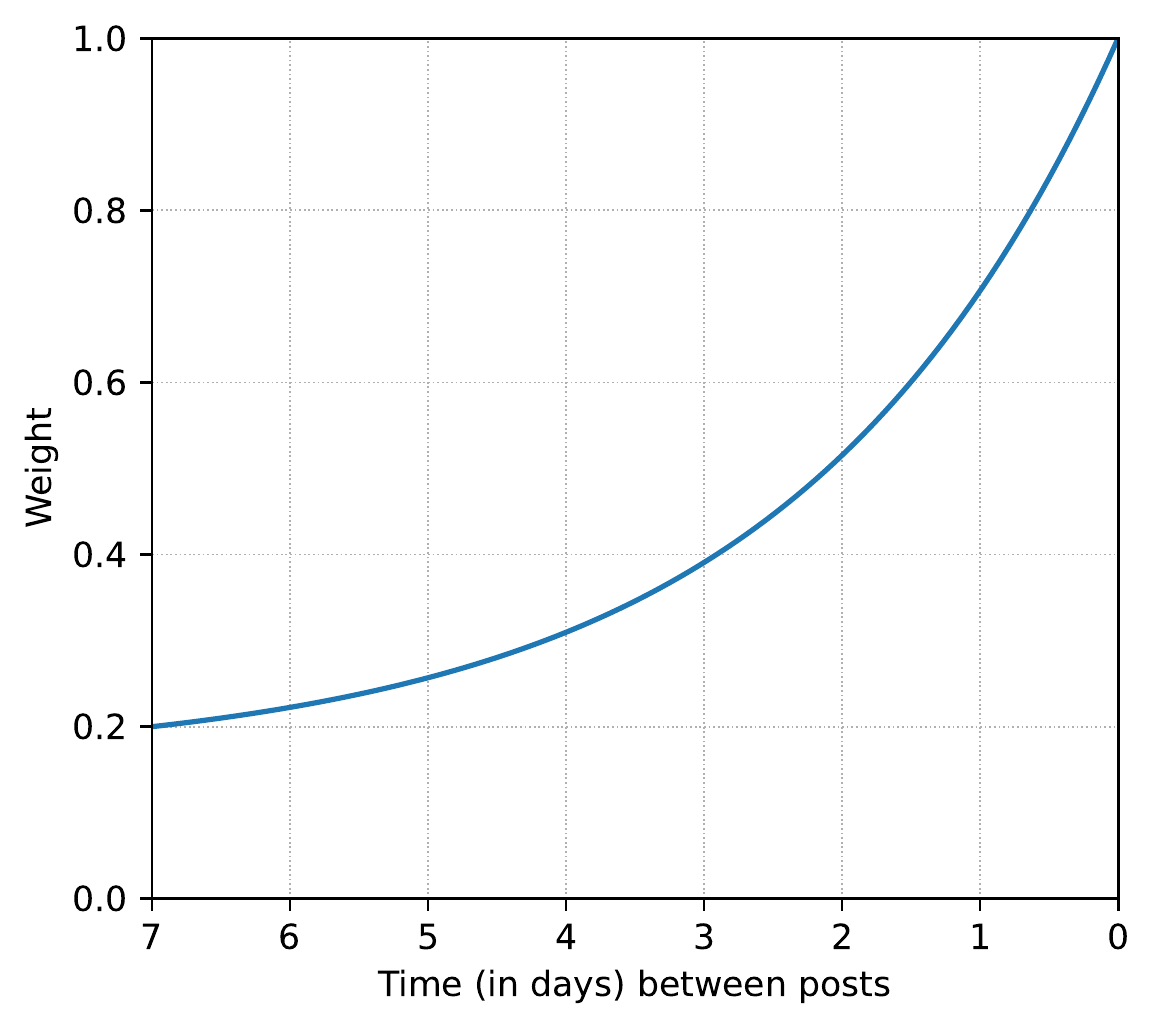}
    \captionof{figure}{Exponential weighting function for $\omega_{lower} = 0.2$ and  $t_{lim} = 7$ days.}
    \label{fig:weight-scheme}
  \end{minipage}
\end{figure}

As mentioned in the Data section, we used the following parameters for network extraction: $\delta_o = 10$, $\delta_t = 1$ month, $\omega_{lower} = 0.2$, $t_{lim} = 7$ days, and $\omega_{first} = 0.5$.
The first two parameters, i.e., $\delta_o = 10$, $\delta_t = 1$ month, set limitations on the existence of an edge.
Specifically, they allow the formation of an edge for users of posts that are ten or fewer posts apart and were placed at most one month apart.
These limitations may lead to some information loss with respect to connections to older posts.
However, the likelihood of links to such older posts representing a meaningful connection is very low, while including them would unfairly favor those posting in long (running) topics due to the sheer number of links they would receive.
The parameters $\omega_{lower} = 0.2$, $t_{lim} = 7$ days, determine the scope and decay of the exponential weighting function applied to ``regular'' edges, i.e., they determine the strength of the implied social tie. 
Specifically, $\omega_{lower}$ sets the minimum weight at $0.2$, while $t_{lim}$ determines that this minimum weight applies for all pairs of posts at least seven days apart.
The resulting exponential weighting function is shown in Figure~\ref{fig:weight-scheme}.
Thus, $\omega_{lower}$ and $t_{lim}$ determine the likelihood that a post was placed in response to or after having at least seen a specific earlier post, while $\delta_o$ and $\delta_t$ determine at what point we consider this likelihood too low to imply a social tie.
The final parameter, $\omega_{first} = 0.5$, sets the weight for all other edges, i.e., edges formed from linking posts to the initial post.
Robustness of our results for these parameters is investigated in Supplementary Material Section S1.

Monthly communication networks were extracted based on all posts up to the end of the given month, thus including posts from previous months.
Additionally, we simplify the networks by merging all parallel edges, i.e., all edges connecting the same two nodes in the same direction, into single edges.
The weights of the resulting edges are exactly the sum of the parallel edges that were merged.
In other words, the resulting weights represent the combined likelihood of a meaningful social tie connecting two users.
As a result, we obtain 15 simplified monthly weighted directed networks $G = (V,E)$, where each node $u \in V$ represents an active user and each weighted edge $(u,v) \in E$ represents the inferred weight of the social tie from user $u \in V$ to user $v \in V$.
It is on these monthly weighted directed networks that the network measures were computed.
\subsection*{Activity indicators}
\label{subsect:methods-baselines}
To evaluate the performance of predicting vendor success using network measures, we compare against activity indicators that can be directly computed from the forum data as a baseline.
Similar to the rationale for our use of network measures, these activity indicators must also adhere to the requirement that we lack knowledge of message content.
We considered three activity indicators in this paper: post activity, topics started and topic engagement.
Each of these indicators can be computed with little computational cost and independent of any knowledge of message content.
Below, we discuss why we believe these are appropriate indicators and how they are computed.
\subsubsection*{Post activity}
Post activity refers to the number of posts a user has posted on the forum up to a given moment in time.
A straightforward link can be made between a user's visibility on a forum and their post activity: the more often someone posts, the more likely it is that another user will come across one of them.
This increased visibility leads to greater name recognition, which has been linked to improved trust~\cite{chen2003interpreting} and market outcomes~\cite{huang2012brand} (e.g., more sales); and trustworthiness has been shown to be a better predictor of vendor selection than product diversity or affordability~\cite{duxbury2018network}.
Therefore, post activity can be used as in indicator of the likelihood of vendor success.
\subsubsection*{Topics started}
Forums that accompany cryptomarkets are intended to allow vendors and their customers to interact.
As such, it is common practice for vendors to promote their products listed for sale by starting a topic promoting their listings~\cite{li2021demystifying}.
The number of topics a user has started is therefore a potential indicator of being a vendor.
As a greater number of topics started may lead to greater visibility, greater name recognition, and simply a greater reach, it may also lead to increased success for vendors~\cite{familmaleki2015analyzing}.
\subsubsection*{Topic engagement}
Topic engagement is the total number of responses to all topics started by a user.
Topic engagement combines the fact that starting topics is a good indicator of being a vendor with the fact that when topics receive a lot of engagement they are naturally also more visible.
Additionally, engagement in any topics about a specific listing is likely to be associated to that listing or the vendor.
For example, a post may concern feedback on the particular listing or on the vendor themselves.
Either way, engagement on these topics is also highly probable to be associated with actual sales.
As such, where the topics started baseline is more likely to be a good indicator of being a vendor or not, topic engagement is more likely to be a good indicator of the success of any such vendor.
\subsection*{Network centrality measures}
\label{subsect:methodsmeasures}
In this subsection we discuss the various network measures utilised in this paper.
We discuss their computation and interpret their meaning within the context of cryptomarket communication networks.
Recall from the Introduction section that a link can represent two types of phenomena, either they represent a certain familiarity or shared interest, i.e., a passive relationship, or they represent the more active relationship of responding to one another (, i.e., communication).
The weighting and direction of edges in our communication networks are closely associated with the latter active relationship. 
After all, the direction of a link implies who posted first and who responded and the weighting indicates how quickly the response was made.
As such, the choice of measure variant is closely related with how we choose to interpret the links in the network.

All network measures were computed using the igraph package~\cite{igraph}.
\subsubsection*{Degree}
The degree of a node is a measure of the number of distinct neighbors connected to that node.
While degree captures this regardless of edge direction, in- and out-degree count only the neighbors connected through incoming and outgoing edges, respectively.
Furthermore, the weighted degree variants sum the weights of the connections with the neighbors.

The degree can be interpreted as the number of different users that a (potential) vendor responds to or receives responses from.
The weighted variant also takes into account how strong the relation to these users is.
Thus, a high in-degree in our networks indicates many different users responding within a relatively short time frame or responding in a topic they started.
Since it is likely that those that respond shortly after a post have seen that post, a high in-degree implies visibility to many different users, thereby improving the aforementioned brand awareness.
As brand awareness promotes trust and sales~\cite{chen2003interpreting, huang2012brand} and trust is a good predictor of vendor selection~\cite{duxbury2018network}, a high in-degree might serve as a good predictor of vendor success.

Unlike incoming edges used for in-degree, outgoing edges do not imply visibility of the user to the neighbors these edges connect to, since those neighbors posted before the user.
Furthermore, in-degree combines the visibility of responses with the visibility of starting topics.
As such, the in-degree is similar to the activity indicators, but focused on the number of individuals engaging with the user rather than the volume of engagement.
For these reasons we focus on the in-degree.
We report results for the unweighted in-degree, as we believe the number of neighbors, i.e., the number of potential customers, to be a better predictor of vendor success than the combined strength of the social ties to these neighbors. Weighted in-degree showed similar results, but with slightly fewer detected vendors that were not found by the activity indicators.
\subsubsection*{Harmonic closeness centrality}
Closeness centrality~\cite{rochat2009closeness} is a measure of how easily a node can reach every other node in the network.
Essentially, it determines whether a node is central based on its distances, i.e. shortest path lengths, to all other nodes.
In other words, where degree was a measure of how well someone is connected locally, closeness is a measure of how well connected a node is globally, i.e., to the entire network.
Harmonic closeness centrality behaves essentially the same as standard closeness centrality and extends properly to directed and disconnected networks, i.e., networks with node pairs that are not connected by any (directed) path~\cite{boldi2014axioms}.

Let $d_G(u,v)$ be the shortest distance connecting nodes $u,v \in V$, where if no path exists $d_G(u,v) = \infty$.
Using $\frac{1}{\infty} = 0$, we can define the harmonic closeness centrality as:
\begin{equation}
hcc_G(u) = \sum\limits_{v \in V} \frac{1}{d_G(u,v)}.
\end{equation}
For bidirectional harmonic closeness centrality, the shortest paths can be determined following edges regardless of their direction.
However, for incoming and outgoing harmonic closeness centrality the paths may follow edges only in one direction, either following the direction of the edges (outgoing) or going against the direction of the edges (incoming).
The weighted variants of these measures use the inverse of the edge weights during shortest distance computation, such that stronger connections equate to shorter distances.

The interpretation of distance more than a single edge away, i.e., a path, with respect to vendor success in cryptomarket communication networks is not straightforward.
After all, other than having a shared author for one of the involved posts, the posts that were responsible for the formation of subsequent links in a path may be wholly unrelated to each other.
For the weighted and directed variant, i.e., links representing responding to one another, the interpretation of paths (and their lengths) that start or end at a specific user are unclear as the path may not represent a single `conversation'.
Perhaps at best, shorter paths would imply closer familiarity. 
Therefore, we are likely better served by relying on the link representation of familiarity directly, i.e., using the undirected and unweighted variant.
Because we also detected the largest share of vendors not found by any of the activity indicators with the unweighted bidirectional variant of harmonic closeness centrality, we report on this variant in the Results section.

For the interpretation of a link as familiarity and shared interest, one can interpret a smaller distance as it being more likely for a user's posts to be visible to other users.
This interpretation would not be dissimilar to that of the `friendship' relation in a social media network such as Facebook.
Even so, it is unknown how the topics that are responsible for forming the edges that make up the connecting paths are related.
They may originate from the same or a highly similar topic, increasing the odds of being visible, or they may differ greatly, making it unlikely that these connections truly form a meaningful path.
As such, a high closeness centrality does not intuitively imply a successful vendor.
Regardless, closeness centrality has often proven to capture users at important positions in a network~\cite{fonhof2018characterizing, chen2012identifying}; its incorporation of global network information in a substantially different manner than betweenness, convinced us that it should be included in our analyses.
\subsubsection*{Betweenness centrality}
Betweenness centrality~\cite{freeman1977set, brandes2001faster} measures the extent to which a node is on shortest paths connecting pairs of nodes in the network.
In other words, it measures how important a node is with respect to connecting various communities in the network.
In the context of cryptomarkets, this makes it a good measure of how well a (potential) vendor reaches different communities of potential buyers.
As such, a vendor with a high betweenness is more likely to have a larger pool of buyers as they may be able to draw from more relatively distinct communities of buyers.
Additionally, betweenness centrality has been shown to perform well in identifying key players in criminal networks~\cite{liu2012criminal, cavallaro2020disrupting}.

The betweenness centrality of node $u \in V$ is determined by computing the sum of the fraction of shortest paths connecting nodes $v,w \in V$ that pass through $u$.
Let $\sigma_{vw}$ indicate the number of shortest paths connecting nodes $v,w \in V$, and let $\sigma_{vuw}$ indicate the number of those shortest paths that pass through node $u \in V$.
Then betweenness centrality can be defined as:
\begin{equation}
bc(u) = \sum\limits_{v,w \in V, u\neq v\neq w} \frac{\sigma_{vuw}}{\sigma_{vw}}
\end{equation}
For directed betweenness centrality, paths must follow the direction of the edges, while undirected betweenness can follow edges in either direction.
Like for harmonic closeness centrality, the weighted variants use the inverse of the edge weights during shortest path computation, such that stronger connections equate to shorter distances.
As direction and weighting can have a large impact on the probability of lying on a shortest path, we use the directed weighted betweenness in this study.
We note that where the interpretation of paths was unclear for harmonic closeness centrality, this is less of an issue for betweenness centrality.
After all, for betweenness centrality we rely on the fact that the user exists on many shortest paths, and is less reliant on its specific position in the path. 
When viewed on the scale of the whole network, existing on many shortest paths implies being a central figure in the various conversations occurring on the forum, regardless of whether the individual paths relate to the same conversation.
For a vendor, this in turn might imply that they or their products are often discussed and that the vendor actively engages with their clientele, all of which may promote their sales.
Thus, a high directed weighted betweenness centrality can be interpreted here as a good indicator of vendor success.
We note that preliminary results showed this variant to also have the best performance.
\subsubsection*{PageRank}
The final measure we consider is PageRank~\cite{page1999pagerank}.
PageRank computes the probability that a random walker, that follows one of the available neighboring edges or jumps to a random node with a particular probability, ends up at a given node.
For the directed variant the choice of edge is restricted to following the direction of the edges and adding weights impacts the odds of following any given edge.
Similar as for betweenness centrality, we report the results for the variant taking both direction and weighting into account as this provides the random walker with more context.
Note that we found this variant of PageRank to indeed have the best performance.

High PageRank values often follow from having paths/edges incoming from (many) other important (i.e., high value) nodes in the network.
As such, we can interpret a high PageRank value as being closely connected to other key players.
As previously stated, Duxbury \& Haynie~\cite{duxbury2018network} found that buyers were more likely to continue ordering with vendors within the same community.
This means that the close connection between users with high PageRank value can be indicative of a boost in their perceived trust and may stimulate their sales.
Thus, a high PageRank value may be able to predict successful vendors.

\subsection*{Evaluation metrics}
In this subsection we discuss the normalization and various evaluation metrics employed in this work.
\subsubsection*{Relative/absolute difference score}
Let $s_{u,t,i}$ indicate the value for user/node $u$, month $t$, and measure $i$.
We first apply min-max normalization to these values, i.e.,
\begin{equation}
    sn_{u,t,i} = \frac{s_{u,t,i} - min_{u_x \in V}\ s_{u_x,t,i}}{max_{u_x \in V}\ s_{u_x,t,i} - min_{u_x \in V}\ s_{u_x,t,i}}.
\end{equation}
Let $V_a, V_b \subset V$ indicate two groups of users, for example, the group of all vendors and all non-vendors for a given month.
The \emph{absolute} and \emph{relative difference score} for a given month $t$ and measure $i$ can then be computed as:
\begin{equation}
    \mathit{abs\_diff\_score}(a, b, t, i) =  \frac{\sum_{u \in V_a} sn_{u,t,i}}{|V_a|} - \frac{\sum_{u \in V_b} sn_{u,t,i}}{|V_b|};
\end{equation}
\begin{equation}
     \mathit{rel\_diff\_score}(a,b,t,i) = \frac{\mathit{abs\_diff\_score}(a, b, t, i)}{\frac{\sum_{u \in V_b} sn_{u,t,i}}{|V_b|}}.
\end{equation}
\subsubsection*{Recall metrics}
Let $TV_{t} \subset V$ indicate the top vendor percentile for a given month $t$ and let $TU_{t,i} \subset V$ indicate the top 20\% users based on measure $i$ for a given month $t$.
Then the \emph{vendor recall} is computed as
\begin{equation}
    \mathit{vendor\_recall}(t, i) = \frac{|TV_{t} \cap TU_{t,i}|}{|TU_{t,i}|} \times 100\%.
\end{equation}
The monthly overlap between detected vendors (for which the mean and standard deviation are presented in Table~\ref{table:overlap}) for given measures $i, j$ and month $t$ is computed as follows:
\begin{equation}
    \mathit{overlap}_{t,i,j} =  \frac{|TV_{t} \cap TU_{t,i} \cap TU_{t,j}|}{|TV_{t} \cap TU_{t,i}|} \times 100\%.
\end{equation}
Thus, $\mathit{overlap_{t,i,j}}$ computes the percentage of (top vendor percentile) vendors detected by measure $i$ that were also found by measure $j$. Note, that in most cases $\mathit{overlap_{t,i,j}} \neq \mathit{overlap_{t,j,i}}$.

Let $pa_{u,t}$ indicate the post activity of user/node $u$ up to and including month $t$; and let $\mathit{sales}_{u,t}$ indicate their sales.
Then we can compute the \emph{post activity recall} and \emph{sales recall} as follows:
\begin{equation}
    \mathit{post\_activity\_recall}(t, i) = \frac{\sum_{u \in TV_{t} \cap TU_{t,i}} pa_{u,t}}{\sum_{u \in TV_{t}} pa_{u,t}} \times 100\%;
\end{equation}
\begin{equation}
    \mathit{sales\_recall}(t, i) = \frac{\sum_{u \in TV_{t} \cap TU_{t,i}} \mathit{sales}_{u,t}}{\sum_{u \in TV_{t}} \mathit{sales}_{u,t}} \times 100\%.
\end{equation}

\section*{Data availability}
The dataset extraction and data quality resolution process as well as the network extraction process, is discussed in the Data Descriptor by Boekhout et al.~\cite{boekhout2023largescale}. The dataset itself is directly available on Zenodo~\cite{boekhout_2023_10171217} at \url{https://zenodo.org/records/10171217}.

\bibliography{bibliography}

\begin{thebibliography}{10}
\urlstyle{rm}
\expandafter\ifx\csname url\endcsname\relax
  \def\url#1{\texttt{#1}}\fi
\expandafter\ifx\csname urlprefix\endcsname\relax\def\urlprefix{URL }\fi
\expandafter\ifx\csname doiprefix\endcsname\relax\def\doiprefix{DOI: }\fi
\providecommand{\bibinfo}[2]{#2}
\providecommand{\eprint}[2][]{\url{#2}}

\bibitem{darkwebhow}
\bibinfo{author}{Egan, M.}
\newblock \bibinfo{title}{What is the dark web, what’s on it \& how to access it}.
\newblock \bibinfo{howpublished}{\emph{Tech Advisor how-to article} \url{https://www.techadvisor.com/article/727316/what-is-the-dark-web-whats-on-it-how-to-access-it.html}} (\bibinfo{year}{2019}).

\bibitem{nadini2022emergence}
\bibinfo{author}{Nadini, M.} \emph{et~al.}
\newblock \bibinfo{journal}{\bibinfo{title}{Emergence and structure of decentralised trade networks around dark web marketplaces}}.
\newblock {\emph{\JournalTitle{Scientific Reports}}} \textbf{\bibinfo{volume}{12}}, \bibinfo{pages}{1--9} (\bibinfo{year}{2022}).

\bibitem{chainanlysis2021}
\bibinfo{author}{Chainalysis}.
\newblock \bibinfo{title}{The chainalysis 2021 crypto crime report}.
\newblock \bibinfo{howpublished}{\url{https://go.chainalysis.com/2021-Crypto-Crime-Report.html}} (\bibinfo{year}{2021}).

\bibitem{martin2014lost}
\bibinfo{author}{Martin, J.}
\newblock \bibinfo{journal}{\bibinfo{title}{Lost on the silk road: online drug distribution and the ‘cryptomarket’}}.
\newblock {\emph{\JournalTitle{Criminology \& Criminal Justice}}} \textbf{\bibinfo{volume}{14}}, \bibinfo{pages}{351--367} (\bibinfo{year}{2014}).

\bibitem{shortis2020drug}
\bibinfo{author}{Shortis, P.}, \bibinfo{author}{Aldridge, J.}, \bibinfo{author}{Monica, J.} \emph{et~al.}
\newblock \bibinfo{title}{Drug cryptomarket futures: structure, function and evolution in response to law enforcement actions}.
\newblock In \emph{\bibinfo{booktitle}{Research Handbook on International Drug Policy}}, \bibinfo{pages}{355--380} (\bibinfo{publisher}{Edward Elgar Publishing}, \bibinfo{year}{2020}).

\bibitem{fonhof2018characterizing}
\bibinfo{author}{Fonhof, A.~M.}, \bibinfo{author}{van~der Bruggen, M.} \& \bibinfo{author}{Takes, F.~W.}
\newblock \bibinfo{title}{Characterizing key players in child exploitation networks on the dark net}.
\newblock In \emph{\bibinfo{booktitle}{International Conference on Complex Networks and their Applications}}, \bibinfo{pages}{412--423} (\bibinfo{publisher}{Springer}, \bibinfo{year}{2018}).

\bibitem{munksgaard2016mixing}
\bibinfo{author}{Munksgaard, R.} \& \bibinfo{author}{Demant, J.}
\newblock \bibinfo{journal}{\bibinfo{title}{Mixing politics and crime--the prevalence and decline of political discourse on the cryptomarket}}.
\newblock {\emph{\JournalTitle{International Journal of Drug Policy}}} \textbf{\bibinfo{volume}{35}}, \bibinfo{pages}{77--83} (\bibinfo{year}{2016}).

\bibitem{van2017new}
\bibinfo{author}{Van~Hout, M.~C.} \& \bibinfo{author}{Hearne, E.}
\newblock \bibinfo{journal}{\bibinfo{title}{New psychoactive substances ({NPS}) on cryptomarket fora: an exploratory study of characteristics of forum activity between {NPS} buyers and vendors}}.
\newblock {\emph{\JournalTitle{International Journal of Drug Policy}}} \textbf{\bibinfo{volume}{40}}, \bibinfo{pages}{102--110} (\bibinfo{year}{2017}).

\bibitem{moeller2017flow}
\bibinfo{author}{Moeller, K.}, \bibinfo{author}{Munksgaard, R.} \& \bibinfo{author}{Demant, J.}
\newblock \bibinfo{journal}{\bibinfo{title}{Flow my {FE} the vendor said: exploring violent and fraudulent resource exchanges on cryptomarkets for illicit drugs}}.
\newblock {\emph{\JournalTitle{American Behavioral Scientist}}} \textbf{\bibinfo{volume}{61}}, \bibinfo{pages}{1427--1450} (\bibinfo{year}{2017}).

\bibitem{armona2018measuring}
\bibinfo{author}{Armona, L.}
\newblock \bibinfo{journal}{\bibinfo{title}{Measuring the demand effects of formal and informal communication: Evidence from online markets for illicit drugs}}.
\newblock {\emph{\JournalTitle{arXiv preprint arXiv:1802.08778}}}  (\bibinfo{year}{2018}).

\bibitem{kamphausen2019digital}
\bibinfo{author}{Kamphausen, G.} \& \bibinfo{author}{Werse, B.}
\newblock \bibinfo{journal}{\bibinfo{title}{Digital figurations in the online trade of illicit drugs: a qualitative content analysis of darknet forums}}.
\newblock {\emph{\JournalTitle{International Journal of Drug Policy}}} \textbf{\bibinfo{volume}{73}}, \bibinfo{pages}{281--287} (\bibinfo{year}{2019}).

\bibitem{li2021demystifying}
\bibinfo{author}{Li, Z.}, \bibinfo{author}{Du, X.}, \bibinfo{author}{Liao, X.}, \bibinfo{author}{Jiang, X.} \& \bibinfo{author}{Champagne-Langabeer, T.}
\newblock \bibinfo{journal}{\bibinfo{title}{Demystifying the dark web opioid trade: content analysis on anonymous market listings and forum posts}}.
\newblock {\emph{\JournalTitle{Journal of Medical Internet Research}}} \textbf{\bibinfo{volume}{23}}, \bibinfo{pages}{e24486} (\bibinfo{year}{2021}).

\bibitem{li2014comparative}
\bibinfo{author}{Li, M.-X.} \emph{et~al.}
\newblock \bibinfo{journal}{\bibinfo{title}{A comparative analysis of the statistical properties of large mobile phone calling networks}}.
\newblock {\emph{\JournalTitle{Scientific reports}}} \textbf{\bibinfo{volume}{4}}, \bibinfo{pages}{1--12} (\bibinfo{year}{2014}).

\bibitem{cavallaro2020disrupting}
\bibinfo{author}{Cavallaro, L.} \emph{et~al.}
\newblock \bibinfo{journal}{\bibinfo{title}{Disrupting resilient criminal networks through data analysis: the case of sicilian mafia}}.
\newblock {\emph{\JournalTitle{Plos one}}} \textbf{\bibinfo{volume}{15}}, \bibinfo{pages}{e0236476} (\bibinfo{year}{2020}).

\bibitem{reisch2022monitoring}
\bibinfo{author}{Reisch, T.}, \bibinfo{author}{Heiler, G.}, \bibinfo{author}{Diem, C.}, \bibinfo{author}{Klimek, P.} \& \bibinfo{author}{Thurner, S.}
\newblock \bibinfo{journal}{\bibinfo{title}{Monitoring supply networks from mobile phone data for estimating the systemic risk of an economy}}.
\newblock {\emph{\JournalTitle{Scientific Reports}}} \textbf{\bibinfo{volume}{12}}, \bibinfo{pages}{13347} (\bibinfo{year}{2022}).

\bibitem{lee2022universal}
\bibinfo{author}{Lee, J.-H.}, \bibinfo{author}{Sato, N.}, \bibinfo{author}{Yano, K.} \& \bibinfo{author}{Miyake, Y.}
\newblock \bibinfo{journal}{\bibinfo{title}{Universal association between depressive symptoms and social-network structures in the workplace}}.
\newblock {\emph{\JournalTitle{Scientific Reports}}} \textbf{\bibinfo{volume}{12}}, \bibinfo{pages}{10170} (\bibinfo{year}{2022}).

\bibitem{wang2023evidence}
\bibinfo{author}{Wang, X.}, \bibinfo{author}{Li, J.}, \bibinfo{author}{Srivatsavaya, E.} \& \bibinfo{author}{Rajtmajer, S.}
\newblock \bibinfo{journal}{\bibinfo{title}{Evidence of inter-state coordination amongst state-backed information operations}}.
\newblock {\emph{\JournalTitle{Scientific Reports}}} \textbf{\bibinfo{volume}{13}}, \bibinfo{pages}{7716} (\bibinfo{year}{2023}).

\bibitem{lu2014network}
\bibinfo{author}{Lu, X.} \& \bibinfo{author}{Brelsford, C.}
\newblock \bibinfo{journal}{\bibinfo{title}{Network structure and community evolution on twitter: human behavior change in response to the 2011 japanese earthquake and tsunami}}.
\newblock {\emph{\JournalTitle{Scientific reports}}} \textbf{\bibinfo{volume}{4}}, \bibinfo{pages}{6773} (\bibinfo{year}{2014}).

\bibitem{jo2023stage}
\bibinfo{author}{Jo, W.}, \bibinfo{author}{Jang, S.~H.} \& \bibinfo{author}{Shin, E.~K.}
\newblock \bibinfo{journal}{\bibinfo{title}{Stage distinctive communication networks of the online breast cancer community}}.
\newblock {\emph{\JournalTitle{Scientific Reports}}} \textbf{\bibinfo{volume}{13}}, \bibinfo{pages}{1726} (\bibinfo{year}{2023}).

\bibitem{liu2012criminal}
\bibinfo{author}{Liu, X.}, \bibinfo{author}{Patacchini, E.}, \bibinfo{author}{Zenou, Y.} \& \bibinfo{author}{Lee, L.-F.}
\newblock \bibinfo{journal}{\bibinfo{title}{Criminal networks: who is the key player?}}
\newblock {\emph{\JournalTitle{FEEM Working Paper}}} \textbf{\bibinfo{volume}{39}} (\bibinfo{year}{2012}).

\bibitem{familmaleki2015analyzing}
\bibinfo{author}{Familmaleki, M.}, \bibinfo{author}{Aghighi, A.} \& \bibinfo{author}{Hamidi, K.}
\newblock \bibinfo{journal}{\bibinfo{title}{Analyzing the influence of sales promotion on customer purchasing behavior}}.
\newblock {\emph{\JournalTitle{International Journal of Economics \& management sciences}}} \textbf{\bibinfo{volume}{4}}, \bibinfo{pages}{1--6} (\bibinfo{year}{2015}).

\bibitem{chen2003interpreting}
\bibinfo{author}{Chen, S.~C.} \& \bibinfo{author}{Dhillon, G.~S.}
\newblock \bibinfo{journal}{\bibinfo{title}{Interpreting dimensions of consumer trust in e-commerce}}.
\newblock {\emph{\JournalTitle{Information technology and management}}} \textbf{\bibinfo{volume}{4}}, \bibinfo{pages}{303--318} (\bibinfo{year}{2003}).

\bibitem{huang2012brand}
\bibinfo{author}{Huang, R.} \& \bibinfo{author}{Sarig{\"o}ll{\"u}, E.}
\newblock \bibinfo{journal}{\bibinfo{title}{How brand awareness relates to market outcome, brand equity, and the marketing mix}}.
\newblock {\emph{\JournalTitle{Journal of Business Research}}} \textbf{\bibinfo{volume}{65}}, \bibinfo{pages}{92--99} (\bibinfo{year}{2012}).

\bibitem{duxbury2018network}
\bibinfo{author}{Duxbury, S.~W.} \& \bibinfo{author}{Haynie, D.~L.}
\newblock \bibinfo{journal}{\bibinfo{title}{The network structure of opioid distribution on a darknet cryptomarket}}.
\newblock {\emph{\JournalTitle{Journal of quantitative criminology}}} \textbf{\bibinfo{volume}{34}}, \bibinfo{pages}{921--941} (\bibinfo{year}{2018}).

\bibitem{evolutionbackground}
\bibinfo{author}{DEEPDOTWEB}.
\newblock \bibinfo{title}{Evolution market background: carding forums, ponzi schemes \& {LE}}.
\newblock \bibinfo{howpublished}{\emph{DEEP.DOT.WEB article} \url{https://www.gwern.net/docs/darknet-markets/evolution/2015-03-18-the_avid-evolutionmarketbackground-cardingforumsponzischemesle.html}} (\bibinfo{year}{2015}).

\bibitem{dnmArchives}
\bibinfo{author}{Branwen, G.} \emph{et~al.}
\newblock \bibinfo{title}{Dark net market archives, 2011-2015}.
\newblock \bibinfo{howpublished}{\url{https://www.gwern.net/DNM-archives}} (\bibinfo{year}{2015}).
\newblock \bibinfo{note}{Accessed: 2021-07-22}.

\bibitem{boekhout2023largescale}
\bibinfo{author}{Boekhout, H.~D.}, \bibinfo{author}{Blokland, A.~A.} \& \bibinfo{author}{Takes, F.~W.}
\newblock \bibinfo{journal}{\bibinfo{title}{A large-scale longitudinal structured dataset of the dark web cryptomarket evolution (2014-2015)}}.
\newblock {\emph{\JournalTitle{arXiv preprint arXiv:2311.11878}}}  (\bibinfo{year}{2023}).

\bibitem{norbutas2020reputation}
\bibinfo{author}{Norbutas, L.}, \bibinfo{author}{Ruiter, S.} \& \bibinfo{author}{Corten, R.}
\newblock \bibinfo{journal}{\bibinfo{title}{Reputation transferability across contexts: Maintaining cooperation among anonymous cryptomarket actors when moving between markets}}.
\newblock {\emph{\JournalTitle{International Journal of Drug Policy}}} \textbf{\bibinfo{volume}{76}}, \bibinfo{pages}{102635} (\bibinfo{year}{2020}).

\bibitem{rochat2009closeness}
\bibinfo{author}{Rochat, Y.}
\newblock \bibinfo{title}{Closeness centrality extended to unconnected graphs: the harmonic centrality index}.
\newblock \bibinfo{type}{Tech. Rep.}, \bibinfo{institution}{ASNA} (\bibinfo{year}{2009}).

\bibitem{freeman1977set}
\bibinfo{author}{Freeman, L.~C.}
\newblock \bibinfo{journal}{\bibinfo{title}{A set of measures of centrality based on betweenness}}.
\newblock {\emph{\JournalTitle{Sociometry}}} \textbf{\bibinfo{volume}{40}}, \bibinfo{pages}{35--41} (\bibinfo{year}{1977}).

\bibitem{brandes2001faster}
\bibinfo{author}{Brandes, U.}
\newblock \bibinfo{journal}{\bibinfo{title}{A faster algorithm for betweenness centrality}}.
\newblock {\emph{\JournalTitle{Journal of mathematical sociology}}} \textbf{\bibinfo{volume}{25}}, \bibinfo{pages}{163--177} (\bibinfo{year}{2001}).

\bibitem{page1999pagerank}
\bibinfo{author}{Page, L.}, \bibinfo{author}{Brin, S.}, \bibinfo{author}{Motwani, R.} \& \bibinfo{author}{Winograd, T.}
\newblock \bibinfo{title}{The pagerank citation ranking: bringing order to the web}.
\newblock \bibinfo{type}{Tech. Rep.}, \bibinfo{institution}{Stanford InfoLab} (\bibinfo{year}{1999}).

\bibitem{kempe2000connectivity}
\bibinfo{author}{Kempe, D.}, \bibinfo{author}{Kleinberg, J.} \& \bibinfo{author}{Kumar, A.}
\newblock \bibinfo{title}{Connectivity and inference problems for temporal networks}.
\newblock In \emph{\bibinfo{booktitle}{Proceedings of the thirty-second annual ACM symposium on Theory of computing}}, \bibinfo{pages}{504--513} (\bibinfo{year}{2000}).

\bibitem{igraph}
\bibinfo{author}{Csardi, G.} \& \bibinfo{author}{Nepusz, T.}
\newblock \bibinfo{journal}{\bibinfo{title}{The igraph software package for complex network research}}.
\newblock {\emph{\JournalTitle{InterJournal, complex systems}}} \textbf{\bibinfo{volume}{1695}}, \bibinfo{pages}{1--9} (\bibinfo{year}{2006}).

\bibitem{boldi2014axioms}
\bibinfo{author}{Boldi, P.} \& \bibinfo{author}{Vigna, S.}
\newblock \bibinfo{journal}{\bibinfo{title}{Axioms for centrality}}.
\newblock {\emph{\JournalTitle{Internet Mathematics}}} \textbf{\bibinfo{volume}{10}}, \bibinfo{pages}{222--262} (\bibinfo{year}{2014}).

\bibitem{chen2012identifying}
\bibinfo{author}{Chen, D.}, \bibinfo{author}{L{\"u}, L.}, \bibinfo{author}{Shang, M.-S.}, \bibinfo{author}{Zhang, Y.-C.} \& \bibinfo{author}{Zhou, T.}
\newblock \bibinfo{journal}{\bibinfo{title}{Identifying influential nodes in complex networks}}.
\newblock {\emph{\JournalTitle{Physica a: Statistical mechanics and its applications}}} \textbf{\bibinfo{volume}{391}}, \bibinfo{pages}{1777--1787} (\bibinfo{year}{2012}).

\bibitem{boekhout_2023_10171217}
\bibinfo{author}{Boekhout, H.}, \bibinfo{author}{Blokland, A.} \& \bibinfo{author}{Takes, F.}
\newblock \bibinfo{title}{{A large-scale longitudinal structured dataset of the dark web cryptomarket Evolution (2014–2015)}}, \doiprefix\url{10.5281/zenodo.10171217} (\bibinfo{year}{2023}).

\end{thebibliography}

\section*{Acknowledgements}
This research was funded by Politie \& Wetenschap.
We would like to acknowledge the advisory committee assigned to this project for its valuable input on relevant law enforcement practices and domain aware baselines.
\section*{Author contributions statement}
All authors aided in conceptualizing the study.
H.B. designed the study, collected and cleaned the data, performed the study and result analysis, and drafted the manuscript.
F.T. and A.B. supervised this process and gave suggestions on the study design and result analysis.
All authors reviewed the report, commented on drafts of the manuscript, and approved the final report.
\section*{Competing interests}
The author(s) declare no competing interests.

\clearpage

\supplementarysection
\section{Robustness of results}
\begin{landscape}
\begin{figure}[t]
  \centering
    \begin{subfigure}[b]{0.25\textheight}
      \centering
      \includegraphics[width=\textwidth]{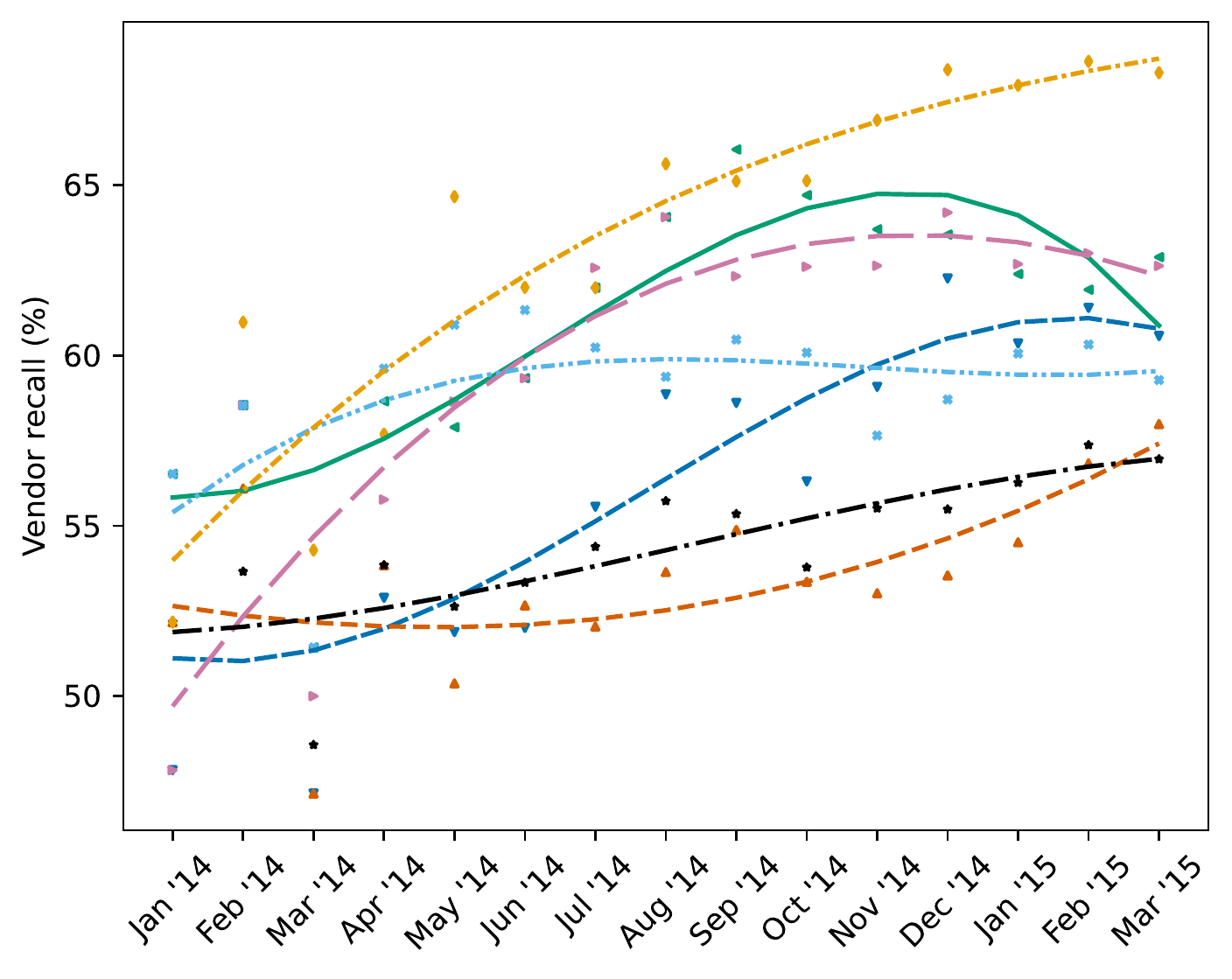}
      \caption{$\delta_o = 2$}
    \end{subfigure}
    ~
    \begin{subfigure}[b]{0.25\textheight}
      \centering
      \includegraphics[width=\textwidth]{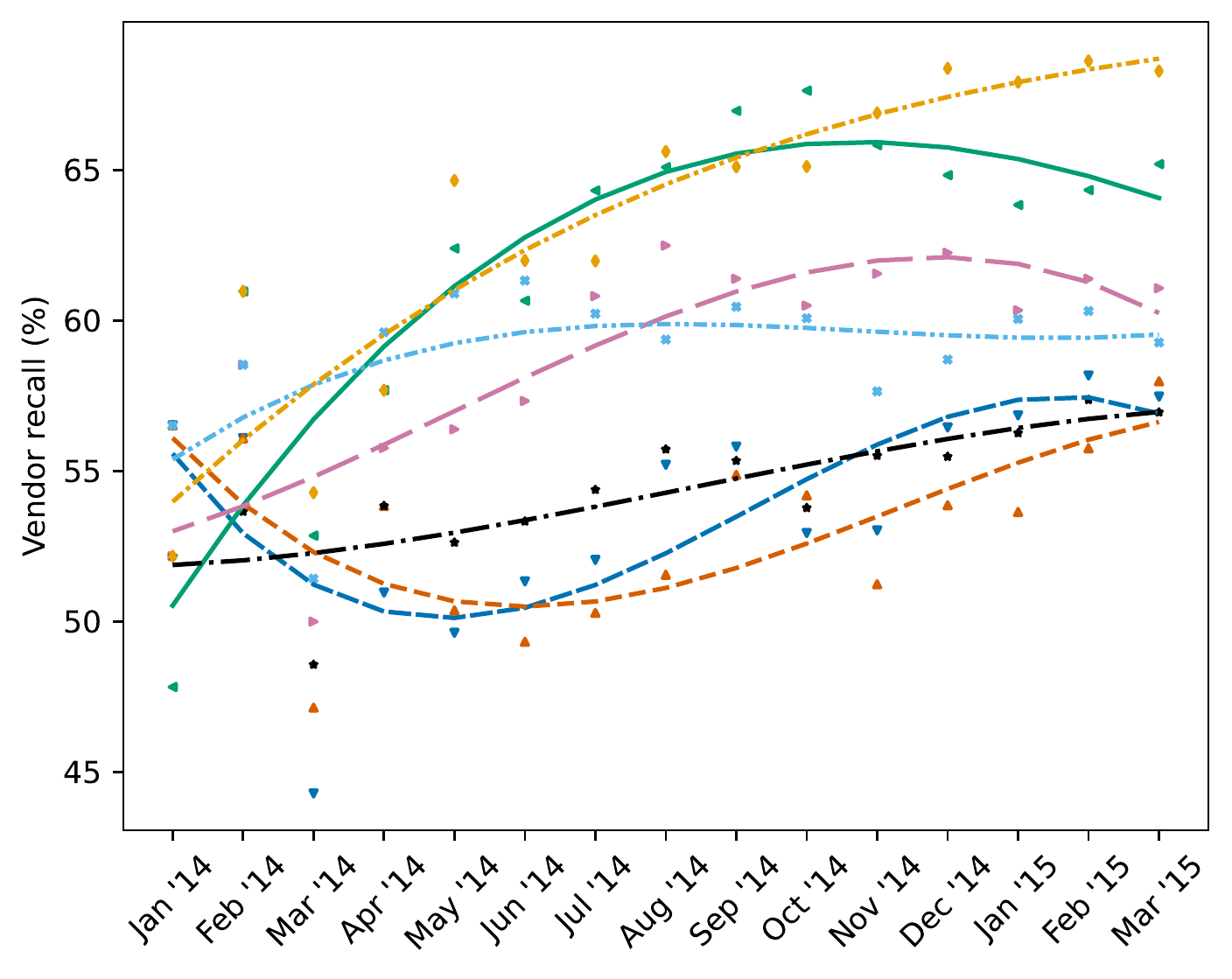}
      \caption{$\delta_o = 5$}
    \end{subfigure}
    ~
    \begin{subfigure}[b]{0.25\textheight}
      \centering
      \includegraphics[width=\textwidth]{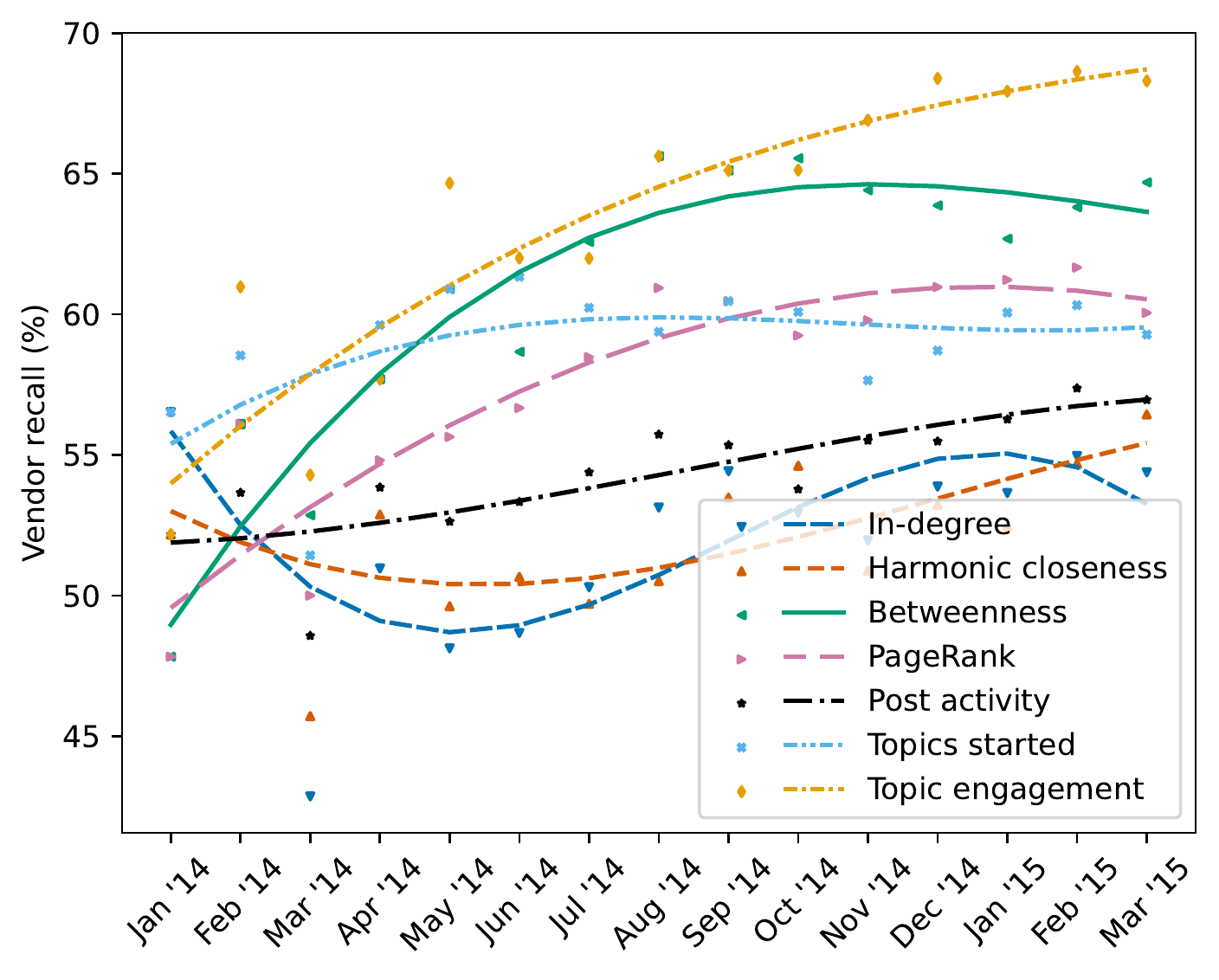}
      \caption{$\delta_o = 10$}
    \end{subfigure}
    ~
    \begin{subfigure}[b]{0.25\textheight}
      \centering
      \includegraphics[width=\textwidth]{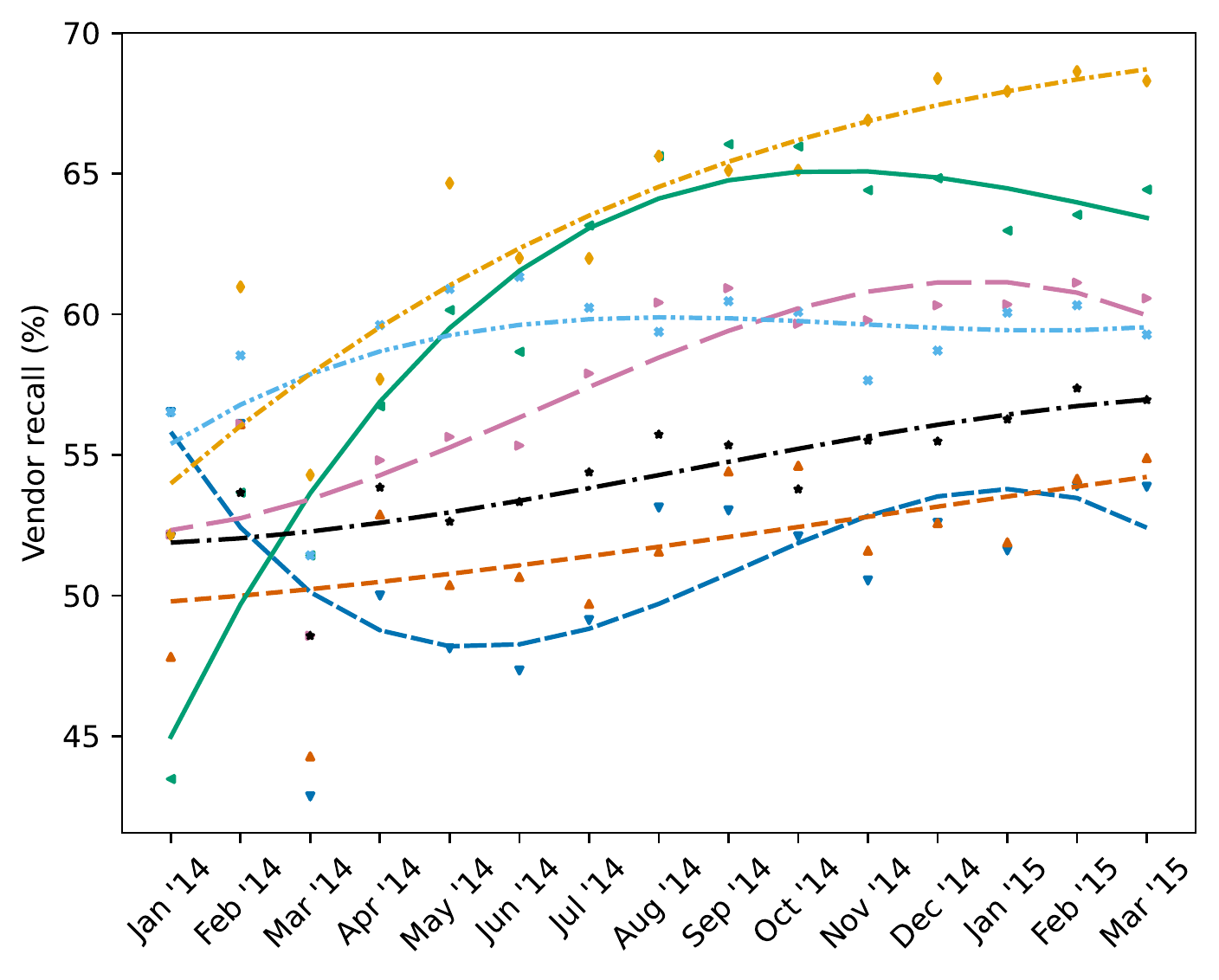}
      \caption{$\delta_o = 15$}
    \end{subfigure}
    ~
    \begin{subfigure}[b]{0.25\textheight}
      \centering
      \includegraphics[width=\textwidth]{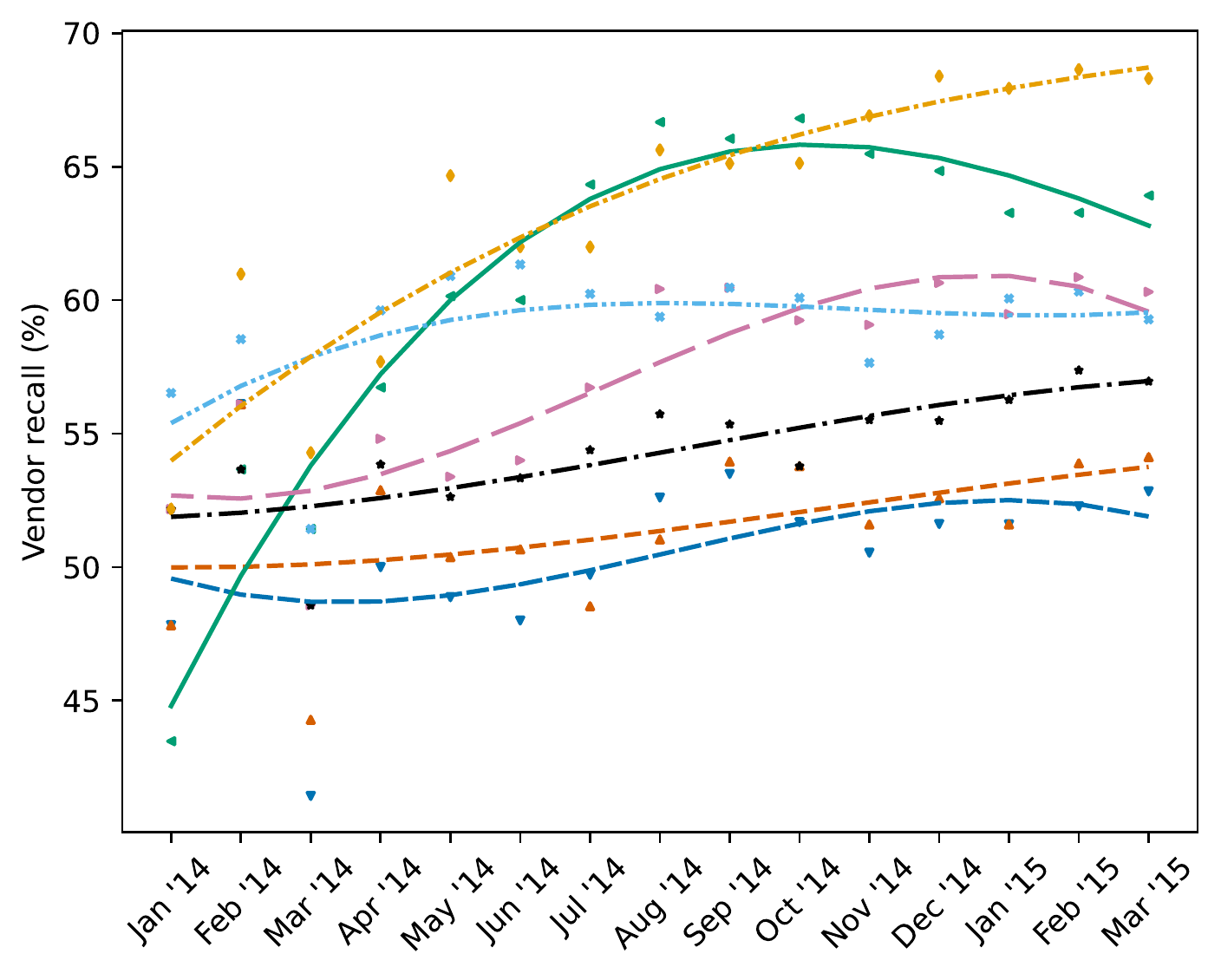}
      \caption{$\delta_o = 20$}
    \end{subfigure}

    \begin{subfigure}[b]{0.25\textheight}
      \centering
      \includegraphics[width=\textwidth]{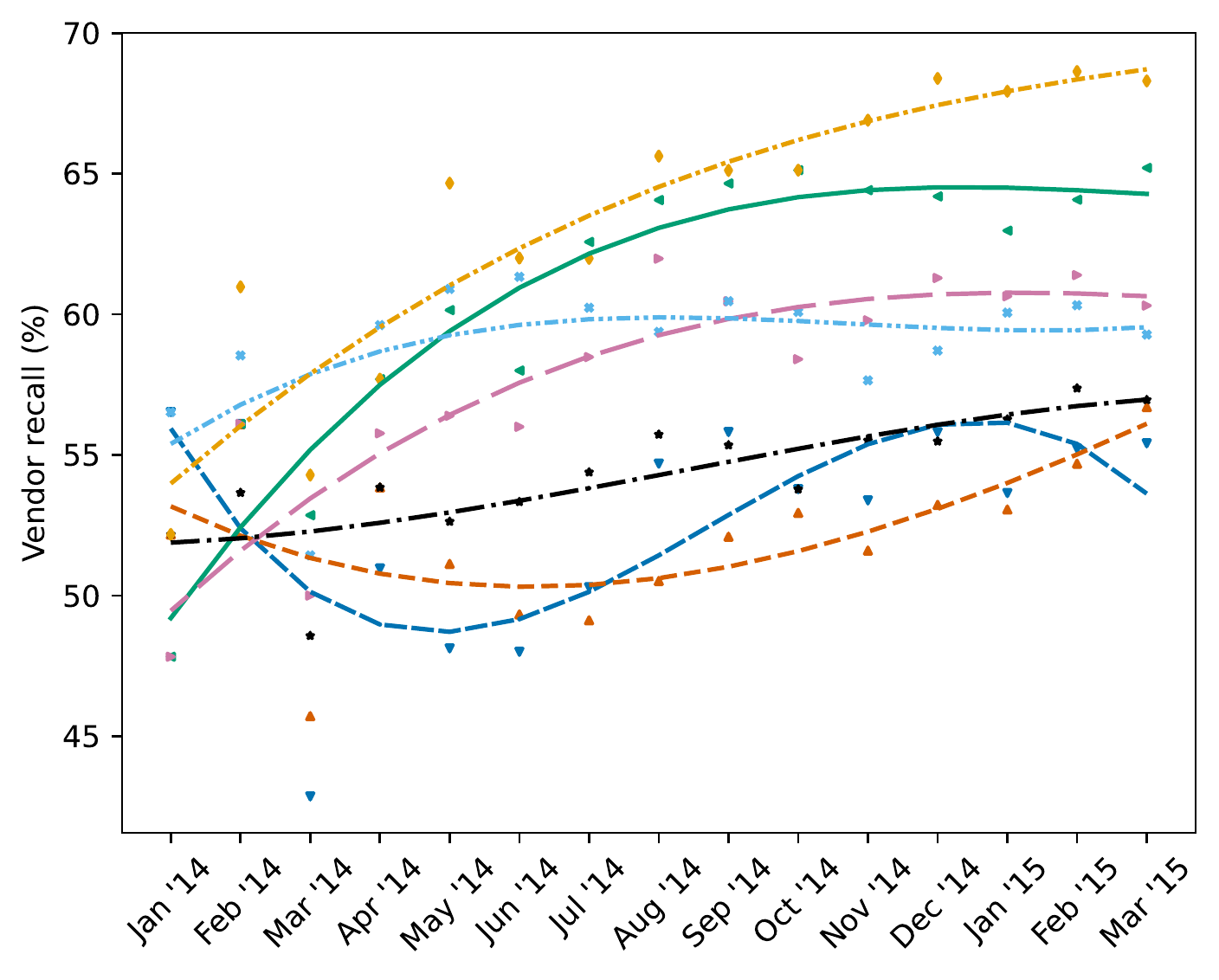}
      \caption{$\delta_t = 7$ days}
    \end{subfigure}
    ~
    \begin{subfigure}[b]{0.25\textheight}
      \centering
      \includegraphics[width=\textwidth]{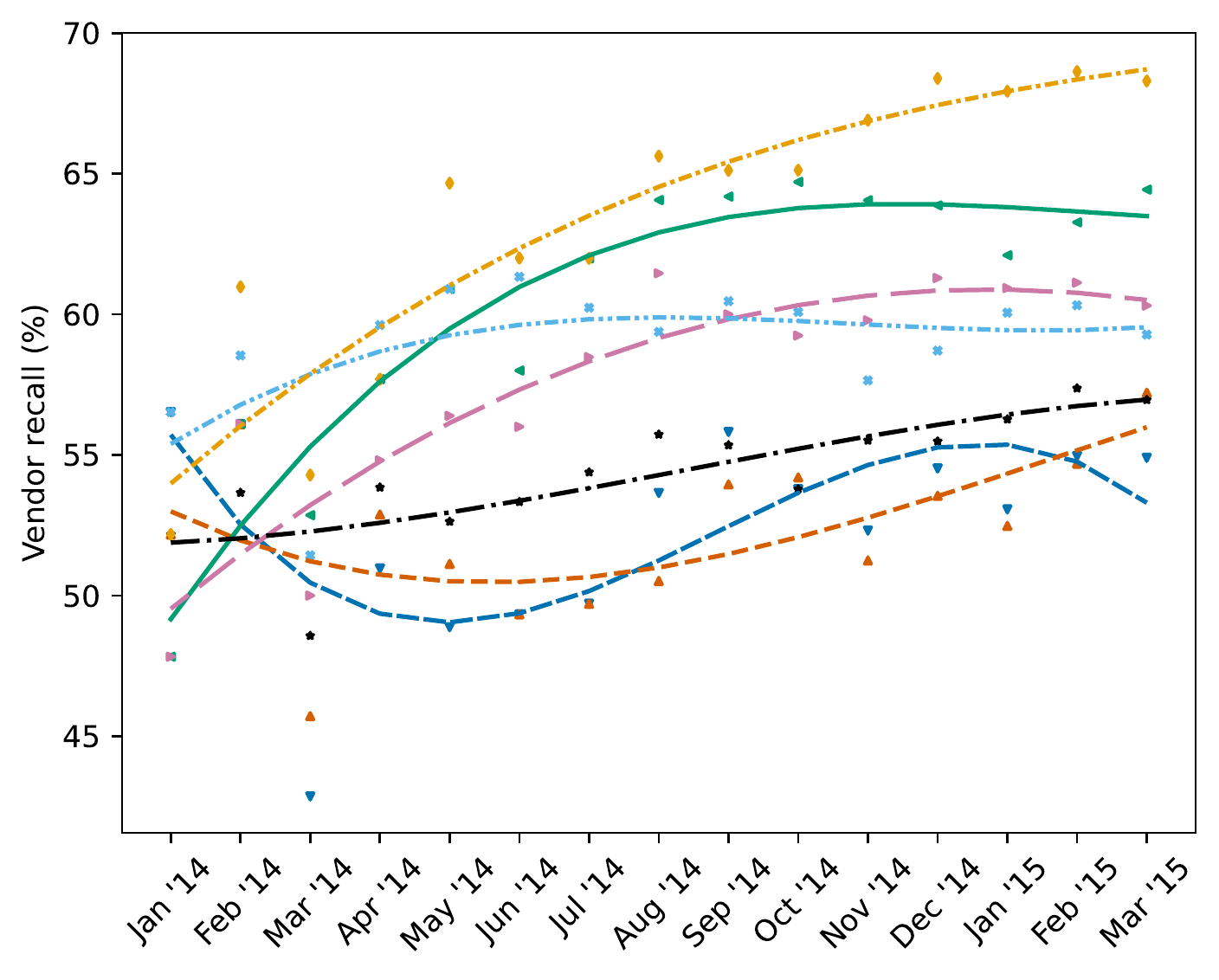}
      \caption{$\delta_t = 14$ days}
    \end{subfigure}
    ~
    \begin{subfigure}[b]{0.25\textheight}
      \centering
      \includegraphics[width=\textwidth]{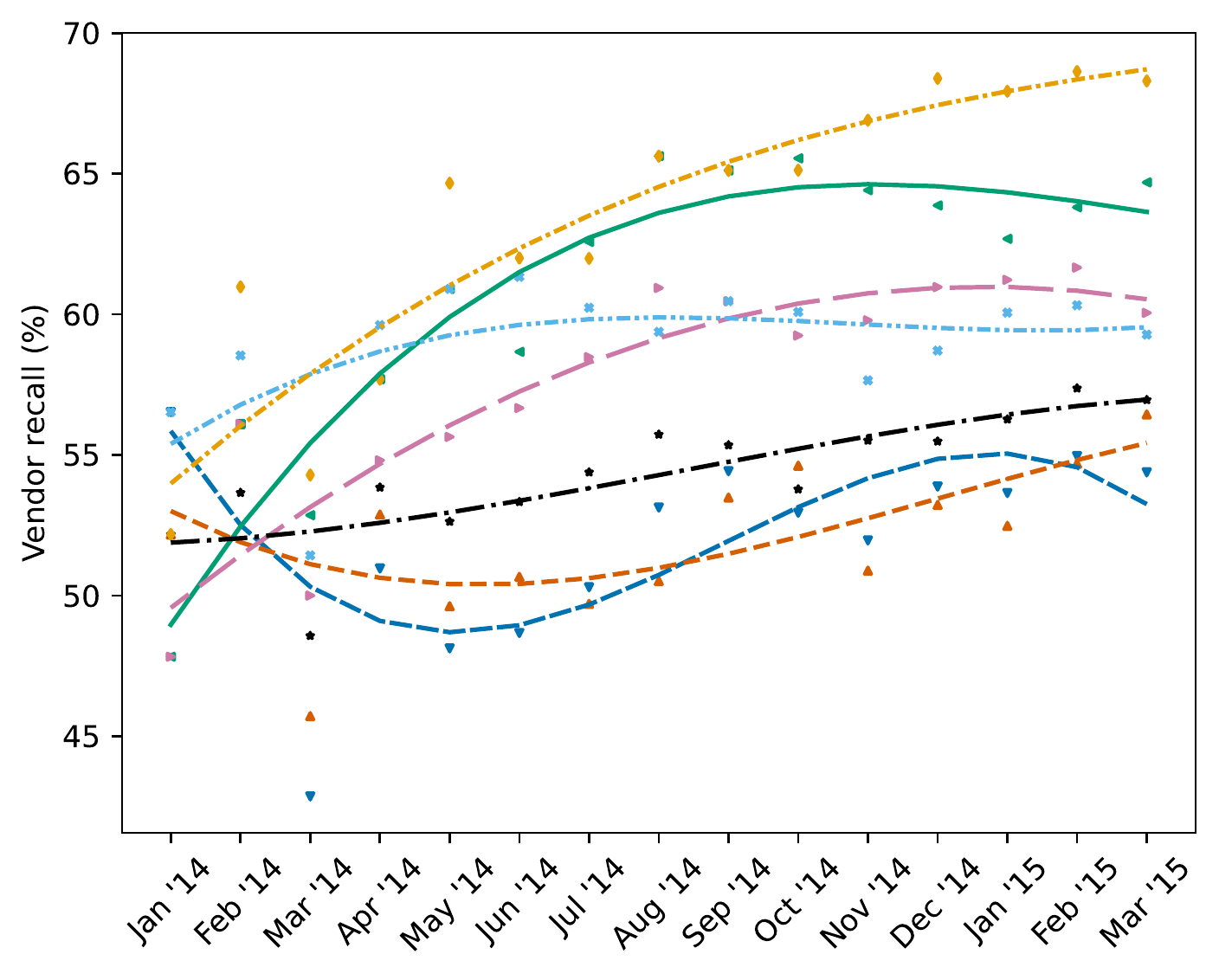}
      \caption{$\delta_t = 1$ month}
    \end{subfigure}
    ~
    \begin{subfigure}[b]{0.25\textheight}
      \centering
      \includegraphics[width=\textwidth]{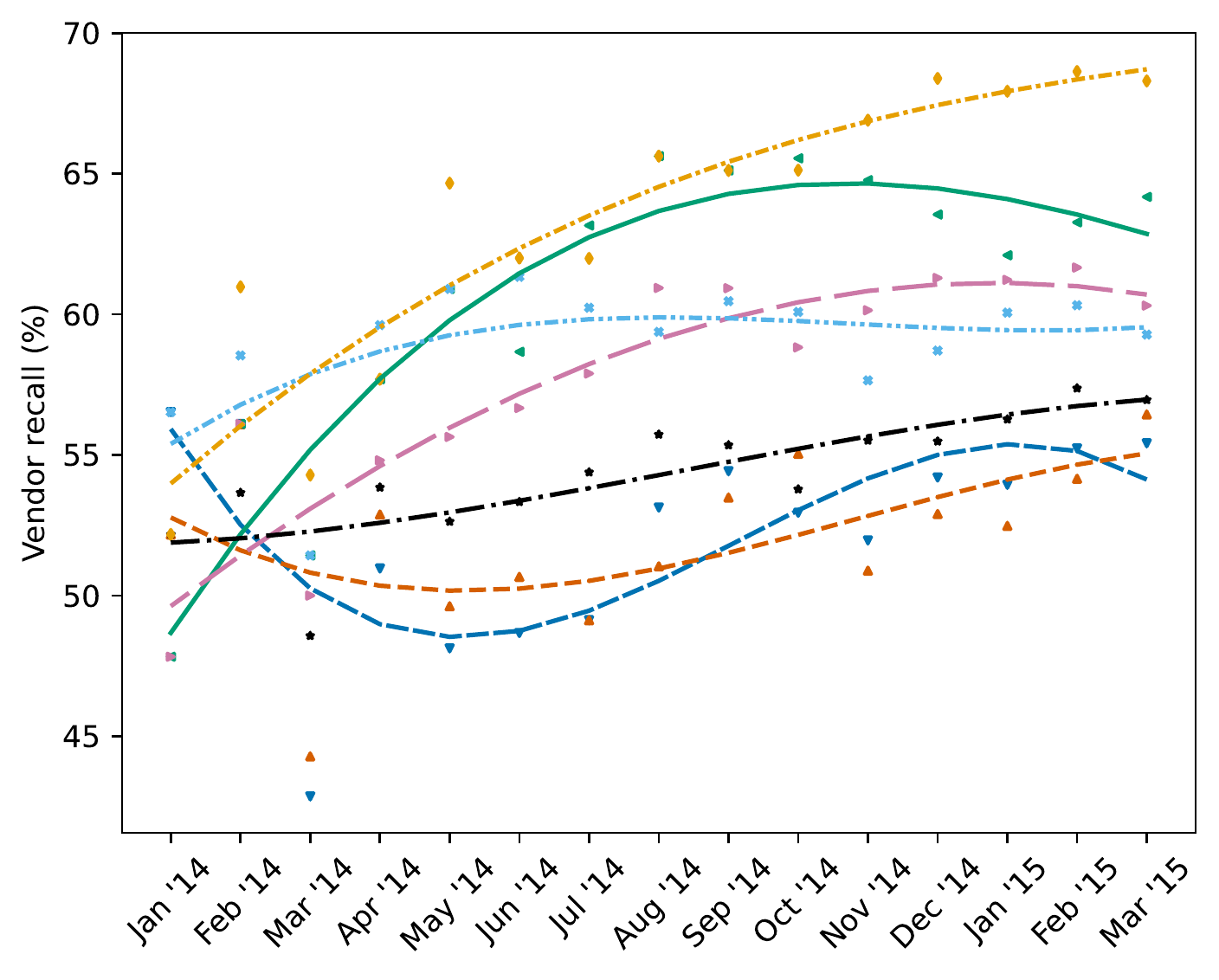}
      \caption{$\delta_t = 2$ months}
    \end{subfigure}
    ~
    \begin{subfigure}[b]{0.25\textheight}
      \centering
      \includegraphics[width=\textwidth]{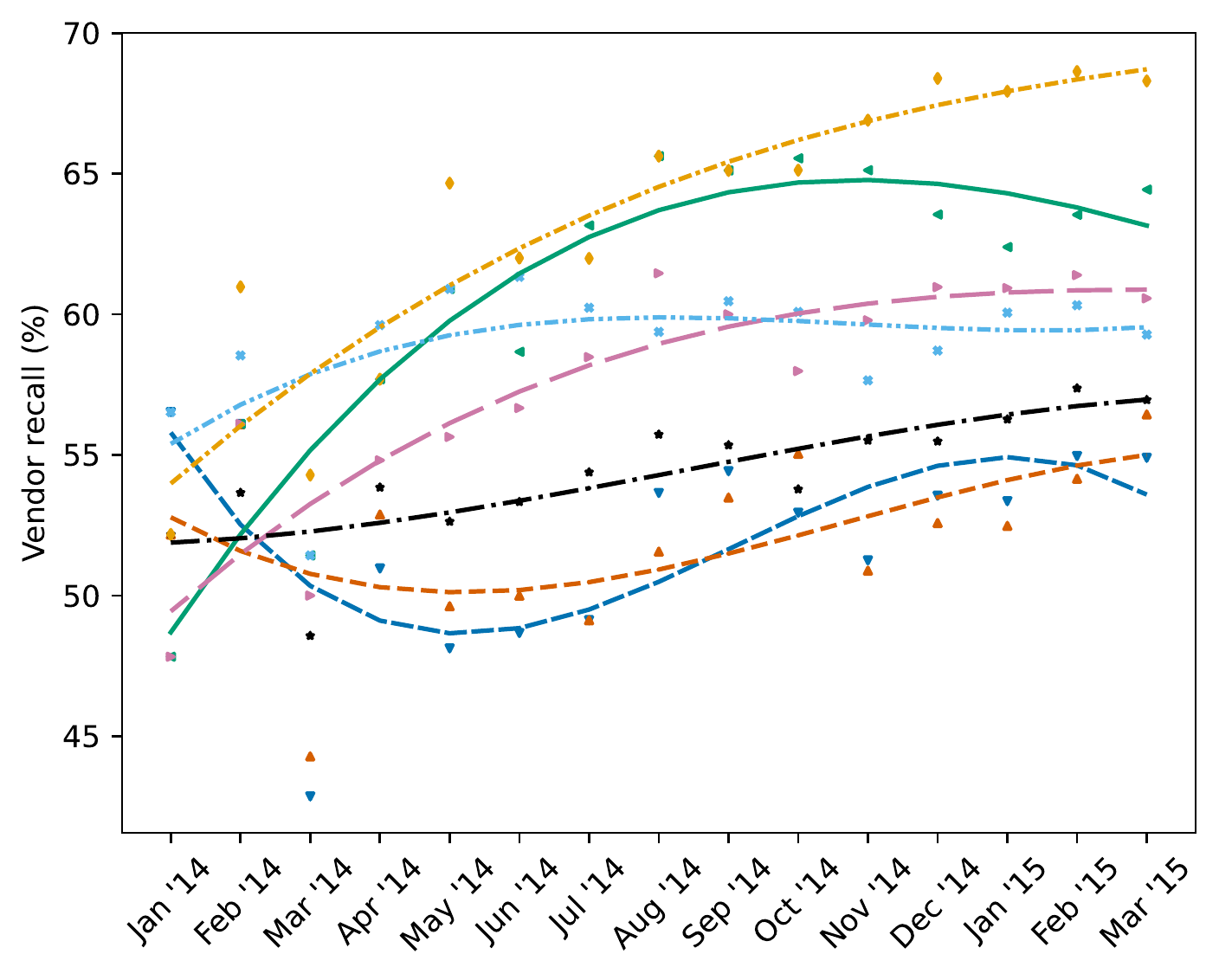}
      \caption{$\delta_t = 3$ months}
    \end{subfigure}
  \caption{Results for varying edge formation parameters. Monthly vendor recall of top vendor percentile (top 0-20\% vendors in terms of sales) among the top 20\% of all users based on the network measures and activity indicators for current success. Each plot displays monthly vendor recall for a different set of parameter values used for generating the network. Each row of plots varies a single parameter influencing edge formation, with the remaining parameters at their default values. The center column always corresponds exactly with the default parameters.} \label{fig:param1}
\end{figure}

\end{landscape}

\begin{landscape}
\begin{figure}[t]
  \centering
    \begin{subfigure}[b]{0.25\textheight}
      \centering
      \includegraphics[width=\textwidth]{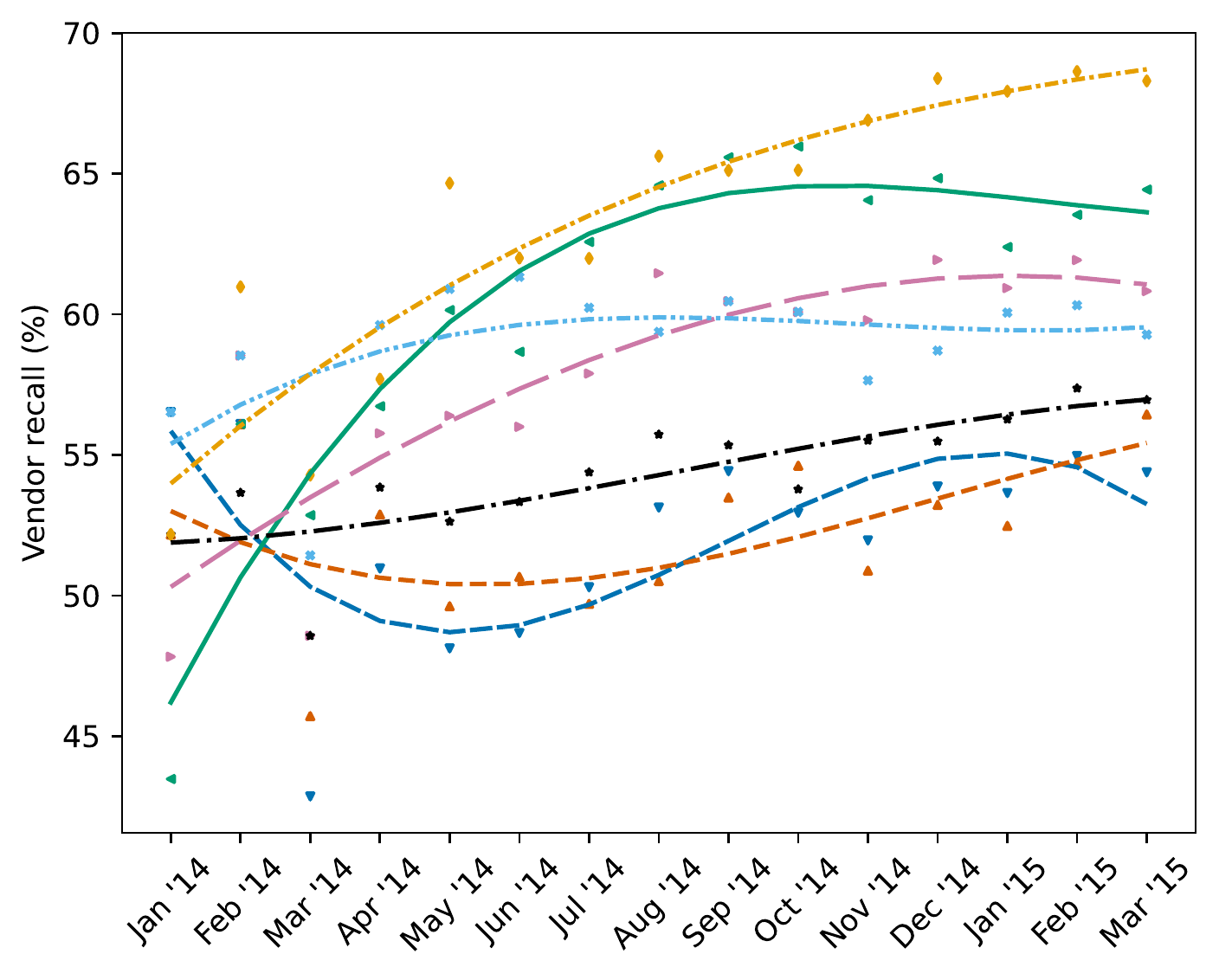}
      \caption{$\omega_{lower} = 0.05$}
    \end{subfigure}
    ~
    \begin{subfigure}[b]{0.25\textheight}
      \centering
      \includegraphics[width=\textwidth]{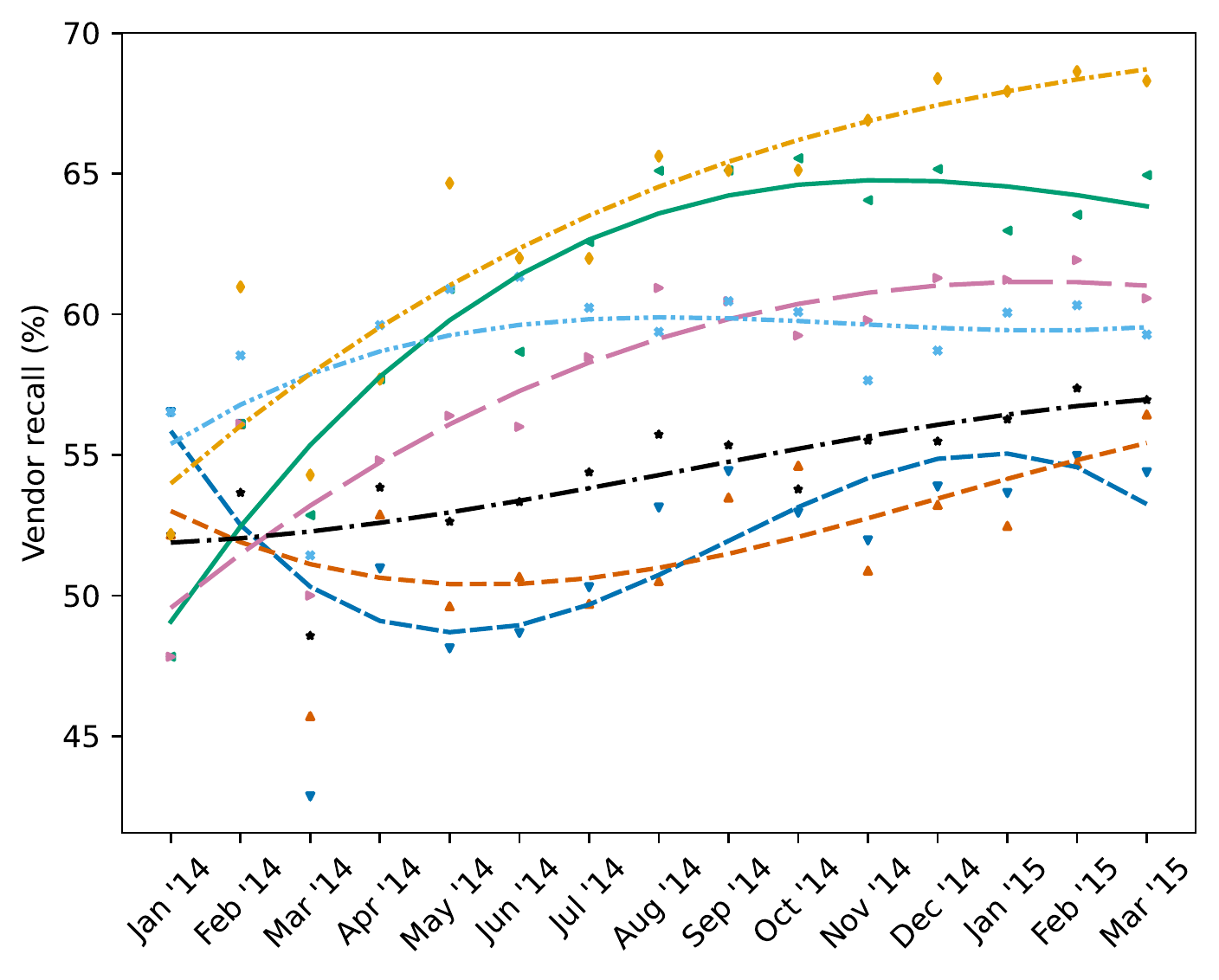}
      \caption{$\omega_{lower} = 0.1$}
    \end{subfigure}
    ~
    \begin{subfigure}[b]{0.25\textheight}
      \centering
      \includegraphics[width=\textwidth]{suplfig-default-recall-current-gr-5-notitle}
      \caption{$\omega_{lower} = 0.2$}
    \end{subfigure}
    ~
    \begin{subfigure}[b]{0.25\textheight}
      \centering
      \includegraphics[width=\textwidth]{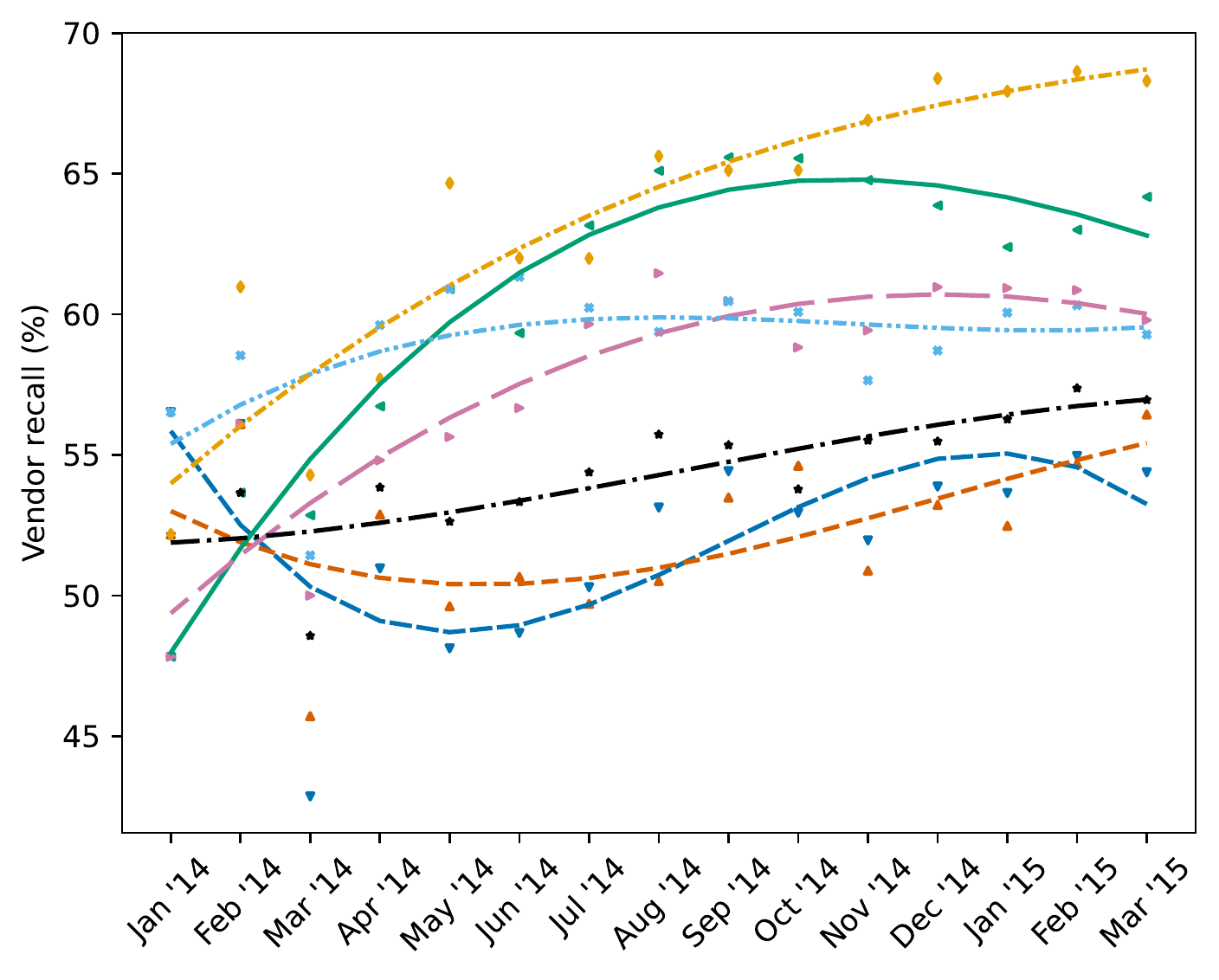}
      \caption{$\omega_{lower} = 0.3$}
    \end{subfigure}
    ~
    \begin{subfigure}[b]{0.25\textheight}
      \centering
      \includegraphics[width=\textwidth]{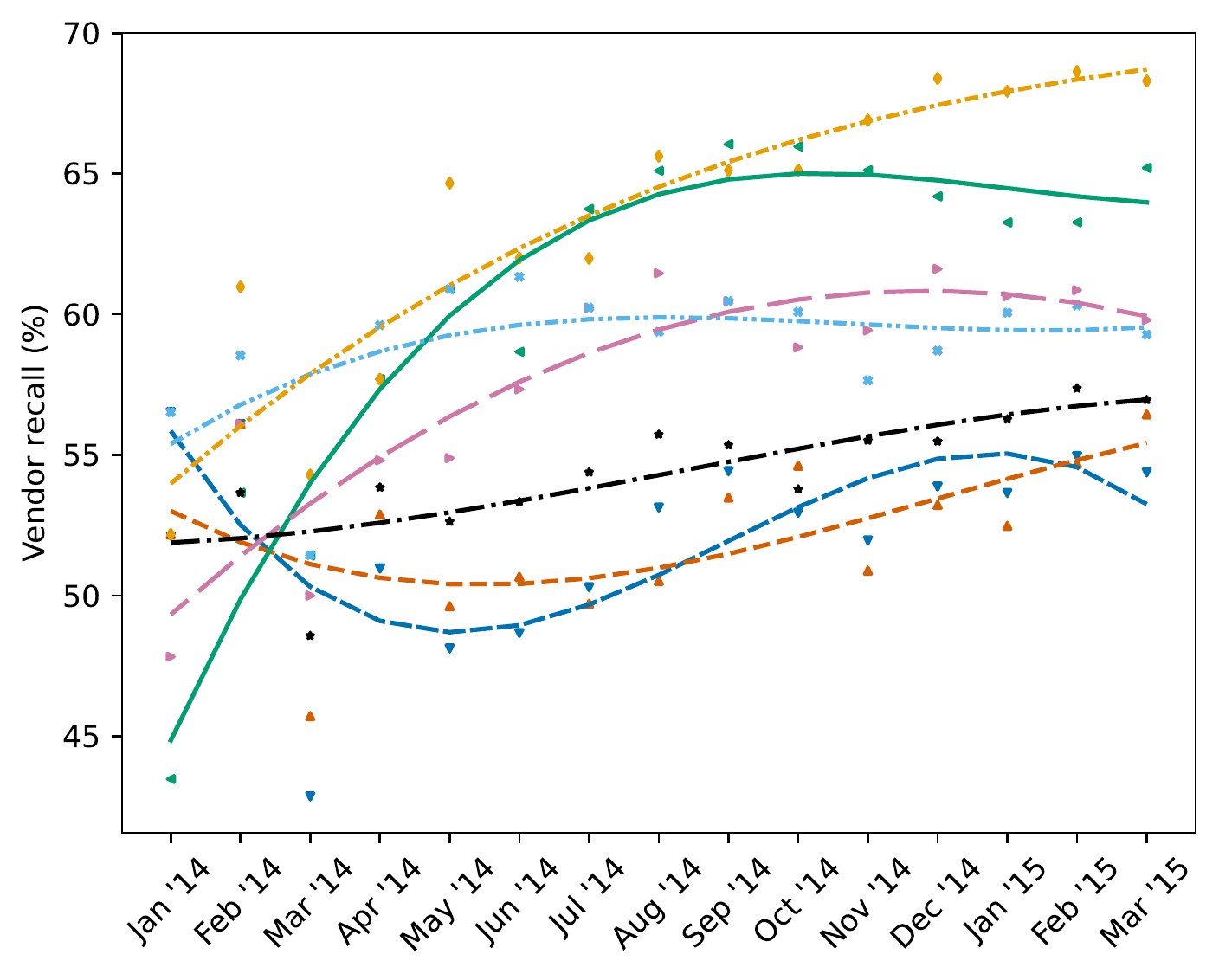}
      \caption{$\omega_{lower} = 0.4$}
    \end{subfigure}
    \begin{subfigure}[b]{0.25\textheight}
      \centering
      \includegraphics[width=\textwidth]{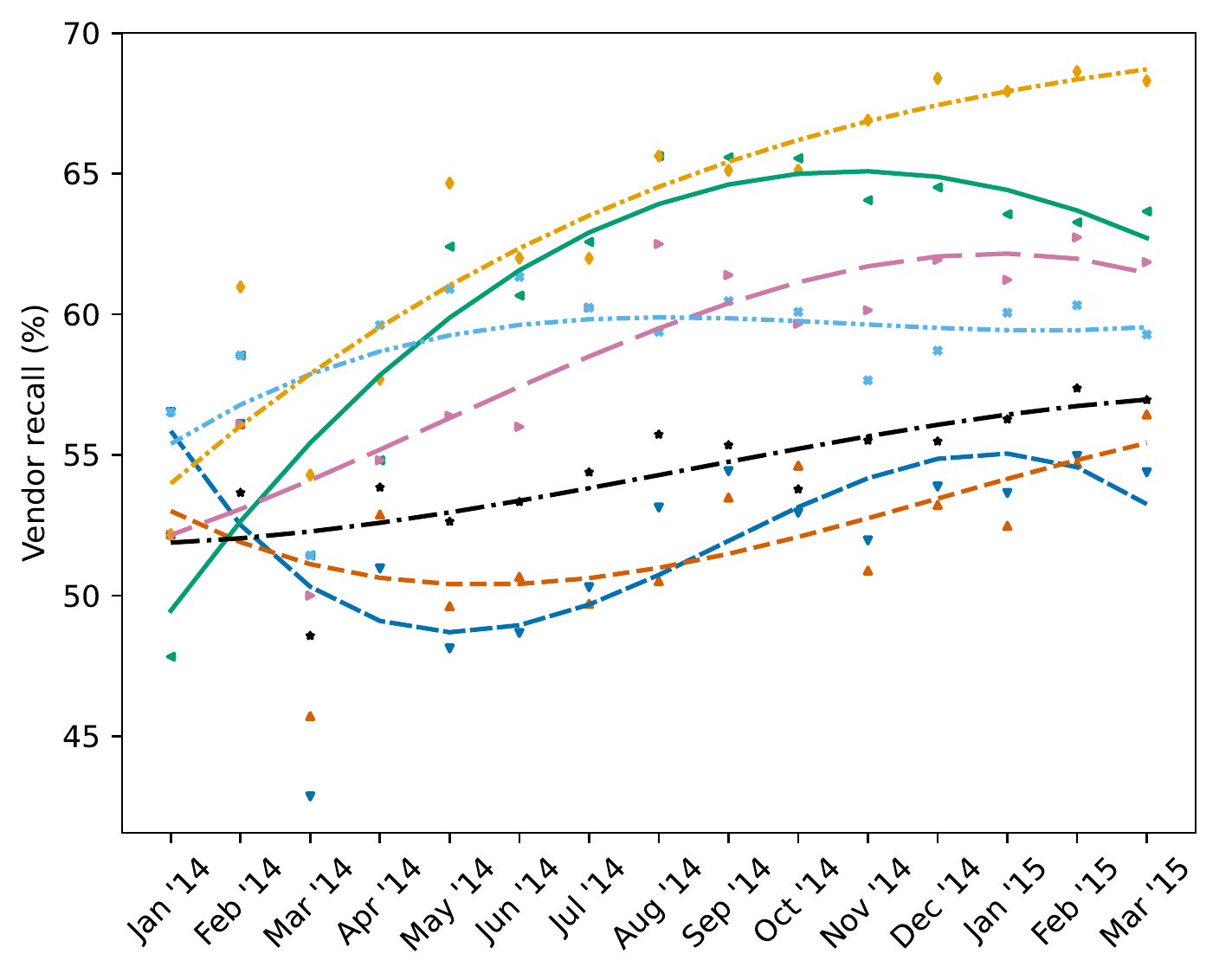}
      \caption{$t_{lim} = 2$ days}
    \end{subfigure}
    ~
    \begin{subfigure}[b]{0.25\textheight}
      \centering
      \includegraphics[width=\textwidth]{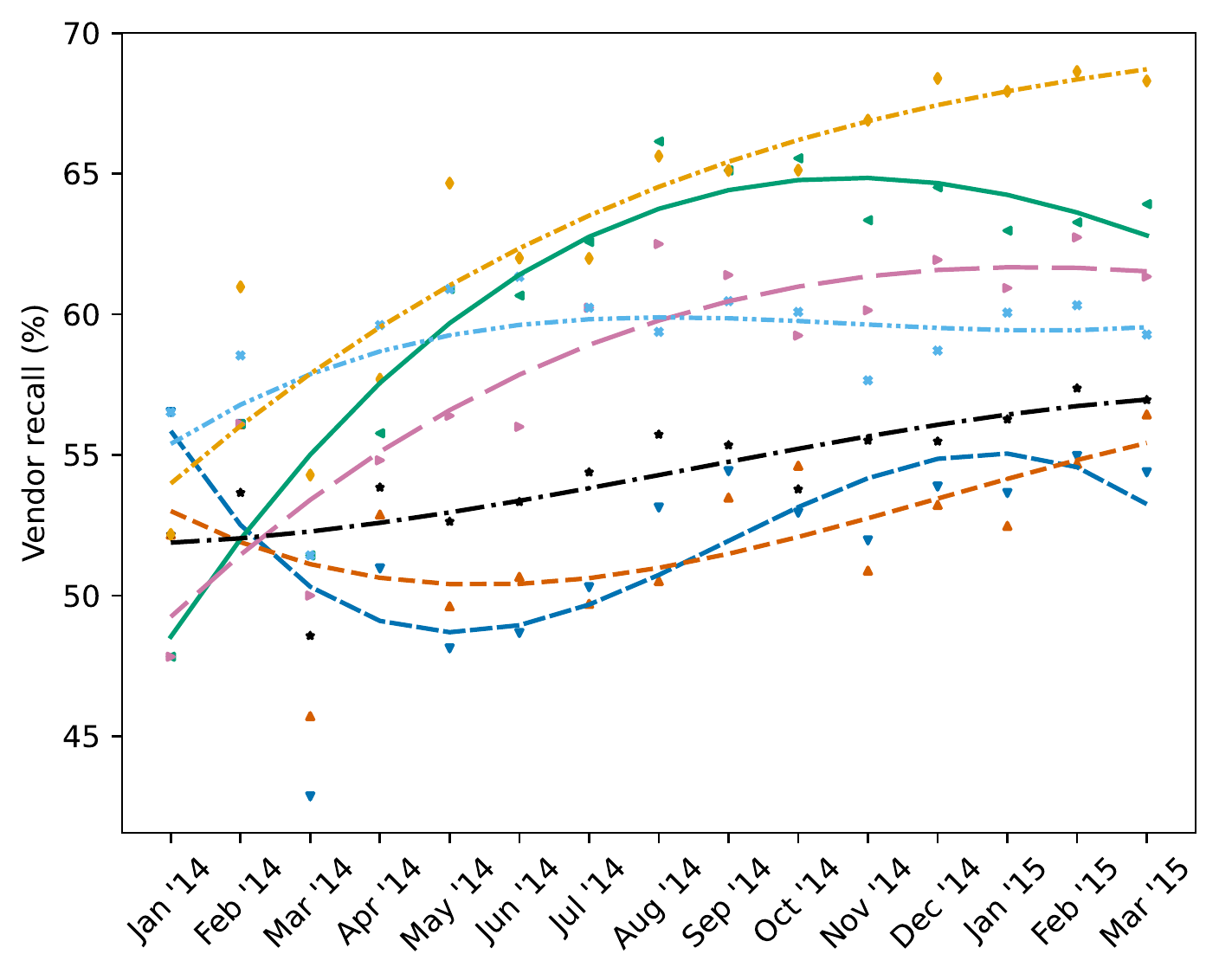}
      \caption{$t_{lim} = 3$ days}
    \end{subfigure}
    ~
    \begin{subfigure}[b]{0.25\textheight}
      \centering
      \includegraphics[width=\textwidth]{suplfig-default-recall-current-gr-5-nolegend-notitle}
      \caption{$t_{lim} = 7$ days}
    \end{subfigure}
    ~
    \begin{subfigure}[b]{0.25\textheight}
      \centering
      \includegraphics[width=\textwidth]{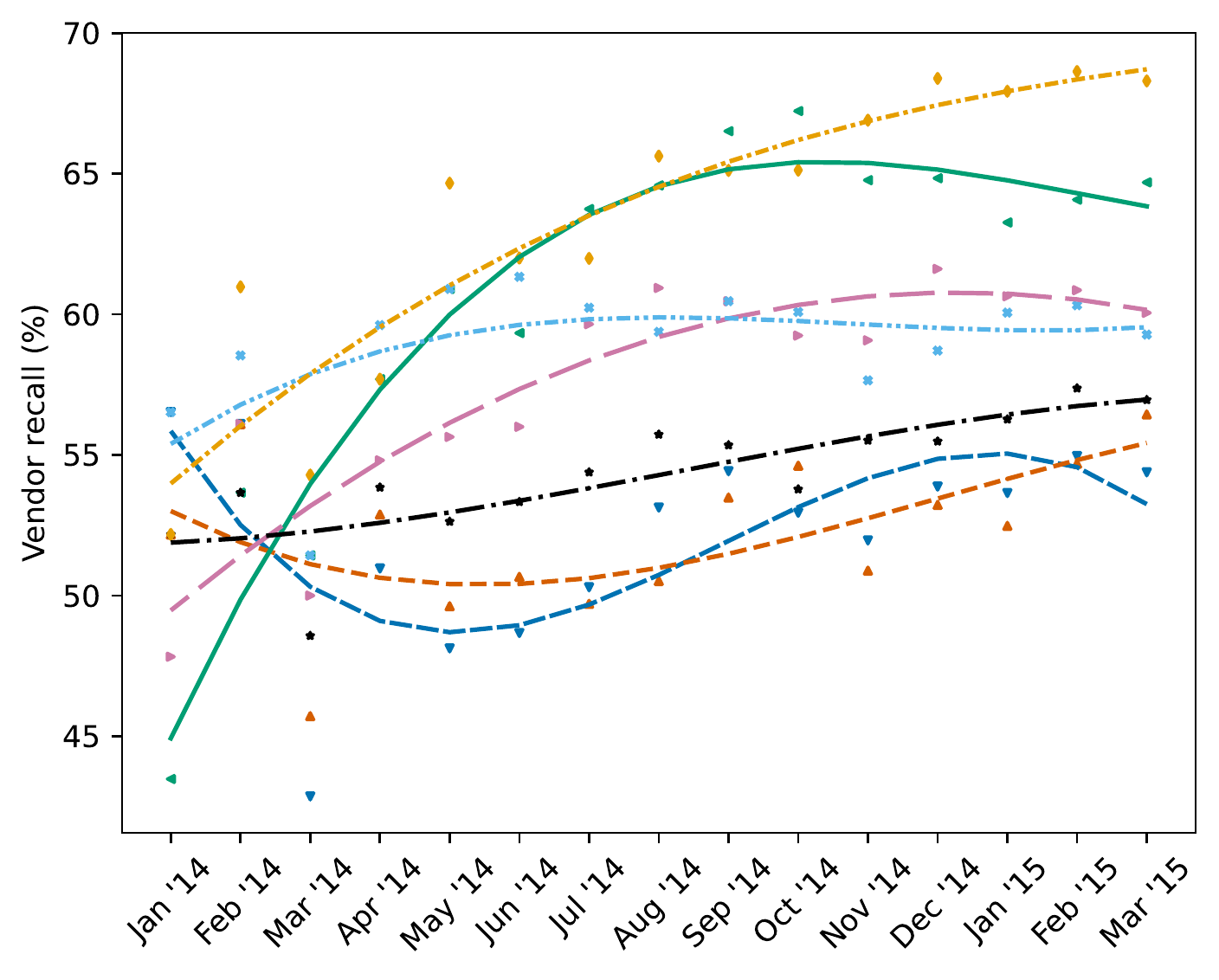}
      \caption{$t_{lim} = 14$ days}
    \end{subfigure}
    ~
    \begin{subfigure}[b]{0.25\textheight}
      \centering
      \includegraphics[width=\textwidth]{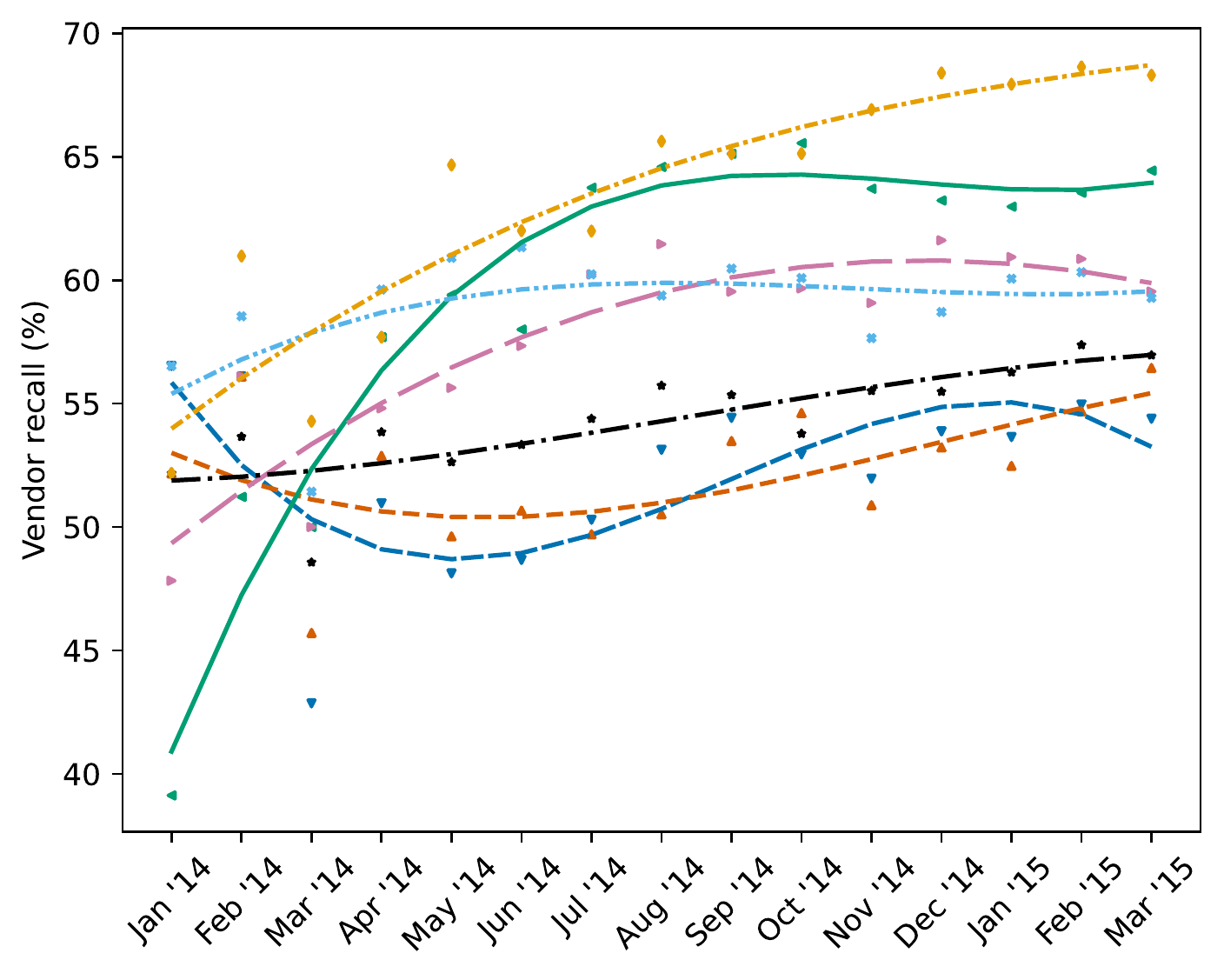}
      \caption{$t_{lim} = 1$ month}
    \end{subfigure}
    \begin{subfigure}[b]{0.25\textheight}
      \centering
      \includegraphics[width=\textwidth]{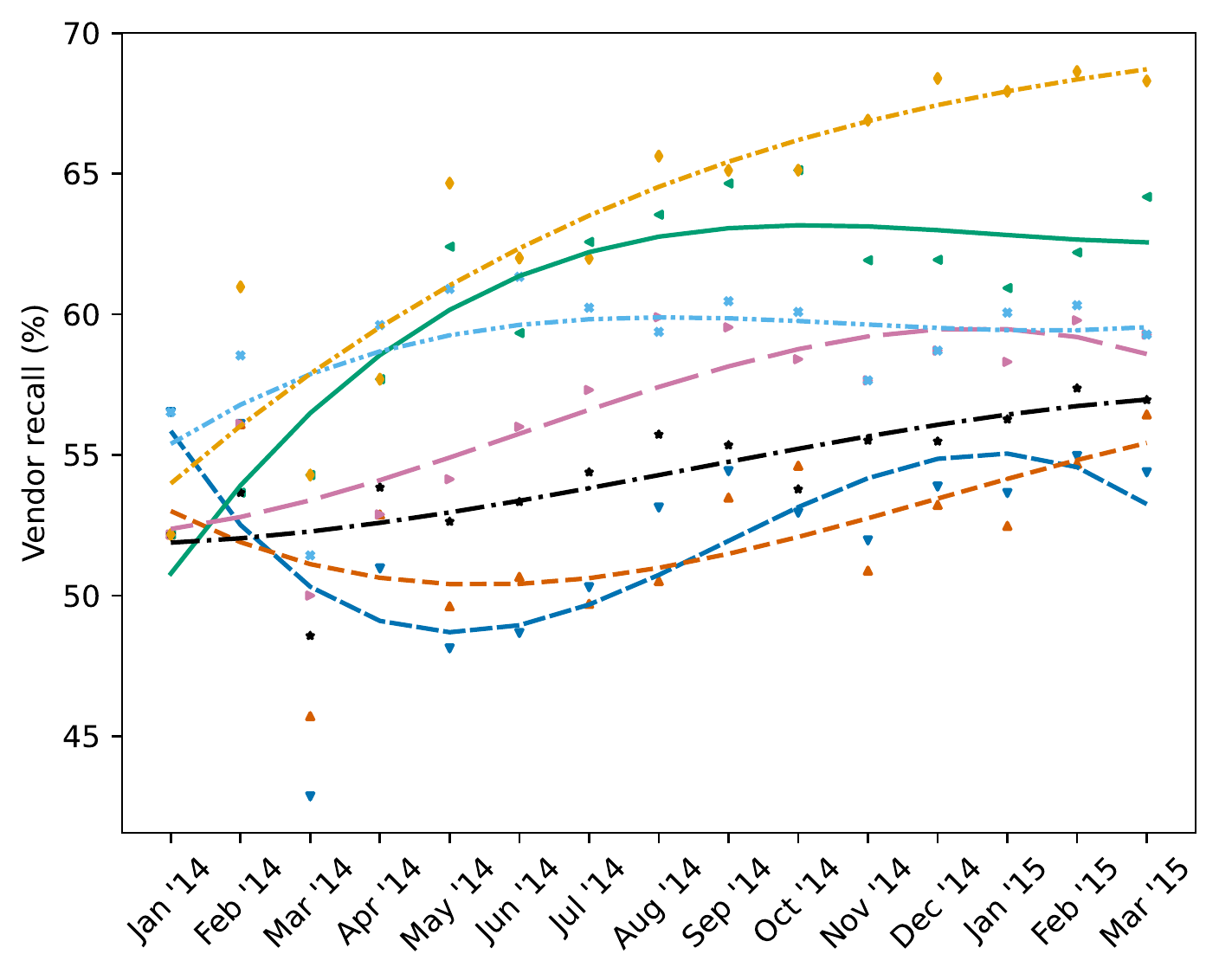}
      \caption{$\omega_{first} = 0.2$}
    \end{subfigure}
    ~
    \begin{subfigure}[b]{0.25\textheight}
      \centering
      \includegraphics[width=\textwidth]{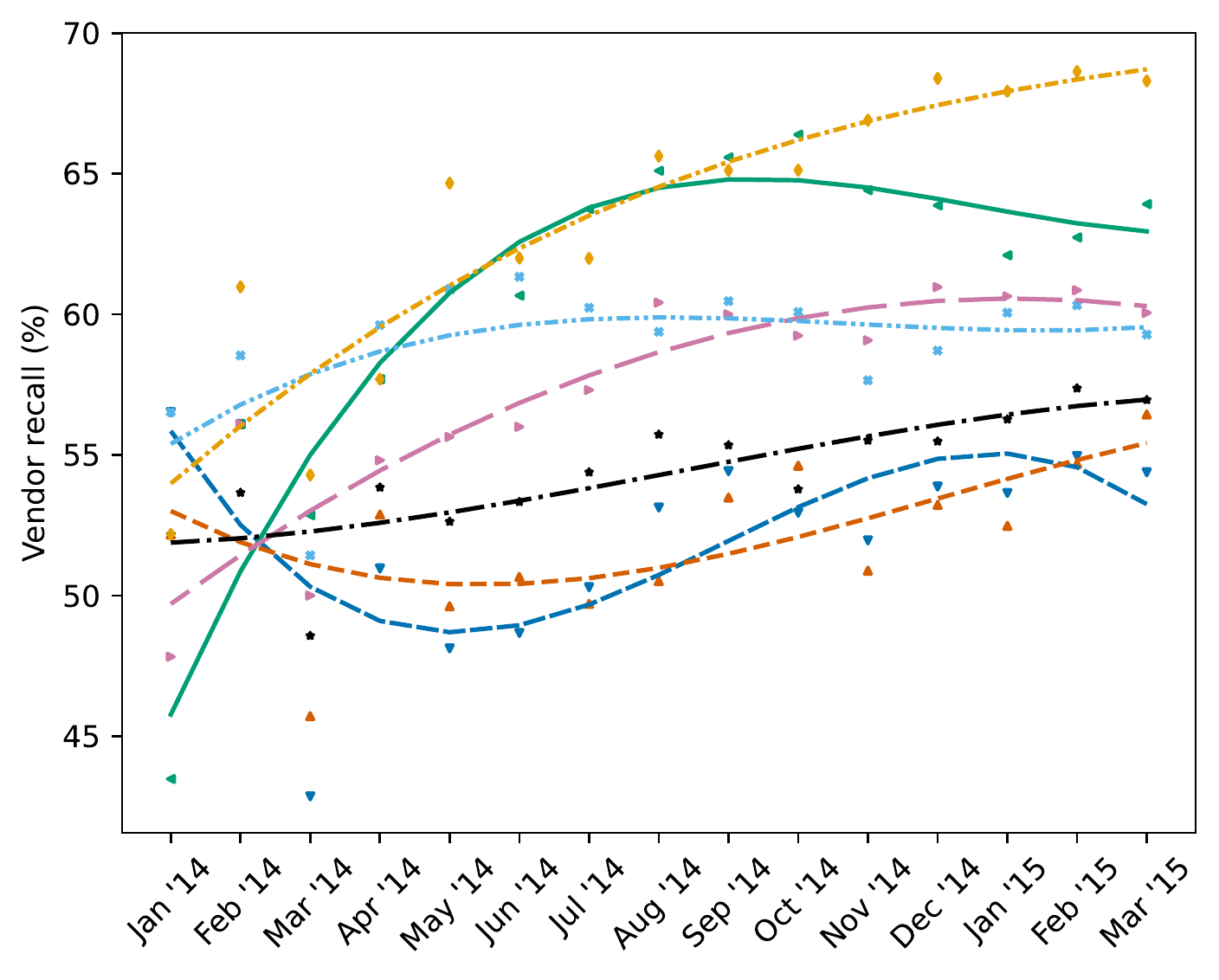}
      \caption{$\omega_{first} = 0.4$}
    \end{subfigure}
    ~
    \begin{subfigure}[b]{0.25\textheight}
      \centering
      \includegraphics[width=\textwidth]{suplfig-default-recall-current-gr-5-nolegend-notitle}
      \caption{$\omega_{first} = 0.5$}
    \end{subfigure}
    ~
    \begin{subfigure}[b]{0.25\textheight}
      \centering
      \includegraphics[width=\textwidth]{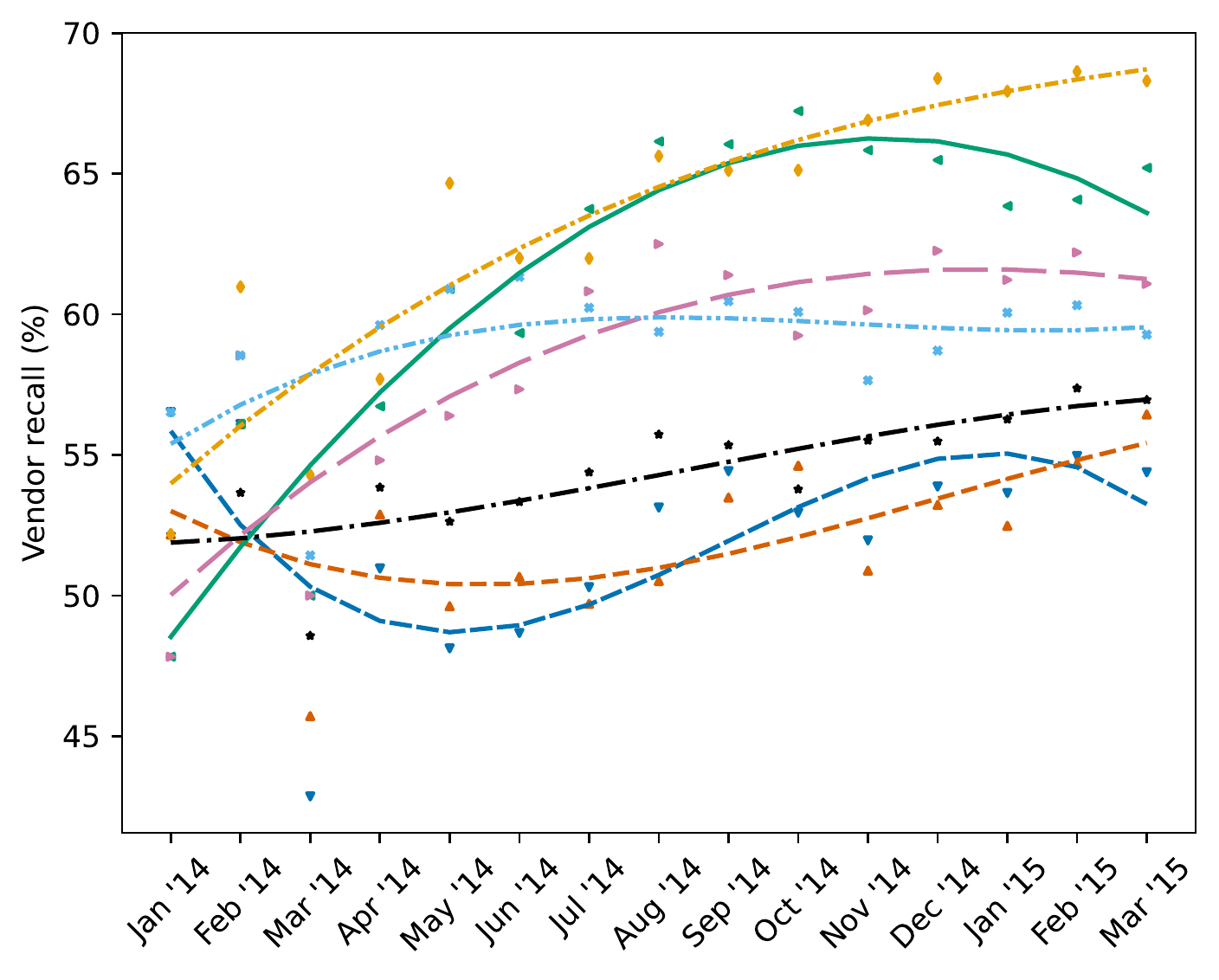}
      \caption{$\omega_{first} = 0.8$}
    \end{subfigure}
    ~
    \begin{subfigure}[b]{0.25\textheight}
      \centering
      \includegraphics[width=\textwidth]{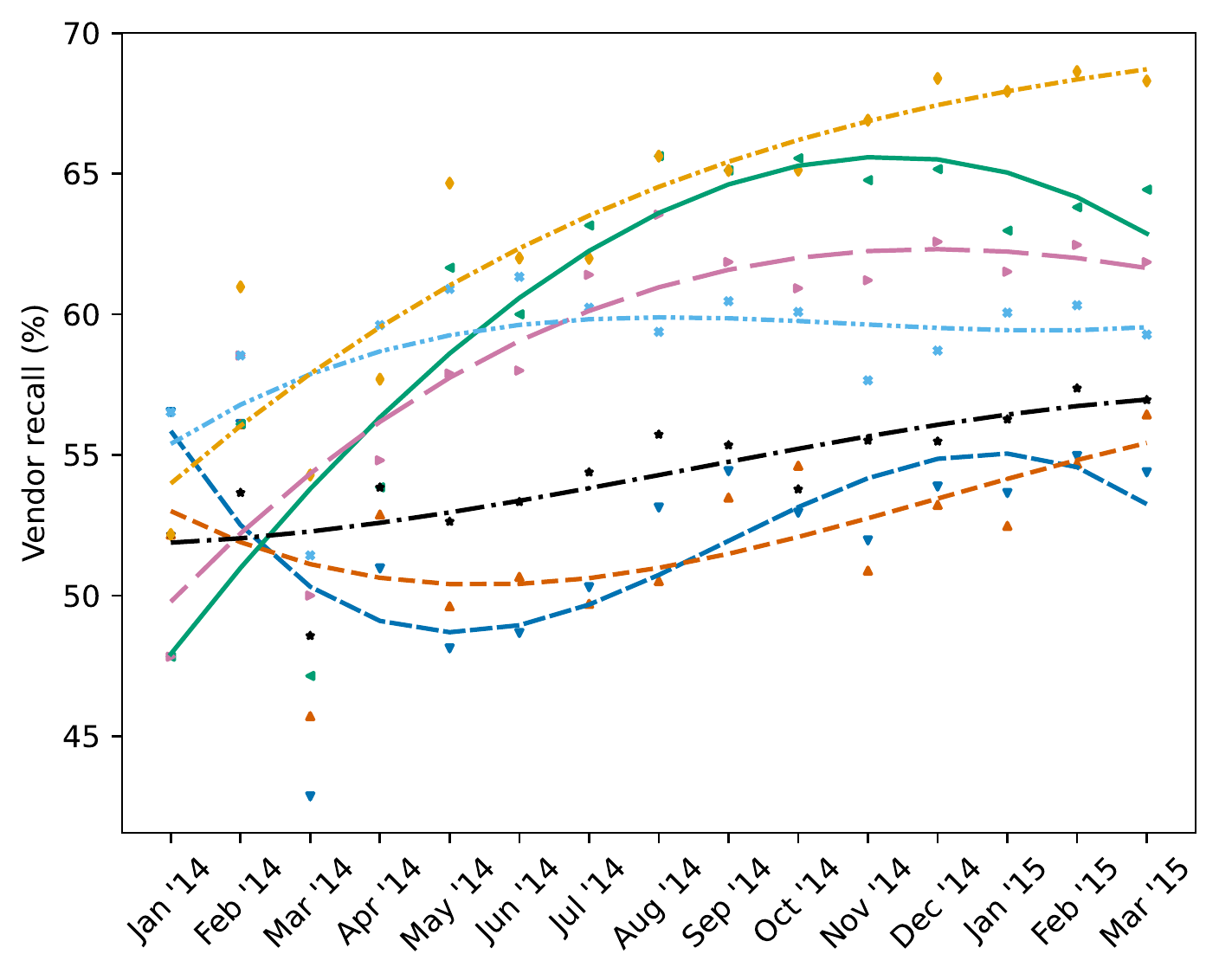}
      \caption{$\omega_{first} = 1$}
    \end{subfigure}
  \caption{Results for varying edge weighting parameters. Monthly vendor recall of top vendor percentile (top 0-20\% vendors in terms of sales) among the top 20\% of all users based on the network measures and activity indicators for current success. Each plot displays monthly vendor recall for a different set of parameter values used for generating the network. Each row of plots varies a single parameter influencing edge weighting, with the remaining parameters at their default values. The center column always corresponds exactly with the default parameters.} \label{fig:param2}
\end{figure}

\end{landscape}

The results presented in the main paper rely on the network extracted based on a single set of parameter values.
Here, we explore the robustness of those results for different parameter values (see Methods for a description of these parameters).
We do so by changing the value of a single parameter at a time while maintaining the same values for all other parameters.
Figures~\ref{fig:param1} and~\ref{fig:param2} shows vendor recall trends (similar to main paper Figure 2a
) for each set of parameter values.
Each row of plots in these figures shows the results for changing one specific parameter, with the middle column corresponding to the `default' values.
Note that since we are only changing the network formation process, the vendor recall results for the forum activity indicators do not change.
As such, they provide a visual aid in analysing the changes in network measure performance.

Figure~\ref{fig:param1} shows vendor recall trends for varying values for the parameters influencing the formation of edges.
First, $\delta_o$ determines how many posts two posts may be apart at most, to still form an edge connecting the users who placed them.
Figures~\ref{fig:param1}a--e show that the in-degree performs better at lower values of $\delta_o$.
Similarly, bidirectional harmonic closeness centrality performs slightly better at lower values.
However, even then neither comes close to outperforming betweenness centrality.
Some variation in performance can be observed for betweenness centrality as well.
Slightly better performance, than default, are observed for $\delta_o = 5$ and $20$.
As such, there is no indication that specifically using a smaller or larger value of $\delta_o$ would improve performance.
Second, $\delta_t$ determines how much time between two posts may have elapsed at most, to still form an edge connecting the users who placed them.
Figures~\ref{fig:param1}f--j show that changing $\delta_t$ between 7 days and 3 months hardly affects the vendor recall at all.
Thus, our findings in the main paper can be considered robust for both $\delta_o$ and $\delta_t$.

Figure~\ref{fig:param2} shows vendor recall trends for varying values for the parameters influencing the weight of edges.
First, $\omega_{lower}$ and $t_{lim}$ determine the scope and rate of decay of the exponential weighting function applied to ``regular'' edges, i.e., the implied social ties.
$\omega_{lower}$ sets a minimum weight, and $t_{lim}$ determines after how much time (between the placement of posts) this minimum weight is reached.
For both of these parameters we can see no meaningful change in the vendor recall performance for either lower or higher values (see Figures~\ref{fig:param2}a--j).
As such, our findings in the main paper can be considered robust for both $\omega_{lower}$ and $t_{lim}$.
Second, $\omega_{first}$ sets the standard weight for all edges formed by linking to the initial post of topics.
Figures~\ref{fig:param2}k--o show slight improvement of vendor recall for weighted directed PageRank as this parameter gets closer to one, the maximum weight.
However, it remains the case that at no point does it outperform betweenness centrality.
We also observe a slight variation in the performance of betweenness centrality.
Specifically, we see movement in when it performs well.
It appears that betweenness centrality performs slightly better in the early months for lower values of $\omega_{first}$, whereas it performs slightly better in the later months for higher values.
As such, it may be advantageous to choose the $\omega_{first}$ value based on the age of the cryptomarket one is dealing with.
Regardless, the change in performance is small and we can conclude that our findings in the main paper can also be considered robust for $\omega_{first}$.

\begin{landscape}
\begin{figure}[t]
  \centering
    \begin{subfigure}[b]{0.25\textheight}
      \centering
      \includegraphics[width=\textwidth]{suplfig-default-recall-current-gr-5-notitle}
      \caption{default}
    \end{subfigure}
    ~
    \begin{subfigure}[b]{0.25\textheight}
      \centering
      \includegraphics[width=\textwidth]{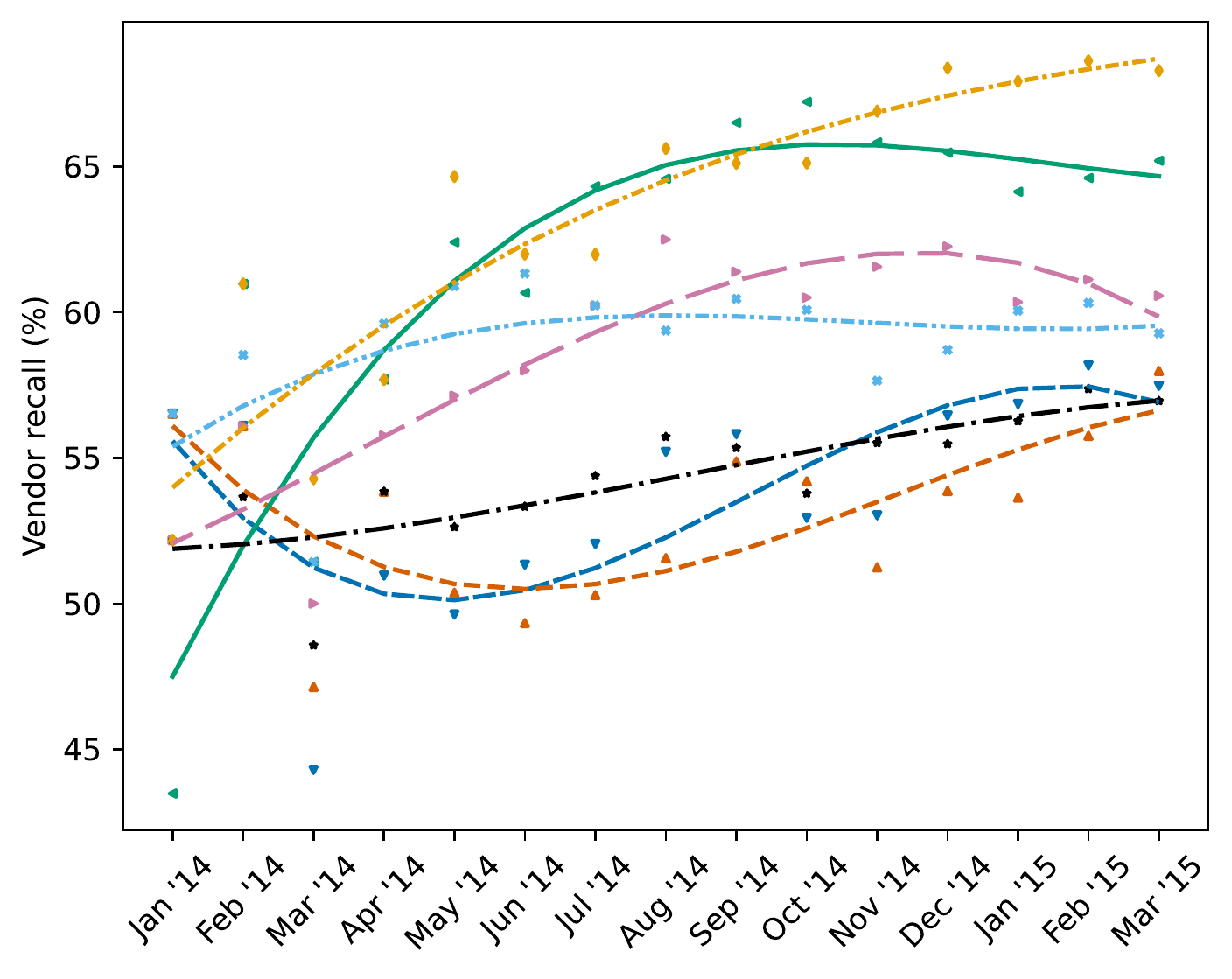}
      \caption{$\delta_o = 5$, $\omega_{lower} = 0.4$}
    \end{subfigure}
    ~
    \begin{subfigure}[b]{0.25\textheight}
      \centering
      \includegraphics[width=\textwidth]{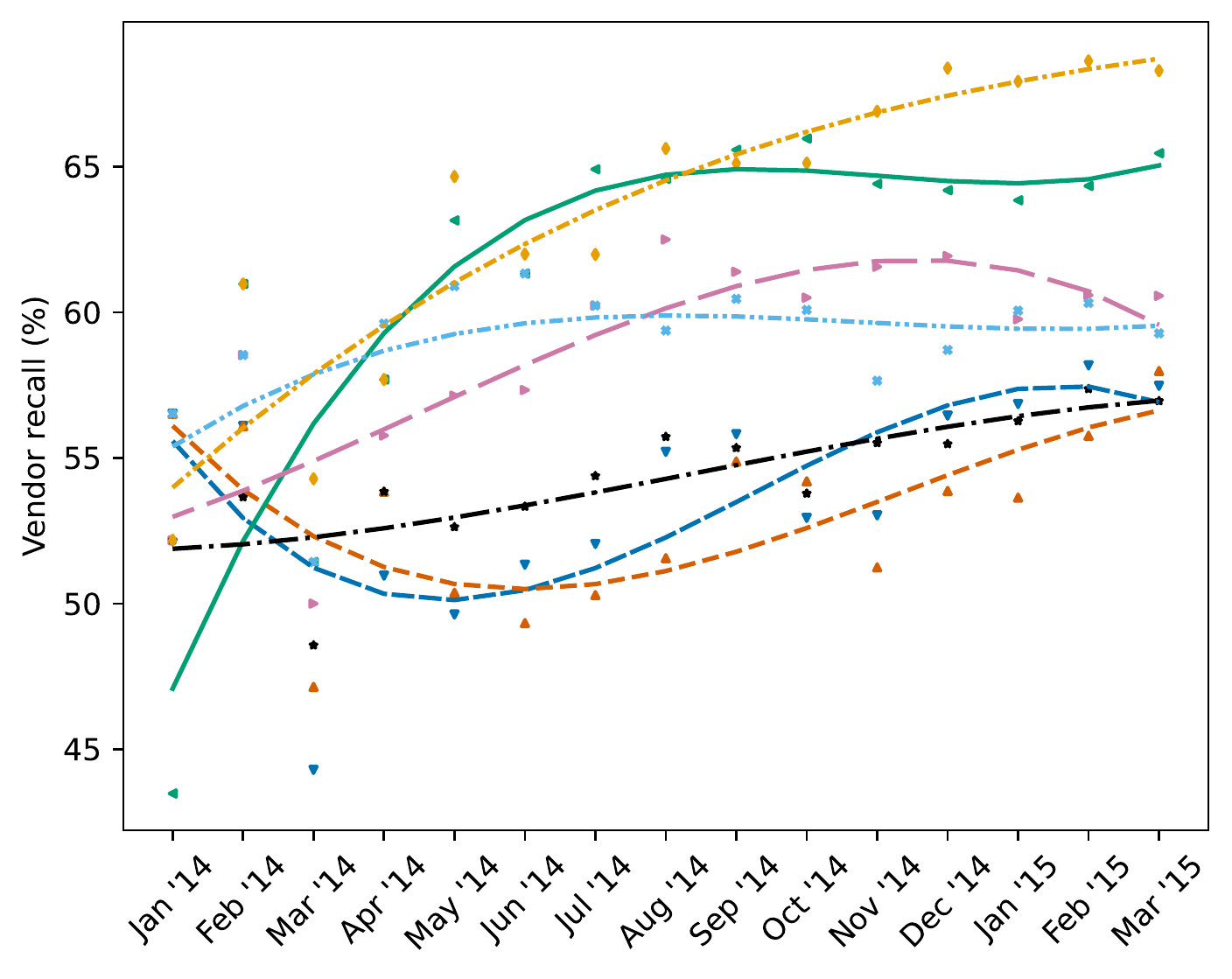}
      \caption{$\delta_o = 5$, $t_{lim} = 14$ days}
    \end{subfigure}
    ~
    \begin{subfigure}[b]{0.25\textheight}
      \centering
      \includegraphics[width=\textwidth]{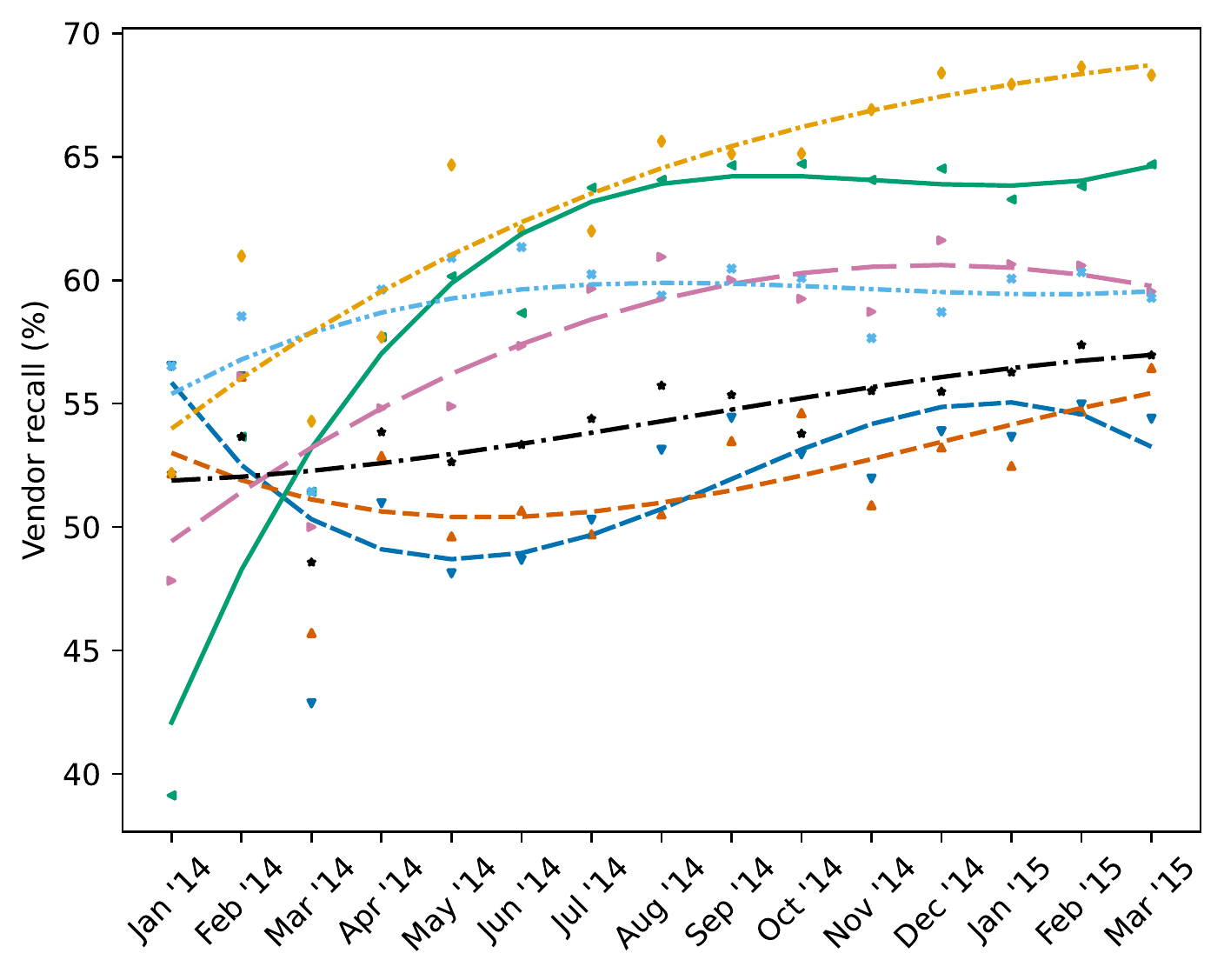}
      \caption{$\omega_{lower} = 0.4$, $t_{lim} = 14$ days}
    \end{subfigure}
    ~
    \begin{subfigure}[b]{0.25\textheight}
      \centering
      \includegraphics[width=\textwidth]{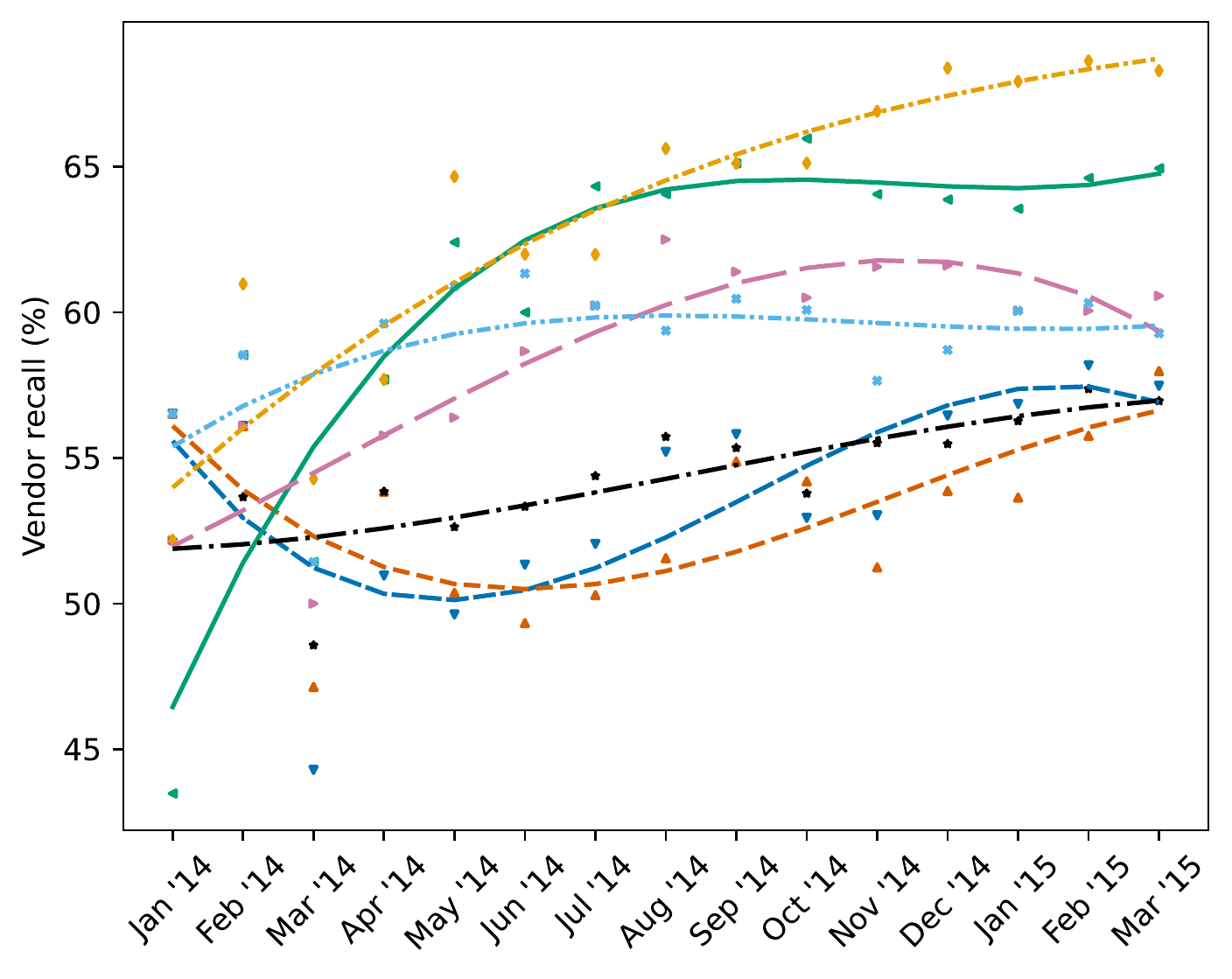}
      \caption{$\delta_o = 5$, $\omega_{lower} = 0.4$, $t_{lim} = 14$ days}
    \end{subfigure}
    \begin{subfigure}[b]{0.25\textheight}
      \centering
      \includegraphics[width=\textwidth]{suplfig-default-recall-current-gr-5-notitle}
      \caption{default}
    \end{subfigure}
    ~
    \begin{subfigure}[b]{0.25\textheight}
      \centering
      \includegraphics[width=\textwidth]{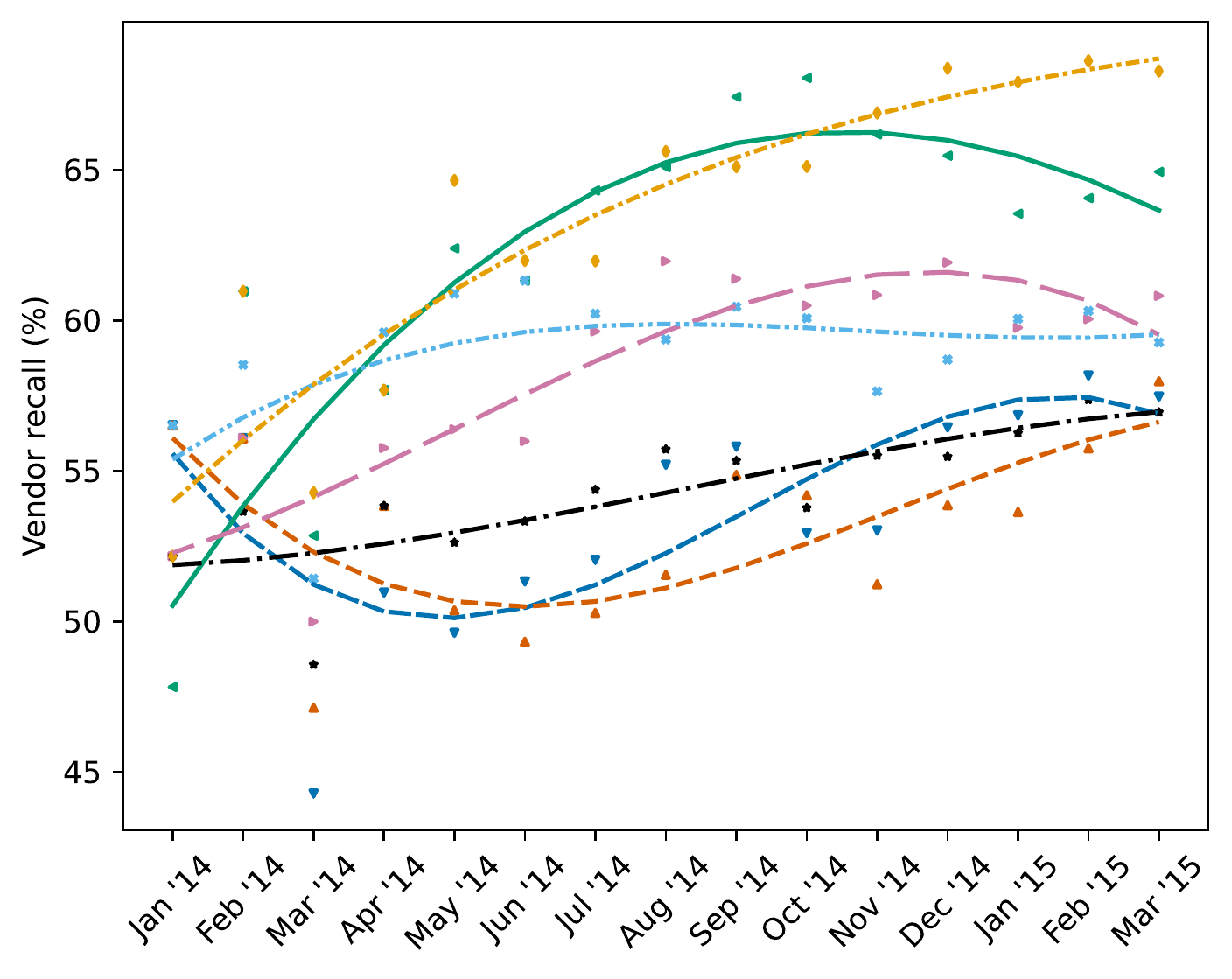}
      \caption{$\delta_o = 5$, $\omega_{first} = 0.4$}
    \end{subfigure}
    ~
    \begin{subfigure}[b]{0.25\textheight}
      \centering
      \includegraphics[width=\textwidth]{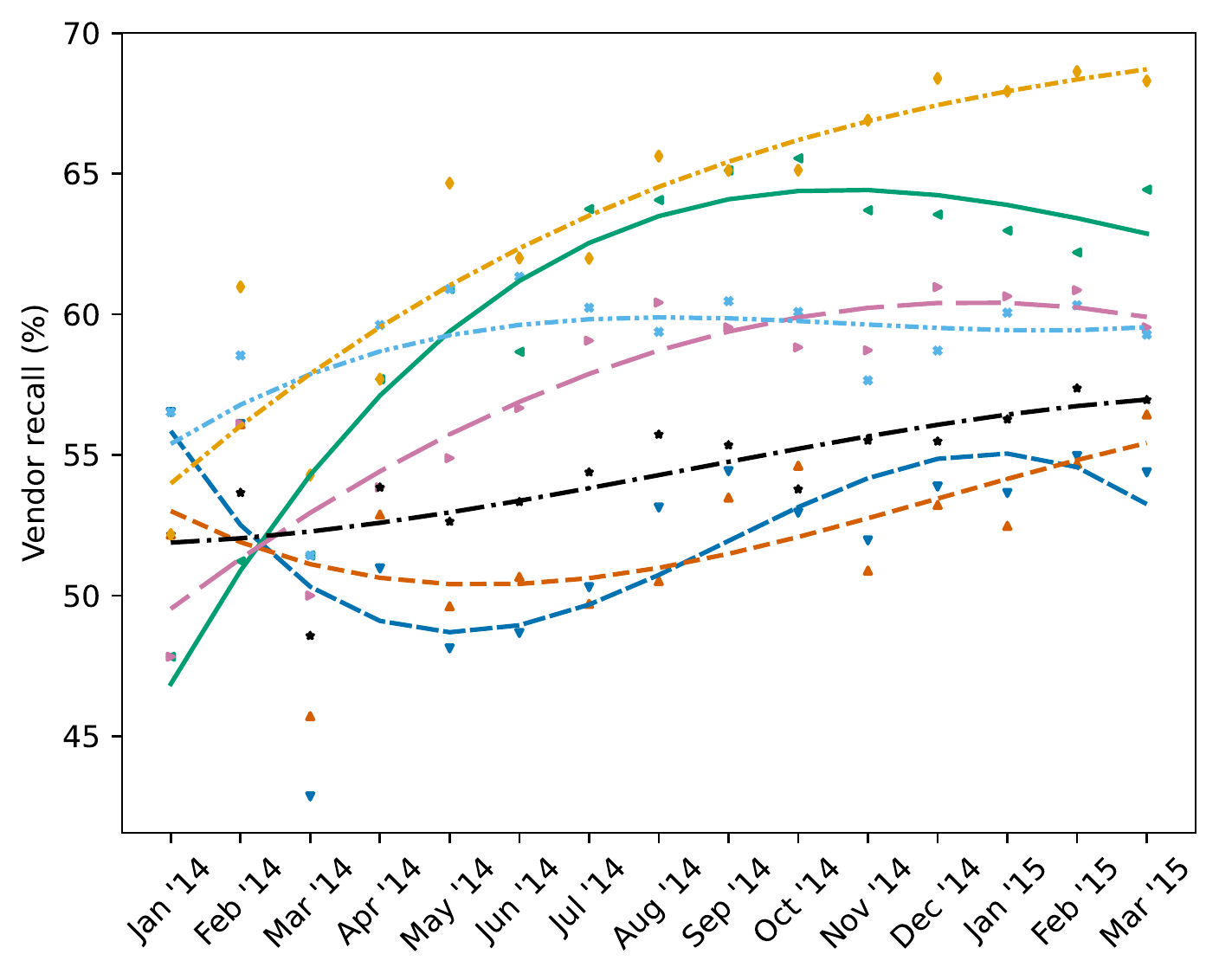}
      \caption{$\omega_{lower} = 0.4$, $\omega_{first} = 0.4$}
    \end{subfigure}
    ~
    \begin{subfigure}[b]{0.25\textheight}
      \centering
      \includegraphics[width=\textwidth]{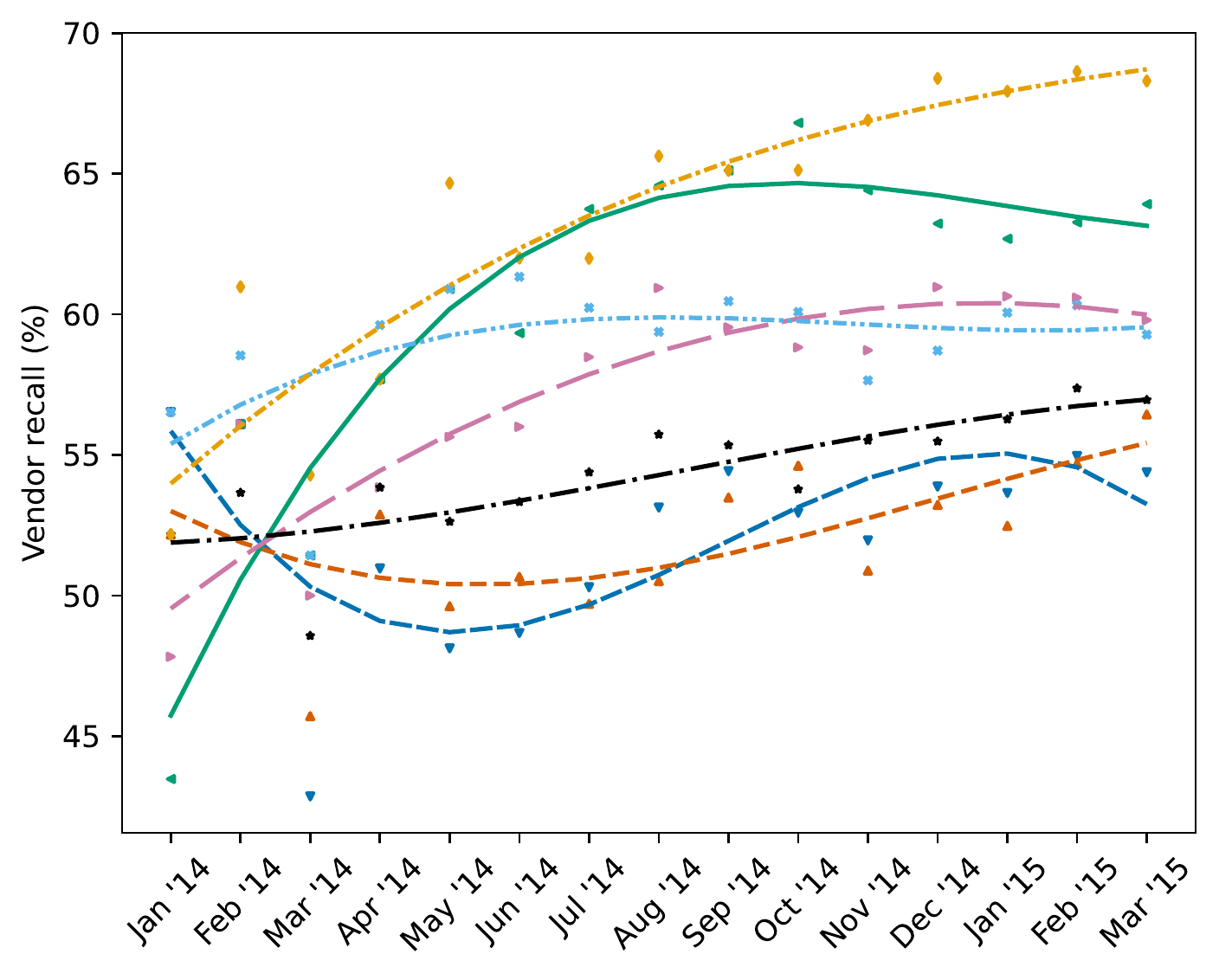}
      \caption{$t_{lim} = 14$ days, $\omega_{first} = 0.4$}
    \end{subfigure}
    ~
    \begin{subfigure}[b]{0.25\textheight}
      \centering
      \includegraphics[width=\textwidth]{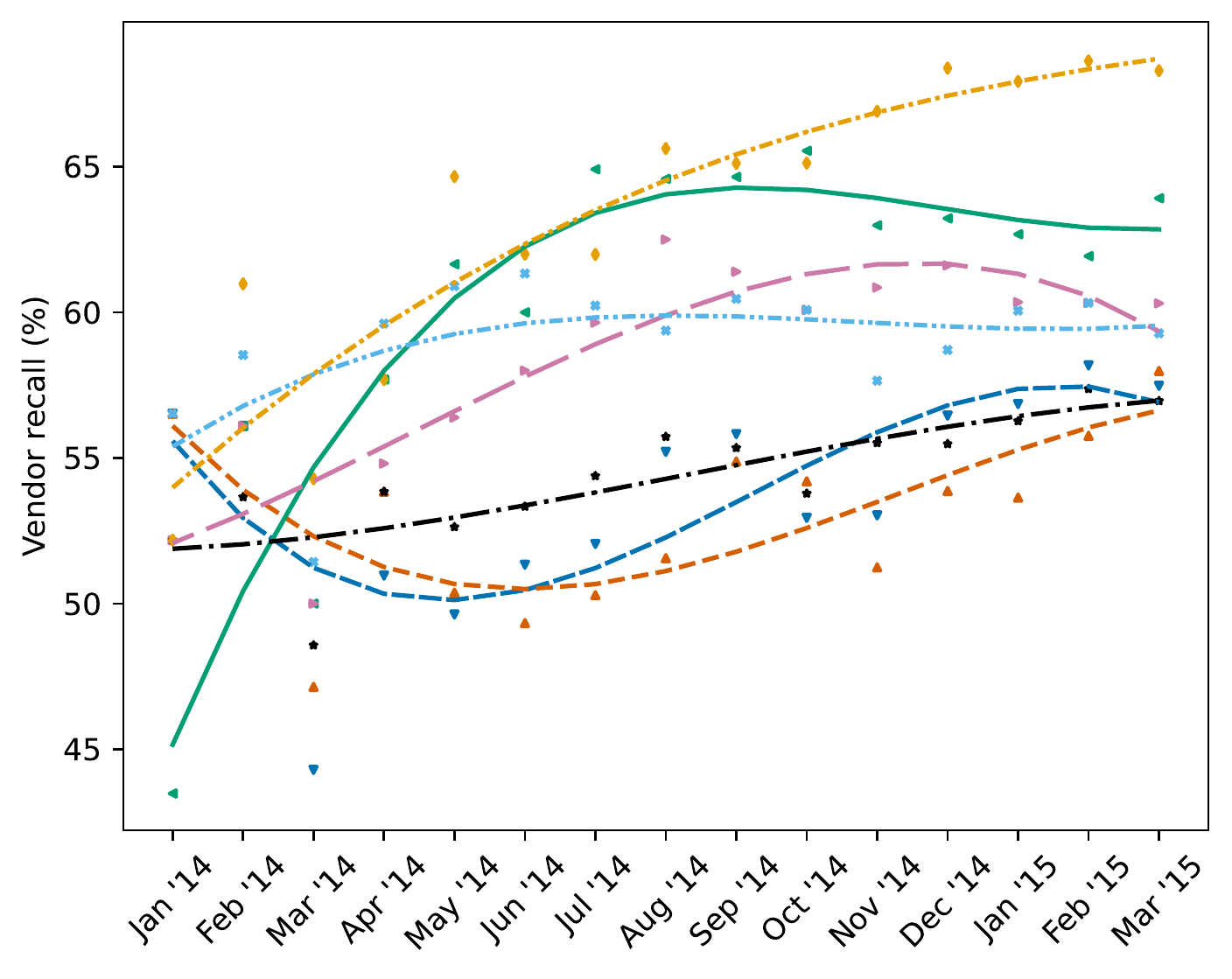}
      \caption{$\delta_o = 5$, $\omega_{lower} = 0.4$, $t_{lim} = 14$ days, $\omega_{first} = 0.4$}
    \end{subfigure}
    \begin{subfigure}[b]{0.25\textheight}
      \centering
      \includegraphics[width=\textwidth]{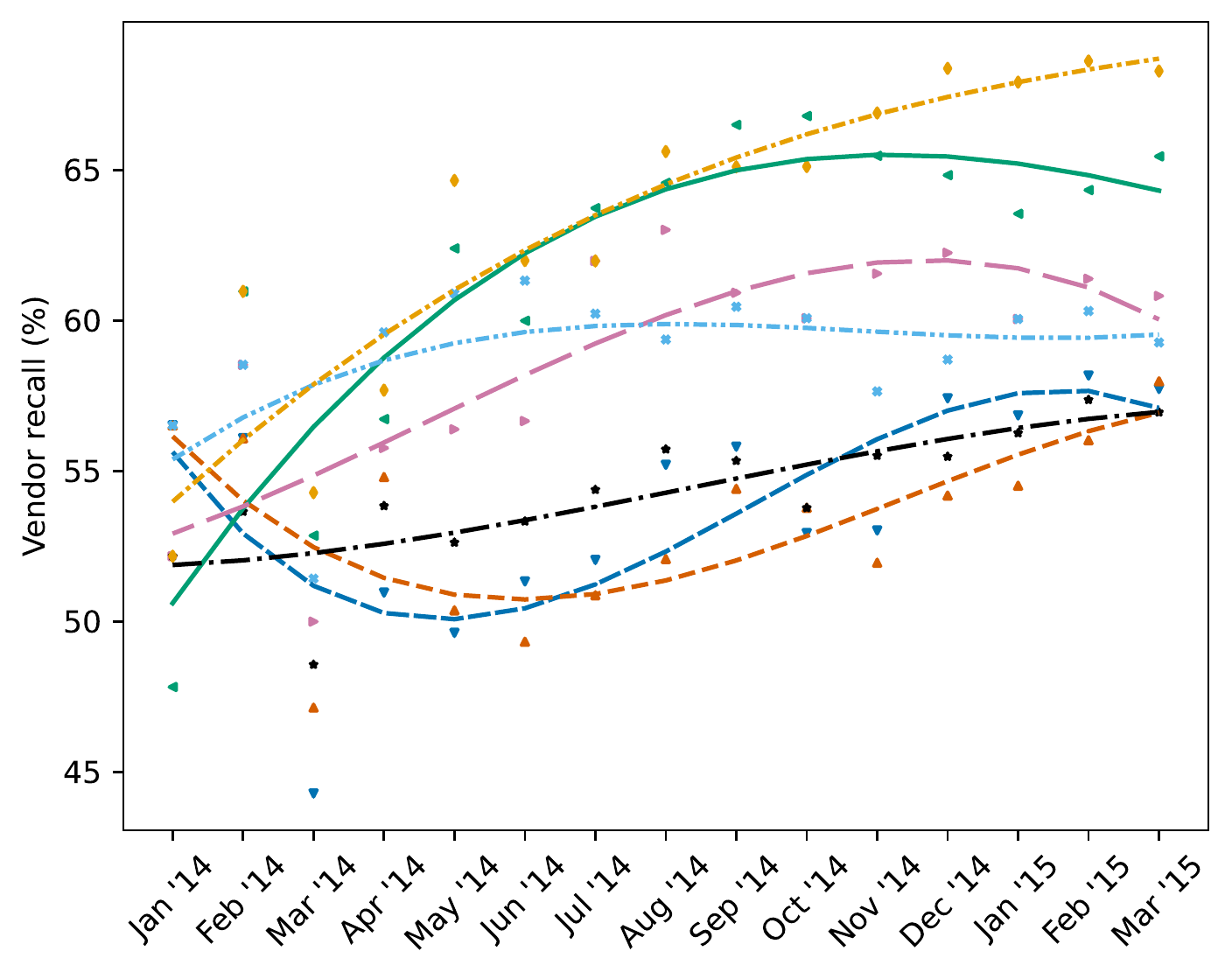}
      \caption{$\delta_o = 5$, $\delta_t = 14$ days}
    \end{subfigure}
    ~
    \begin{subfigure}[b]{0.25\textheight}
      \centering
      \includegraphics[width=\textwidth]{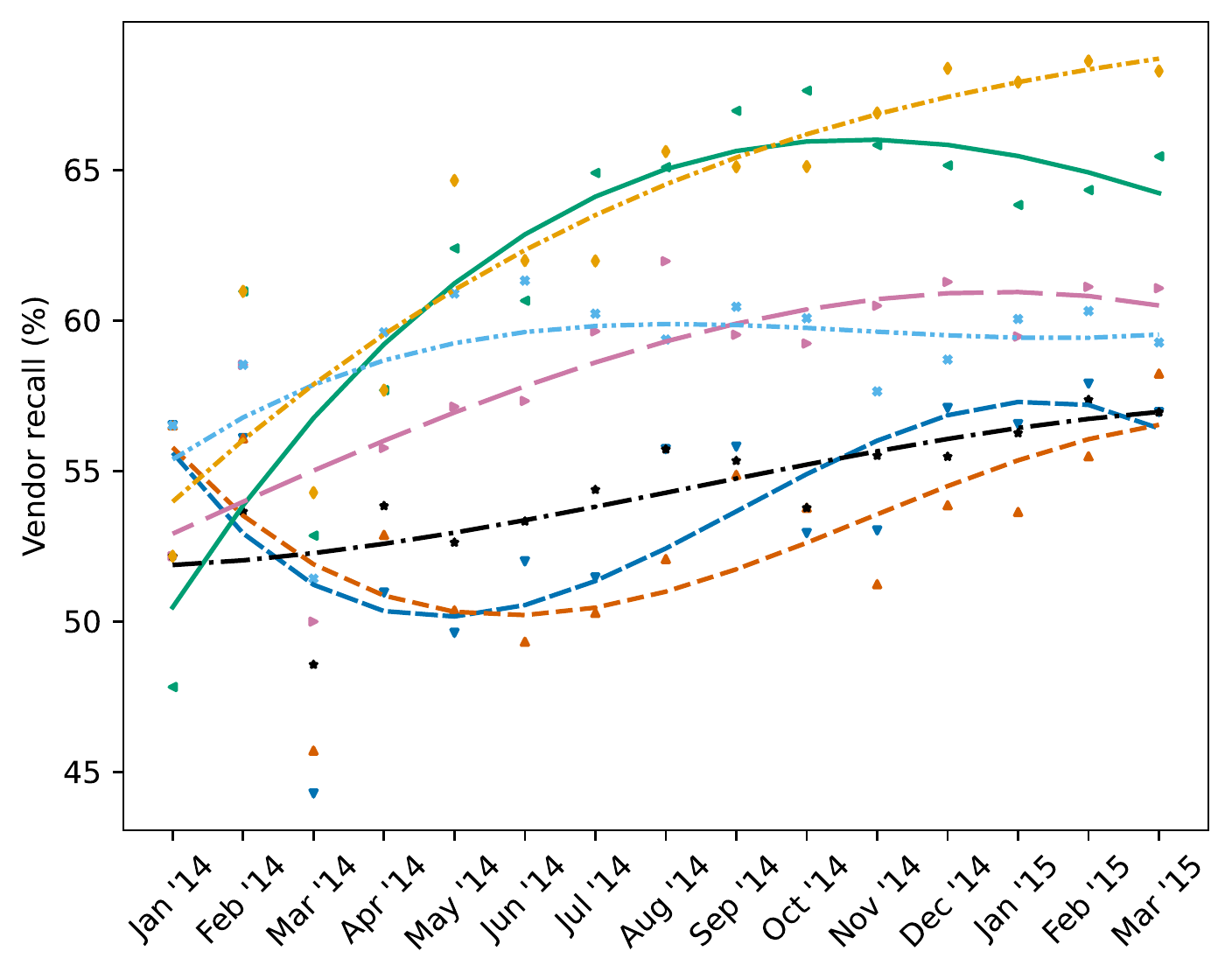}
      \caption{$\delta_o = 5$, $\delta_t = 3$ months}
    \end{subfigure}
    ~
    \begin{subfigure}[b]{0.25\textheight}
      \centering
      \includegraphics[width=\textwidth]{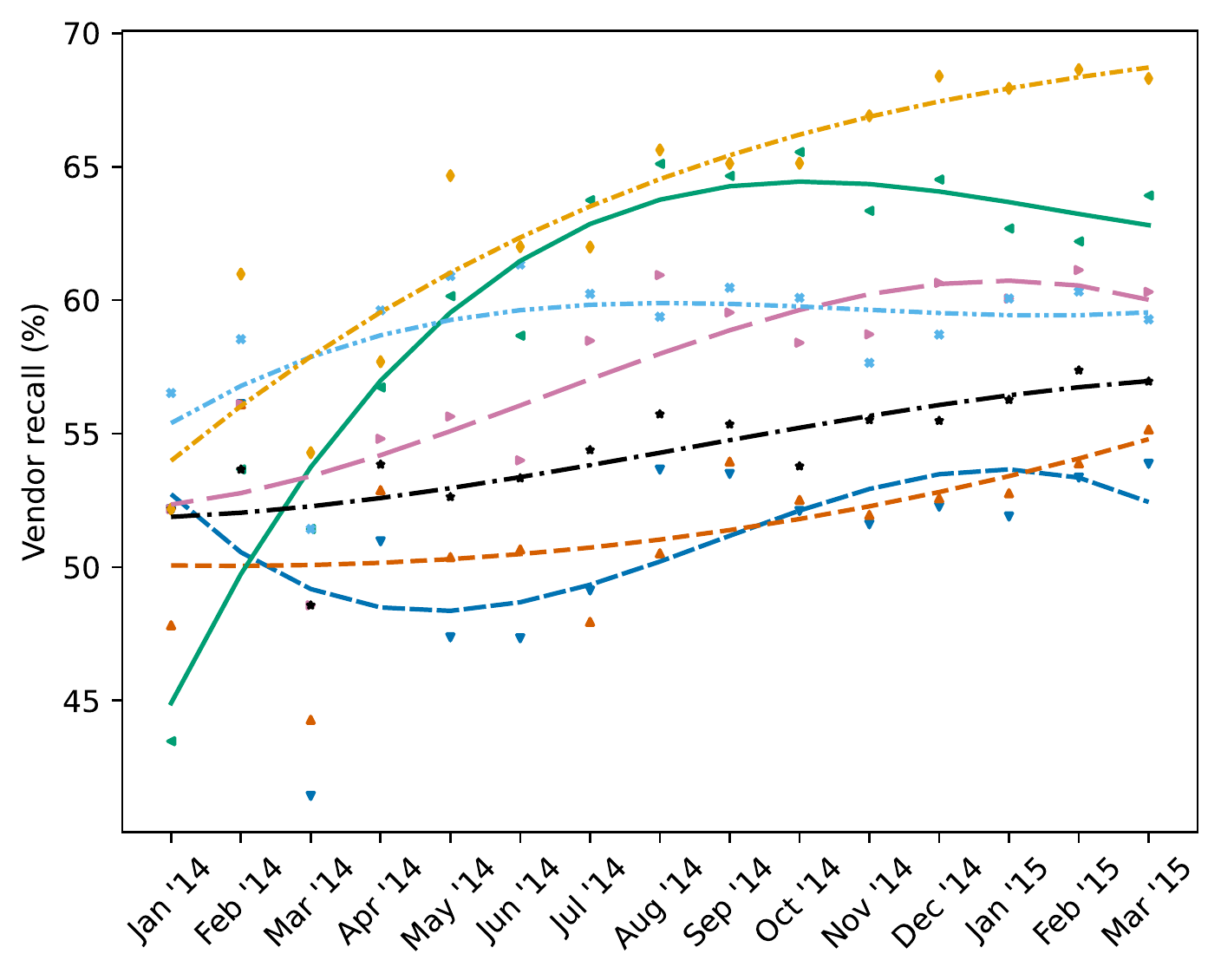}
      \caption{$\delta_o = 20$, $\delta_t = 14$ days}
    \end{subfigure}
    ~
    \begin{subfigure}[b]{0.25\textheight}
      \centering
      \includegraphics[width=\textwidth]{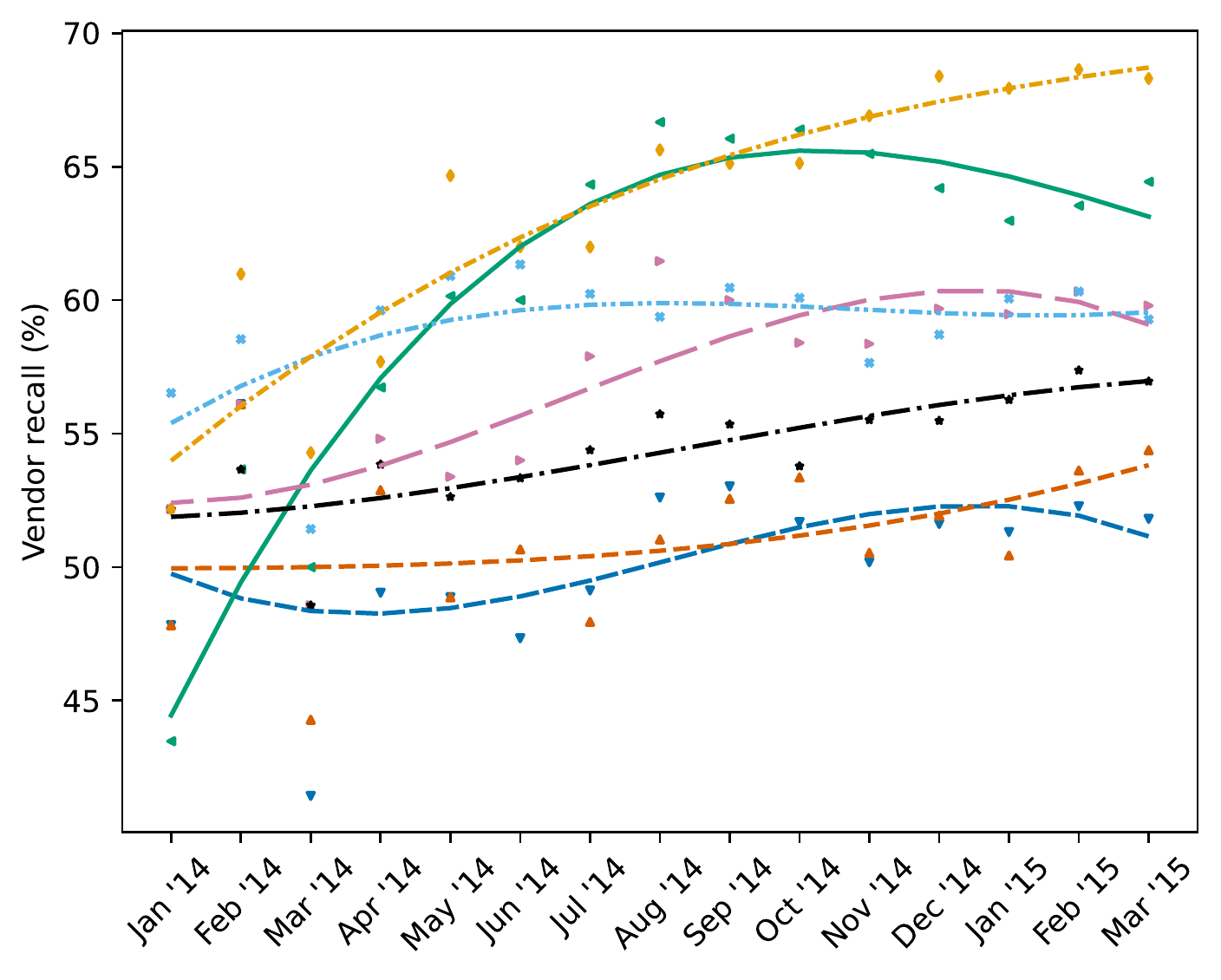}
      \caption{$\delta_o = 20$, $\delta_t = 3$ months}
    \end{subfigure}
  \caption{Results for varying multiple edge weighting parameters. Monthly vendor recall of top vendor percentile (top 0-20\% vendors in terms of sales) among the top 20\% of all users based on the network measures and activity indicators for current success. Each plot displays monthly vendor recall for a different set of parameter values used for generating the network. For the top two rows, each set of varied parameters consist of those that showed some improvement in Figures~\ref{fig:param1} and~\ref{fig:param2}. The bottom row specifically compares different variations for $\delta_o$ and $\delta_t$, to investigate the effects for different scopes of information loss.} \label{fig:param3}
\end{figure}
\end{landscape}

Having established the robustness of changing a single parameter at a time, we finally consider changing several parameters simultaneously.
For this we first combined parameter values which showed slight improvements compared to default parameter values in Figures~\ref{fig:param1} and~\ref{fig:param2} and plot the vendor recall trends for various combinations of these values in Figure~\ref{fig:param3}a--j.
Although some combinations show relatively improved performance for the centrality measures, none show sufficient improvement to become inconsistent with our findings.

Next, we investigate how potential loss of information caused by parameters $\delta_o$ and $\delta_t$, as described in the ``Network extraction'' section of the main paper, impacts vendor recall performance if chosen differently in Figures~\ref{fig:param3}k--n (i.e., the bottom row of Figure~\ref{fig:param3}).
Specifically, we consider each variation of including fewer or more posts and allowing for less or more maximum time between posts.
The figures show that the performance changes are largely dominated by the $\delta_o$ parameter, showing the same performance changes as those we observed for the default $\delta_t$.
Specifically, we see a slight improvement in the performance of betweenness centrality at both a lower and higher value of $\delta_o$ than default.
Thus, there does not seem to be a clear connection between the performance and the information loss caused by smaller $\delta_o$ values.
On the contrary, Figures~\ref{fig:param3}k,m show that when we decrease $\delta_t$ to 14 days, the performance drops slightly for all tested $\delta_o$ parameter values.
However, Figures~\ref{fig:param3}l,n show that when we extend $\delta_t$ to 3 months, we observe the same performance as at 1 month for the same $\delta_o$ parameter value.
As this was also the case for the default value of $\delta_o$, we may conclude that although performance can be compromised by the information loss caused by $\delta_t$, there is a point beyond which extending $\delta_t$ no longer improves performance.
In other words, there is a value of $\delta_t$ at which the performance plateaus.
It appears that for the Evolution cryptomarket this aligns with our chosen default value of 1 month.
However, note that even at a $\delta_t$ as low as 7 days (Figure~\ref{fig:param1}f), the performance drop observed is very small.
Thus, it appears that the vast majority of the added-value to performance lies in the responses that are written within days of a post.

In short, we observe that in general the vendor recall performance varies little for changes in parameter values determining the formation of the network.
Thus, we can consider our findings in the main paper to be robust for these parameters.

\section{Post activity and Sales recall}
\begin{figure}[h]
  \centering
      \begin{subfigure}[b]{.44\textwidth}
        \centering
        \includegraphics[width=\textwidth]{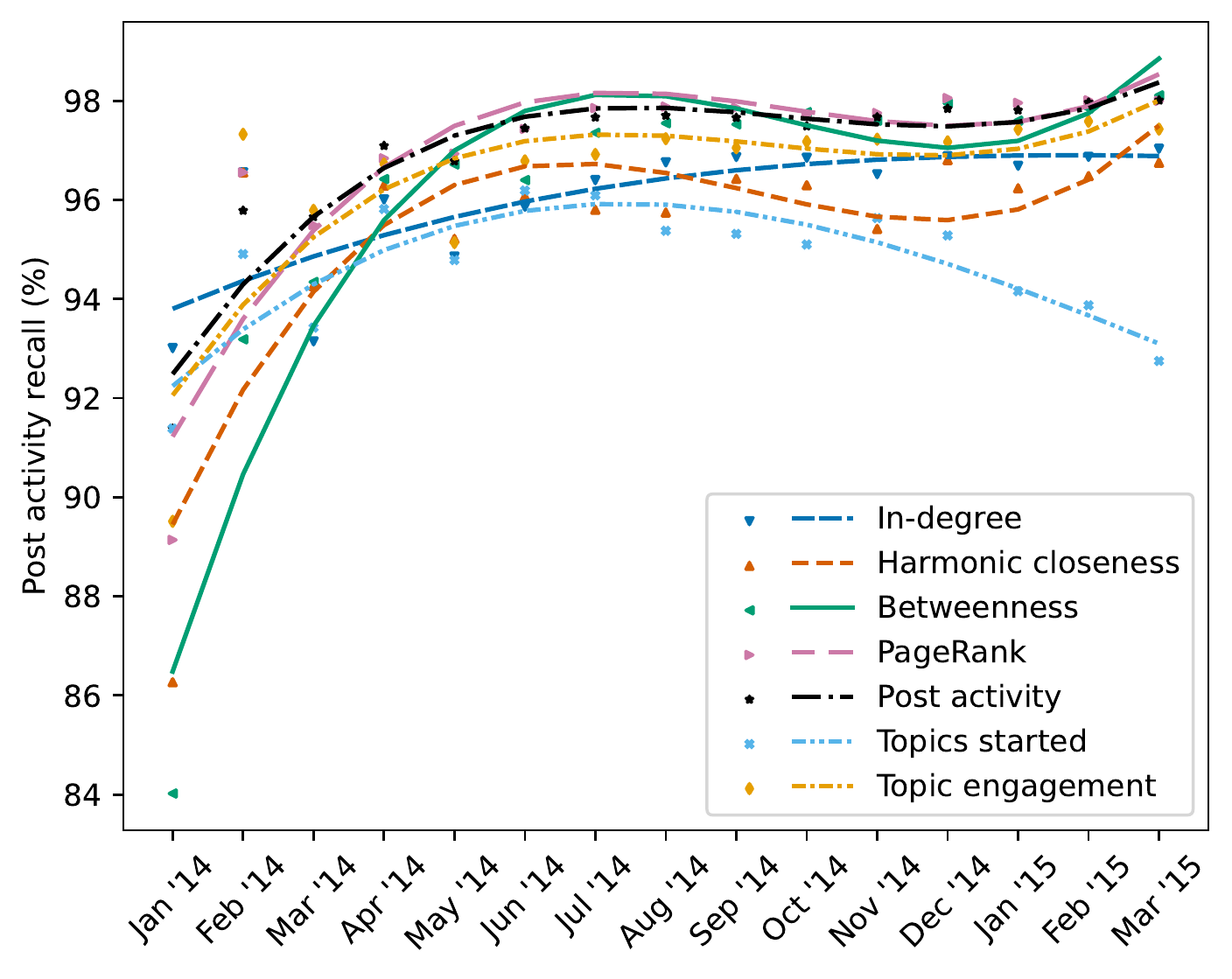}
        \caption{Current success}
      \end{subfigure}
      \begin{subfigure}[b]{.44\textwidth}
        \centering
        \includegraphics[width=\textwidth]{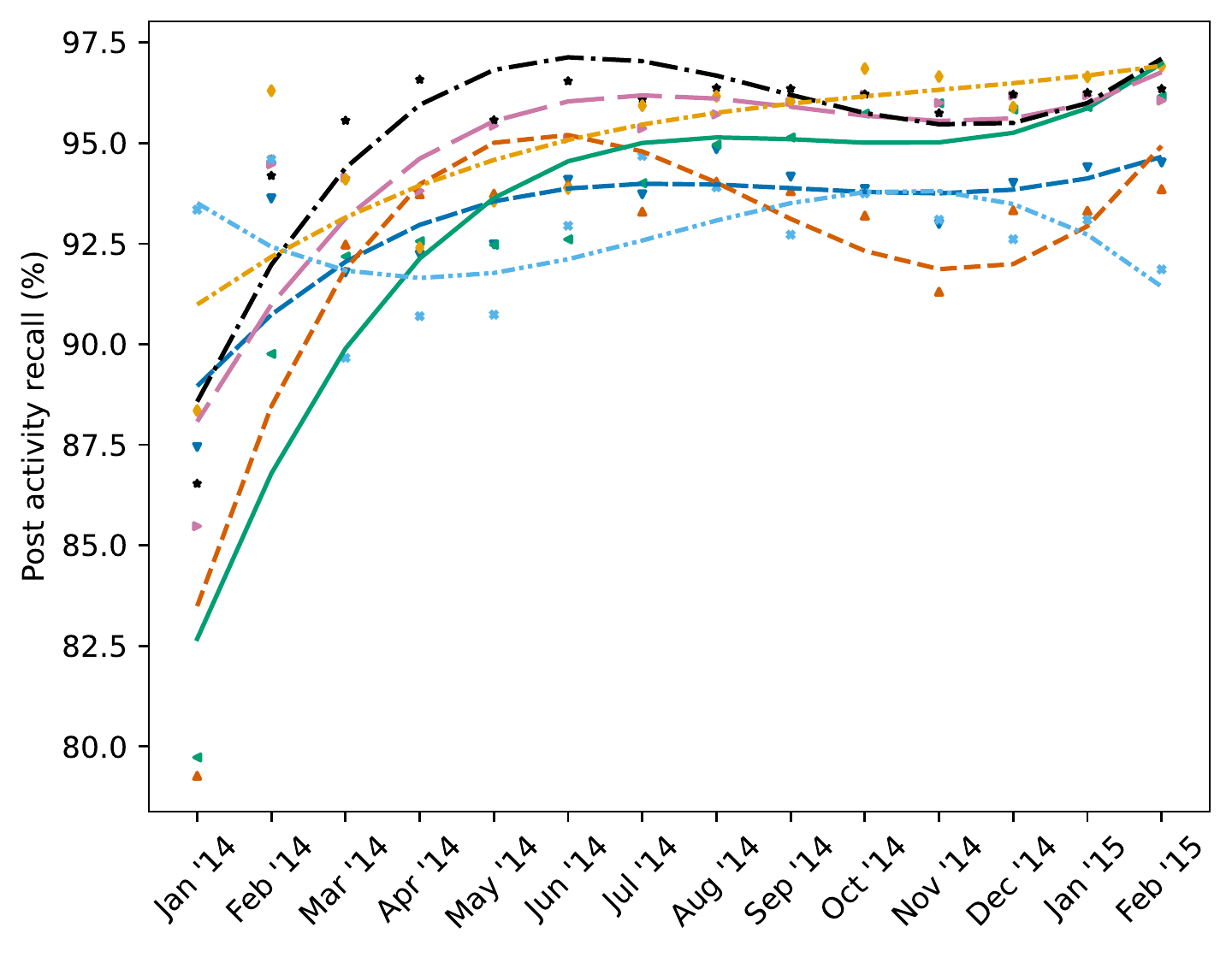}
        \caption{Future success}
      \end{subfigure}
      \caption{Monthly post activity recall for both current (a) and future success (b). Higher post activity recall indicates that recalled vendors placed a relatively larger share of the top vendor percentile's total post activity.} \label{fig:recall-activity}
\end{figure}

\begin{figure}[h]
  \centering
    \begin{subfigure}[b]{0.44\textwidth}
      \centering
      \includegraphics[width=\textwidth]{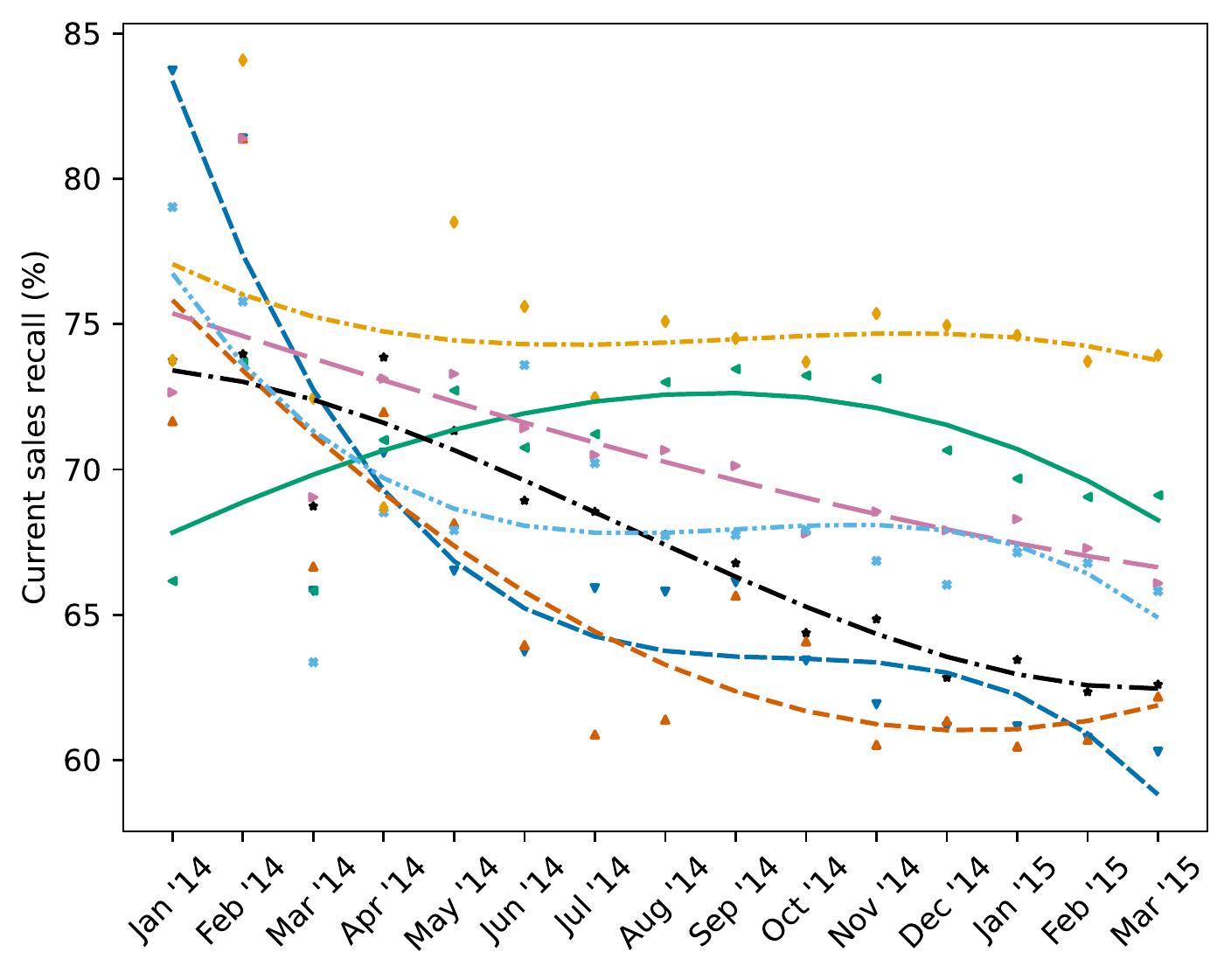}
      \caption{Current success}
    \end{subfigure}
    ~
    \begin{subfigure}[b]{0.44\textwidth}
      \centering
      \includegraphics[width=\textwidth]{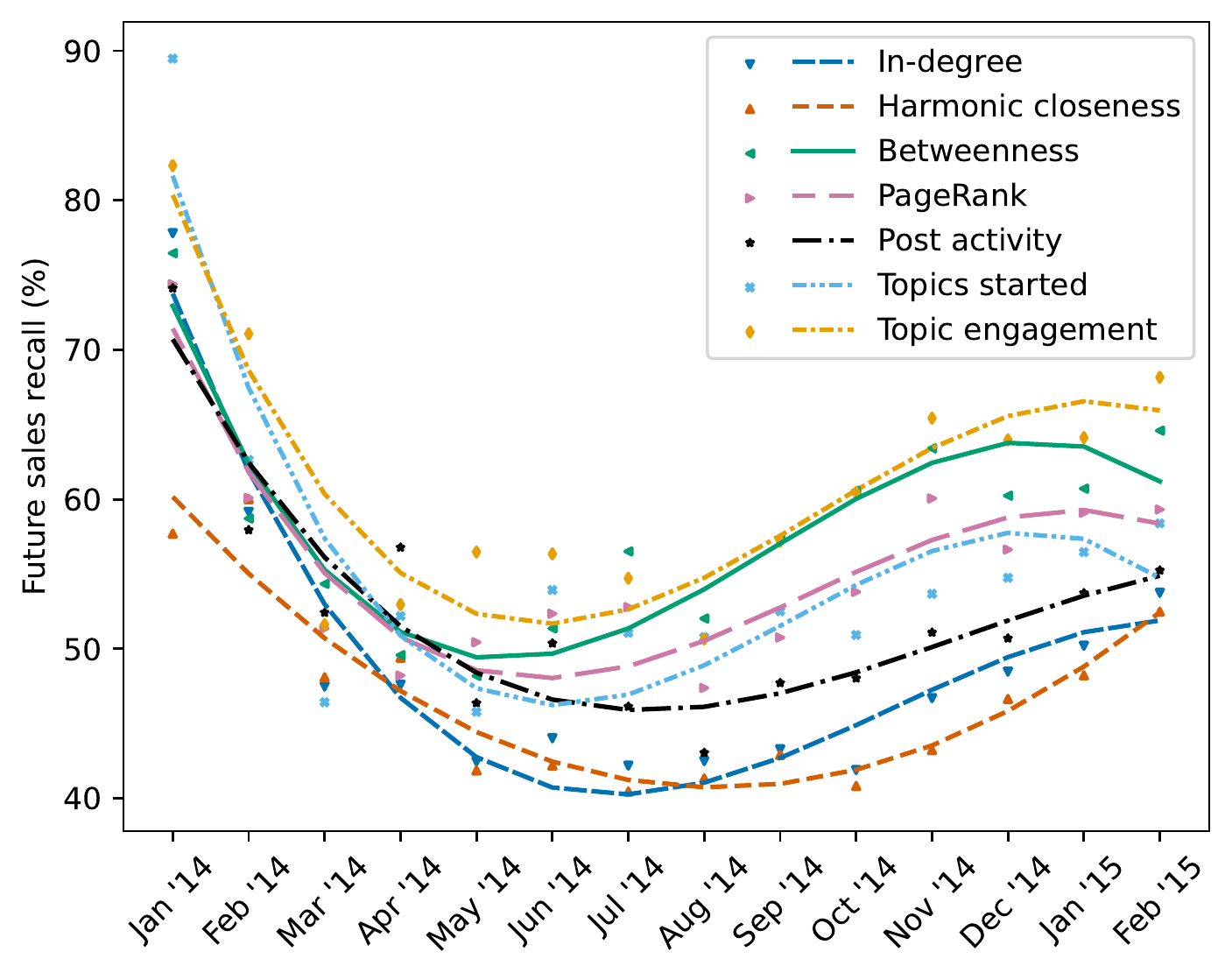}
      \caption{Future success}
    \end{subfigure}
  \caption{Monthly sales recall for both current (a) and future success (b). Higher sales recall indicates a greater portion of the top vendor percentile's total sales was attributed to the recalled vendors.} \label{fig:recall-sales}
\end{figure}

In this section, we consider the \emph{post activity recall} and \emph{sales recall}, which measure respectively what percentage of post activity and sales of the entire top percentile the detected vendors are responsible for.
We plot the monthly post activity recall in Figure~\ref{fig:recall-activity} and current and future sales recall in Figure~\ref{fig:recall-sales}.
These figures show the recall in terms of what percentage of post activity/sales the recalled vendors are responsible for.

From Figure~\ref{fig:recall-activity} we see that the vast majority of post activity by the top vendor percentile is associated with vendors with high network centrality and activity indicators.
As such, many of the non-recalled vendors are likely to be those with very few posts.
Not unexpectedly, the post activity indicator often has the highest post activity recall, while the topics started indicator often captures the least post activity.

Figure~\ref{fig:recall-sales} shows that for current success most of our observations for vendor recall hold up.
Perhaps the most significant change is that the differences between PageRank and topics started and between betweenness centrality and topic engagement are more prominent.
Similarly, for future success PageRank now outperforms the topics started baseline more consistently.
For both current and future success, we observe that the sales recall is generally between 10-20\% higher than the corresponding vendor recall.
This indicates that the detected vendors are, on average, the more successful vendors among the top percentile.
\section{Overviews of sales and post activity of recalled and non-recalled users}
\begin{figure}[h]
  \centering
    \begin{subfigure}[b]{0.44\textwidth}
      \centering
      \includegraphics[width=\textwidth]{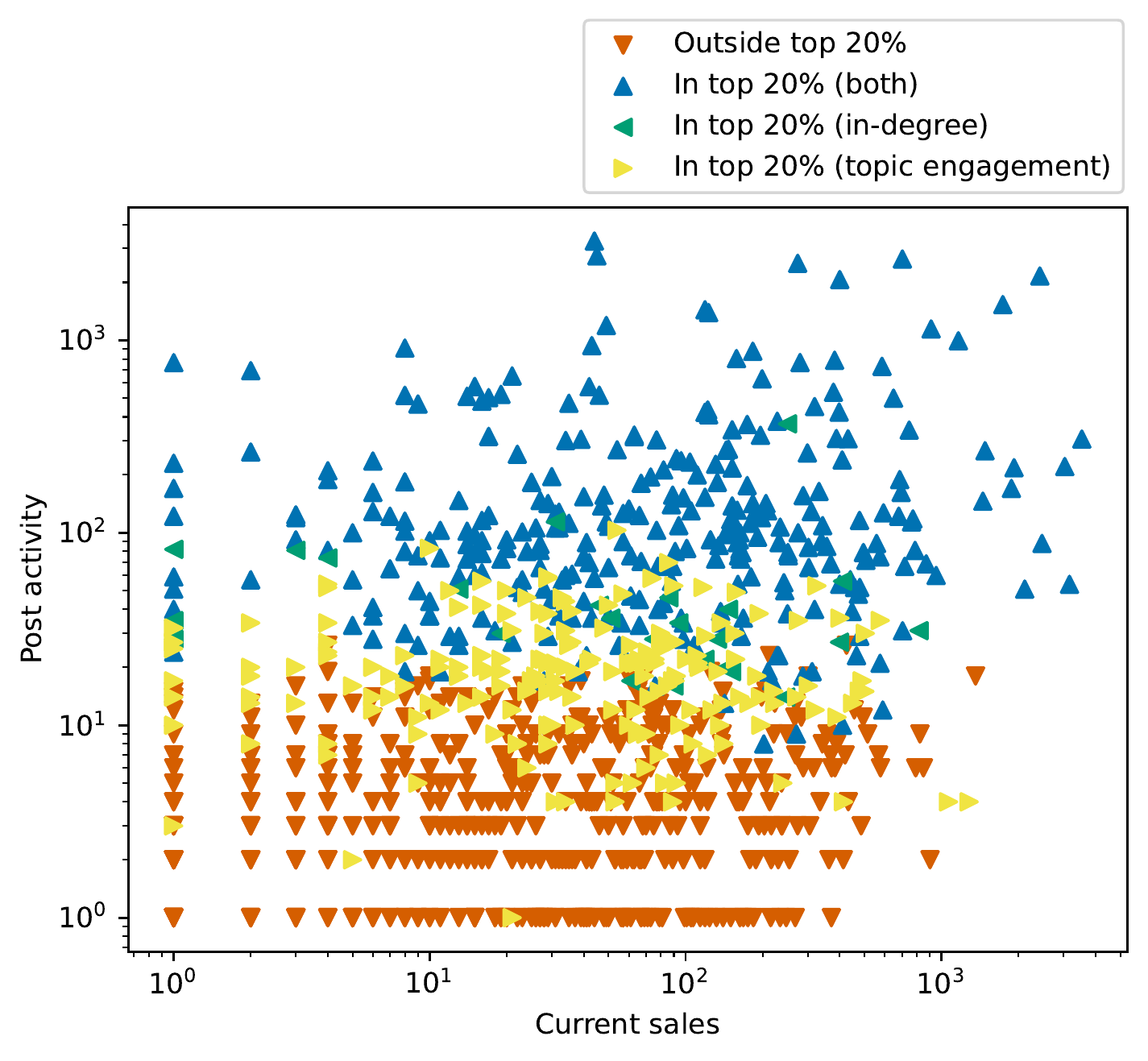}
      \caption{In-degree}
    \end{subfigure}
    ~
    \begin{subfigure}[b]{0.44\textwidth}
      \centering
      \includegraphics[width=\textwidth]{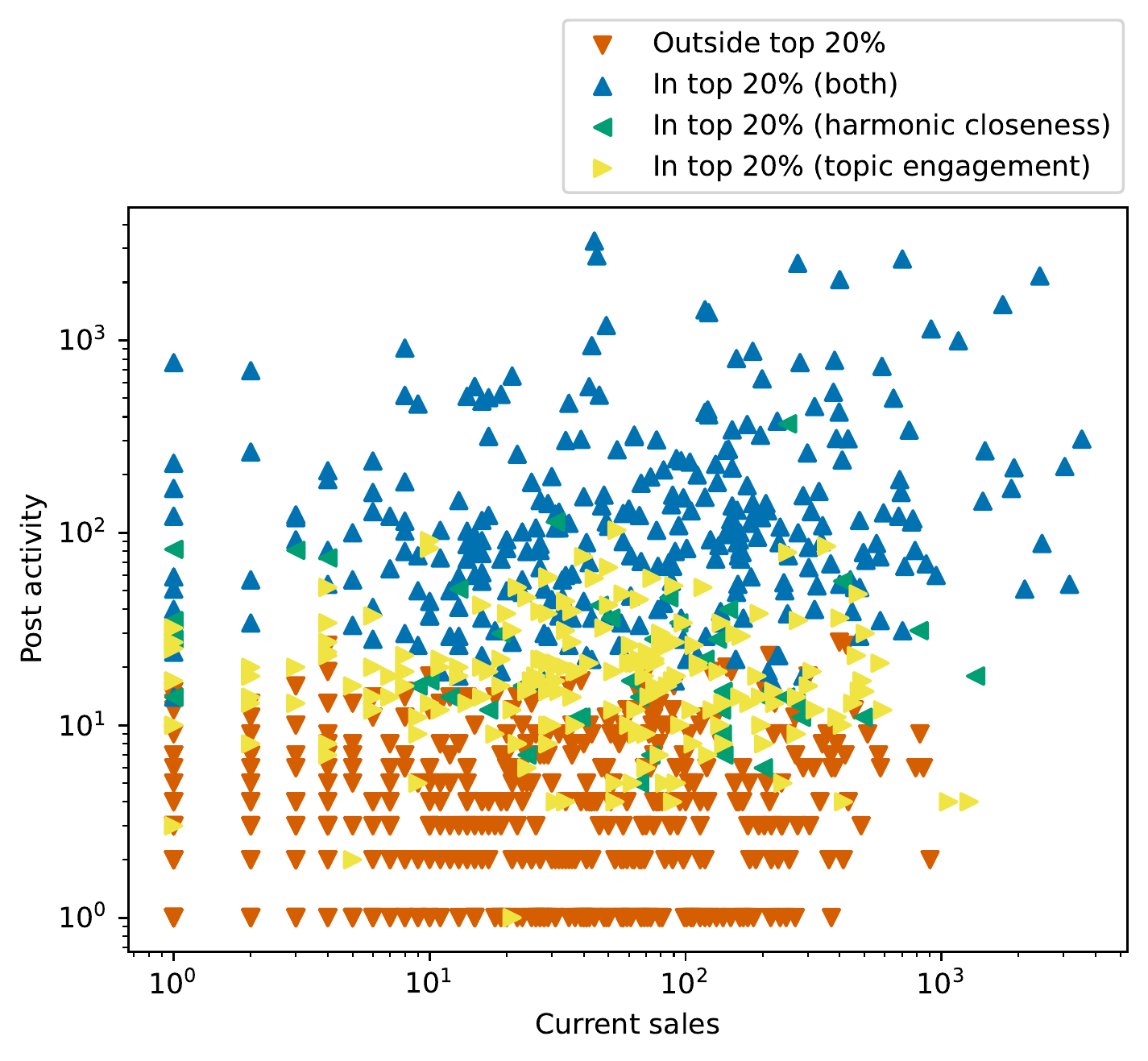}
      \caption{Harmonic closeness centrality}
    \end{subfigure}
    \begin{subfigure}[b]{0.44\textwidth}
      \centering
      \includegraphics[width=\textwidth]{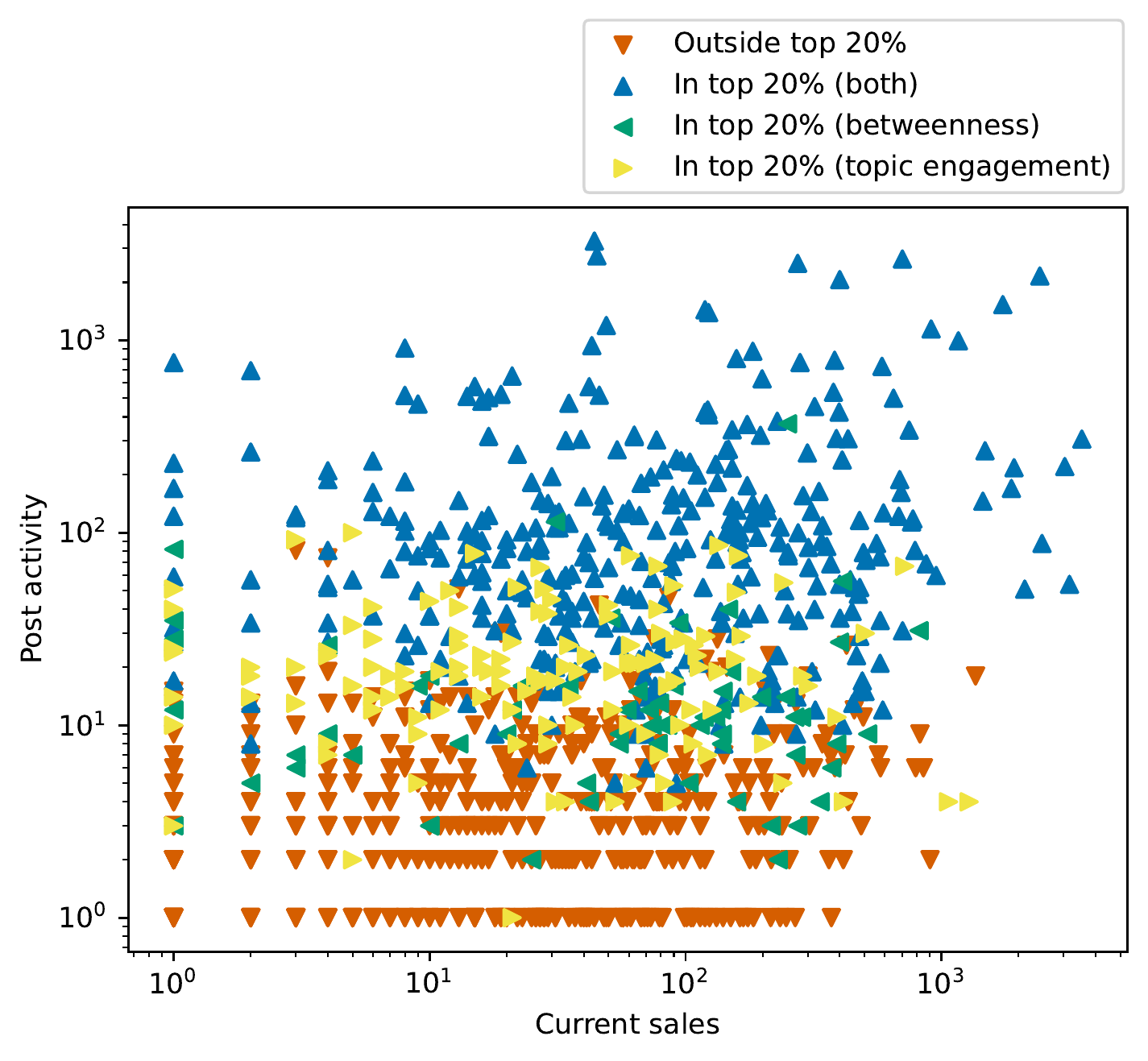}
      \caption{Betweenness centrality}
    \end{subfigure}
    ~
    \begin{subfigure}[b]{0.44\textwidth}
      \centering
      \includegraphics[width=\textwidth]{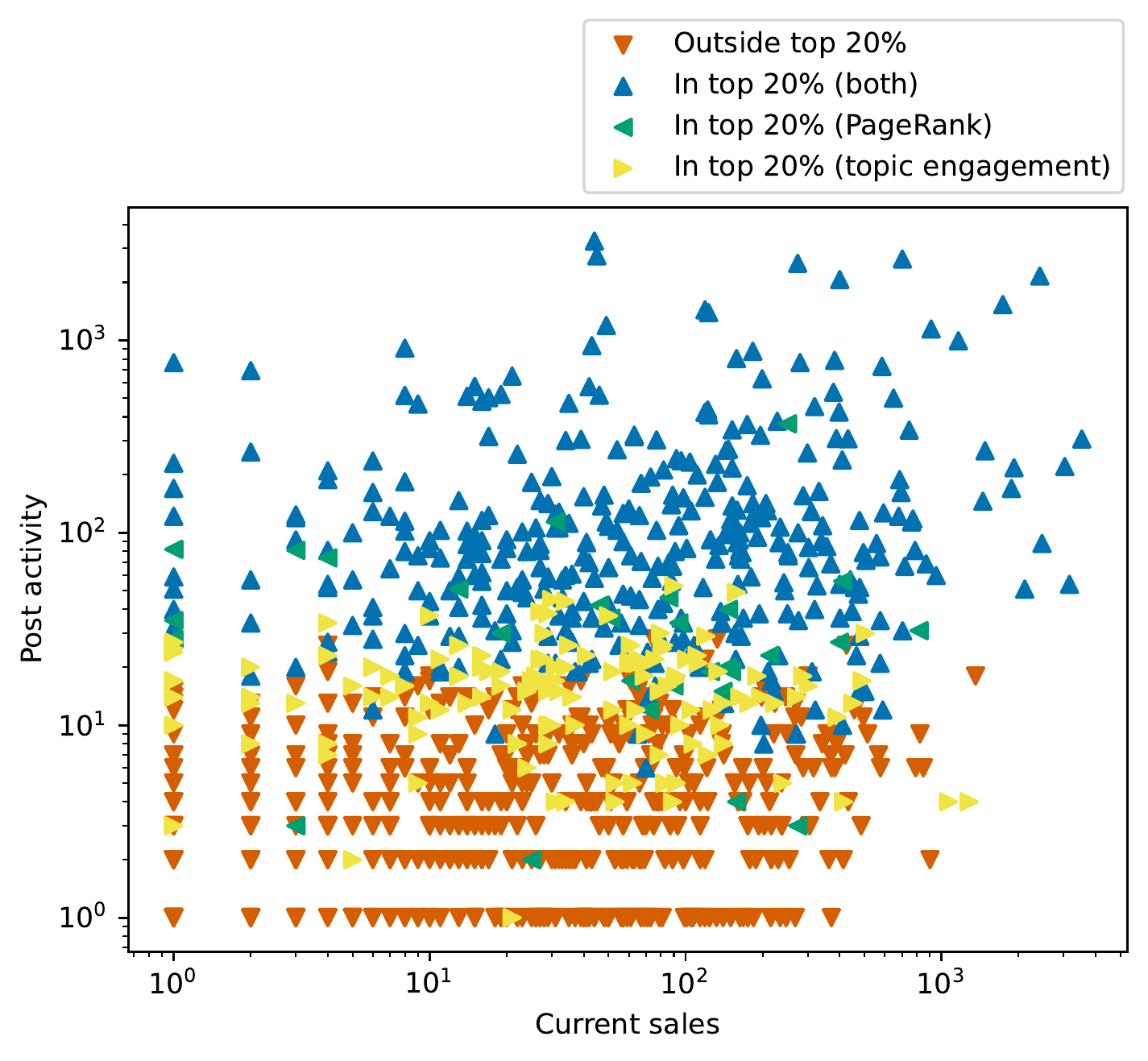}
      \caption{PageRank}
    \end{subfigure}
  \caption{Sales and post activity of recalled (in top 20\%) and non-recalled (outside top 20\%) users for September 2014, for each of the network centrality measures compared with topic engagement and their intersection, respectively.} \label{fig:overview1}
\end{figure}

\begin{landscape}
\begin{figure}[t]
  \centering
    \begin{subfigure}[b]{0.25\textheight}
      \centering
      \includegraphics[width=\textwidth]{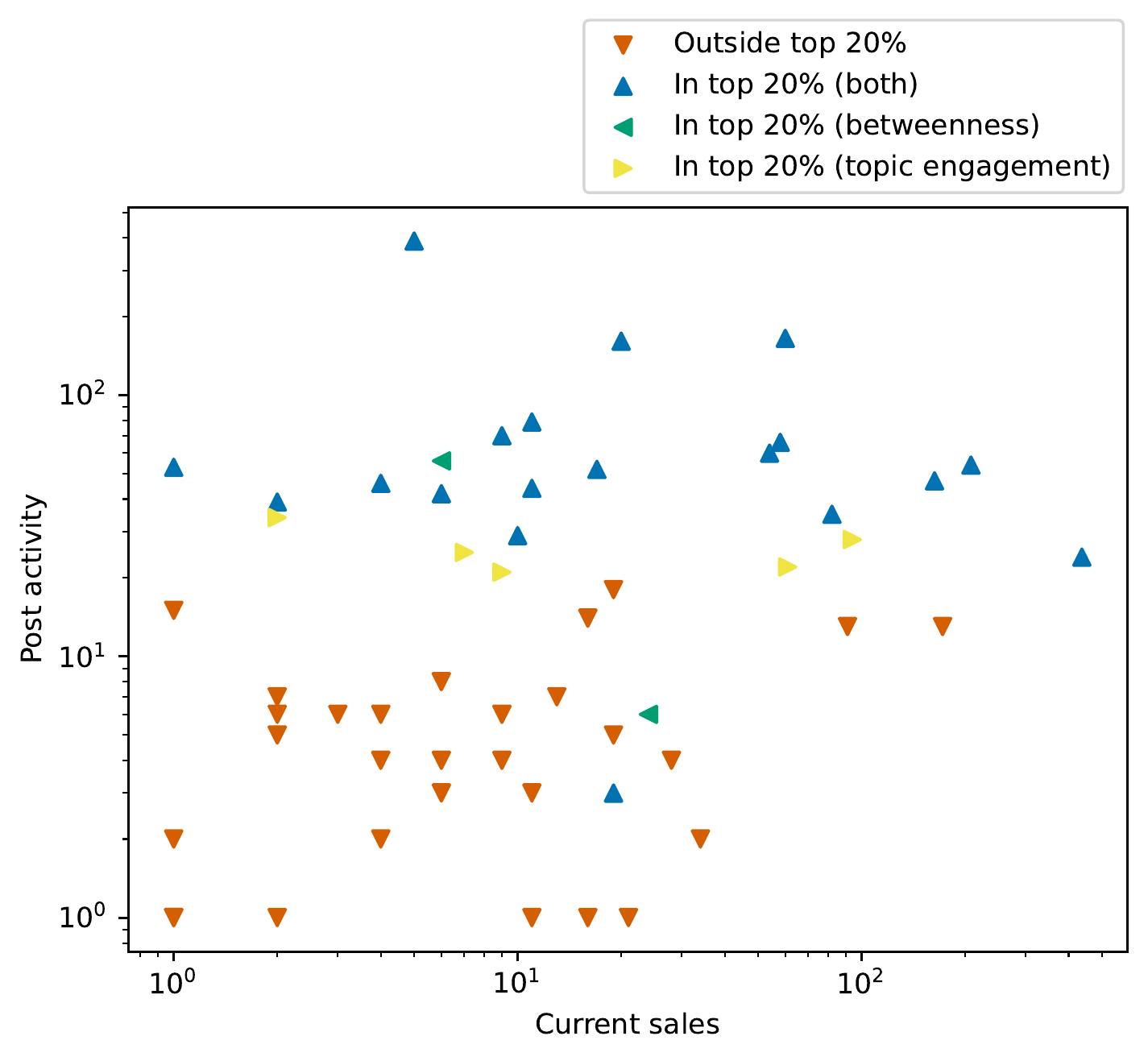}
      \caption{January 2014}
    \end{subfigure}
    ~
    \begin{subfigure}[b]{0.25\textheight}
      \centering
      \includegraphics[width=\textwidth]{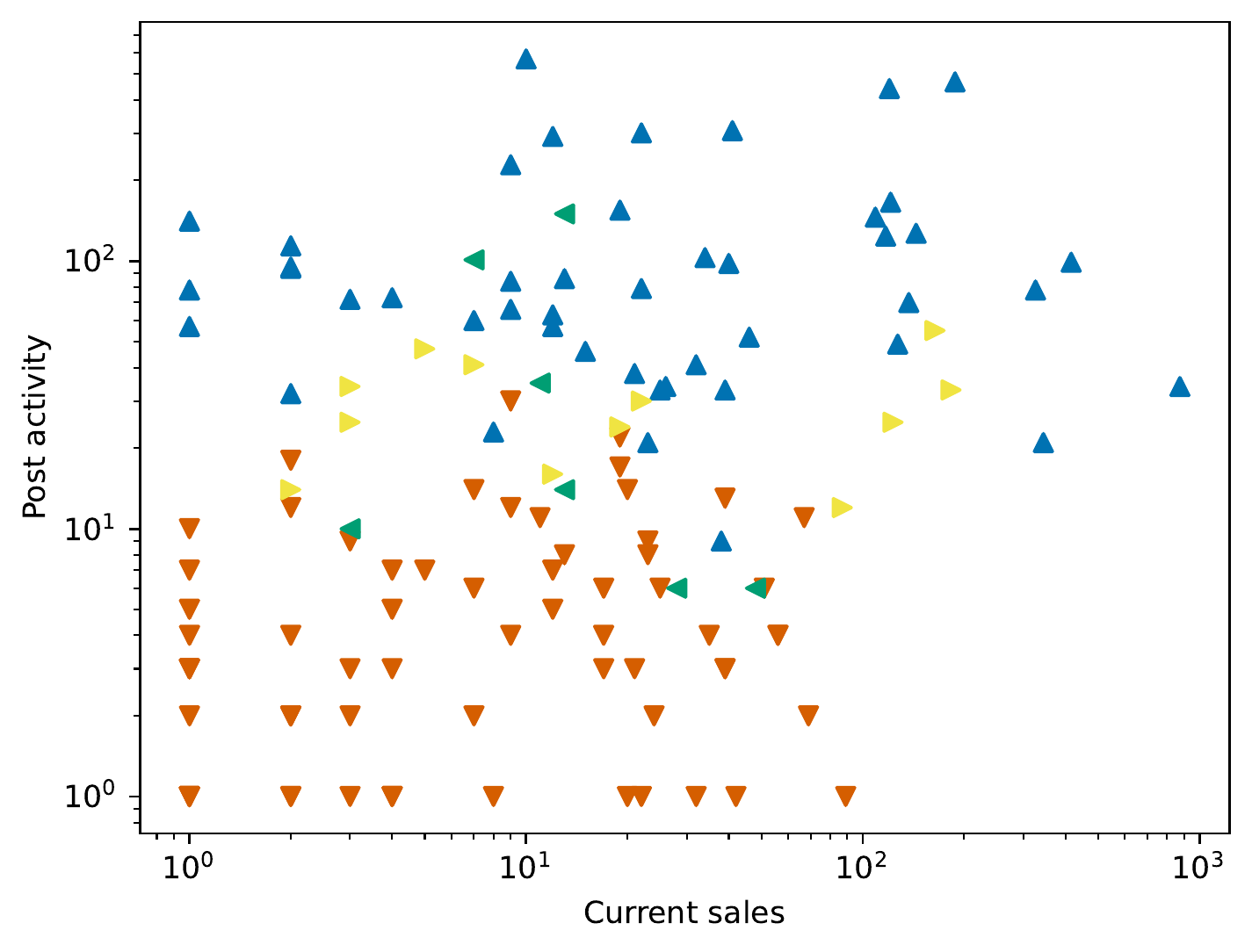}
      \caption{February 2014}
    \end{subfigure}
    ~
    \begin{subfigure}[b]{0.25\textheight}
      \centering
      \includegraphics[width=\textwidth]{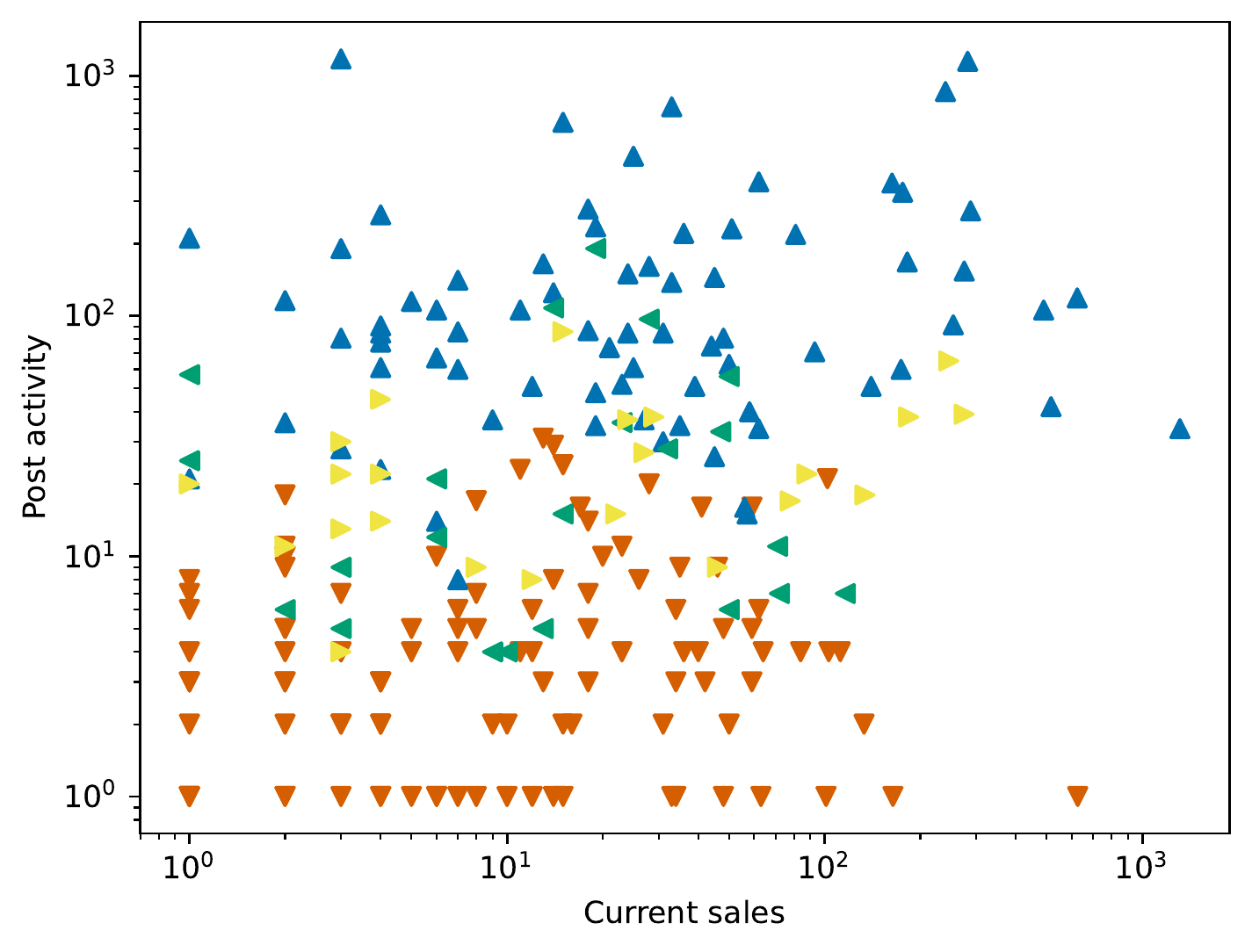}
      \caption{March 2014}
    \end{subfigure}
    ~
    \begin{subfigure}[b]{0.25\textheight}
      \centering
      \includegraphics[width=\textwidth]{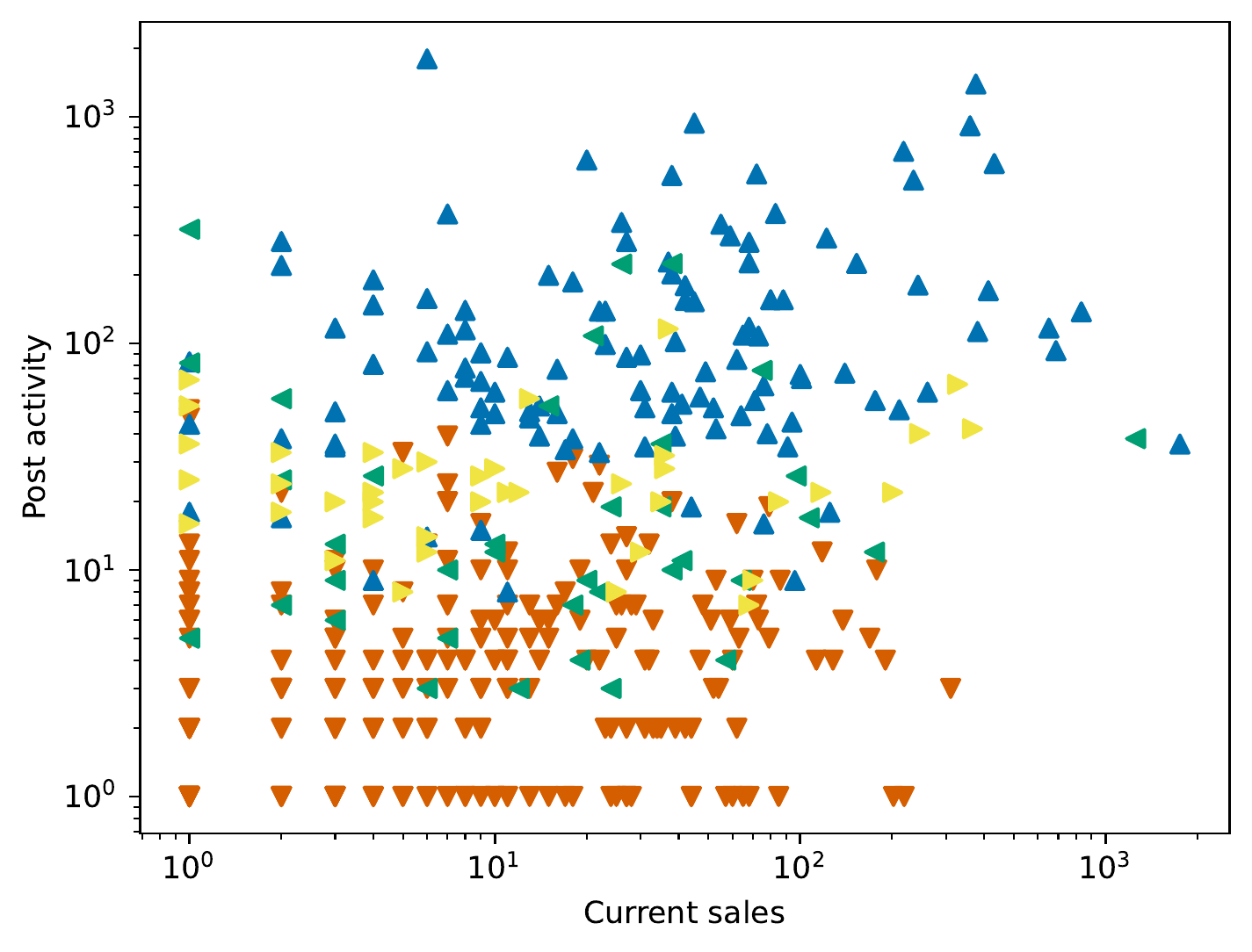}
      \caption{April 2014}
    \end{subfigure}
    ~
    \begin{subfigure}[b]{0.25\textheight}
      \centering
      \includegraphics[width=\textwidth]{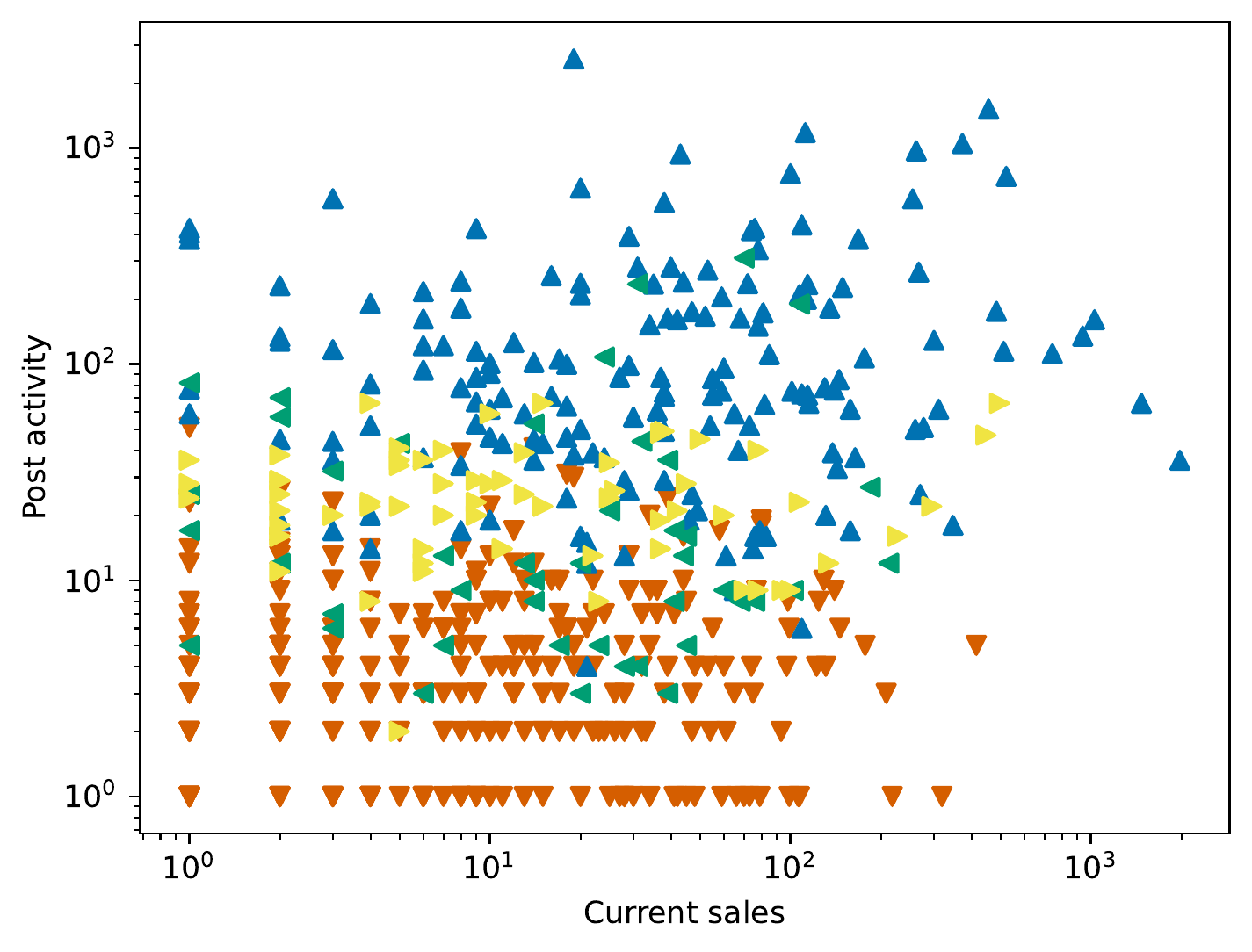}
      \caption{May 2014}
    \end{subfigure}
    \begin{subfigure}[b]{0.25\textheight}
      \centering
      \includegraphics[width=\textwidth]{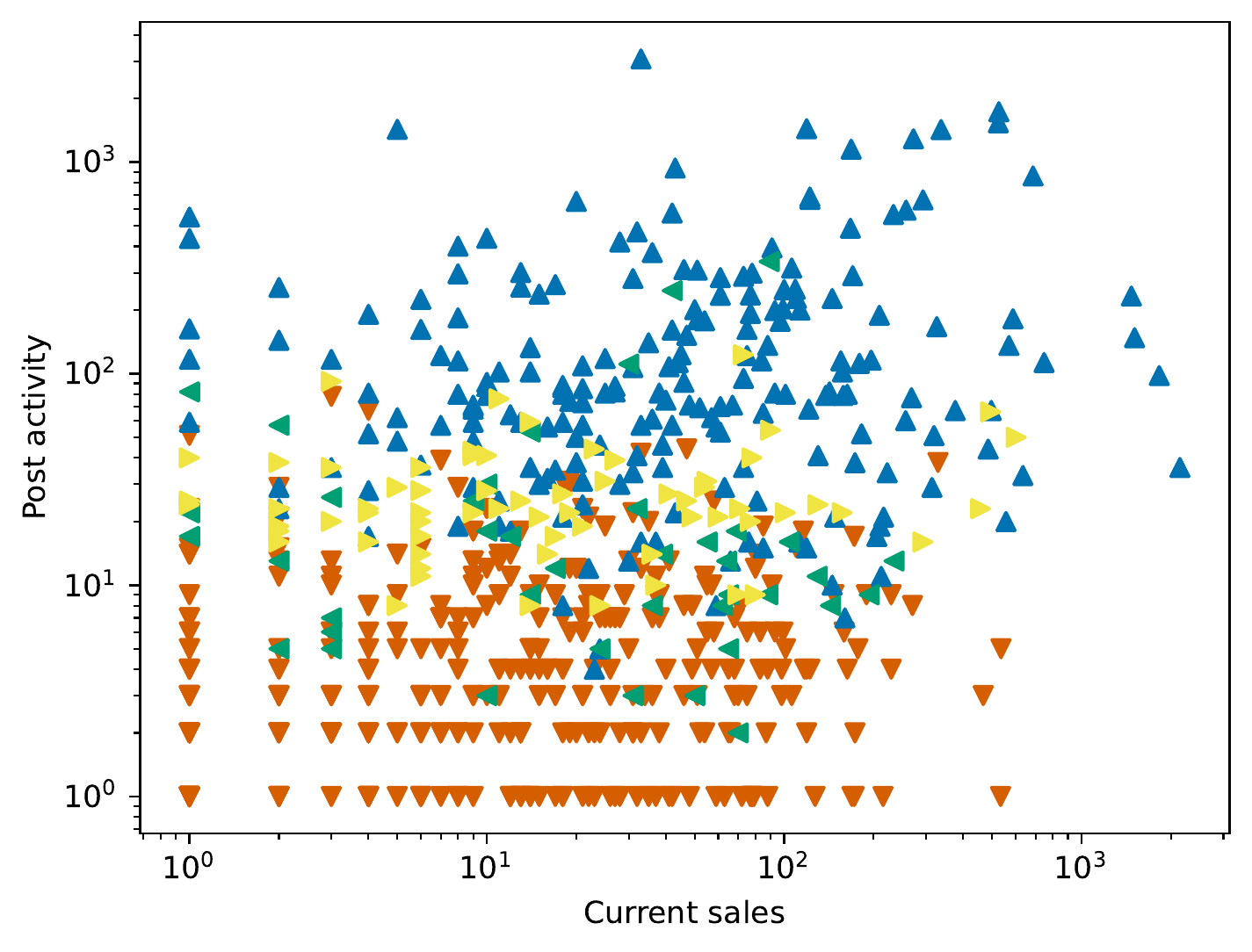}
      \caption{June 2014}
    \end{subfigure}
    ~
    \begin{subfigure}[b]{0.25\textheight}
      \centering
      \includegraphics[width=\textwidth]{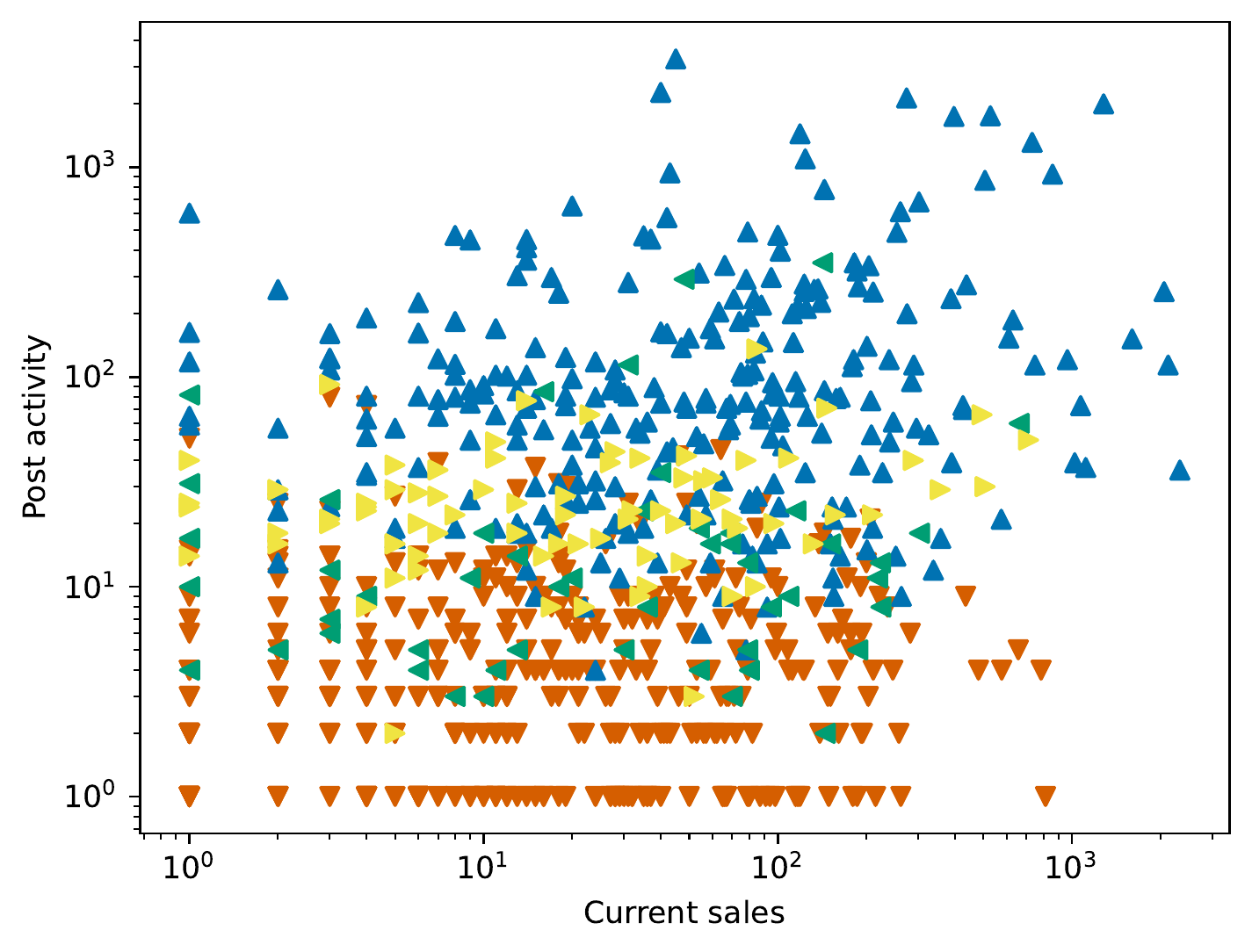}
      \caption{July 2014}
    \end{subfigure}
    ~
    \begin{subfigure}[b]{0.25\textheight}
      \centering
      \includegraphics[width=\textwidth]{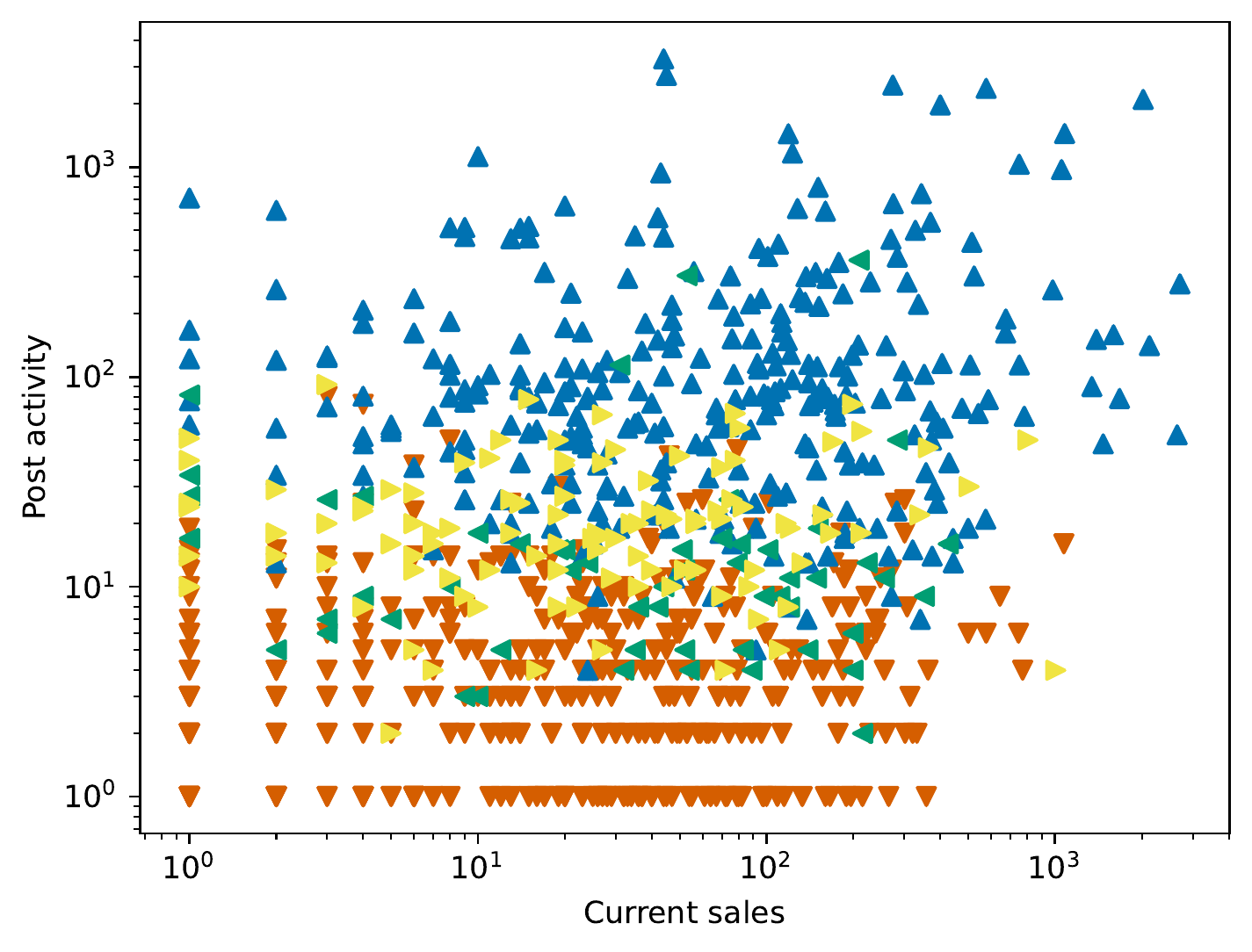}
      \caption{August 2014}
    \end{subfigure}
    ~
    \begin{subfigure}[b]{0.25\textheight}
      \centering
      \includegraphics[width=\textwidth]{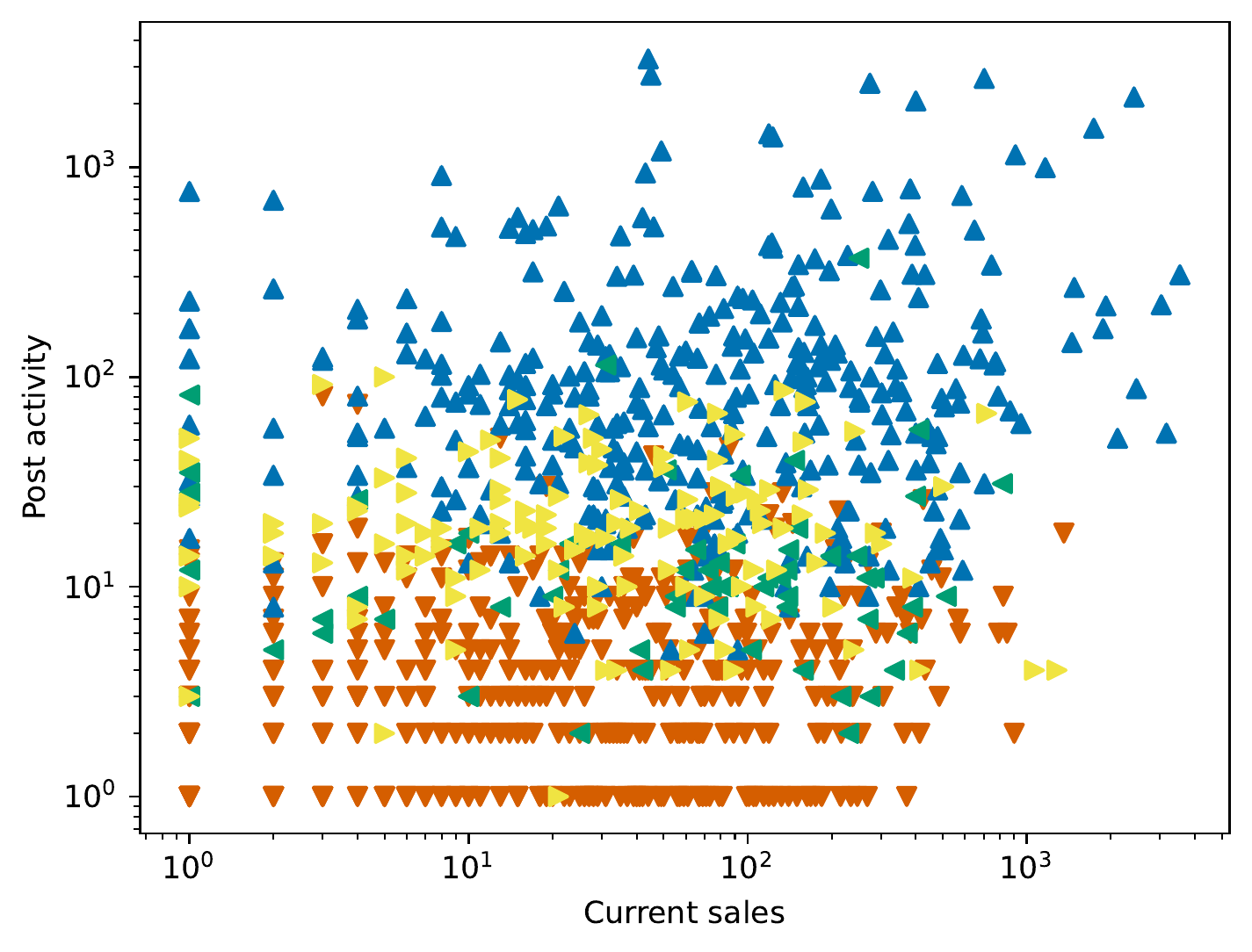}
      \caption{September 2014}
    \end{subfigure}
    ~
    \begin{subfigure}[b]{0.25\textheight}
      \centering
      \includegraphics[width=\textwidth]{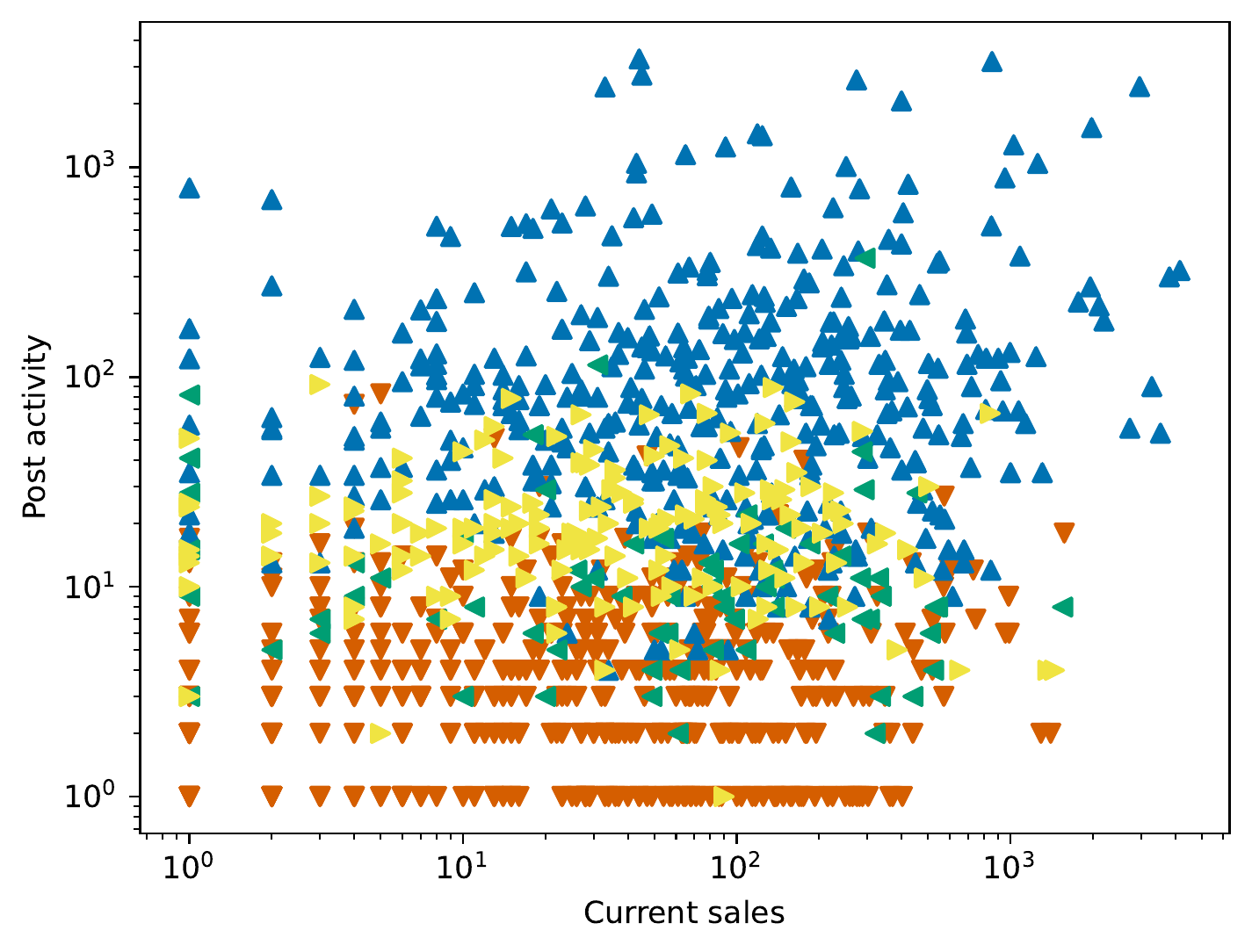}
      \caption{October 2014}
    \end{subfigure}
    \begin{subfigure}[b]{0.25\textheight}
      \centering
      \includegraphics[width=\textwidth]{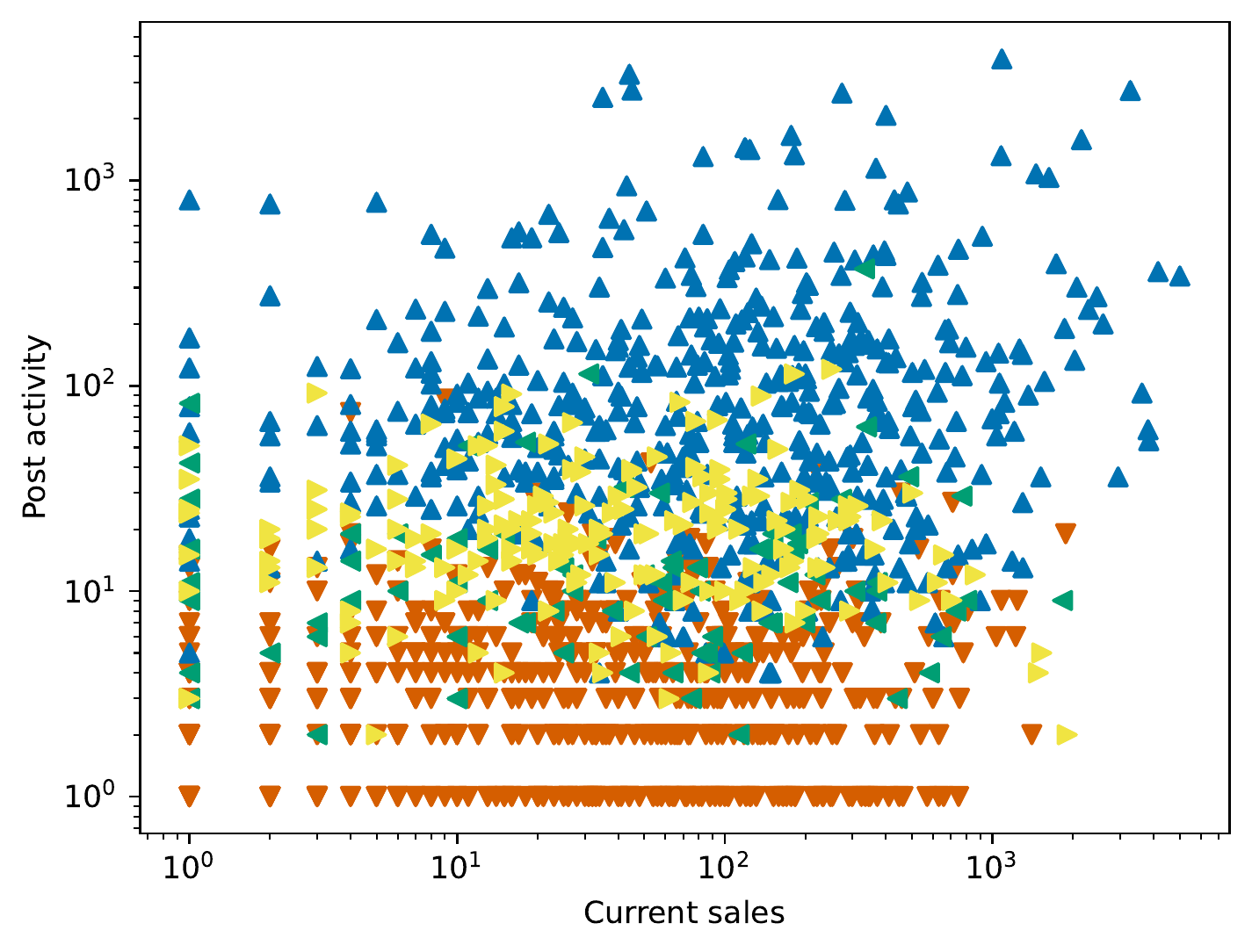}
      \caption{November 2014}
    \end{subfigure}
    ~
    \begin{subfigure}[b]{0.25\textheight}
      \centering
      \includegraphics[width=\textwidth]{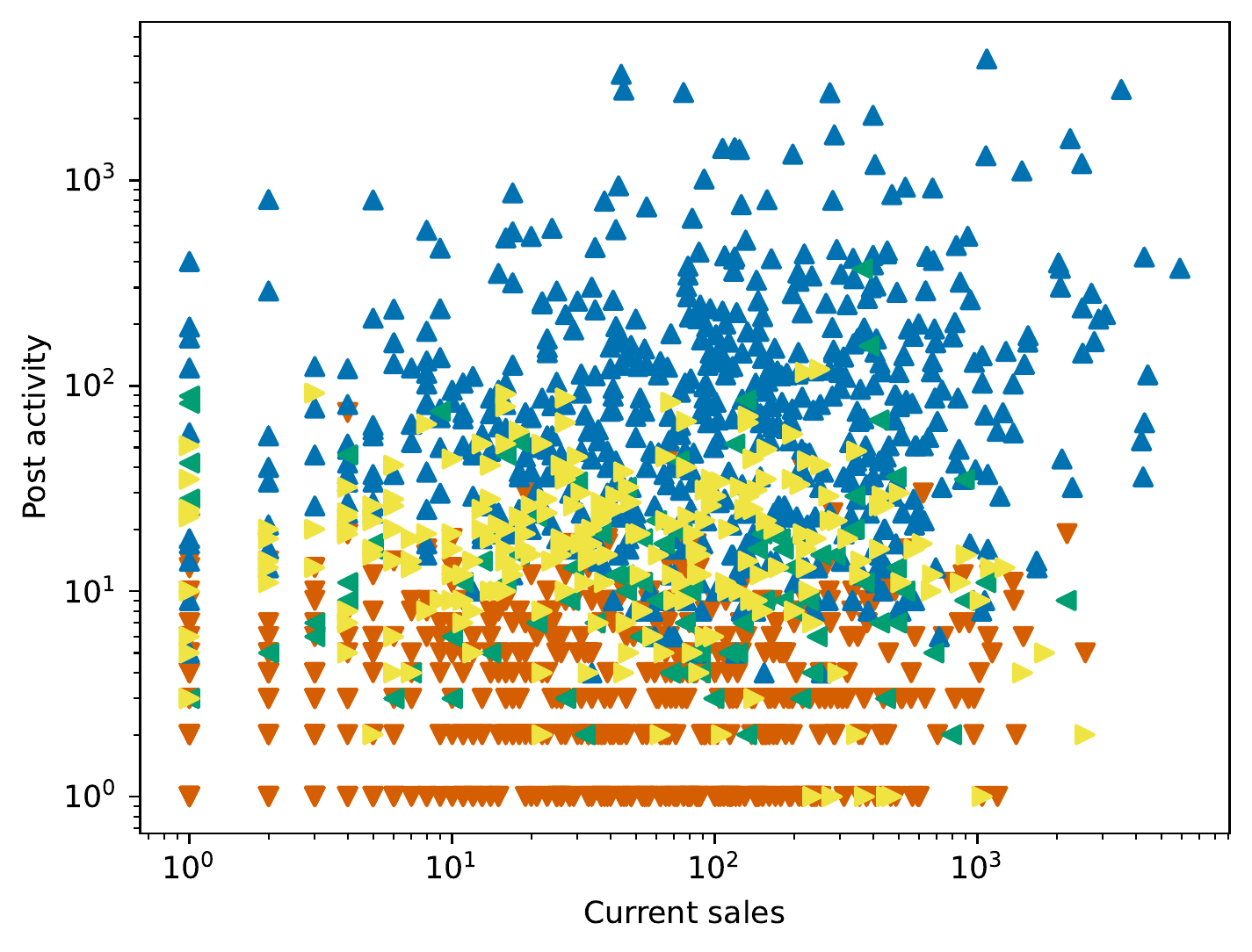}
      \caption{December 2014}
    \end{subfigure}
    ~
    \begin{subfigure}[b]{0.25\textheight}
      \centering
      \includegraphics[width=\textwidth]{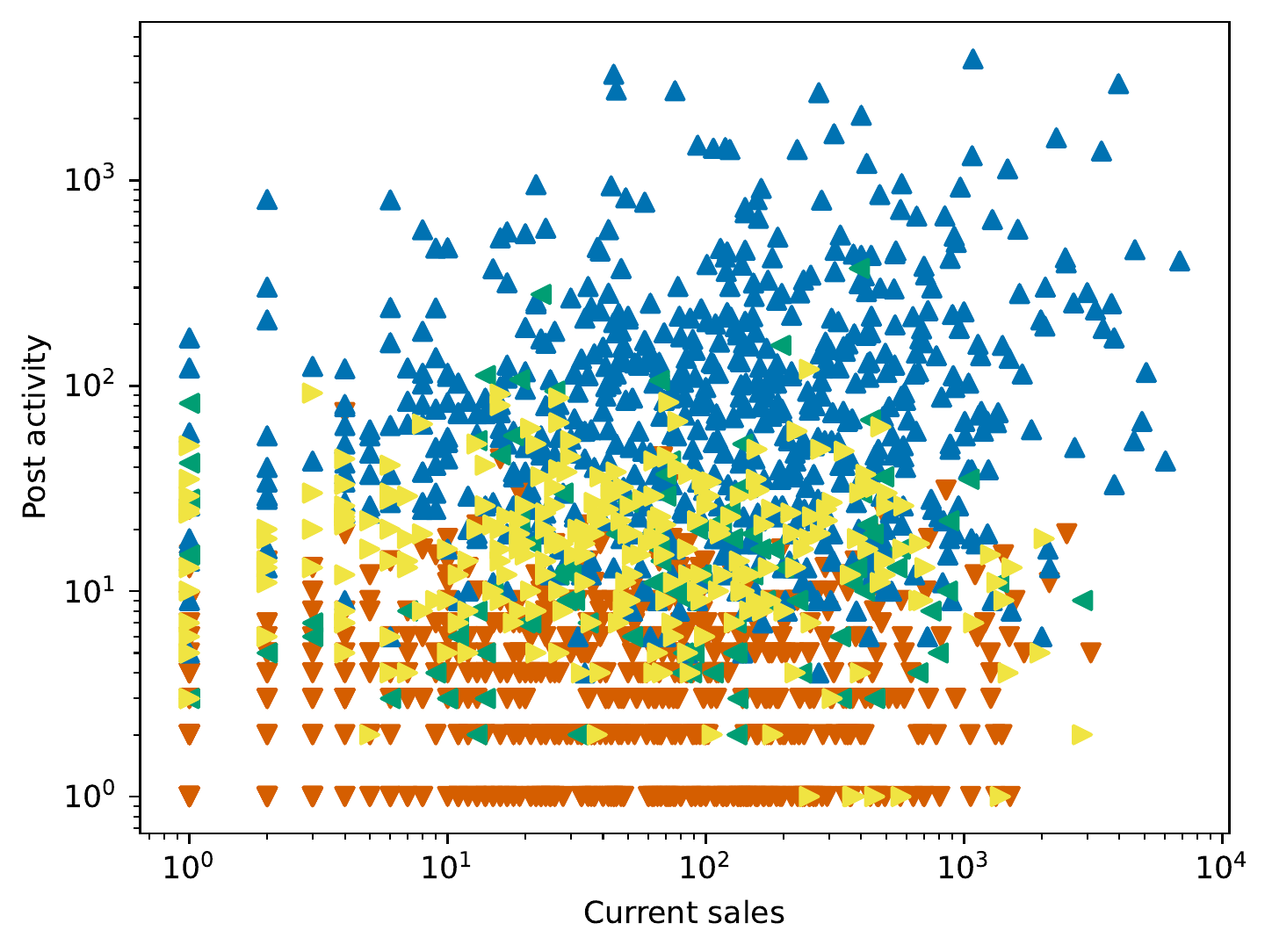}
      \caption{January 2015}
    \end{subfigure}
    ~
    \begin{subfigure}[b]{0.25\textheight}
      \centering
      \includegraphics[width=\textwidth]{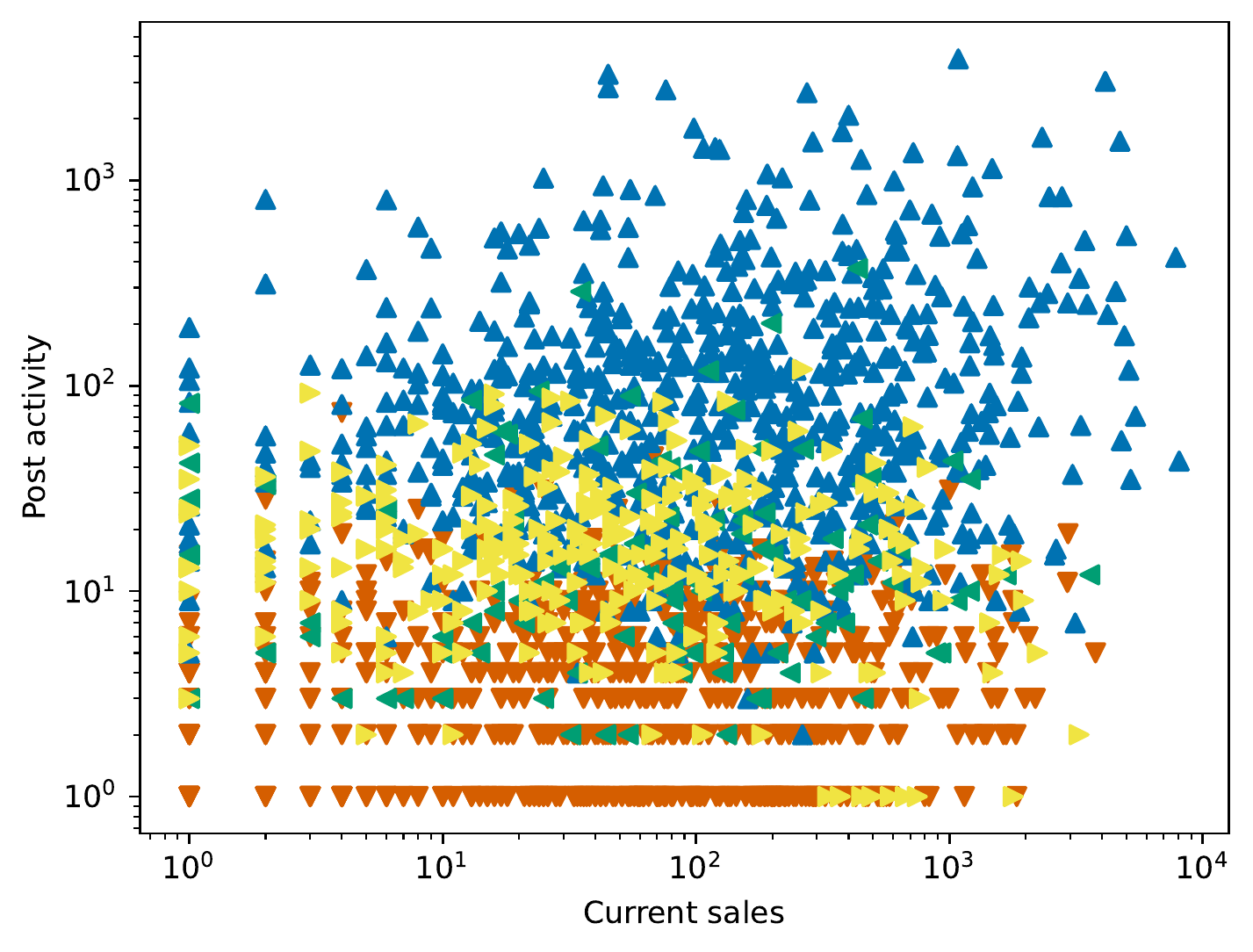}
      \caption{February 2015}
    \end{subfigure}
    ~
    \begin{subfigure}[b]{0.25\textheight}
      \centering
      \includegraphics[width=\textwidth]{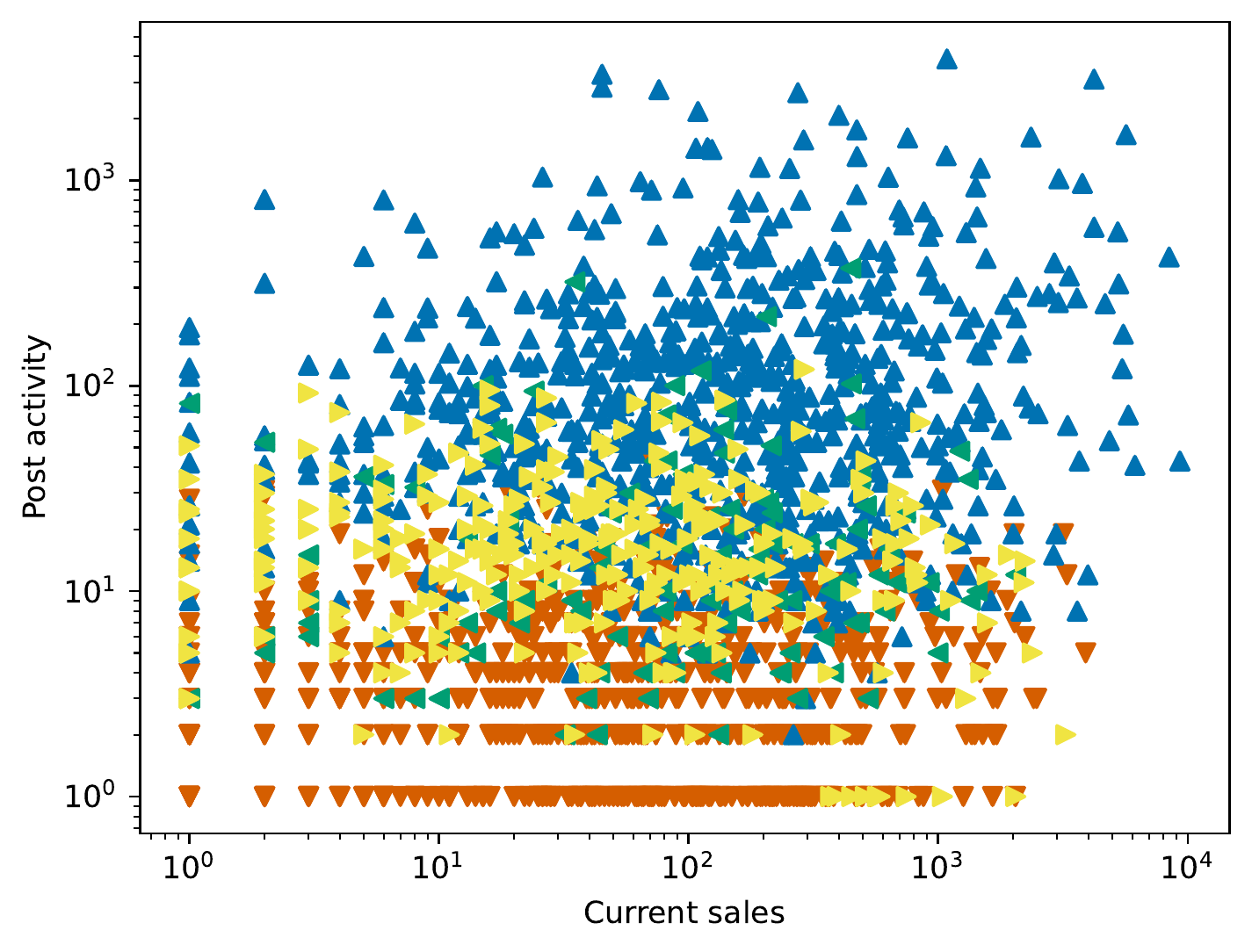}
      \caption{March 2015}
    \end{subfigure}
  \caption{Current sales and post activity of recalled (in top 20\%) and non-recalled (outside top 20\%) users for topic engagement, betweenness centrality, and their intersection, for each month.} \label{fig:overview2}
\end{figure}
\end{landscape}

\begin{landscape}
\begin{figure}[t]
  \centering
    \begin{subfigure}[b]{0.25\textheight}
      \centering
      \includegraphics[width=\textwidth]{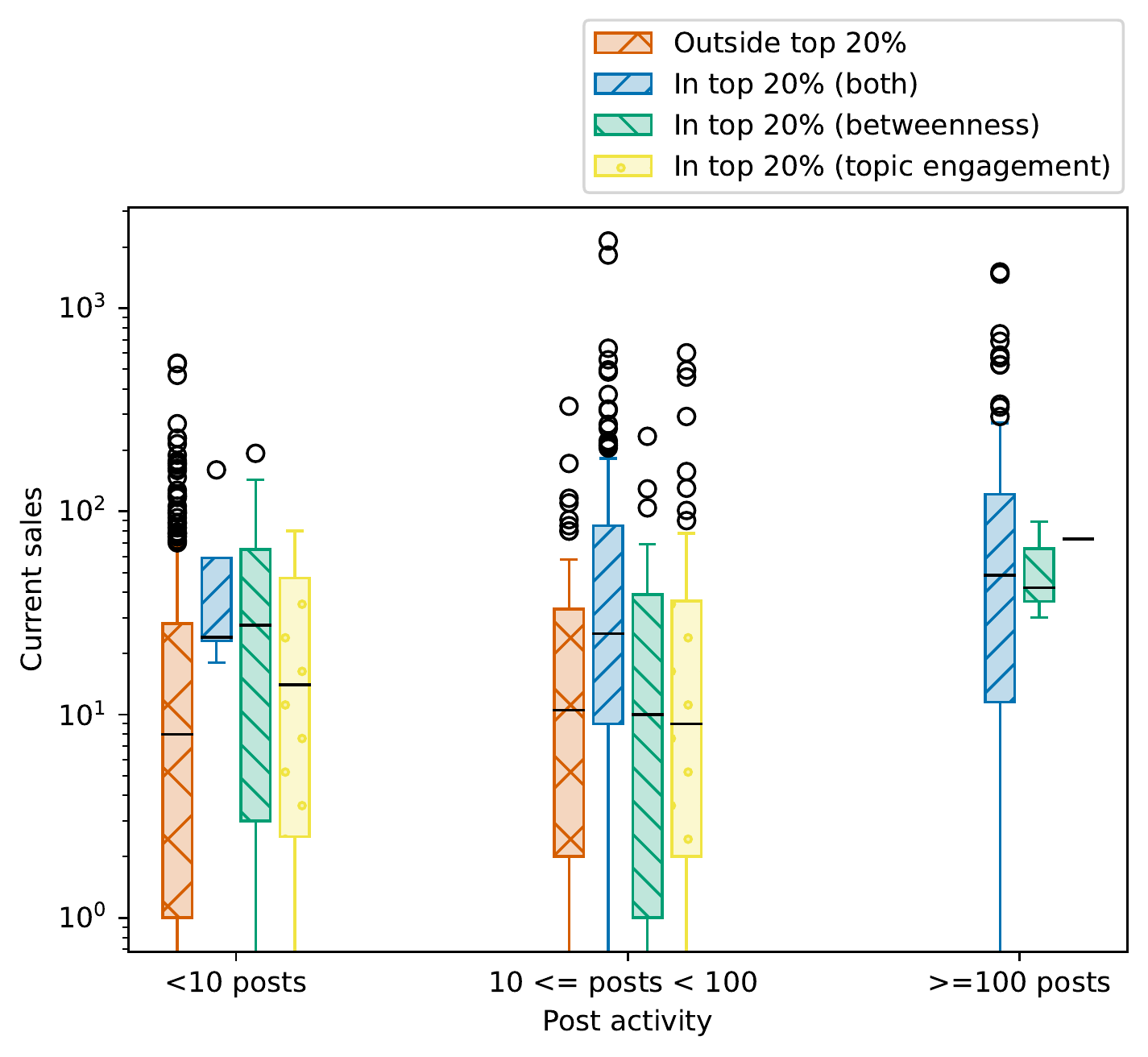}
      \caption{June 2014}
    \end{subfigure}
    ~
    \begin{subfigure}[b]{0.25\textheight}
      \centering
      \includegraphics[width=\textwidth]{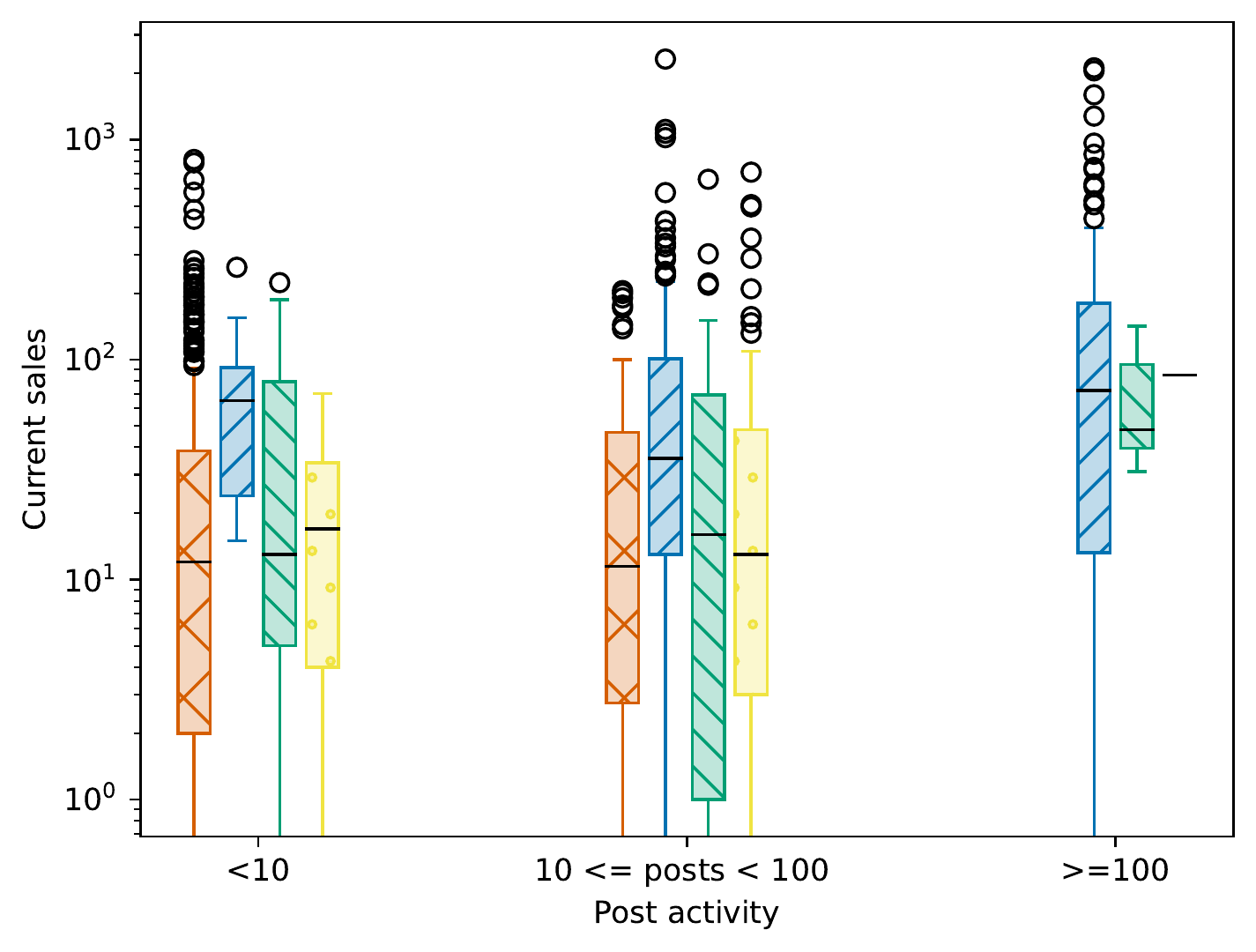}
      \caption{July 2014}
    \end{subfigure}
    ~
    \begin{subfigure}[b]{0.25\textheight}
      \centering
      \includegraphics[width=\textwidth]{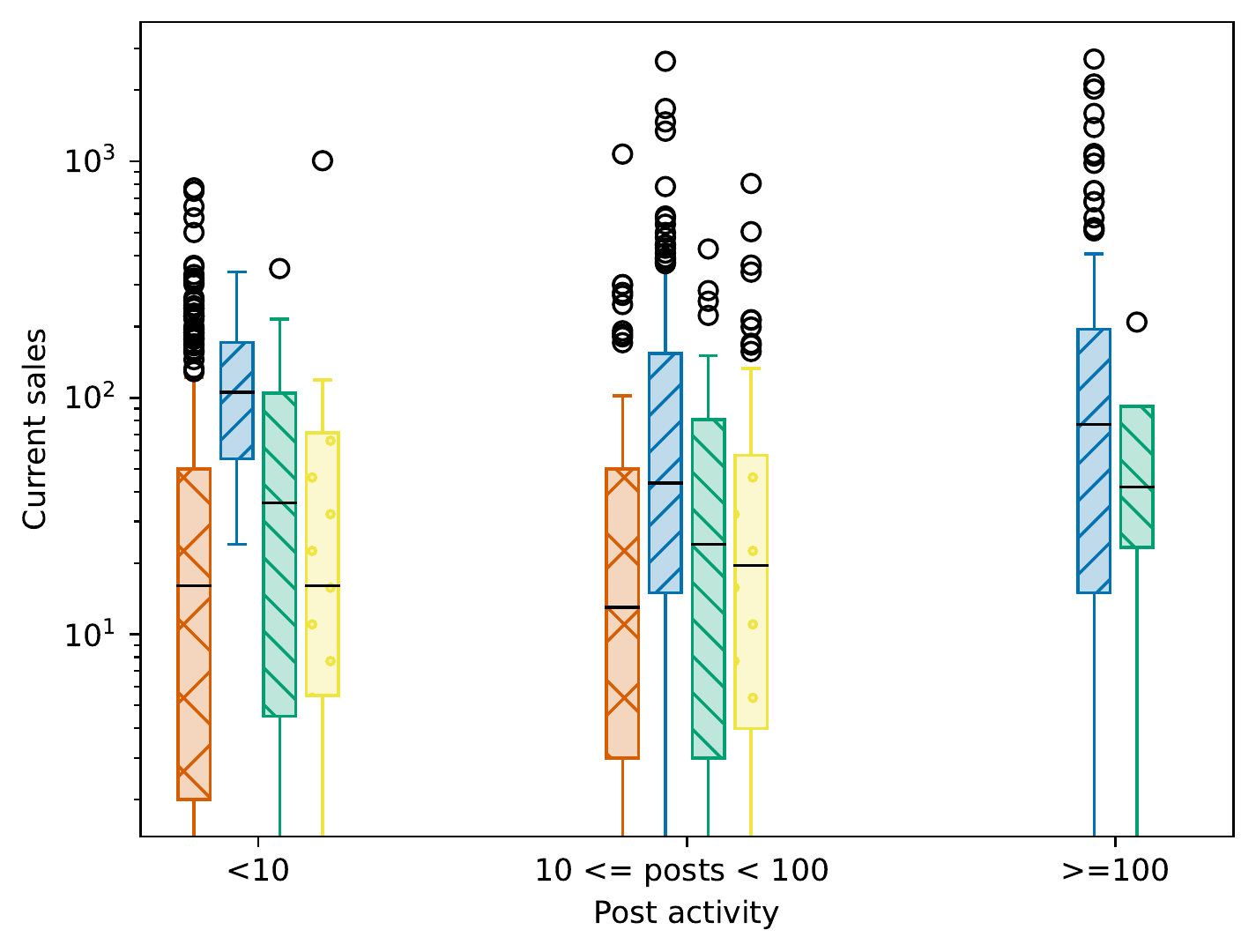}
      \caption{August 2014}
    \end{subfigure}
    ~
    \begin{subfigure}[b]{0.25\textheight}
      \centering
      \includegraphics[width=\textwidth]{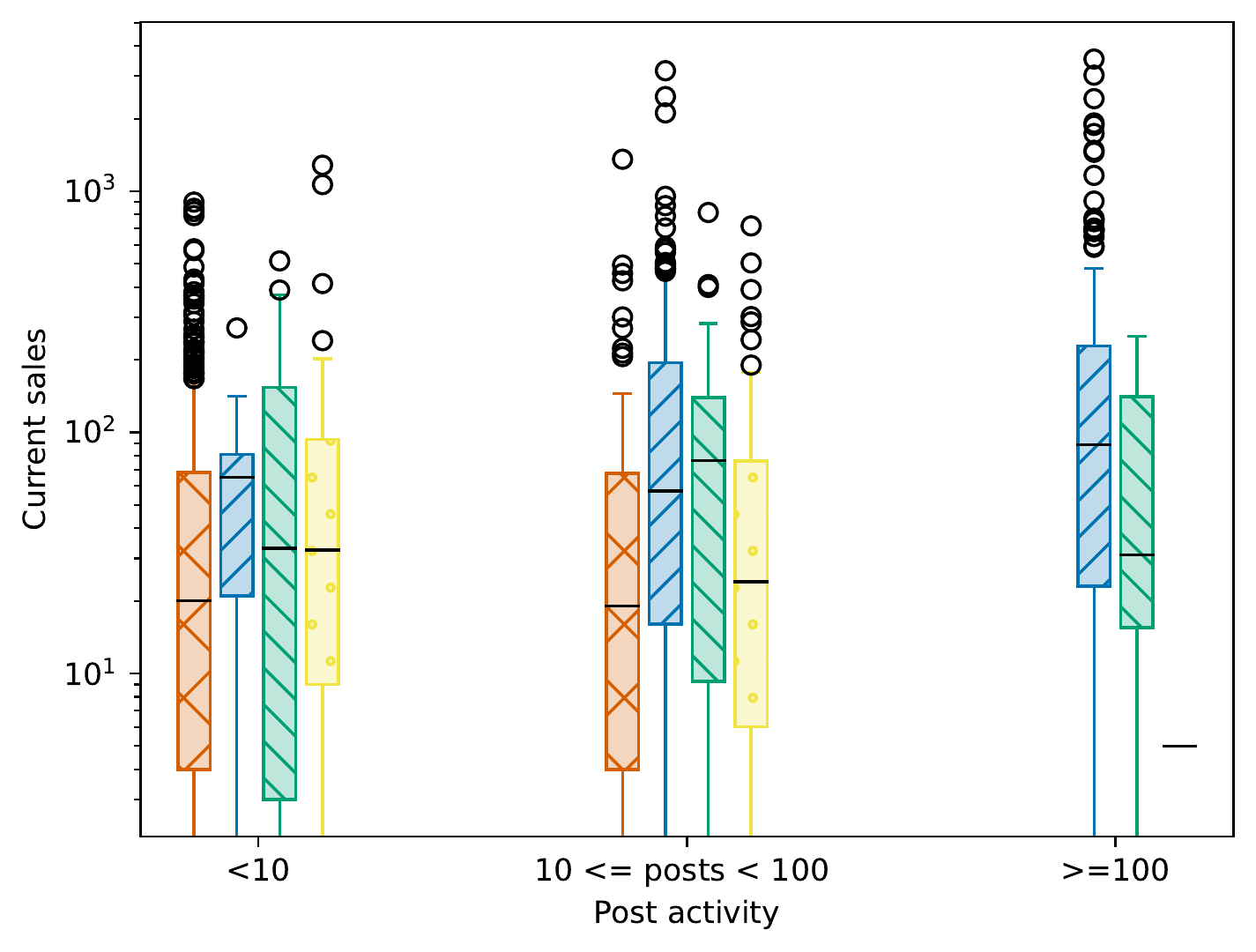}
      \caption{September 2014}
    \end{subfigure}
    ~
    \begin{subfigure}[b]{0.25\textheight}
      \centering
      \includegraphics[width=\textwidth]{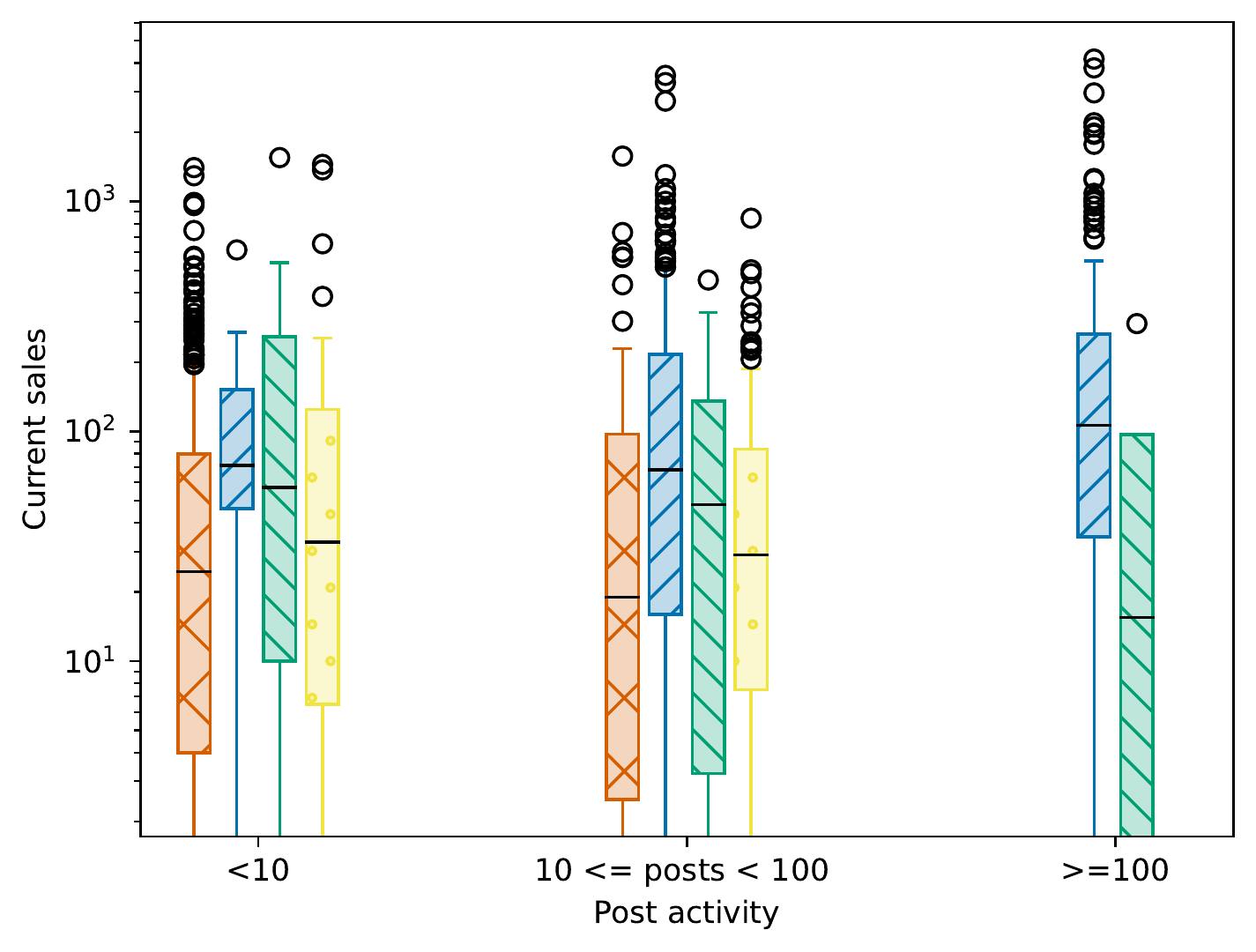}
      \caption{October 2014}
    \end{subfigure}
    \begin{subfigure}[b]{0.25\textheight}
      \centering
      \includegraphics[width=\textwidth]{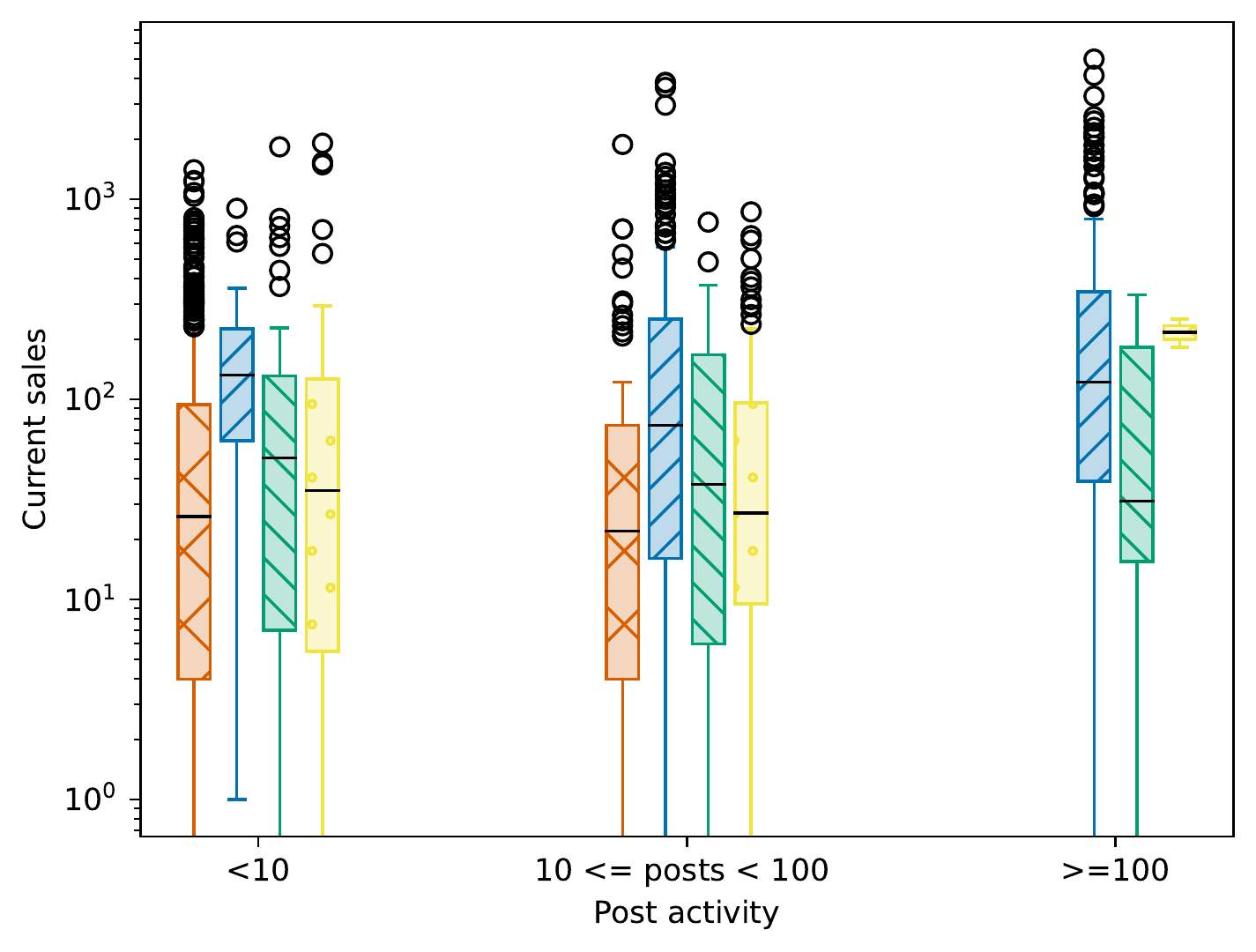}
      \caption{November 2014}
    \end{subfigure}
    ~
    \begin{subfigure}[b]{0.25\textheight}
      \centering
      \includegraphics[width=\textwidth]{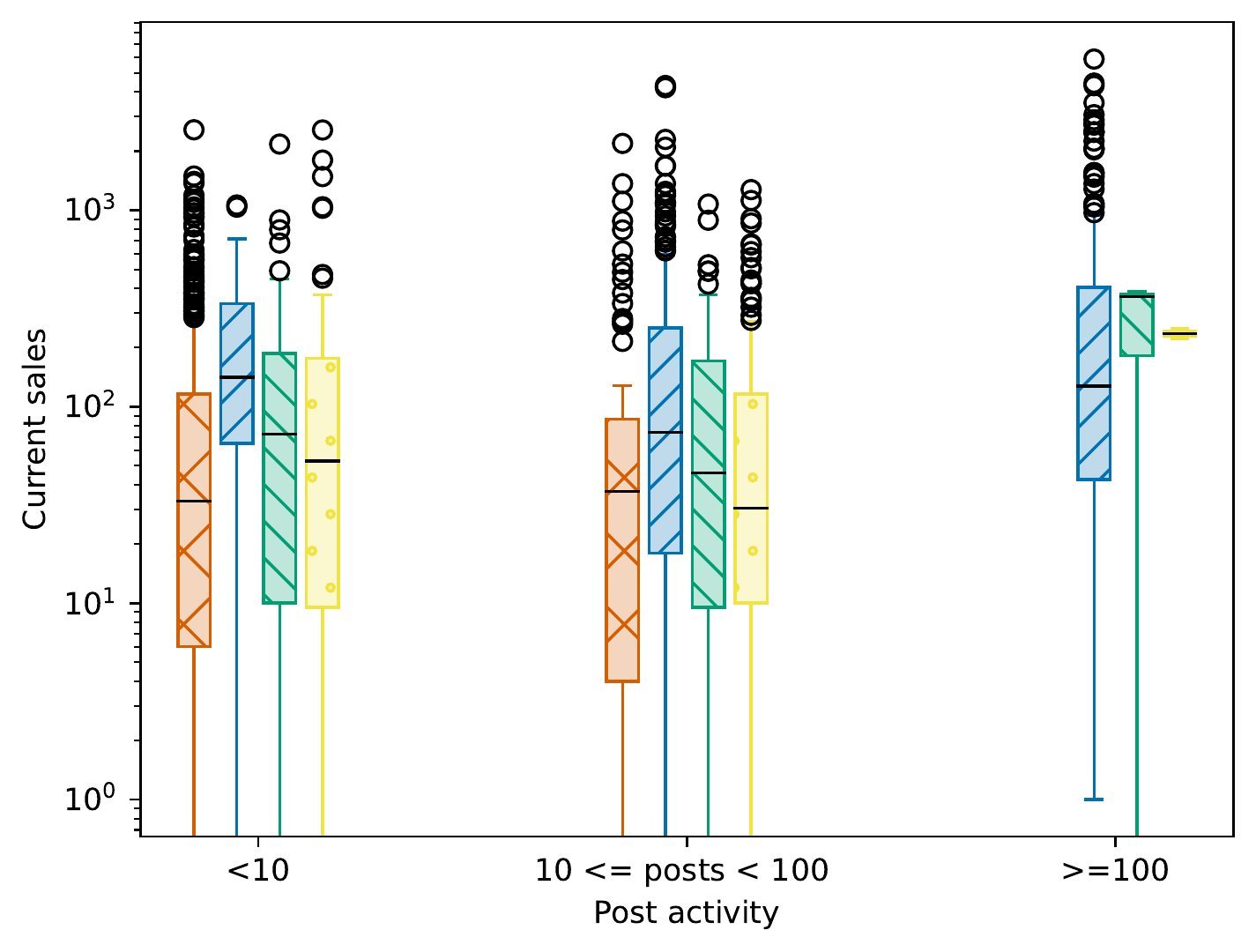}
      \caption{December 2014}
    \end{subfigure}
    ~
    \begin{subfigure}[b]{0.25\textheight}
      \centering
      \includegraphics[width=\textwidth]{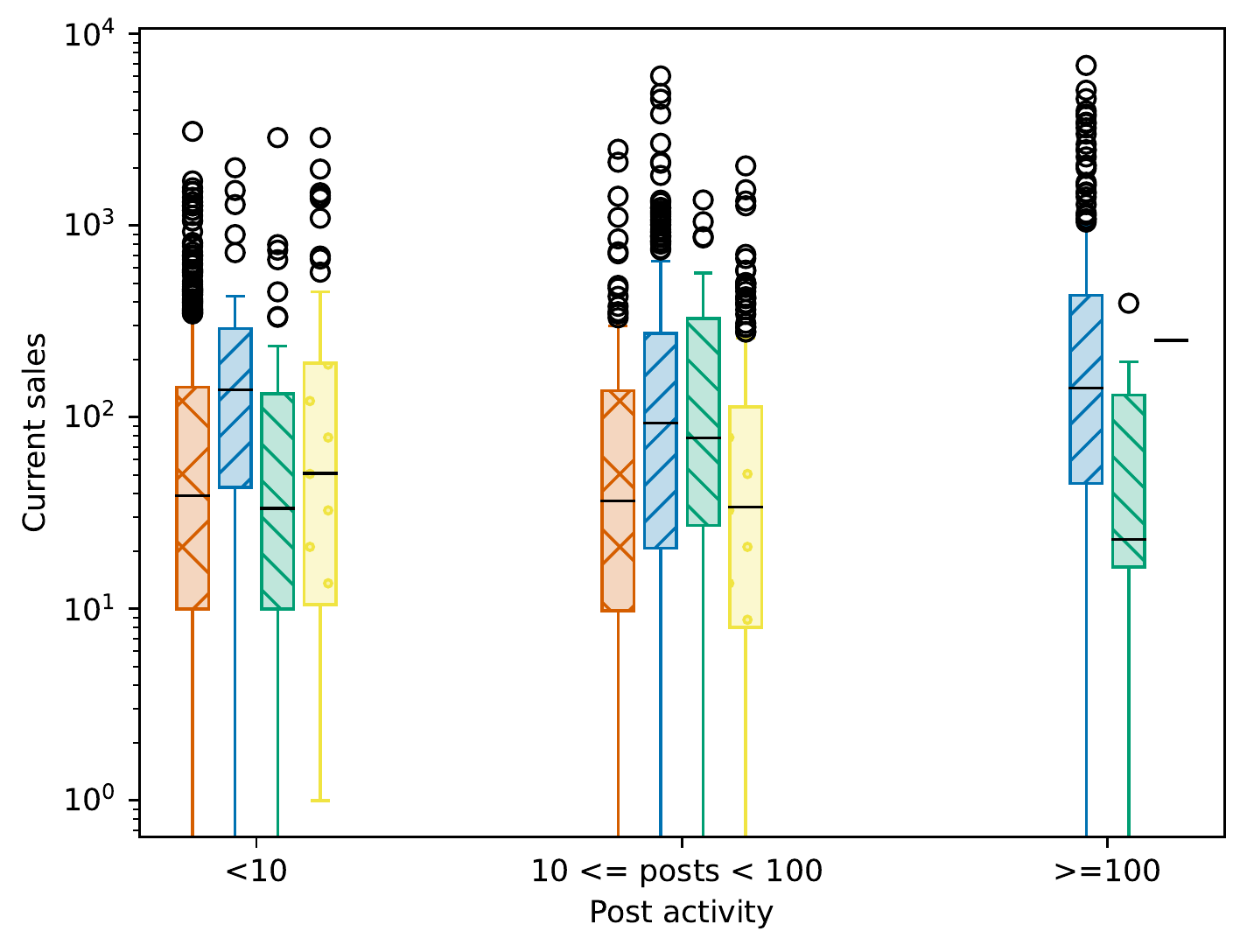}
      \caption{January 2015}
    \end{subfigure}
    ~
    \begin{subfigure}[b]{0.25\textheight}
      \centering
      \includegraphics[width=\textwidth]{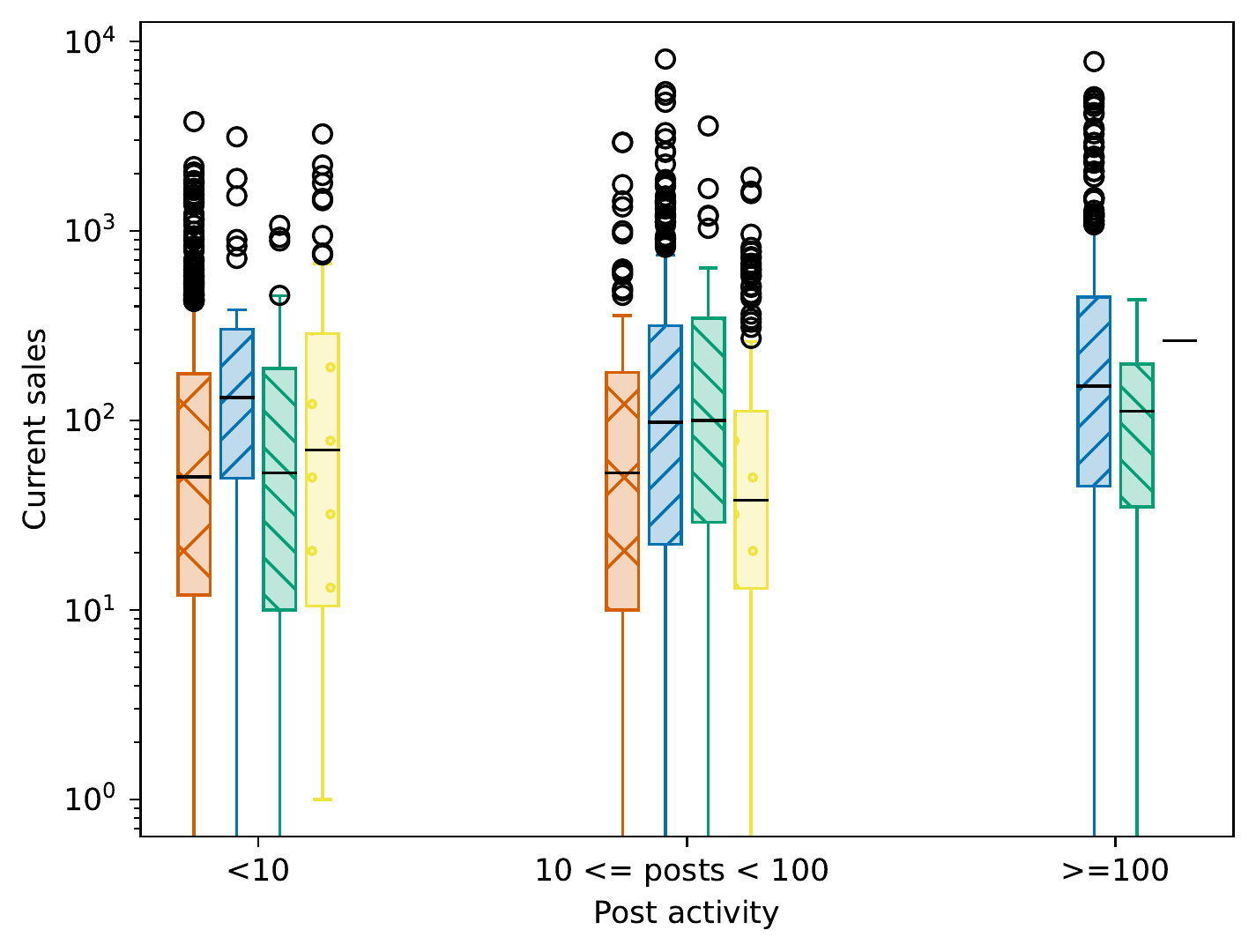}
      \caption{February 2015}
    \end{subfigure}
    ~
    \begin{subfigure}[b]{0.25\textheight}
      \centering
      \includegraphics[width=\textwidth]{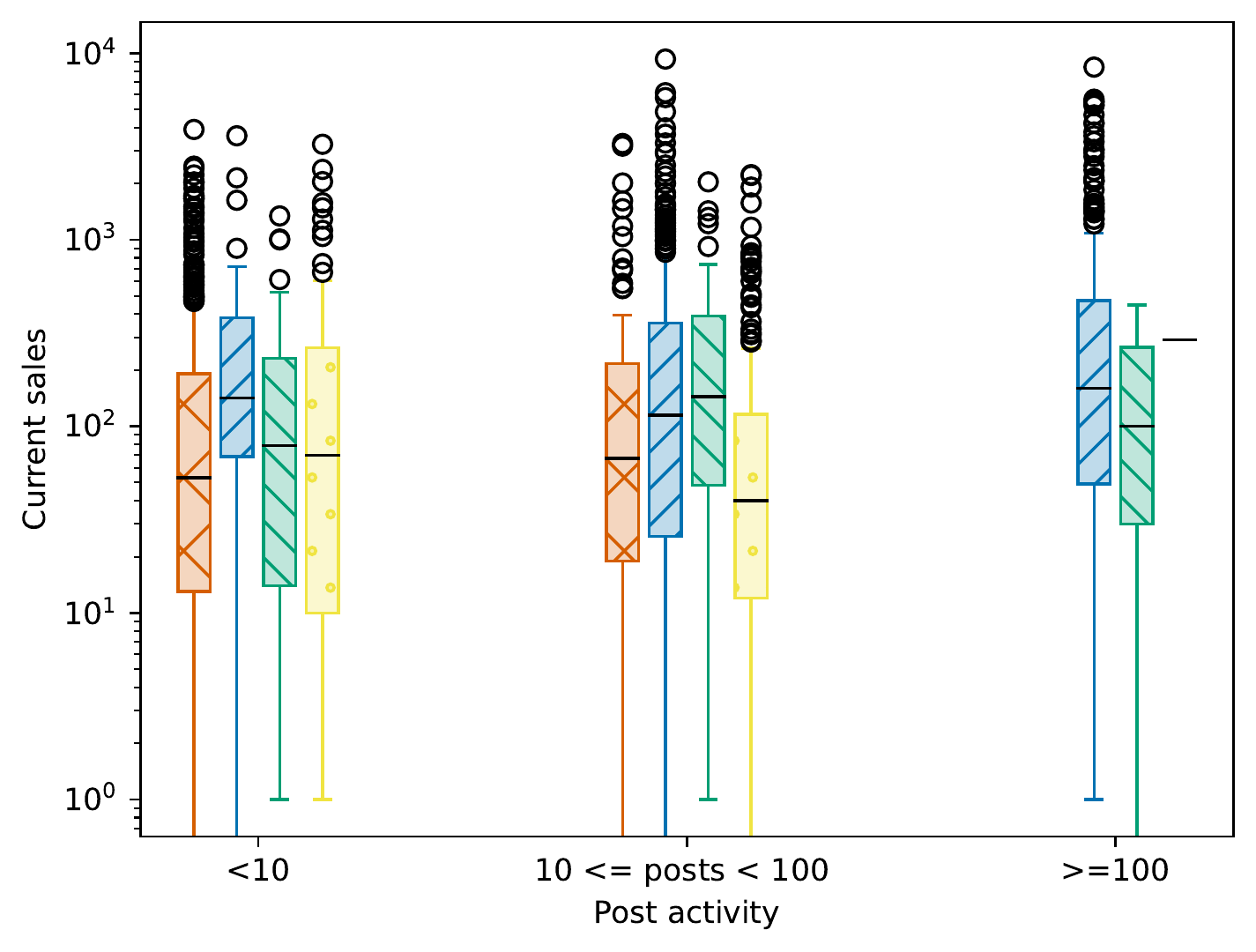}
      \caption{March 2015}
    \end{subfigure}
  \caption{Current sales and post activity of recalled (in top 20\%) and non-recalled (outside top 20\%) users for topic engagement, betweenness centrality, and their intersection, for each month.} \label{fig:overview3}
\end{figure}
\end{landscape}

\begin{landscape}
\begin{figure}[t]
  \centering
    \begin{subfigure}[b]{0.25\textheight}
      \centering
      \includegraphics[width=\textwidth]{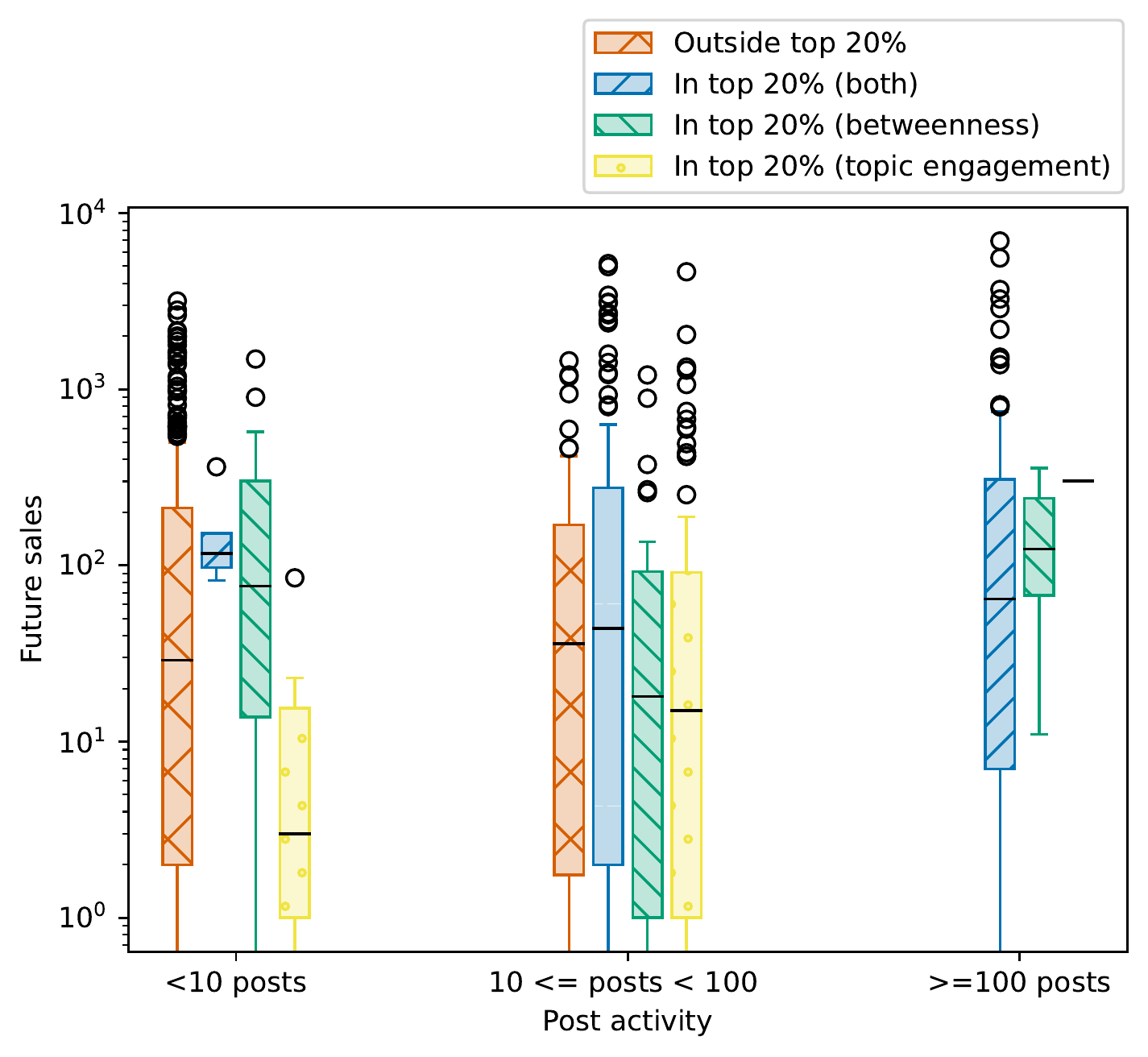}
      \caption{June 2014}
    \end{subfigure}
    ~
    \begin{subfigure}[b]{0.25\textheight}
      \centering
      \includegraphics[width=\textwidth]{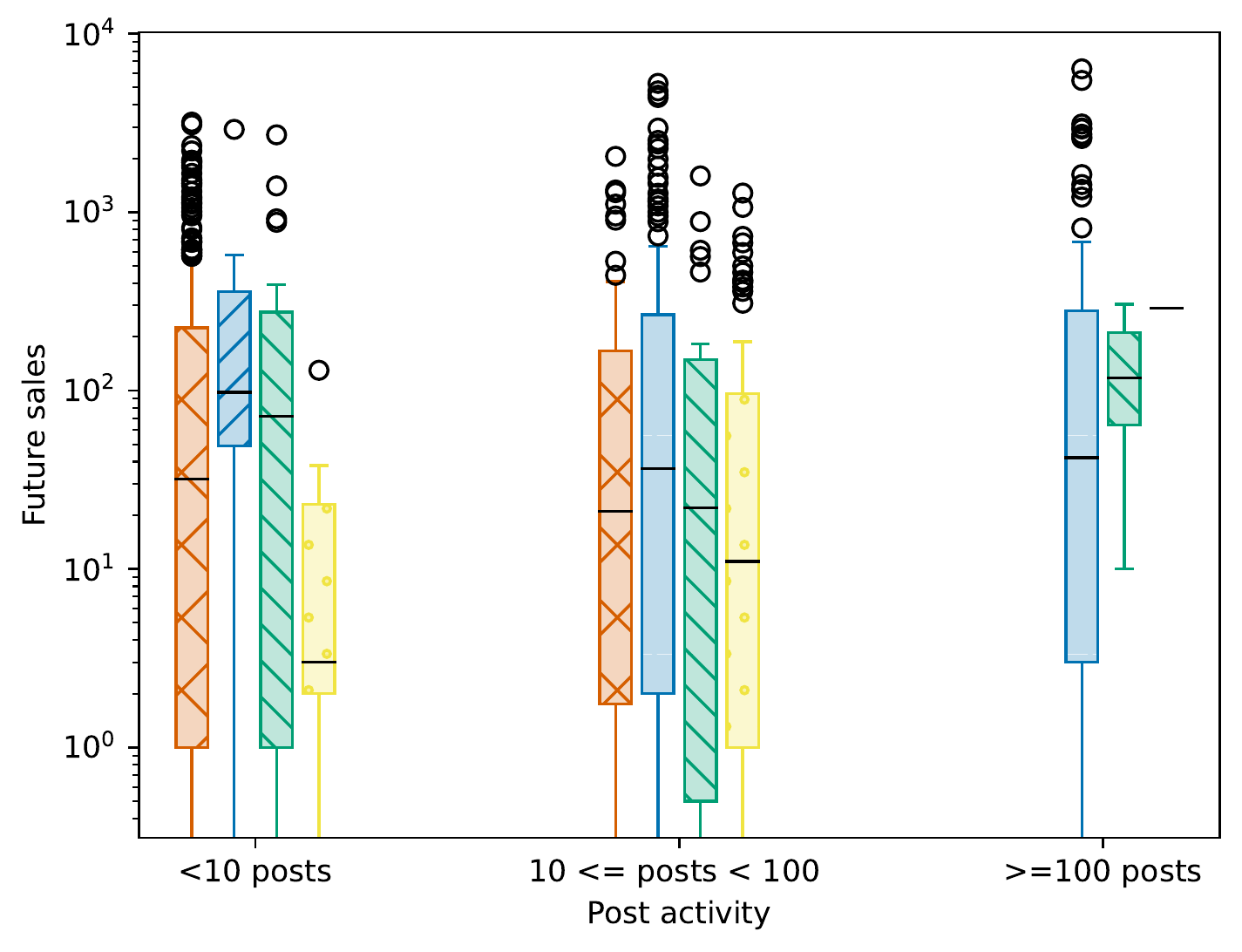}
      \caption{July 2014}
    \end{subfigure}
    ~
    \begin{subfigure}[b]{0.25\textheight}
      \centering
      \includegraphics[width=\textwidth]{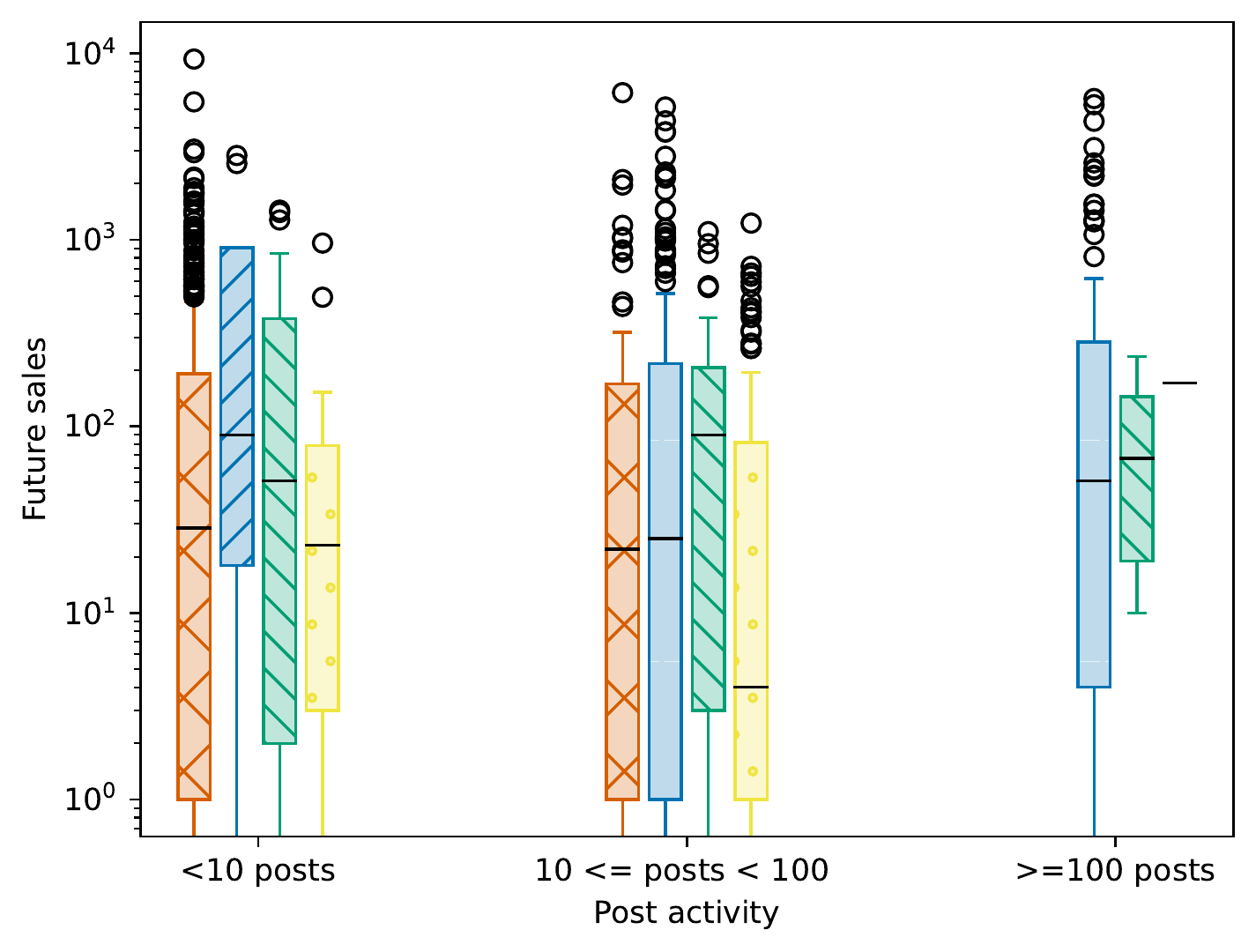}
      \caption{August 2014}
    \end{subfigure}
    ~
    \begin{subfigure}[b]{0.25\textheight}
      \centering
      \includegraphics[width=\textwidth]{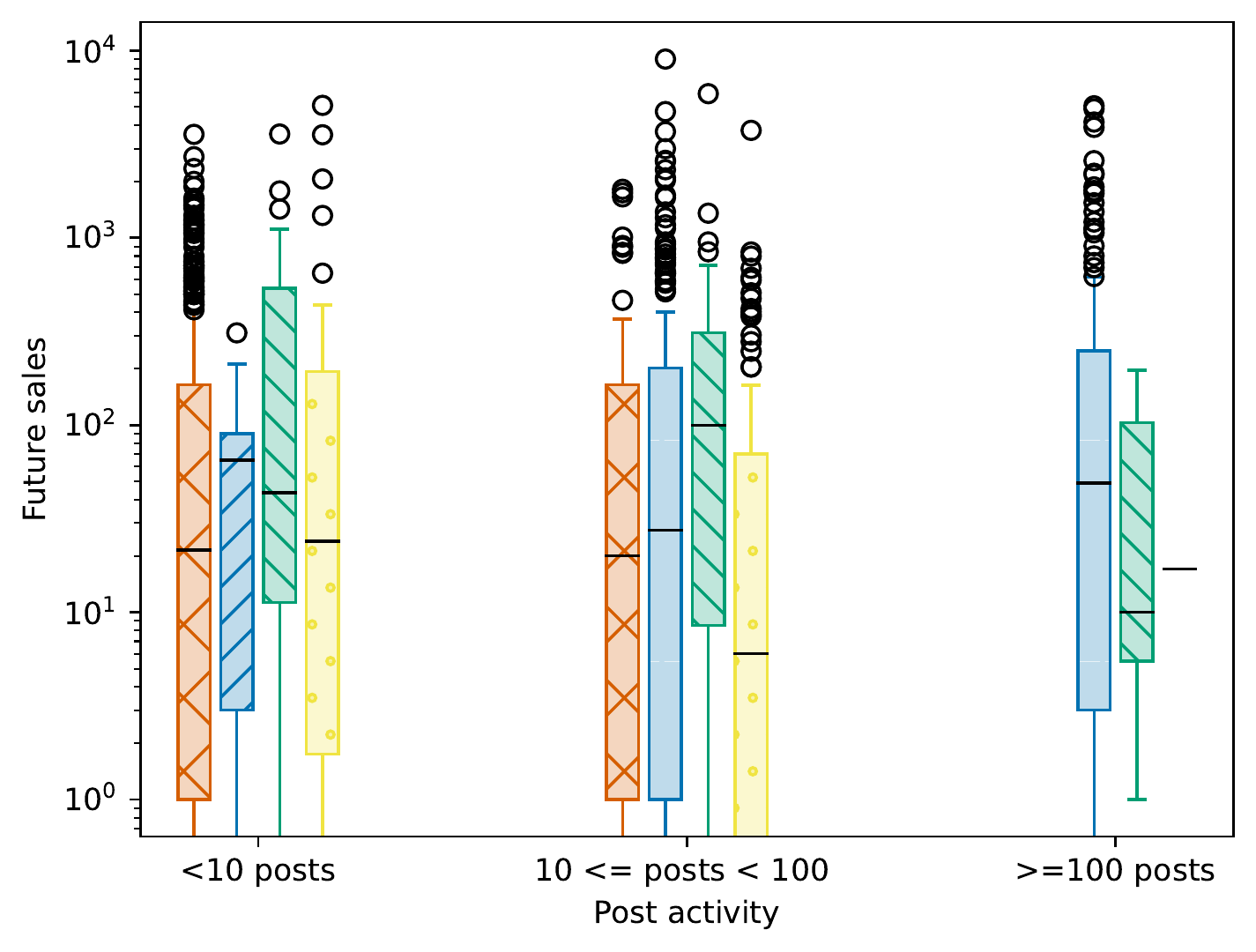}
      \caption{September 2014}
    \end{subfigure}
    ~
    \begin{subfigure}[b]{0.25\textheight}
      \centering
      \includegraphics[width=\textwidth]{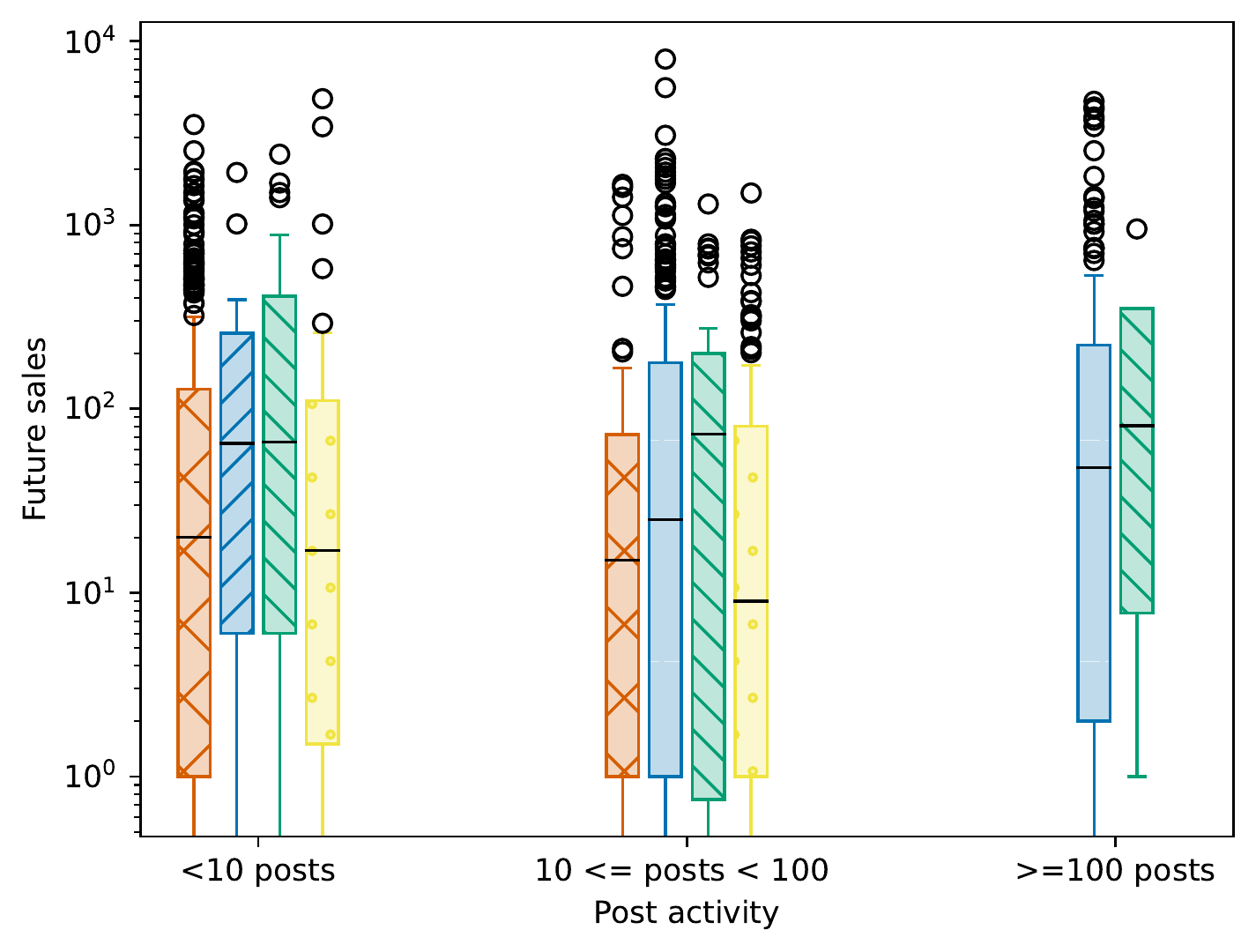}
      \caption{October 2014}
    \end{subfigure}
    \begin{subfigure}[b]{0.25\textheight}
      \centering
      \includegraphics[width=\textwidth]{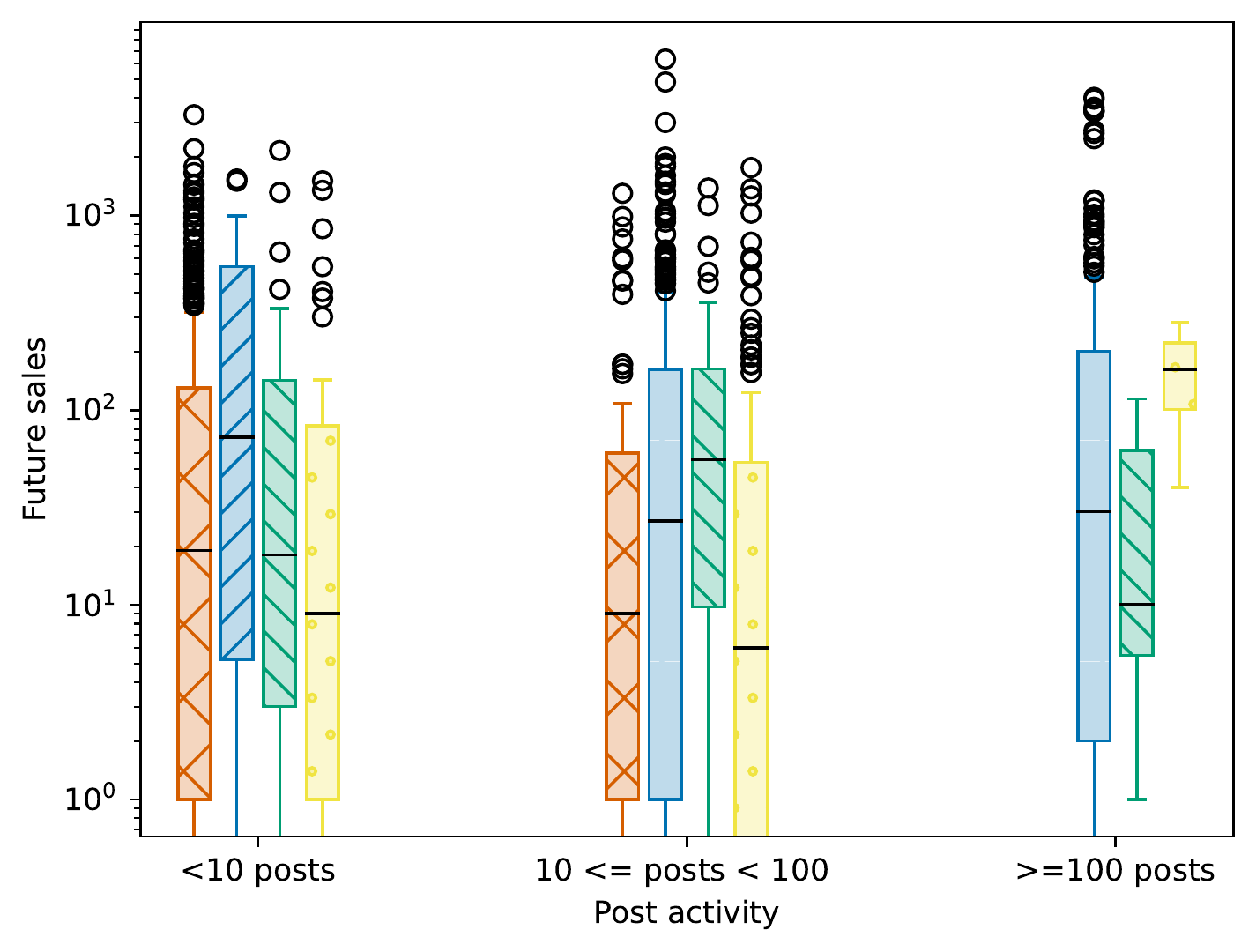}
      \caption{November 2014}
    \end{subfigure}
    ~
    \begin{subfigure}[b]{0.25\textheight}
      \centering
      \includegraphics[width=\textwidth]{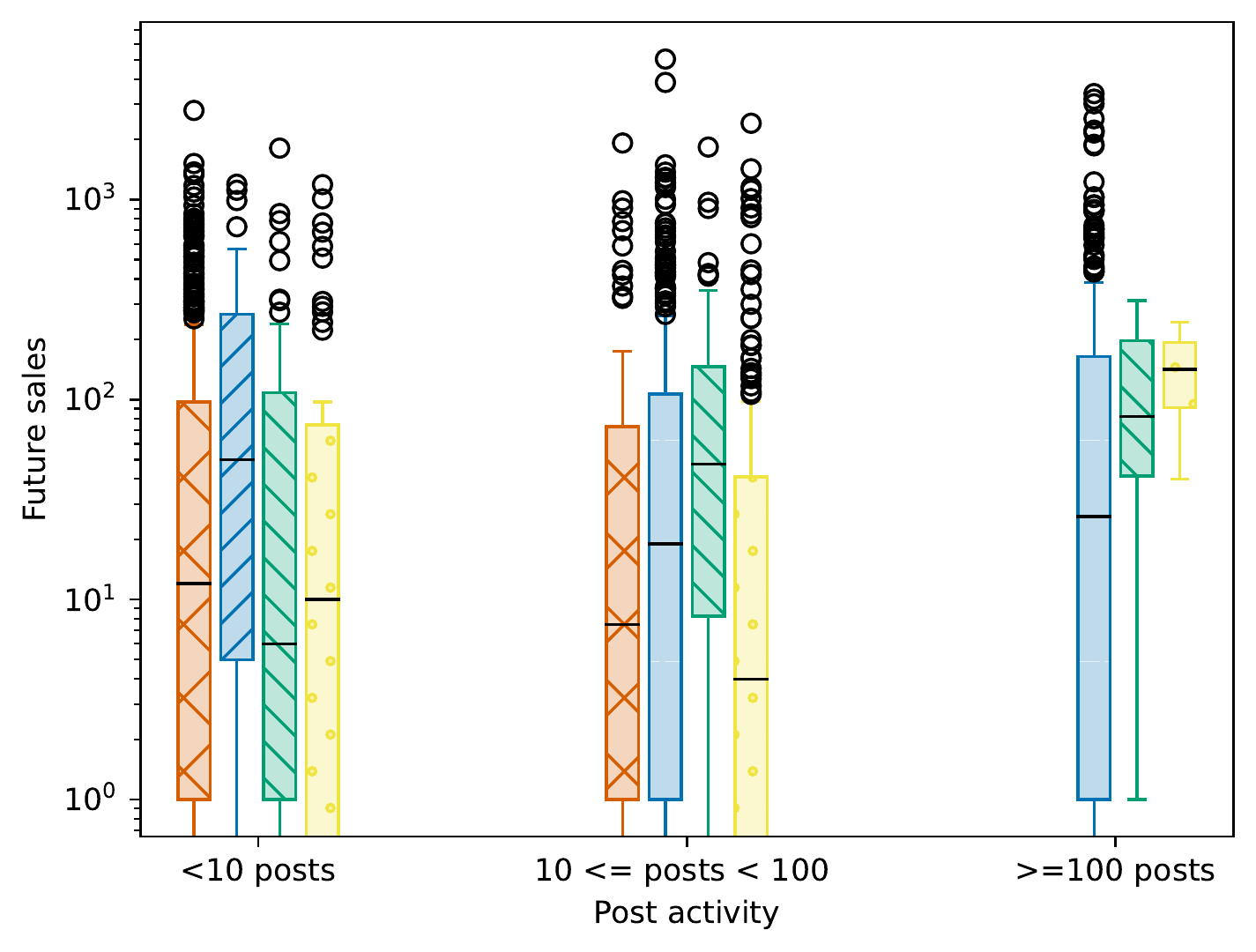}
      \caption{December 2014}
    \end{subfigure}
    ~
    \begin{subfigure}[b]{0.25\textheight}
      \centering
      \includegraphics[width=\textwidth]{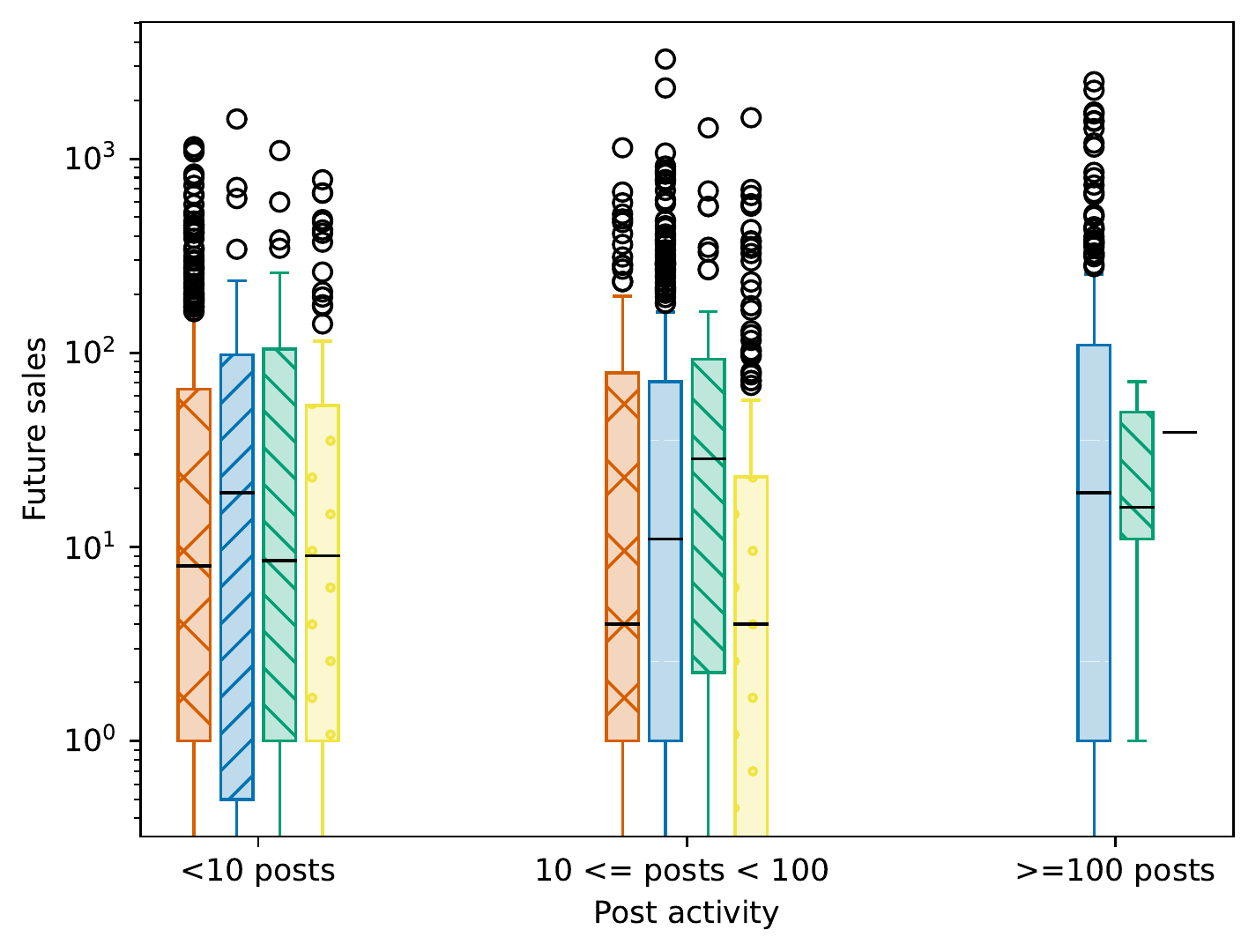}
      \caption{January 2015}
    \end{subfigure}
    ~
    \begin{subfigure}[b]{0.25\textheight}
      \centering
      \includegraphics[width=\textwidth]{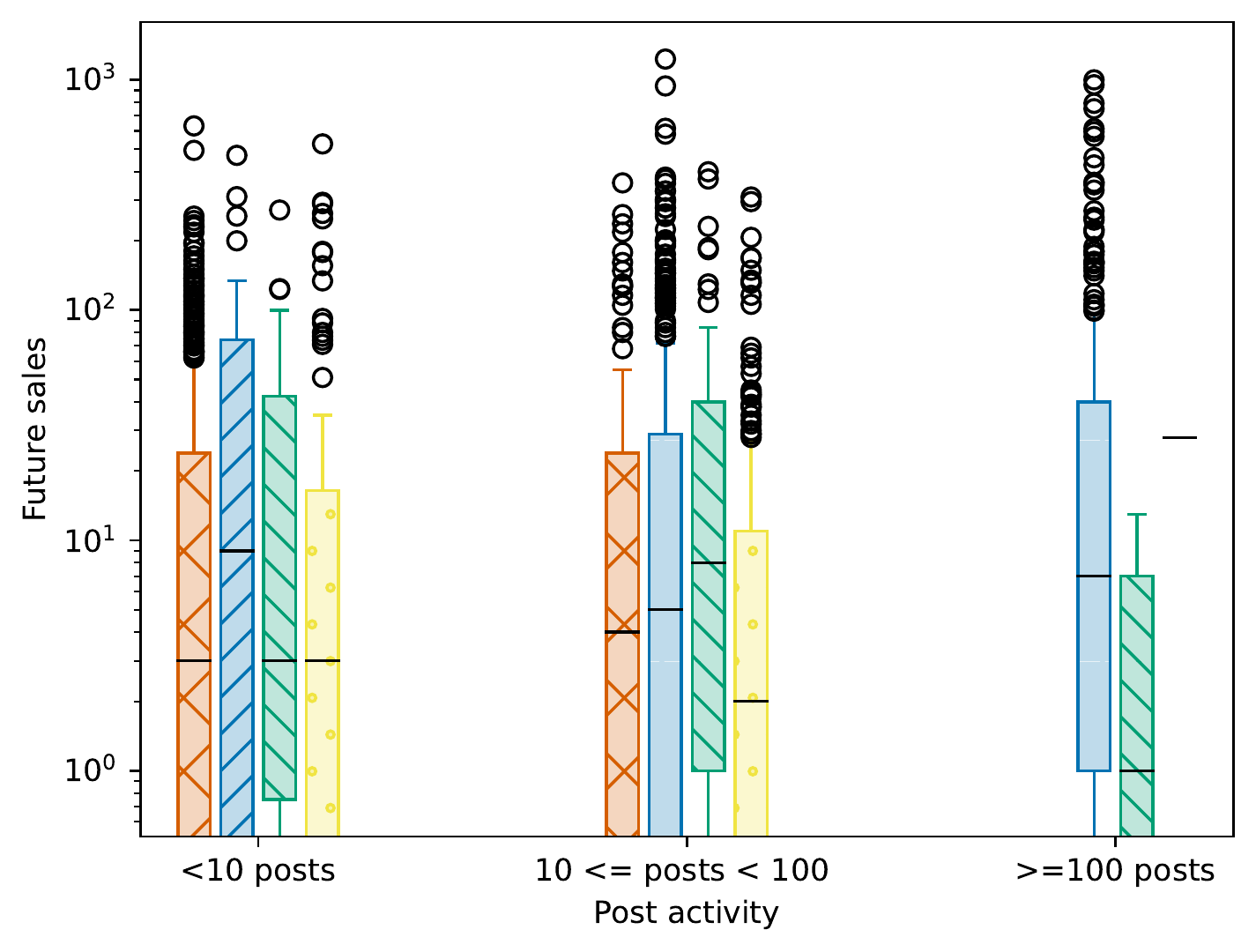}
      \caption{February 2015}
    \end{subfigure}
  \caption{Current sales and post activity of recalled (in top 20\%) and non-recalled (outside top 20\%) users for topic engagement, betweenness centrality, and their intersection, for each month.} \label{fig:overview4}
\end{figure}
\end{landscape}

Here, we explore the sales and post activity of recalled and non-recalled vendors for the various network centrality measures (see Figure~\ref{fig:overview1}) and months (see Figures~\ref{fig:overview2},~\ref{fig:overview3} and~\ref{fig:overview4}).
We do so, to check if the conclusions drawn from the case study included in the main paper (Figure 3) are robust for the other months and if those conclusions also apply to the other centrality measures.

Figure~\ref{fig:overview1} confirms that, for all centrality measures, there exists a threshold of activity (around 100 posts for September 2014) beyond which users are always included.
Furthermore, we observe that in-degree, harmonic closeness centrality, and PageRank are far less inclined to include low activity vendors within the top 20\% of their rankings.
Although less obvious, for these centrality measures it does still appear to hold that greater success leads to greater odds of being included at lower activity.
Whereas for betweenness centrality we found that the vendors uniquely found were relatively less active and more successful, we find that only the latter holds for the other measures.
This lines up with our earlier finding from Table 1, that in-degree, harmonic closeness, and PageRank uniquely find very few vendors that are not likely to be found through their forum activity.

Figure~\ref{fig:overview2} shows that, throughout all months, a threshold of activity exists beyond which users are always included.
This threshold seems to slowly increase as the forum and the overall activity increases.
However, after October 2014 this increase seems marginal at best.
Furthermore, each month we observe the tendency for moderate activity vendors to be more likely to be included the more successful they are.
We confirm this in Figure~\ref{fig:overview3} by grouping vendors by their activity.
Here, we have excluded the first few months which we in the main paper found to be a development period for the network, before achieving reliable vendor recall.
We observe that indeed each month, though to a less significant extent for some, the average success of vendors uniquely found by betweenness exceeds that of those uniquely found by topic engagement.
Furthermore, Figure~\ref{fig:overview4} confirms our finding that this difference in average success is larger for future success.
Thus, we can conclude that our findings in the main paper (with regards to Figure 3) are robust beyond only September 2014.

\section{Performance at different thresholds}

The main paper presented evaluation metric results for a specific threshold for the measure rankings, namely 20\% of all users.
Here, we use ROC curves (receiver operating characteristic curve) to investigate the performance of the activity indicators and network measures at different thresholds.
Specifically, we computed statistics at each interval of 5\% in the range 0--100\%, checking the performance for predicting groups of vendors for each of our measures.
We consider the entire top vendor percentile, ``less active'' vendors among the top vendor percentile, all vendors, and all ``less active'' vendors.
We set the activity threshold for ``less active'' at fewer than 100 posts, above which all vendors were shown to be found for all centrality measures.
The resulting ROC curves, for September 2014, are shown in Figure~\ref{fig:roc-curves}.

\begin{figure}[h]
    \centering
    \includegraphics[width=0.8\textwidth]{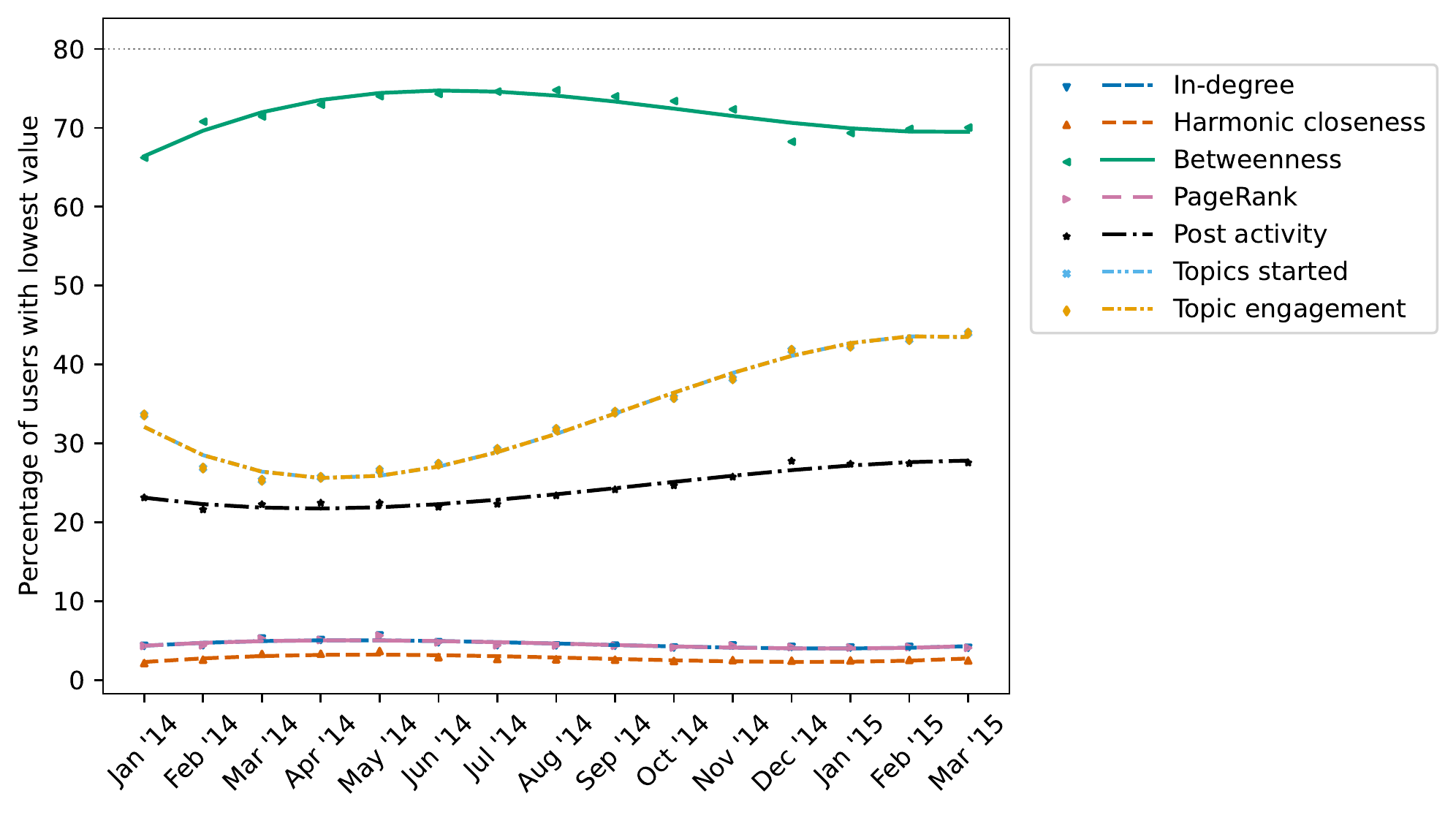}
    \caption{Percentage of users with the lowest value. Identical values indicate these users may have any random ordering, thus providing no additional predictive power. The dotted line at 80\% indicates the threshold that would need to be exceeded for these random orderings to impact our standard threshold for the measure rankings of 20\% of all users. None of the measures exceed this threshold for any of the snapshots.}
    \label{fig:perc-low}
\end{figure}

\begin{figure}[h]
  \centering
    \begin{subfigure}[b]{0.44\textwidth}
      \centering
      \includegraphics[width=\textwidth]{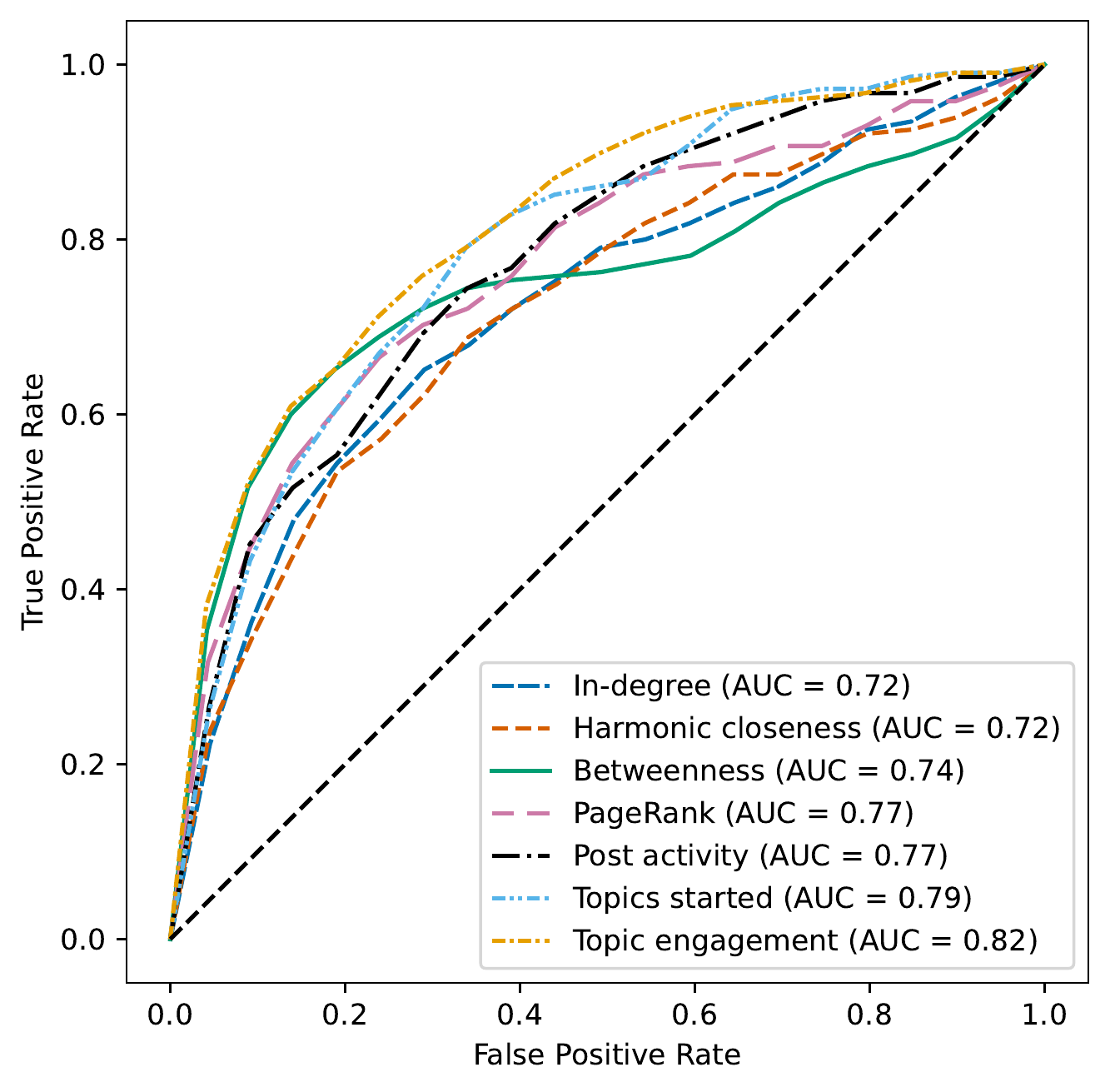}
      \caption{Top percentile vendors regardless of activity.}
    \end{subfigure}
    ~
    \begin{subfigure}[b]{0.44\textwidth}
      \centering
      \includegraphics[width=\textwidth]{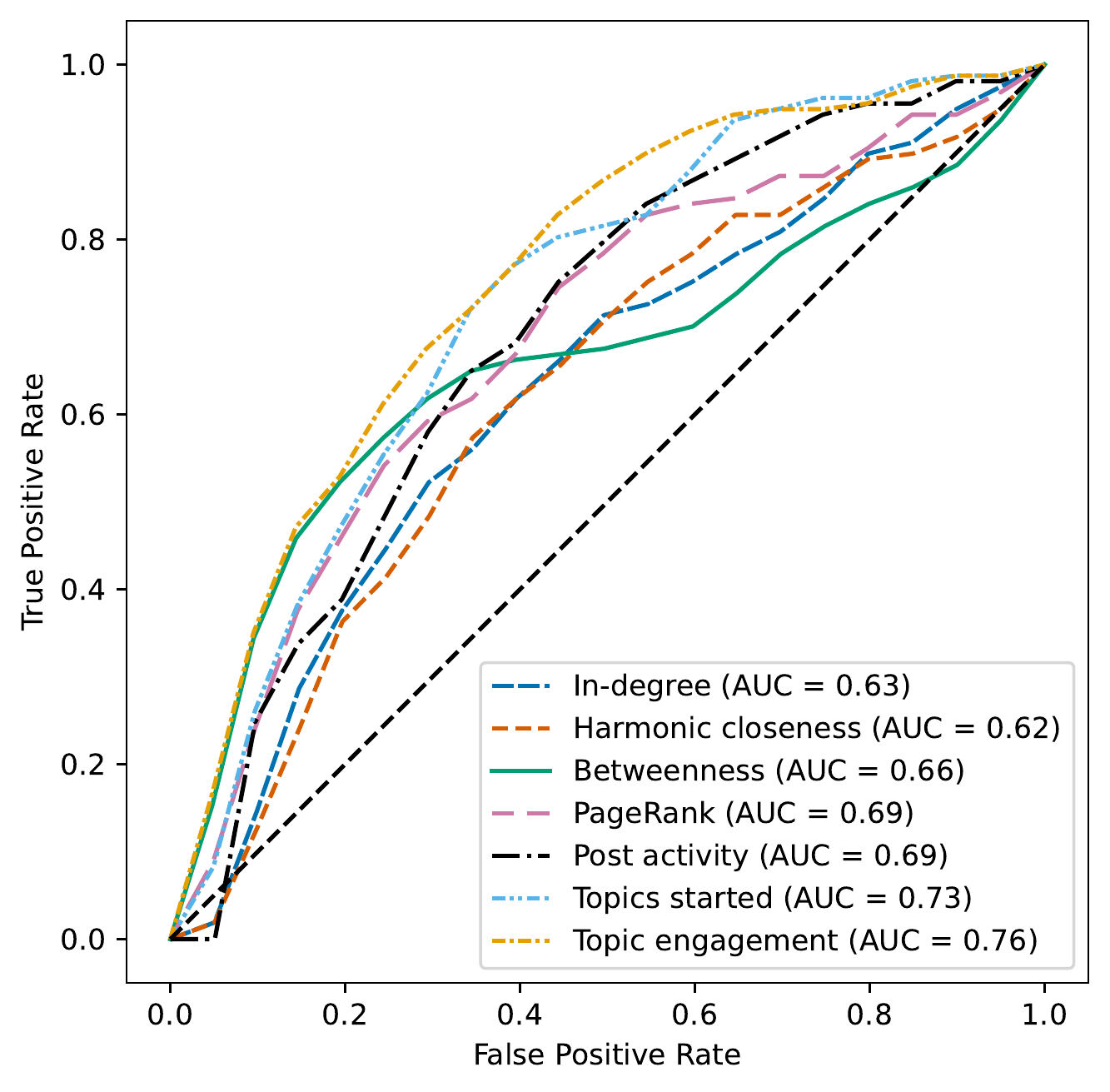}
      \caption{Top percentile vendors with fewer than 100 posts}
    \end{subfigure}
    \begin{subfigure}[b]{0.44\textwidth}
      \centering
      \includegraphics[width=\textwidth]{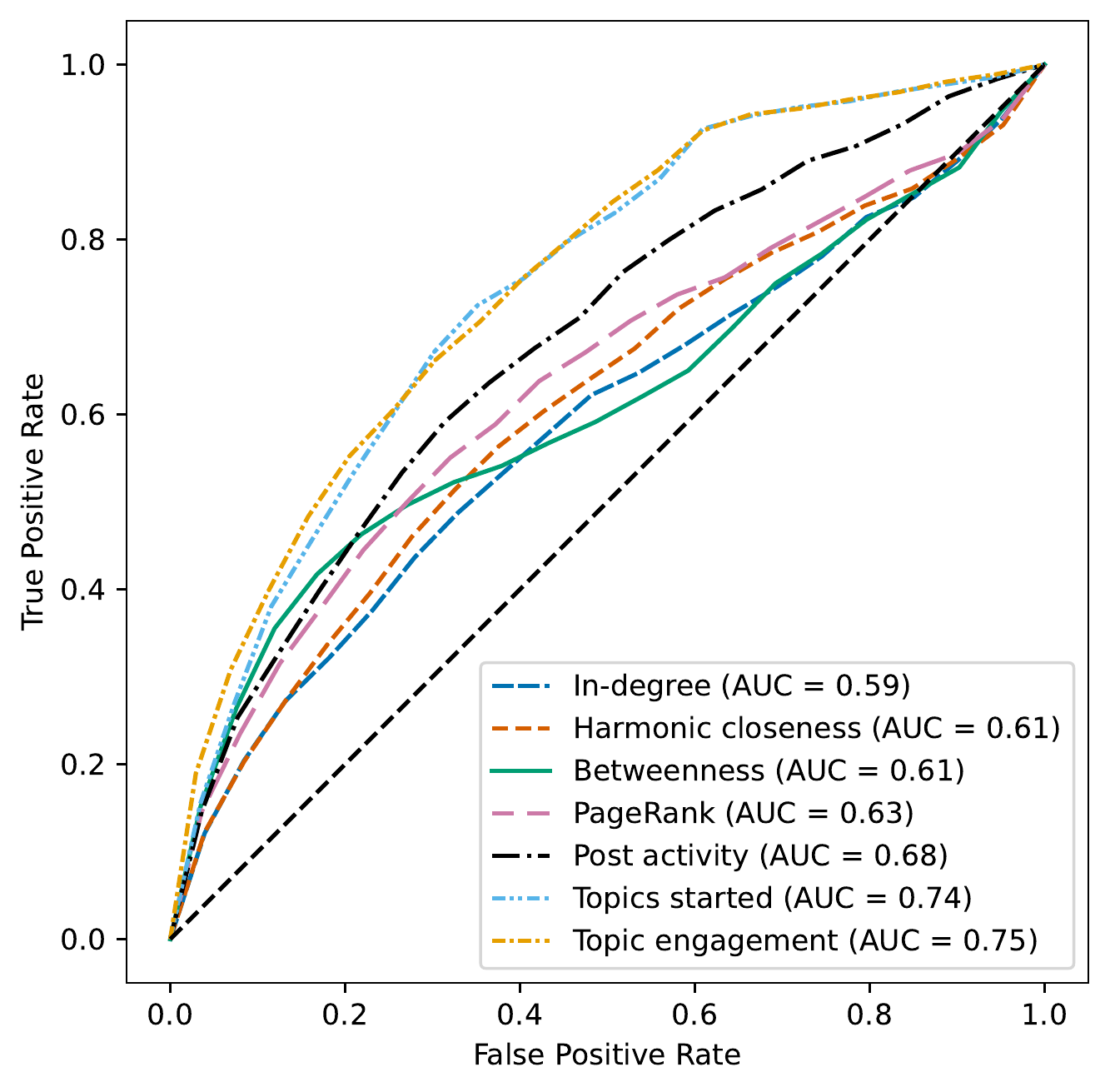}
      \caption{All vendors regardless of activity.}
    \end{subfigure}
    ~
    \begin{subfigure}[b]{0.44\textwidth}
      \centering
      \includegraphics[width=\textwidth]{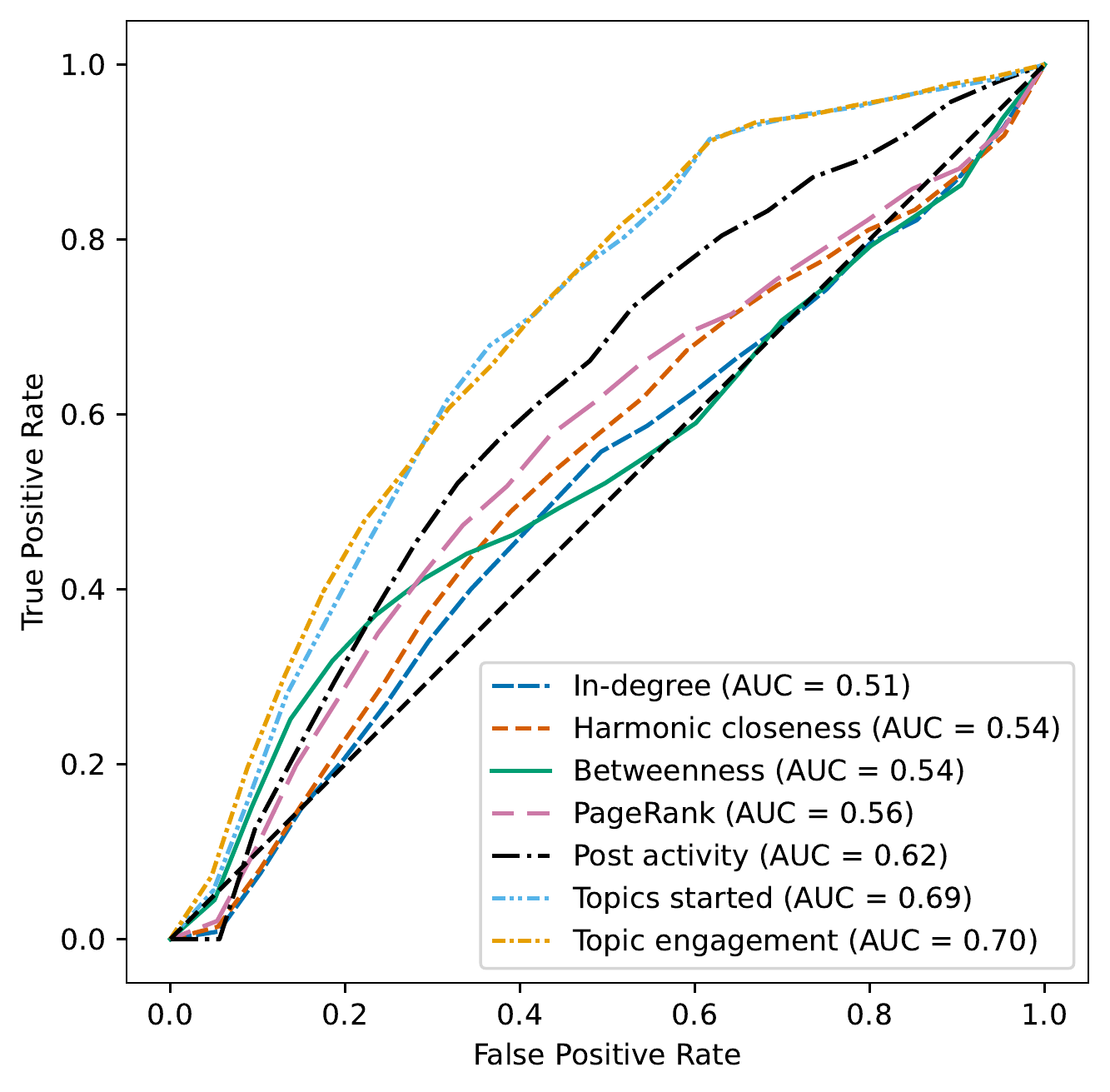}
      \caption{All vendors with fewer than 100 posts}
    \end{subfigure}
    \caption{ROC curves  for September 2014, predicting groups of vendors.}
    \label{fig:roc-curves}
\end{figure}

For the top vendor percentile (Figures~\ref{fig:roc-curves}a,b), we see that topic engagement and betweenness centrality achieve a similar true positive rate up to a false positive rate of 20\%.
Afterwards, the topic engagement clearly outperforms all centrality measures.
One factor contributing to the poor performance of betweenness centrality at false positive rates above 20\%, is that many users get the same lowest betweenness value of zero.
Since identically scoring users are essentially randomly ordered, their ordering does not add any predictive power.
In Figure~\ref{fig:perc-low} we show that for each monthly snapshot around 70\% of users end up with the same lowest value for betweenness centrality, while these percentages are far lower for the remaining measures.
As such, betweenness is not suited for higher thresholds as they would include increasingly more users that are essentially randomly ordered.
However in practice, given the limited resources of law enforcement, we are far more interested in the performance at low false positive rates.
After all, the higher the false positive rate, the more resources would be wasted on non-vendors.
As such, Figures~\ref{fig:roc-curves}a,b confirm that at least for any lower thresholds, our findings with regards to vendor recall hold up.

When we consider all vendors, not just the top vendor percentile, Figures~\ref{fig:roc-curves}c,d paint a different picture.
Again, topic engagement provides the best overall performance, but now closely followed by the topics started indicator.
On the contrary, though still outperforming the remaining measures at low thresholds, betweenness centrality has clearly worse performance.
However, this is is in line with our conclusions, drawn from the results in the ``Detecting vendors in the user base" section in the main paper, that topic engagement is the best overall predictor and that betweenness centrality performs particularly well for the more successful vendors.
The performance of the topics started indicator for all vendors compared to for successful vendors, indicates that starting topics on the Evolution forum was quite indicative of being a vendor in general.


\end{document}